\documentclass[a4paper,12pt]{report}
\usepackage{amssymb,amsmath,mathbbol,mathrsfs}
\usepackage[usenames,dvipsnames]{color}
\usepackage{hyperref}
\usepackage{xcolor}
\usepackage{stmaryrd}
\usepackage{authblk}
\usepackage{framed}
\usepackage{empheq}
\usepackage{slashed}
\usepackage{hyphenat}
\usepackage{cite}
\usepackage{apacite}
\usepackage{natbib}
\usepackage{dsfont}
\usepackage{chngcntr}

\usepackage{marginnote}    

\usepackage[left=.8in,right=.8in,top=.8in,bottom=.8in]{geometry}                  

\usepackage{sectsty} 
\allsectionsfont{\sffamily\mdseries\upshape} 
\usepackage{tocloft}

\makeatletter
\renewenvironment{abstract}{%
    \if@twocolumn
      \section*{\abstractname}%
    \else 
      \begin{center}%
        {\bfseries\sffamily\abstractname\vspace{\z@}}
      \end{center}%
      \quotation
    \fi}
    {\if@twocolumn\else\endquotation\fi}
\makeatother

\numberwithin{equation}{section}
\counterwithout*{footnote}{chapter}

\hypersetup{
	colorlinks=true,         
	linkcolor=MidnightBlue,          
	citecolor=BrickRed,        
	urlcolor=MidnightBlue            
}

\newcommand{\tocite}[1]{[{\color{magenta}\texttt{#1}}]}
\newcommand{\be}{\begin{equation}}
\newcommand{\ee}{\end{equation}}

\newcommand{\F}{{{\cal M}}}
\renewcommand{\d}{{\mathrm{d}}}
\newcommand{\D}{{\mathrm{D}}}

\newcommand{\SU}{{\mathrm{SU}}}

\newcommand{\G}{{\mathcal{G}}}

\newcommand{\pp}{{\partial}}

\newcommand{\fG}{{\mathrm{Lie}(\G)}}

\newcommand{\Diff}{{\mathrm{Diff}}}

\renewcommand{\bar}{\overline}
\newcommand{\dd}{{\mathbb{d}}}

\renewcommand{\hat}{\widehat}
\newcommand{\rad}{{\text{rad}}}

\newcommand{\La}{\mathcal{L}}
\newcommand{\RR}{\mathds{R}} 

\newcommand{\WZ}{\mathsf{WZ}}
\newcommand{\wz}{\mathsf{wz}}
\newcommand{\cs}{\mathsf{cs}}
\newcommand{\CS}{\mathsf{CS}}

\newcommand{\ch}{\mathsf{ch}}
\newcommand{\Ch}{\mathsf{Ch}}
\renewcommand{\t}{\mathfrak{t}}

\newtheorem{defi}{Definition}

\newcommand{\cint}{{\int\kern-.87em{<}}}
\newcommand{\sint}{{\int\kern-.75em{\sim}}}
\newcommand{\fint}{{\int\kern-1.00em{\int}}}

\newcommand{\bb}{\mathbb}

\newcommand{\order}[1]{\ensuremath{\mathcal{O}(#1)}}

\newcommand{\tr}{\text{tr}}

\renewcommand{\#}{\sharp}

\let\oldmarginpar\marginpar
\renewcommand\marginpar[1]{\oldmarginpar{\color{red}\raggedright\footnotesize #1}}

\newcommand{\old}{\color{red}}

\title{
\begin{figure}[h]
  \centering
  \includegraphics[width= 0.5\textwidth]{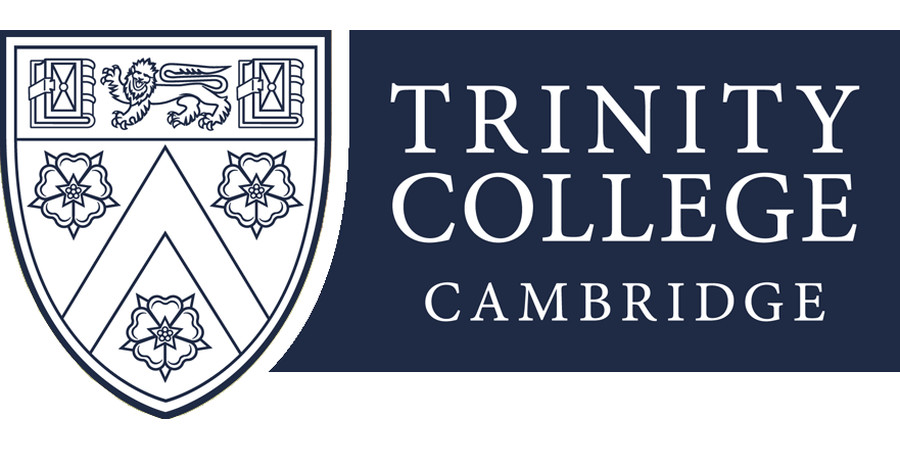}
\end{figure}\medskip

{\bf{\huge{Why gauge? Conceptual Aspects of Gauge theories}}}

}

\author{{\large{Henrique de A. Gomes}}\\~
\\
under the supervision of Dr. Jeremy Butterfield.\\
~\\
~\\
This thesis is submitted for the degree of Doctor of Philosophy\\
 at the  University of Cambridge\\
~\\
~\\
Date of Submission: December 2021
}

\date{}
 
\begin{document}

\maketitle

This thesis is the result of my own work and includes nothing which is the outcome of work done in collaboration except as declared in the preface and specified in the text.\\

It is not substantially the same as any work that has already been submitted before for any degree or other qualification except as declared in the preface and specified in the text.\\
\begin{itemize}
\item Chapters \ref{ch:sd_1} and \ref{ch:sd_2} are partly based on H. Gomes, ``\emph{Same-Diff?: Part I: Conceptual similarities (and one difference) between gauge transformations and diffeomorphisms},'' and H. Gomes, ``\emph{Same-Diff?: Part II: A compendium of similarities between gauge transformations and diffeomorphisms} (2021, unpublished);
\item
 Chapter \ref{ch:Noether} is based on my contributions to: H. Gomes,  J. Butterfield, and B. Roberts,  ``\emph{The Gauge Argument: A Noether Reason
}'', Forthcoming in `The Physics and Philosophy of Noether's Theorems', Cambridge University Press, (2021);
\item
 Chapter \ref{ch:Coulomb} is based on joint work: H. Gomes and J. Butterfield: ``\emph{How to Choose a Gauge? The case of Hamiltonian Electromagnetism 
}'', (2021, unpublished);
\item
 Chapter \ref{ch:subsystems} is partly based on H. Gomes, ``\emph{Holism as the empirical significance of symmetries}'', European Journal of the Philosophy of Science (2020); H. Gomes,  ``\emph{Noether charges: the link between empirical significance of symmetries and non-separability''}, Forthcoming in `The Physics and Philosophy of Noether's Theorems', Cambridge University Press, (2021);  H. Gomes, ``\emph{Gauging the Boundary in Field-space}'', Studies in the History and Philosophy of Science, (2019); and H. Gomes, ``\emph{The role of representational conventions in assessing the empirical significance of symmetries}'', (2020, unpublished).
\end{itemize}


\section*{Acknowledgements}

I would like to thank:\\
$\bullet$ The Cambridge International Trust, for the complete funding of the studies that have led to this thesis.\\
 $\bullet$ My supervisor and personal hero, Jeremy Butterfield, who patiently guided me through this transition.  His generosity of spirit is only comparable to his encyclopedic knowledge of the field; all I can say is that I tried my best to learn on both accounts.\\
 $\bullet$  My co-supervisor, Hasok Chang, for  ushering me into the wonderful topic of philosophy of science; and Gordon Belot, Jim Weatherall, Aldo Riello, and Bryan Roberts, for many interesting and helpful conversations on the topic of gauge. 

\newpage

\begin{abstract}
This thesis is about conceptual aspects of gauge theories. 

Gauge theories lie at the heart of modern physics: in particular, they constitute the standard model of particle physics.  
At its simplest, the idea of gauge is that nature is best described using a descriptively redundant language; the different descriptions are said to be related by a gauge symmetry. The over-arching question the thesis aims to answer is: how can descriptive redundancy be fruitful for physics? This question embraces many important topics in the philosophical literature on gauge theory, which I will address.  

This thesis has two main Parts.  

Part \ref{part:I}  provides technical and conceptual background. It relates the redundancies of gauge theories with the redundancies in the foundations of spacetime physics. In particular, to those of Einstein's theory of general relativity,  that are more familar to the average  philosopher. This Part provides a perspicuous, geometrical understanding of the physics of gauge theory, on a par with the chronogeometric understanding of general relativity. 

In Part \ref{part:II} I will assess two surprising uses of 
and one contentious question about gauge symmetry. First,  I will provide  one answer to the question: ``Why gauge theory?'', that is: why introduce redundancies in our models of nature in the first place? This type of answer is  pragmatic: because such redundancies are useful for model-building, in a particular way; and they allow us to focus our mathematical apparatus on different aspects of the same phenomena. Second, I present  a choice of gauge that is related  to a physically natural, and general, splitting of the electric field; which undermines the way one usually thinks of a choice of gauge as motivated by calculational convenience, or as completely arbitrary. Last, I will  assess arguments and counter-arguments for  the direct physical significance of gauge symmetries.  The conclusion provides a second type of answer to the question of ``Why gauge?''. Namely:  because we need it to couple subsystems. 

 \end{abstract}
 
\newpage

\tableofcontents

\chapter{Introduction}\label{ch:syms}

In this introductory Chapter, I will set the scene. First, I must  delineate a  formal definition of symmetries, in Section \ref{sec:syms}. This may seem like a straightforward task, but it is far from it. The intuitions we commonly have about symmetries clash with most attempts of formalization (as described in \cite{Belot_sym}). So we tread carefully, and define symmetries more flexibly than is usually done. Symmetry here will be taken as a formal preservation of some selected quantities in a given theory. 
The brief treatment already allows us to ask interesting questions, about the interpretation of symmetries, and about symmetry-related models.  With the scene set, in Section \ref{sec:method}  I will describe the methodological morals of the thesis: essentially, I want to close conceptual gaps that exist between philosophers and physicists in the subject of gauge theory.


\section{General remarks on dynamical symmetries}\label{sec:syms}
In its broadest terms, a symmetry is a transformation of a system which preserves the values of a relevant (usually large) set of physical quantities. Of course, this broad idea is made precise in various different ways: for example as a map on the space of states, or on the set of quantities; and as a map that must respect the system's dynamics, e.g.  by mapping solutions to solutions or even by preserving the value of the Lagrangian functional on the states. In Section \ref{sec:syms_tech} I will  provide the definitions about symmetries that we will be using throughout this thesis. 
Section \ref{sec:soph_elim} briefly discusses the  doctrine of \emph{structuralism} and its relation to a  reductive understanding of symmetry-related models, called \emph{eliminativism}.

\subsection{Technical considerations about symmetries}\label{sec:syms_tech}
Technically, the general gloss on  dynamical symmetries is that they are transformations acting on the models of a given theory such that the  models that they relate are empirically indiscernible according to that theory. This gloss provides important intuition, but  nailing down symmetry more precisely  is a  challenge. For instance: defining a dynamical symmetry as any transformation that takes each solution of the equations of motion of a theory to another solution is far too weak: it would imply that any solution is related by a symmetry to any other. And there are other problems. For instance:  models which we would intuitively take to depict physically distinct situations may  nonetheless be symmetry-related, depending on the notion of symmetry; and it is also false that  empirically identical situations are always symmetry-related according to every account of symmetry.  \cite{Belot_sym} gives an exposition of the obstacles  to a general definition. Different authors have risen to Belot's challenge, of providing a general account of symmetry that is   coherent and yet non-circular (see e.g. \cite{Wallace2019, Fletcher2018}). 
 For now, I give what I believe to be a plausible definition of symmetries, that disallows some but likely not all of \cite{Belot_sym}'s counter-examples.

   Let $\cal M$ be the space of models of the theory. Models are supposed to be complete descriptions of the world, according to the given theory. Here the word `world' is purposefully ambiguous: it can refer to an  instantaneous state or to an entire history.    
 And `instantaneous state' is also ambiguous: one may understand an instantaneous description to include or not include information about rates of change---theories whose models are  states in phase space include this information and those whose models are complete instantaneous configurations do not.  In both cases of models of instantaneous states of affairs, they will here be dubbed \emph{states} of the \emph{universe} (not of the world); and I will keep using the word `model' and `world' as the more inclusive terms, that apply also to descriptions of entire histories.

Now, each physical theory will postulate some mathematical structure for its models. 
For example, in non-relativistic mechanics, we could have each model be a configuration of $N$ point particles in Euclidean space, $\RR^3$. So each model is endowed with both the differentiable and vector space structure of $\RR^3$, which can be used in formal manipulations. Now this mathematical structure of each model is reflected in a different level of mathematical structure for the space of models, $\cal M$. In the non-relativistic mechanics example, the space of models is configuration space, which is isomorphic to $\RR^{3N}$. So, while the linear and smooth structure of $\RR^3$ is part of each model, and we use it for important operations such as taking derivatives, we also require the smooth structure of configuration space to do variational calculus. Or similarly, if we are employing a Hamiltonian formalism for mechanics, the symplectic structure can be seen as a structure on the state space $\cal M$; it does not inhere in each model (which, in this case,  would be a phase space point). 

 In field theories, the space of models ${\cal M}$ is usually endowed with a topological structure that allows definitions of neighborhoods of models,  differentiable one-parameter families of models, etc. Indeed, we will usually endow it with further structure: smooth, symplectic, etc.\footnote{An important question here is: in what sense does the mathematical structure of the models constrain or determine the mathematical structure of $\cal M$? For example, in \cite[Ch. 10]{Ringstrom_book}, it is argued that other criteria, such as stability of solutions of the theory, have the power to largely determine the appropriate topology of $\cal M$. But I do not aim to answer this complicated question in general. } 
Of course, using these further, e.g. topological, structures, $\cal M$ becomes an infinite-dimensional manifold. But I would like to reassure the concerned reader on this point: infinite and finite dimensional geometries may differ in certain details, but much of the abstract geometrical reasoning that we are familiar with in the finite case extends to the infinite one.\footnote{\cite{Michor} have a general approach to geometry that is based on curves and their differentiability as embedded in arbitrary spaces; and for many of the geometrical objects and intuitions of the finite-dimensional case,  the approach builds  bridges towards the infinite-dimensional. Another useful source, that develops differential geometry in the infinite-dimensional case by replacing $\RR^n$ as the image of local charts of manifolds (cf. Chapter \ref{ch:sd_1}) by more general Hilbert or Banach vector spaces, is \cite{Lang_book}. One useful rule of thumb about generalizing mathematical theorems is the following: theorems of finite-dimensional geometry whose proof requires some sort of integration are not straightforwardly extendible, whereas those that do not require integration are relatively easily extendible.\label{ftnt:inf_dim}}

Thus, in sum,  each of the models and also ${\cal M}$ are endowed with some mathematical structures (e.g. symplectic, differential, topological,  vector space,  set-theoretic, etc) and the mathematical structures relevant for the models and for ${\cal M}$ need not be the same.

\begin{defi}[$S$-symmetry]\label{def:sym} Let $S$ be some quantity on the system, represented as a real function on ${\cal M}$ that respects these structures (e.g. is   smooth, linear, etc.).  Then a transformation $\Theta:{\cal M}\rightarrow {\cal M}$ is an $S$-symmetry iff $\Theta$:\\
 \noindent (i) respects the structure of ${\cal M}$ (e.g. is  smooth, linear, etc.);  \\
 \noindent (ii) is definable without fixed parameters from ${\cal M}$, i.e. all models enter as free variables in the  transformation $\Theta$; and\\
 \noindent (iii) $\Theta$ preserves the values of $S$: for any model $m\in \cal M$, $S(\Theta(m))=S(m)$. 
\end{defi}
 
  Note  that a transformation $\Theta$ that only preserves the value of $S$ at a subset of models is not an  $S$-symmetry. A symmetry transformation respects   the  structure of ${\cal M}$ \emph{and} preserves the value of a function on ${\cal M}$. So, for example, given some structure, such as e.g. a symplectic form, $\Omega$ (in which case ${\cal M}$ is a smooth manifold, infinite-dimensional in the case of field theories and finite-dimensional for  particle mechanics), and a Hamiltonian $H$ that is a real-valued function on ${\cal M}$,  then item (i) implies $\Theta^*\Omega=\Omega$, and item (iii) implies $\Theta^*H:=H\circ \Theta=H$.

\paragraph{Infinitesimal symmetries}
  In this thesis I will only be interested in symmetries that are  continuous and connected to the identity transformation, and in the case where ${\cal M}$ has at least a topological structure. Thus, for field theories, ${\cal M}$ is an infinite-dimensional manifold, as I mentioned above. 
  
   Specializing to those cases,  let $S$ be some quantity on ${\cal M}$, as above.  
   \begin{defi}[Infinitesimal $S$-symmetry]\label{def:inf_sym}  I will take a vector field $\mathcal{X}$ on ${\cal M}$ to generate an infinitesimal $S$-symmetry, iff:\\
 \noindent (i) $\mathcal{X}$ respects the structure of ${\cal M}$ (e.g. whose flow is symplectic, smooth, continuous,  etc.);  \\
 \noindent (ii) $\mathcal{X}$ is definable without fixed parameters from ${\cal M}$, i.e. all models enter as free variables in the  argument of $\mathcal{X}$; and\\
 \noindent (iii) $\mathcal{X}$ preserves the values of $S$: for any model $m$, $\mathcal{X}[S](m)=0$.\footnote{Here, for any 1-parameter curve of models $m(t)$ such that $\frac{d}{dt}_{|t=0}m(t)={\cal X}_m$ and $m(0)=m$, this is taken as $\mathcal{X}[S](m):=\frac{d}{dt}_{|t=0}(S(m(t))$.}
\end{defi}

  When an infinitesimal $S$-symmetry can be integrated for parameter time $t$, we have finite symmetries generated by the flow of $\cal X$:  $\Theta_{\mathcal{X}}^t:{\cal M}\rightarrow {\cal M}$, such that  (omitting $\cal X$ and $t$): $S(\Theta(m))=S(m)$.  
 
 Infinitesimal symmetries are generically much more tractable than the full group of symmetries; and, even in field theory, given $S$, they can often be found algorithmically, e.g. as kernels of certain integro-differential operators (cf. \cite[Sec. 3]{Lee:1990nz}).

These definitions downplay the role of the dynamical equations of motion of a given theory. Indeed, we can include statements about dynamics by equating $S$ with an \emph{action functional}.  Such an action functional provides a more complete characterization of the dynamics of a given theory, since it can be used as a starting-point for quantization within either the Lagrangian or Hamiltonian formalisms, and it also yields the classical equations of motion in a straightforward manner. So, for almost the entirety of this thesis, the quantity $S$ for the $S$-symmetries will be identified with the action functional. And so $m\in \cal M$ is a history, for which I will suppress boundary conditions in the elementary notation, and I will write $\varphi$ for $m$, to match field theory notation (which will be my focus).


In sum,  we endow $\mathcal{M}$ with a (infinite-dimensional) manifold-like structure of its own; and take dynamics to be obtained from a variational principle. That is, given an action functional on this space: $S:\mathcal{M}\rightarrow \RR$, the extremization requirement $S[\varphi+\delta \varphi]-S[\varphi]=0$ for all directions (or vector fields) $\delta \varphi\in T_{\varphi}\mathcal{M}$, gives rise to the equations of motion, as conditions on the `base point' $\varphi$. Moreover, certain vector fields on $\mathcal{M}$ may leave $S$ invariant,  e.g.  $S[\varphi+\widehat{\delta \varphi}(\varphi)]-S[\varphi]=0$, for all $\varphi$, where $\widehat{\delta \varphi}:\varphi\rightarrow T_\varphi\mathcal{M}$ is a smooth vector field on this infinite-dimensional field space, $\mathcal{M}$, that, importantly, obeys supposition (ii) from Definition \ref{def:inf_sym}.

 \paragraph{Empirical unobservability}

An $S$-symmetry relates empirically indistinguishable models if $S$ captures all the empirically accessible quantities. Theories are their own arbiters of empirical (in)discernibility (cf. \citep{ReadMoller}),\footnote{Einstein made this very point to Heisenberg. Here is how  \cite[p.63]{Heisenberg_dial} described the interaction: 
\begin{quote} I said ``We cannot observe electron orbits inside the atom...Now, since a good theory must be based on directly observable magnitudes, I thought it more fitting to restrict myself to these, treating them, as it were, as representatives of the electron orbits." But Einstein protested: ``But you don't seriously believe that none but observable magnitudes must go into a physical theory?". In some surprise, I asked ``Isn't that precisely what you have done with relativity?". 
``Possibly I did use this kind of reasoning," Einstein admitted, ``but it is nonsense all the same....In reality the very opposite happens. It is the theory which decides what we can observe." \end{quote}} so different theories may have different  $S$'s being sufficient for empirical indiscernibility. But for  all theories of modern physics,  taking $S$ as the Hamiltonian or the action functional will be enough for our purposes.\footnote{Boundary conditions are here taken as features of ${\cal M}$, jointly with  the other mathematical structure delineated above. One of the most notable counter-examples of \cite{Belot_sym} is the Lenz-Runge symmetry, which preserves the equations of motion of a Newtonian two-body problem, but does not preserve features we take to be observable, such as the orbit eccentricity. We could disallow these symmetries by including eccentricity as one of our quantities $S$, but, in this case, this is not necessary. For Lenz-Runge is not an $S$-symmetry when $S$ is the action functional, since the action is not preserved by that symmetry (it is only preserved up to a boundary term that is non-vanishing). Such a definition would also disallow Galilean boosts. A milder definition of $S$, which allows arbitrary boundary terms, is also a possibility, and indeed it is necessary to account for the infinitesimal diffeomorphisms as symmetries if $S$ is taken as the Einstein-Hilbert action of general relativity. } 

To be more precise: it is not that I believe that the action or Hamiltonian somehow encompasses all physical quantities for a given theory: it is rather that I endorse \emph{the unobservability thesis} of \cite{Wallace2019}. Namely, the preservation of these quantities---the Hamiltonian or the action functional---ensures that the dynamics of the theory are preserved by the set of transformations. Moreover,  empirical access, in particular a physical process of observation,  is itself a dynamical notion. Thus a dynamical symmetry can have consequences for what is observable  when that symmetry encompasses
the physical processes involved in a  measurement. From these two suppositions, it is not far-fetched to conclude that quantities or properties that are symmetry-variant for the action functional or Hamiltonian have no grip on, or relevance to, a dynamical process such as a measurement. Or put differently:  the values of such quantities cannot be inferred from dynamical processes; and in particular, by certain types of observation.  That is,  under certain assumptions about the measurement process,
 the unobservability thesis states  that quantities or properties that have values that are not invariant under a dynamics-preserving  symmetry transformation of the
system are unobservable.
 
\paragraph{Symmetries as isomorphisms}

But, as will be discussed at length in Chapter \ref{ch:sd_1}, if we are to judge symmetry-related models as representing the same physical possibility, it makes sense to seek a type of physical and mathematical structure that reflects the quantities that are symmetry-invariant. This brings the category theoretic framework (cf. footnote \ref{ftnt:category} below) into the discussion: we identify symmetries with the isomorphisms of some structure, as represented in a category in which the objects are the models of the theory. This strategy is closest to what \cite[p. 3]{Wallace2019} dubs the `representational strategy', which ``builds the representational equivalence of symmetry-related models into the definition [of symmetry], usually by requiring that symmetries
are automorphisms of the appropriate mathematical space of models (hence
preserve all structure, and thus all representation-apt features, of a model)''.

 That is, since item (iii) implies that symmetries can be composed,  
 we  demand that symmetries  form a groupoid, i.e. a category in which every arrow is a morphism, with the objects of the category being the models, i.e. the elements of   ${\cal M}$.\footnote{The most important characteristic for category theory is its focus on morphisms and transformations between
mathematical objects that preserve (some of) their internal structure. For instance, these morphisms could be group homomorphisms in the category of groups, or linear maps in the category of
vector spaces. More precisely, given a
category $\mathcal{C}$, a morphism $f:A\rightarrow B$ is an isomorphism between objects $A$ and $B$ if and only if there is another morphism $f^{-1}:B\rightarrow A$ such that $f\circ f^{-1}=\mathsf{Id}_A$ and $f^{-1}\circ f=\mathsf{Id}_B$. And a property $P$ is structural, just in case $P(A)$ iff $P(f(A))$ for all isomorphisms $f$.  Another important type of mapping are the functors between different categories. This is, essentially, a mapping of objects to objects and arrows to arrows that preserves 
the categorical properties in question. Such functors  are crucial  for comparing
the objects of different mathematical categories. A groupoid is a category in which every arrow has an inverse in the above sense. \label{ftnt:category} } 

 Here it will prove useful to make a further assumption: that  symmetries are represented as groups (which could be infinite-dimensional), denoted $\G$, such that, given the space  of models of a theory, ${\cal M}$, there is an action  of $\G$ on ${\cal M}$, a map ${\cal M}:\G\times {\cal M}\rightarrow {\cal M}$, that preserves the action functional.\footnote{Such an assumption---that symmetries are represented by the action of an infinite-dimensional group---holds for the covariant Lagrangian version of both Yang-Mills theories and general relativity, and it holds for the Hamiltonian version (in which ${\cal M}$ is phase space) of Yang-Mills theory, but it does not hold for the Hamiltonian version of general relativity; there we have only a groupoid structure (see \cite{Blohmann_2013}).}
 More formally: there is  a structure preserving map on $\F$ that can be characterized  element-wise,  for $g\in \G$ and $\varphi\in \F$, as follows:
 \begin{eqnarray}\mu:\G\times \F&\rightarrow&\F\nonumber\\
(g,\varphi)&\mapsto&\mu(g, \varphi)=: \varphi^g. \label{eq:group_action}
\end{eqnarray}
Since each $g\in \G$ is a symmetry, the action is such that, as per item (i) in the Definitions above, $S(\varphi^g)=S(\varphi)$, for all $\varphi$ and $g$. 

The symmetry group partitions the state space into equivalence classes in accordance with an equivalence relation, $\sim$, where $\varphi\sim \varphi'$ iff for some $g$, $\varphi'=\varphi^g$. We denote the equivalence classes under this relation by square brackets $[\varphi]$ and  the orbit of $\varphi$ under $\mathcal{G}$ by $\mathcal{O}_\varphi:=\{\varphi^g, g\in \mathcal{G}\}$. Though there is a one-to-one correspondence between $[\varphi]$ and $\mathcal{O}_\varphi$, the latter is rather seen as an embedded manifold of $\cal M$, whereas the former exists abstractly, outside of $\cal M$. More mathematically, were we to write the canonical projection operator onto the equivalence classes, $\mathsf{pr}:{\cal M}\rightarrow {\cal M}/\G$, taking $\varphi\mapsto [\varphi]$, then  the orbit ${\cal O}_\varphi$ is the pre-image of this projection, i.e.  $\mathcal{O}_\varphi:=\mathsf{pr}^{-1}([\varphi])$.



 Tacitly endorsing these extra assumptions about symmetries,  we call  $[\varphi]$ the  \textit{physical state}, and $\varphi'\in \mathcal{O}_\varphi$  its \textit{representative} (when there is no need to  emphasise that $\varphi$ involves a choice of representative, we call it just `the state'  for short). We call the collection of equivalence classes, $[{\cal M}]:=\{[\phi], \phi\in{\cal M}\}$,  \emph{the physical state space}. As written, this is an abstract space, i.e. defined implicitly by an equivalence relation, or as  certain classes of isomorphic models, under the appropriate notion of isomorphism.

 \subsection{Structuralism in physics, summarized}\label{sec:soph_elim}
 In this formulation of our theory, there is an important distinction  between the objects---represented by the models---and the structure: represented by the isomorphism classes. 
 

 In  physics, the distinction becomes more salient in the context of \emph{determinism}.  
In the case of theories with `time-dependent' symmetries---such as Yang-Mills theory and general relativity---determinism can only be secured for the equivalence classes, $[\varphi]\in [\F]$,  not for the states $\varphi\in \F$ (see e.g. \cite{Wallace_LagSym, Earman_det}). But, as in pure  mathematics, we usually cannot explicitly express the structure encoded by $[\varphi]$ (at least not without significant pragmatic or explanatory deficit); we can  do so only implicitly, by pointing to the isomorphism  classes, or by selecting representatives of those classes. Thus we enter debates about structuralism within physics.

 \emph{Eliminativism} about symmetries is the position that  seeks an  \emph{intrinsic parametrization of $[\F]$} that makes no reference to the elements of $\F$. In other words, eliminativism seeks to render the  structure of the old theory  as the primary objects of a new theory, thus securing   physical determinism by jettisoning  representational redundancy.

\emph{Sophistication} is, in broad terms, the position that rejects eliminativism while maintaining a commitment to structuralism as an abstract---often higher-order, in the logic sense of requiring quantification over properties and relations---characterization of the ontology, often under the label of \emph{Leibniz equivalence} (see \cite{EarmanNorton1987}). This position claims an intrinsic parametrization of $[\F]$ is \emph{not} required for an ontological commitment only to members of $[\F]$ (see \cite{Dewar2017}). We will discuss this position at length in Chapter \ref{ch:sd_1}.

\section{Morals of this thesis}\label{sec:method}

 In the absence of intrinsic characterizations of structure, i.e. if wholesale eliminativism fails,  we must instead, perhaps provisionally, endorse sophistication: retaining isomorphisms but denying that they relate distinct physical possibilities.

One of the main methods I will use for investigating  gauge theory is through comparison with general relativity. And for general relativity, in both the older and the more recent philosophical literature about isomorphisms  (cf. e.g. \cite{Butterfield_hole, Brighouse_hole,  Hoefer_hole, Weatherall_hole}, and many more), it has been claimed that, ever since Einstein's own discussion of the hole argument (see \cite{Janssen, EarmanNorton1987} for a description),  sophistication is, in effect  if not in name,  how the majority of  theoretical physicists see isomorphic models: as reported by e.g.  \cite[p. 438]{Wald_book}, \cite[p. 5]{Oneill}, \cite[p. 68]{HawkingEllis}. 

But there are dissonant voices: as \cite{Belot50} points out, in certain sectors of the theories, certain isomorphisms \emph{are}  taken to relate different physical possibilities. Moreover,   a blanket endorsement of sophistication may be only a half-way house: as described in \cite{belot_earman_1999, belot_earman_2001}, in practice, theoretical physicists---especially those working in quantum gravity---aim to develop a more perspicuous characterization of the structure that is common to the isomorphic models, i.e a more perspicuous characterization of $[\F]$. 

So theoretical physicists do not form a single monolithic block, and some, in some circumstances, would question the familiar or standard view of Leibniz equivalence. Another distinction amongst physicists is between those who work at a more abstract and those who work at a more concrete  level.  And the views of the  latter also run up against the conceptual understanding of the majority of philosophers of physics: contrary to the picture of isomorphisms relating distinct models of the universe---which elicits the debate about structuralism---many practicing physicists tend to see only a conceptually harmless redundancy of choices of coordinates with which we describe physical systems. That is, these physicists tend to construe redundancy `passively', whereas philosophers and the more abstract-minded physicists tend to construe them `actively'. 

One of the central aims of this thesis  is to close these gaps between the philosophy of gauge theory and how gauge theory  is in practice used by different types of physicists. 

Due to a lack of space, I will not in this thesis be able to fully describe my response to the first gap mentioned above: that the existence of sectors of the theory where Leibniz equivalence fails can be reconciled with sophistication and Leibniz equivalence.  But I will give the upshot in Section \ref{sec:subsystems} and complement it in Chapter \ref{ch:subsystems}: the main idea is an elaboration of a reply considered by \citet[Sec. 4.4]{Belot50} and by many others, including Einstein himself. The idea will be that we distinguish what the theory says about the world as a whole, from how the theory can be used to model particular kinds of subsystem. For the world as a whole, we maintain that isomorphic models represent the same physical possibility. And we construe sectors in which isomorphisms relate different physical possibilities as representing subsystems, wherein those  isomorphisms change the relationship between subsystem and environment. But cashing out this idea  requires a careful examination of subsystems in the context of gauge theories: this is done in \cite{Gomes_new, DES_gf} and in more summarized form in Chapter \ref{ch:subsystems} (but not in detail in this thesis). 

I will leave the the gap between how the practical and the abstract minded physicists construe symmetries to Chapter \ref{ch:sd_1}. More specifically, in Sections \ref{sec:gr_re} and \ref{sec:soph_antiq} I will, among other things,  develop a  link, at least for infinitesimal symmetries, between the active and passive view of isomorphisms in general relativity and Yang-Mills theory. 

     In order for Chapter \ref{ch:sd_1} to also satisfactorily close the remaining gap left by the blanket endorsement of sophistication---that we owe a more perspicuous characterization of the symmetry-invariant structure---I will need  the arguments of Section \ref{sec:rep_conv}, below. Here  I lay out the core of that reply: my argument is that the  more perspicuous characterizations of $[\F]$---yearned for by the quantum gravitist in particular---are furnished by   what I will dub \emph{representational conventions}. Such conventions employ intra-theoretic resources to  provide perspicuous, and yet choice-dependent, characterizations of structure. They allow us to understand the transformation from domain to codomain of an isomorphism of a theory as a change in the relations and properties that we select to  represent a given physical structure. 

This relation will be made more explicit once we relate these choices to notational ones, and isomorphisms to notational changes, through the passive-active correspondence, to be elaborated on in Chapter \ref{ch:sd_1}.

 Further along the thesis, these representational conventions will again be used. In Section \ref{sec:antiheal}, I use   them to answer  a claim by Healey that diffeomorphisms of general relativity are different from the gauge symmetries of Yang-Mills theories in an important way. In  Chapter \ref{ch:Coulomb}, I will provide the details of how a particular representational convention can be anchored to intra-theoretic choices; and in Chapter \ref{ch:subsystems}, I will use conventions to discuss the composition of subsystems, when I complement my response to \cite{Belot50}'s concern about sectors in which isomorphisms do not correspond to symmetries.

Thus, since the concept of representational conventions threads together the topics of this thesis, I will now devote the remaining of this Chapter to elaborating  its general features and the application to subsystems.

\subsection{Representational conventions: general definitions}\label{sec:rep_conv}

Physical facts and representational facts come to us highly entangled. This is of course, a common theme. It occurs in  the logical positivists’ aim of presenting physical theories with a once-and-for-all division of fact and convention; and it was the center of a dispute between Carnap and Quine. I reject this once-and-for-all distinction, both in gauge theory and in the broader philosophical context (for familiar reasons, that I take to be  best articulated by \cite{Putnam_analytic}). But I judge that we can nonetheless assess matters of physical fact. The trick is to anchor these facts to an analogue of a Carnapian framework, that I will call a \emph{representational convention}. 
 Each representational convention will have a unique representation of the physical facts. And as long as we stick with a single convention---whatever that is---we can compare and count different physical possibilities unambiguously. Like any good anchor, it will only serve its function if it doesn't move about.



To emphasize what I said in the preamble to this Section, it bears repeating: in the present context, representational conventions employ intra-theoretic resources to  provide perspicuous, and yet choice-dependent, characterizations of structure; and isomorphisms of a theory can then be understood as  changes in the relations and properties we select to  represent a given physical structure.

\begin{defi}[Representational convention] A representational convention is an injective map 
\begin{align}
\label{eq:rep_conv}
\sigma:[\F]&\rightarrow \F\\
[\varphi]&\mapsto \sigma([\varphi])\nonumber
\end{align}
 that  respects the required mathematical structures of ${\cal M}$, e.g. smoothness or differentiability and is such that $\mathsf{pr}(\sigma([\varphi]))=[\varphi]$, where $\mathsf{pr}$ is the canonical projection map onto the equivalence classes (cf. end of Section \ref{sec:syms_tech}).
\end{defi}

Armed with such a choice of representative for each orbit, a generic state $\varphi$ could be written uniquely as some doublet $\varphi=([\varphi], g)_\sigma:=\sigma([\varphi])^g$, which of course satisfies: $\varphi^{g'}=(\sigma([\varphi]))^{gg'}=([\varphi], gg')_\sigma$.\footnote{Our notation is slightly different than \citet[p. 9]{Wallace2019}'s, who denotes these doublets as $(O, g)$ (in our notation $([\varphi], g)$), and  labels the choice of representative (or gauge-fixing) as $\varphi_O$ (our $\varphi_\sigma$). We prefer the latter notation, since it makes it clear that there is a choice to be made. As with coordinate systems, the interesting quantities will be invariant under these choices; nonetheless, we need to keep them fixed. This requirement  becomes nuanced when we are comparing different subsystems, with each other and with the joint system, as we will do in Section \ref{sec:subsystems}.\label{ftnt:Wallace_equi}}    Thus we identify ${\cal M}\simeq [{\cal M}]\times \G$ via the diffeomorphism:
\begin{align}\label{eq:doublet} \bar\sigma: [{\cal M}]\times \G&\rightarrow \cal M\\
([\varphi], g)&\mapsto \sigma([\varphi])^g\nonumber
\end{align}

 Given just the state, $\varphi$, we cannot discern any symmetry transformation that has been  applied to it. But armed with a choice of representative as in \eqref{eq:doublet}, we can do exactly that.
Thus, as a general principle, \emph{any}  significance that we attribute to group elements, or functions of group elements, must make reference to such a choice.
 Thus representational conventions are crucial in the discussion about the observability of symmetries, as we will see in more detail in Chapter \ref{ch:subsystems}.


Now, as I mentioned, the space $[\F]$ is abstract, or only defined implicitly. Since we cannot usually represent elements $[\varphi]$ of $[\F]$ intrinsically, we in practice replace $\sigma$ by an equivalent projection operator that takes any element of a given orbit to the image of $\sigma$, as in: 
\begin{defi}[Projection operator]
A map 
\begin{align}
h_\sigma:\F&\rightarrow\F\nonumber\\
\varphi&\mapsto  h_\sigma (\varphi)=\sigma([\varphi])\label{eq:proj_h}
,\end{align}
is called the projection operator for the representational convention, $\sigma$.  \end{defi}
Since $[\varphi^g]=[\varphi]$, we must have
\be\label{eq:h_equiv} h_\sigma(\varphi^g)=h_\sigma(\varphi).\ee
 Moreover, $h_\sigma(\varphi)$  can be seen explicitly, as a map on $\F$, i.e. as a function of $\varphi$. As such,  an $h_\sigma$ uniquely and concretely represents structural content; and so projection operators provide a valuable mathematical tool with which we can discuss physical possibility in terms of structure. In other words, two given representatives, $\varphi, \varphi'$, that are in principle unrelated, are physically the same, i.e. give the same value for \textit{all} symmetry-invariant quantities iff $h_\sigma(\varphi)=h_\sigma(\varphi')$. 
Thus a projection resolves problems of  identity of physical states. 

\subsubsection{Unobservability and other theses about symmetry}\label{sec:unobs}
With these definitions in place, we can briefly address the relation between symmetries of the whole universe and empirical significance, or the observability, of those symmetries, as discussed in the last two subsections of Section \ref{sec:syms_tech}.

 To see this, suppose for simplicity that, given some notion of dynamical evolution of the states, $U$, then evolution by $U$ commutes with some group action. So $\varphi(t)$ satisfies the evolution equation, $U(t')(\varphi(s))=\varphi(t'+s)$, if and only if $\varphi(t)^g$ also satisfies it.  Once we assume a well-defined representational convention exists, we can write   $\varphi_o=\sigma([\varphi_o])^{g_o}$, for  $\varphi(0)=\varphi_o$ and some initial $g_o\in \G$. Then from $\varphi(t)=U(t)(\varphi_o)$, we obtain:  
\be (U(t)(\sigma([\varphi_o]))^{g_o}= U(t)(\sigma([\varphi_o])^{g_o}). 
\ee Applying the projection $h_\sigma$ to both sides, we get:
\be h_\sigma((U(t)(\sigma([\varphi_o]))^{g_o})= h_\sigma(U(t)(\sigma([\varphi_o]))=h_\sigma(U(t)(\sigma([\varphi_o])^{g_o})),
\ee
where we applied \eqref{eq:h_equiv} in the first equality. So the projected evolution along the image of $\sigma$ is indifferent to the initial $g_o$ (an extension to the time-dependent case is rather trivial). And since the map $[{\cal M}]\times \mathrm{Id}\rightarrow \mathrm{Im}(\sigma)$ is a diffeomorphism, we translate these statements into ones about the equivalence classes: the future evolution of $[\varphi]$ depends only on the present value of $[\varphi]$---which is how it is is stated by \citet[p. 10]{Wallace2019} (where this last step of translation from Im$(\sigma)$ to $[\F]$ is omitted). 
 
So there is ``a self-contained dynamics
for the invariant degrees of freedom of the system that is quite independent of
the $\G$-variant features'' \citep[p. 10]{Wallace2019}. Wallace then assumes that ``the system under investigation is rich enough to model
its own dynamics, and that the system is measuring itself rather than being
observed from outside,'' and takes this to demonstrate 
\noindent \textit{the unobservability thesis}: that given a family of models of a global system which
are related by a symmetry transformation, it is impossible to determine
empirically---i.e through dynamical, self-contained measurements---which model in fact represents the system.


\subsubsection{Representational conventions in field theories: relation to gauge-fixing}\label{subsec:gf}

 In the type of field theories we will focus on in this thesis, the procedure for fixing  a representational convention  is intimately related to a procedure called \emph{gauge-fixing}.\footnote{And in quantum gravity, they are also intimately related to \cite{Rovelli_partial}'s \emph{partial observables}, and also to other tools used by that community, such as \emph{intrinsic time parametrizations} \cite{Isham_POT}.}

 To illustrate the relationship between a representational convention and a gauge-fixing, we require some of the mathematical tools of fiber bundles, to be described in detail in Section \ref{sec:PFB}. In practice, the gauge-fixing procedure relies on the given representational convention $\sigma([\varphi])$ satisfying some auxiliary condition. In that language, a representational convention, like a gauge-fixing, requires   a choice of  \textit{section} of the  configuration space, seen as a (possibly infinite-dimensional) principal bundle. 
But in \eqref{eq:doublet}, we oversimplified: there is no such global product form, and even a local product form is not guaranteed to exist, as it is in the finite-dimensional case (as we will see in Chapter \ref{ch:sd_1}, in Section \ref{sec:PFB_formalism}).

 Nonetheless, in the spirit of footnote \ref{ftnt:inf_dim}, it is in fact true that the space of models $\cal M$ is very similar to a principal fiber bundle, with $\G$ as its structure group. But there are important differences between the infinite-dimensional and the finite-dimensional case. As we will see in Section \ref{sec:PFB}, in the finite-dimensional case, it suffices that the action of a group $G$ on the given manifold $P$ be free (and proper) for that manifold to have a principal ${G}$-bundle structure, usually written as ${G}\hookrightarrow P\rightarrow P/{G}$.  In the infinite-dimensional case, these properties of the group action are not enough to guarantee the necessary fibered, or local product structure: one has to construct that structure by first defining a \emph{section}.\footnote{ In fact, in the case of field theories, such as general relativity and Yang-Mills: no such local product structure exists:  $\mathcal{M}$ is not an infinite-dimensional principal $\mathcal{G}$-bundle, i.e. $\mathcal{G}\hookrightarrow\mathcal{M}\rightarrow \mathcal{M}/\mathcal{G}$.  The obstacle is that there are special states---called \emph{reducible}---that have stabilizers, i.e. elements $\tilde g\in \G$ such that  $\tilde\varphi^{\tilde g}=\tilde \varphi$. Indeed, it is easy to show that for  some $\tilde\varphi':=\tilde\varphi^g\in\mathcal{O}_{\tilde \varphi}$, 
$$\tilde\varphi'^{g^{-1}\tilde g g}=\tilde\varphi^{\tilde g g}=\tilde\varphi^g=\tilde\varphi'$$
 and so all the elements of the orbit are also reducible (with stabilizers related by the co-adjoint action of the group). And so entire orbits are of `different sizes': in the language of Section \ref{sec:PFB}, they are not all isomorphic to the structure group. Nonetheless, there is a generalization of a section, called a slice, that provides a close cousin of the required product structure. As has been shown using different techniques and at different  levels of mathematical rigour,  \cite{Ebin, Palais, Mitter:1979un, isenberg1982slice, kondracki1983, YangMillsSlice, Slice_diez} both the Yang-Mills configuration space and the configuration space of Riemannian metrics (called $\mathrm{Riem}(M)$), admit a \emph{stratified} product structure. That is, the state space is stratified into orbits of states that possess more and more stabilizers; with the orbits with more stabilizers being at the boundary of the orbits of states with fewer stabilizers. For each stratum, we can find a section and form a product structure as in the standard picture of the principal bundle.

 Unfortunately, the space of Lorentzian metrics is not known to have such a structure: it has only been shown for the space of Einstein metrics that admit a constant-mean-curvature (CMC) foliation (see also footnote \ref{ftnt:CMC}). 
For both general relativity and non-Abelian gauge theories,   reducible configurations   form a meagre set.   \textit{Meagre}  sets are those that arise as countable unions of nowhere dense sets. In particular, a small perturbation will get you out of the set (and this is true of the reducible states in the state spaces of those theories, according to the standard field-space metric topology  (the Inverse-Limit-Hilbert topology cf. e.g. \cite{kondracki1983, fischermarsden}). In this respect, Abelian theories, such as electromagnetism, are an exception: {\it all} their configurations are reducible,  possessing the constant gauge transformation as a stabilizer. 

And apart from this obstruction to the product structure---i.e. even if we were to restrict attention to the generic configurations in the case of non-Abelian field theories---one can have at most a \emph{local} product structure: no representational convention, or section, giving something like \eqref{eq:doublet}, is global (this is known as the  \emph{Gribov obstruction}; see \cite{Gribov:1977wm, Singer:1978dk}).\label{ftnt:stab} }

A choice of section is essentially an embedded submanifold on the state space $\F$ that intersects each orbit exactly once. 
 That is, we impose further functional equations that the state in the aimed-for representation must satisfy; this is like defining a submanifold indirectly, through the regular value theorem: E.g. defining a co-dimension one surface $\Sigma\subset N$ for some manifold $N$,  as $\mathcal{F}^{-1}(c)$, for $c\in \RR$, and $\mathcal{F}$ a smooth and regular function, i.e.  $\mathcal{F}:N\rightarrow \RR$ such that $\d \mathcal{F}\neq 0$.\footnote{In the infinite-dimensional case, both the dimension and the co-dimension of a regular value surface can be infinite, and it becomes trickier to construct  a section: roughly, one starts by endowing $\cal M$ with some  $\G$-invariant (super)metric, and then finds the orthogonal complement to the orbits, $\cal{O}_\varphi$, with respect to this supermetric. But here the intersection of the orbit with its orthogonal complement cannot be assumed to vanish, as it does in the finite-dimensional case. Nonetheless, in the cases at hand, that intersection is given by the kernel of an elliptic operator, and one therefore can invoke the `Fredholm Alternative' (see \cite[Sec. 5.3 and 5.9]{Trudinger}) to show that that intersection is at most finite-dimensional, but generically is zero, and thus the generic orbit has the `splitting' property: the total tangent space decomposes into a direct sum of the tangent space to the orbit and its orthogonal complement. Then one extends the directions transverse to the orbit by  using the Riemann normal exponential map with respect to the supermetric, and thus concludes, by the Gauss lemma, that  this submanifold is transverse to the neighboring orbits and  for a sufficiently small radius has no caustics; (it is a bit harder to show that this `section' is not only transverse to the orbits, but indeed that it intersects neighboring orbits only once cf. \cite{Ebin}). 
\label{ftnt:slice}}

Once the surface is defined, $\sigma$ can be seen as the embedding map with range  $\sigma([{\cal M}])=\mathcal{F}^{-1}(0)\subset {\cal M}$.  The next step is to find a gauge-invariant projection map, $h_\sigma$, that projects any configuration  to this surface.


In general, the real-valued function $\mathcal{F}$ that defines the section (surface) $\sigma([\F])$ as its level surfaces $\mathcal{F}^{-1}(0)$ must satisfy two conditions: \\

 \noindent$\bullet$\,\,\textit{Universality} (or existence):~ 
For all $\varphi\in \F$, the equation $\mathcal{F}(\varphi^g)=0$ must be solvable by a functional $g_\sigma:\cal M\rightarrow \G$. Here, $g_\sigma(\varphi)$ is a gauge transformation required to transform $\varphi$ to a configuration $\varphi^{g_\sigma(\varphi)}$ which belongs to the gauge-fixing section $\sigma$. That is: 
\be\label{eq:gauge-fixing} g_\sigma:\F\rightarrow \G ,\,\,\text{is such that}  \,\, \mathcal{F}(\varphi^{g_\sigma(\varphi)})=0, \,\, \text{for all}\,\, \varphi\in \F.
\ee
This condition ensures that $\mathcal{F}$ doesn't forbid certain states, i.e. that each orbit possesses at least one intersection with the gauge-fixing section.

\smallskip

\noindent$\bullet$\,\,\textit{Uniqueness}:~ If  $g_\sigma$ as above satisfies  $\mathcal{F}(\varphi^{g_\sigma(\varphi)})=0$, then     $\varphi^{g_\sigma(\varphi)}=\varphi'{}^{g_\sigma(\varphi')}$ if and only if $\varphi\sim \varphi'$, meaning that only isomorphic models have the same projection. From this condition, we get $\varphi^{g_\sigma(\varphi)}=\varphi{}^{g_\sigma(\varphi^g)g}$, and obtain the following equivariance property for $g_\sigma$:
\be\label{eq:cov_g} g_\sigma(\varphi^g)=g_\sigma(\varphi)g^{-1}.
\ee
This equation is, schematically identical to one we will find when we look at sections of principal fiber bundles in the finite-dimensional case, namely, \eqref{eq:g_equi},  and so we refer to Section \ref{subsec:active_passiveYM} for more details. (The section in the space of models is essentially a state-dependent section of the finite-dimensional bundle). 

It is convenient to rewrite $h_\sigma$ of \eqref{eq:proj_h} explicitly including $g_\sigma$:
\begin{align}
h_\sigma:\F&\rightarrow\F\nonumber\\
\varphi&\mapsto  h_\sigma (\varphi):=\varphi^{g_\sigma(\varphi)}
\label{eq:h_proj_F}\end{align}
  And, as expected, from \eqref{eq:cov_g},   $h_\sigma (\varphi)$ is a gauge-invariant functional, in the sense that $h_\sigma(\varphi^{g})=h_\sigma(\varphi)$.\footnote{But note that reducible states, i.e. states that have stabilizers as per footnote \ref{ftnt:stab} are not ``wrinkly enough'', do not have features that vary enough, to completely fix the representation. Stabilizers inevitably produce degeneracies in the representational convention: they foil uniqueness, but for physical reasons.\label{ftnt:stab2}}

\subsubsection{Isomorphisms relate choices of the physical relations that are used for description}\label{subsec:isos_descs}

Of course, we can still change the representational convention  itself, i.e. act with the group  on the image of $h_\sigma$:  since $h_\sigma(\varphi)\in \F$, we could consider $(h_\sigma(\varphi))^g$. That is because $h:\F\rightarrow\F$ is a projection, as opposed to a reduction ($\mathsf{pr}:\F\rightarrow [\F]$). 
 And indeed, given two representational conventions $\sigma, \sigma'$, we obtain a relation: 
\be\label{eq:transition_h} h_\sigma(\varphi)=(h_{\sigma'}(\varphi))^{\mathfrak{t}_{\sigma\sigma'}(\varphi)},
\ee
where $\mathfrak{t}_{\sigma\sigma'}(\varphi):=g_{\sigma'}(h_\sigma(\varphi))$  is a state-dependent isomorphism (the analog of the transition map between sections of a principal bundle, given in more detail in \eqref{eq:transition}). Note also that since $h_\sigma(\varphi^g)=h_\sigma(\varphi)$, the transition map depends only on the orbit $\mathcal{O}_\varphi$ of $\varphi$. So the `translations' between the physical quantities written as $h_\sigma$ to those written as $h_\sigma'$ recover the original isomorphisms. This idea gives us a new gloss on symmetries: we know the symmetries because they relate \emph{choices of relations} with which we describe  physics. 

Or rather: each choice of $\sigma$ corresponds to a certain intra-theoretic choice of relations and quantities with which we want to describe the physical, i.e. symmetry-invariant, features of the given system. Then each such description, depicted as a function $h_\sigma$ on the space of models $\cal M$, is fully symmetry-invariant: it reports the physical goings-on. Nonetheless, given two choices of relations or features we want to use to describe the system,  we can `translate' or `reverse engineer' from  the  invariant quantities  according to one to those according to the other.  The transition map $\mathfrak{t}_{\sigma\sigma'}(\varphi)$ of  \eqref{eq:transition_h} is the `reverse engineering'.

Examples of $\mathcal{F}$ (and $\sigma$, $h_\sigma$, and $g_\sigma$) in field theory will be given in Section  \ref{sec:antiheal} and Chapter \ref{ch:Coulomb}, along with their physical interpretations. A simple example for particle mechanics: let $\mathcal{ M}:=T^*\cal Q$ be a phase space, with configuration space $\cal Q$ and let $\G$ be, say,  a group of boosts; then the definition $\mathcal{F}(q, p):=\sum_\alpha p_\alpha=0$, where $p_\alpha$ are the momenta conjugate to the configurations $q_\alpha$ (with $\alpha$ labeling the particles; cf Chapter \ref{ch:Coulomb} and Appendix \ref{sec:Ham})  i.e. the choice of center of mass coordinates, satisfies all the criteria above.   So here $g_\sigma$ would be the boost required for the system to have zero total linear momentum.  

Or even simpler: I could choose a particular particle  to be my center of coordinates, and represent all  symmetry-invariant quantities with respect to this choice. Of course, examples abound: beyond the case of translations and boosts,  we could fix a frame in $\RR^3$  by diagonalizing the moment of inertia tensor around the center of mass; in this case $g_\sigma$ would be a state-dependent rotation (see \cite[Sec. 4]{DES_gf} for details).\footnote{Configurations that are collinear, or are spherically symmetric, etc. would be unable to fix the representation: as per footnotes \ref{ftnt:stab} and \ref{ftnt:stab2}, these are \emph{reducible} configurations; they have stabilizers that cannot be fixed by any feature of the state. } 

 Similar kinds of choices can be made in any theory with symmetries. For instance,  in the more familiar case of special relativity, we may choose Lorentz frames that are adapted to some phenomenon under study: e.g.  ``co-moving with a rocket''. Every Lorentz-invariant fact can be described in this frame, and one can see the frame as `physical', because it is anchored on  features of the state. And singling out some choice of frame does not imply that another frame is less capable of describing the goings-on in the laboratory inside the rocket, just that it may be more cumbersome to do so.\footnote{ To tie this discussion in with the topic of Section \ref{sec:PFB}, and in particular with the view of gauge transformations as changes of bases in the frame bundle (associated to the tangent spaces in Section \ref{subsec:PFB_ex}, and to internal spaces in Section  \ref{subsec:internal_basis}), we could think of a smoothly varying choice of coordinate systems about each $x\in M$, with which we scaffold the tangent bundle $TM$, and thereby represent any vector field in a basis. \label{ftnt:frame_conv}}




 \subsection{Subsystems, symmetry and equivalence}\label{sec:subsystems}


The second topic I want to broach in this introductory Chapter, at a very general level, is the issue that \cite{Belot50} raises, discussed in the preamble of Section \ref{sec:method}.

The issue will be elaborated in slightly more detail in the case of general relativity at the end of Section \ref{subsec:drag}, but it extends well beyond this superficial gloss, and occurs in gauge theory as well.\footnote{ See e.g. \cite{Giulini_asymptotic} for similar remarks about gauge theory. There is today a healthy theoretical physics sub-discipline that focuses on understanding the different conditions on and formulations of asymptotic structure. 
See e.g.  \citep{Henneaux:2018gfi, Henneaux_rev} for extensive references and a treatment of asymptotic spatial infinity for both general relativity and electromagnetism,  \citep{strominger2018lectures, AshtekarNapoli} for textbooks treating asymptotic null infinity and \cite{AshtekarHansen} for how the treatments can be unified). There is also an extensive literature on asymptotically de Sitter (cf. \citep{AshtekarBonga_I}) and Anti-de Sitter spacetimes \citep{Asymp_ads, Ashtekar_1984}.}  Here I do not aim to introduce  the subject in any comprehensive form. I will only use a well-known response to Belot's challenge as a motivation to formalise a notion of `subsystems': a notion that is focused on systems with symmetries. 

In a few words, the general version of Belot's challenge can be put as follows: suppose we have a mature theory, in which we have identified symmetries with a notion of  isomorphism of some mathematical structure, as at the end of Section \ref{sec:syms_tech}.  Can there still be sectors that contain successful applications of the theory and yet harbor a mismatch between  isomorphisms and  symmetries?  As \cite{Belot50} points out, this is indeed the case  for some formulations of general relativity in the asymptotically flat  sector.\footnote{The issue is that some of these formulations are coordinate-dependent. Namely, one assumes that there is a certain type of spacelike coordinate, $r$, which represents radial distance, and in coordinates that employ $r$,  asymptotically flat metrics are defines as those that can be decomposed as $g_{\mu\nu}=\eta_{\mu\nu}+\order{r^{-1}}$ (with derivatives of the metric decaying faster). This type of definition was used originally  to allow the definition of conserved global charges of  spacetime as a whole, such as mass, energy, and linear and angular momentum  (see \cite{ADM, ReggeTeitelboim1974}). \label{ftnt:Belot_ADM} }

The topic is deep and wide-ranging, and finds applications both in  more speculative questions of  black hole evaporation (and its related information paradox) and in  more concrete investigations about gravitational waves. Indeed, as \citet[p.  2]{AshtekarBonga_I} point out, it was a confusion about asymptotic coordinate transformations---or rather, the differences between isomorphisms and asymptotic symmetries---that led to the controversy about the physical reality of gravitational waves. This confusion was only resolved when  Bondi and others (Sachs, Metzner, Van der Berg, Penrose, and Newman), using a precursor of Penrose compactification (cf. \cite[Ch. 11]{Wald_book}), formulated the theory of asymptotic radiation in an invariant form, thereby untangling physical issues from notational ones.

In fact, although Einstein himself had already derived the quadrupole formula (that was indicative of gravitational radiation) in 1917, he himself was rather suspicious of some features of the formalism he had used to do so. In a letter to Mie in 1918, Einstein
remarks that a locally generally covariant theory in which isolated systems correspond to asymptotically
flat solutions “is a monstrosity”, \cite[vol. 8, doc. 470]{CPAE} since they “pre-suppose a definite choice of the system of reference, which is
contrary to the spirit of the relativity principle”   \cite[vol. 6, doc. 43]{CPAE}. 

Though less pessimistic,  \citet[p. 182]{Penrose_gr_problems} also acknowledges the (often hidden) assumption that asymptotically flat  sectors  ``are interesting  not because they are thought to be realistic models for the entire universe, but because they describe the physics of isolated systems''.

Along these lines, I believe  that the asymptotic discrepancy between isomorphism and symmetry arises when we forget that these target sectors of any theory are used to represent subsystems, and that, at the asymptotic boundary, they implicitly carry some fixed structure that simulates our measuring apparata.\footnote{The first to actually endow the asymptotic coordinate system with some dynamics of their own, in an attempt to realize this proposal, were \cite[Sec. 5]{ReggeTeitelboim1974}. See also \cite{BeigOMurchadha} for a treatment in the constrained Hamiltonian formalism of Chapter \ref{ch:Coulomb} and Section \ref{sec:syms_consts}. } \citet[p. 970]{Belot50} considers this type of response:
\begin{quote}
But wait! Surely here we are undeniably talking about subsystems of the universe! As
Penrose was quoted saying above: asymptotically flat solutions provide idealized
models of relatively isolated self-gravitating subsystems of our universe. So it may
well seem that in this context it is safe to fall in with the sort of account [...] on which symmetries are to be understood as generating new
possibilities when they act on the state of a subsystem of the universe, but not when
they act on global histories. And then we can set aside these funny asymptotic
symmetries as irrelevant, and go back to flat-out denying that generalized shifts
generate new possibilities.
\end{quote}
 Belot goes on to deny this argument, but I believe it is  essentially right. 
 Belot's denial says that a cosmological model, viz. de Sitter spacetime, admits a similar discrepancy between asymptotic isomorphisms and symmetries: 
\begin{quote} it is not just isolated systems that tend to look
asymptotically like de Sitter spacetime: the same is true for a wide variety of models
of the universe as a whole. In this context, asymptotic boundary conditions and
the distinction between isometry and gauge equivalence that travels in their wake
are not just for subsystems. [...And although] there are several competing characterizations of the asymptotically de Sitter sector and
its asymptotic symmetries [...] they each embody the principle that [the isomorphisms that asymptote to the identity] should be treated as gauge symmetries, and others as physical symmetries (Ibid, 970, 977)
\end{quote} 

To back his claim,  he cites \citep{Kelly_2012} and \citep{AshtekarBonga_I}, but it seems to me he misinterprets them.\footnote{One often looks at flat  patches of de Sitter spacetime (those that are `slanted'  just so that they stretch out so as to obtain something like spatial infinity) and there imposes structure at the asymptotic regions; but for what is called `the global slicing', which assumes space to be a three-dimensional sphere, no such conditions are required.}
Here is what they say \cite[p. 47]{Kelly_2012}:  
\begin{quote}Since $S^{(d-1)}$ is compact, there is no need to impose further boundary conditions. The constraints then imply that all gravitational charges vanish identically. \emph{All diffeomorphisms are gauge symmetries} and the asymptotic symmetry group is trivial.
On the other hand, it is natural in cosmological contexts to consider pieces of de Sitter space which may be foliated by either flat or hyperbolic Cauchy surfaces. [...] These Cauchy surfaces are non-compact, and boundary conditions are required in the resulting asymptotic regions. \end{quote}
 And here are \cite[p. 3]{AshtekarBonga_I}: \begin{quote}What is the situation for $\Lambda> 0$ [de Sitter spacetime]? Penrose's
construction of null infinity naturally generalizes; he showed this already in his first papers.
[The conformal boundary] is again a boundary of the physical space-time within its conformal completion. But it
is now space-like. Consequently, as we will see in detail, the asymptotic symmetry group [...] \emph{is now the the full diffeomorphism group} of the [conformal boundary]. [my italics]\end{quote}

In the spirit of these responses, in what follows I will examine the relation between symmetries and subsystems: 
I think it is worthwhile to pursue one idealization at a time; to divide and conquer, if you will. Thus I will  set aside asymptotic issues and focus instead on what conditions we expect physical subsystems to satisfy, especially as regards their symmetries. In this respect, I follow \cite{Wallace2019, Wallace2019b}.

 Section \ref{subsec:sub_rec} gives a first gloss on the idea of a subsystem as reflecting important kinematical features of the larger system of which it is a part. Section \ref{subsec:promise} concludes with an upshot of these reflections, and with one promisory note about another use of representational conventions, that will be cashed  in by Chapter \ref{ch:subsystems}.

 
 
 
 \subsubsection{Kinematical subsystem recursivity}\label{subsec:sub_rec}

 A relatively strong requirement on subsystems, formulated and endorsed by \cite[p. 5]{Wallace2019a}, is that they satisfy \emph{subsystem recursivity}, so that the theories
\begin{quote} have the remarkable and
underappreciated feature of being able to reinterpret subsystems of their models, when dynamically isolated, as other models of the same theory. [... in these cases]  any model can be interpreted [...as a]
dynamically isolated subsystem under certain idealizations about its environment
and where, if we want to remove those idealizations, we can embed the
model in a model of a larger system within the same theory---and where that
larger system in turn is interpretable in the first instance as a subsystem of a
still-larger system.
\end{quote}
As Wallace argues, `dynamical isolation' is a term of art in physics, but we will not need to be more precise about this, except that we need to assume  that isolation entails a weak form of dynamical autonomy. 
 A subsystem is dynamically autonomous when  its dynamical equations, up to the level of approximation required by the situation at hand,  does not depend on the details of the rest of the system, except insofar as the rest of the system defines \emph{initial} boundary conditions for the subsystem. 

In the field theoretic context, Wallace interprets conditions of dynamical isolation as asymptotic boundary conditions. But there is here a worry that local field theories, by respecting relativistic causality and being  local, would be able to `reinterpret subsystems of their models, as other models of the same theory', even without being completely dynamically isolated. That is, field theories are already to a certain extent `separable': a condition, according to \cite[p. 321]{Einstein}, that  ``things claim an existence independent
of one another, insofar as these things ``lie in different parts of space''.'' Thus, for example, take a point $x$ in the future of  a Cauchy surface $\Sigma$,  and define $R$ as the intersection of the past causal cone of $x$ (called $J^-(x)$) and $\Sigma$. Then the state on $R$  fully determines the states in the region bounded by $\pp J^-(x)$ and $R$ (see \cite{Geroch1970}; this topic also plugs into much of the philosophy of determinism, cf. \cite[Ch. 4]{Earman_det}). It thus seems that we could find weaker conditions for subsystem recursivity that would still provide a rich universe of applications. 

Here I am only interested in the behavior of symmetries of the laws, at both subsystem and global levels. As Marc Lange argues extensively, there is a sense in which symmetries can be seen as `laws on laws', or `metalaws' (see e.g. \cite{Lange_meta}, and my discussion about points off  the constraint surface in Chapter \ref{ch:Coulomb}). In this spirit, my requirement about dynamical isolation and subsystem recursivity can  be weakened in two senses. 

First, I will only be interested in whether the subsystem enjoys the `same type' of symmetries as the larger system in which it is embedded. This will be labeled \emph{downward consistency of the symmetries}. It is required if (1) we define the state space by restriction to a subsystem, and (2) we want the isomorphisms of these restricted states to match the dynamical symmetries.  If applied to gravity, this condition would exclude the type of definition used in e.g. spatially asymptotically flat spacetimes  as described by \cite{Belot50}; but I will at most apply it to Yang-Mills theories. The requirement allows evolving boundary conditions, if they are symmetry-invariant; it is in general  a weaker condition of isolation that allows us to treat more general types of non-asymptotic subsystems. 

Other than (2) above,  there are two further motivations for the requirement of downward consistecy of symmetries. First, (a): since, as argued by \cite{Wallace2019} empirical observations are tied to subsystems, our evidence for symmetries---both of the direct and the indirect sort, such as charge conservation---must ultimately arise from the behavior of subsystems. And (b): any consistent theory owes an account of what happens to the larger system's symmetries when they are restricted to a subsystem, and, from (a), downward consistency would be an explanatory  account: we think the larger system has certain symmetries because we see their action on subsystems and infer they must extend to our own sector, seen as a subsystem.  Moreover, a conflict with downward consistency for local field theories would reflect a type of incompatibility between an inside and an outside perspective of the  boundary of a subsystem. How does  the environment, i.e. the entire universe, `see' the symmetries of the subsystem? For the symmetries of the theory (that act far away from the asymptotic boundary) are unconstrained: they are not pared down. So how should observers from the environment construe a definition of subsystem---a sector of the theory, in Wallace's nomenclature---that does not, \emph{can not}, support the full action of the local symmetries? 

The second sense in which my requirements about dynamical isolation and dynamical recursivity will be weakened is that, in the case of instantaneous states,  in this thesis (Chapter \ref{ch:subsystems}), I will focus on the relation between system and subsystem symmetries \emph{for initial states}. Borrowing the label from physics, we can call this \emph{kinematical} isolation. Here the motivation is that some given isolation condition may only hold   for a certain interval of time, $I\subset \RR$. But I do not want to focus  on the loss of autonomy over time, and so I will only require some small $I\neq \emptyset$.   Thus, differently from Wallace, I assume only that some interval $|I|=\delta> 0$ exists in which downward consistency is satisfied. 

In the case of gauge theories, the arguments of this thesis (more completely made in \cite{DES_gf}) will require only such kinematical isolation considerations. 

As  we will see in Chapter \ref{ch:subsystems},  there are mathematical obstacles on the way to  implementing downward consistency for  local symmetries in gauge theory. This is due, essentially, to other,   non-local aspects of physical observables in gauge theories (as will be  discussed in Chapter \ref{ch:Coulomb} and briefly in Section \ref{par:active_passive} and \ref{sec:syms_consts}).
But there are also local aspects of gauge theory: in particular, `interactions' should be local and the theory should be causal. This is enough to ensure that my weakened form of kinematical isolation, and my condition of downward consistency can  be satisfied by  subsystems defined through a partition of space. 
For local field theories such as  Yang-Mills gauge theories and general relativity, certain subsystem whose boundaries do not break the symmetries of the larger system will respect downward consistency.\footnote{ In the case of general relativity, downward consistency would require us to demarcate subsystems using diffeomorphism-invariant conditions; such as Komar-Bergmann scalars \citep{Bergmann_Komar}. And there are also many characterizations of black holes that are diffeomorphism-invariant in this way (see e.g. \cite[Chs. 5, 8 and 9]{Hayward_book}, which moreover employ a notion of dynamical isolation that is not necessarily asymptotic).
}  

In the particle theory case, due to action-at-a-distance, the assumption that the subsystem dynamics inherits the symmetries of the larger universe requires stronger isolation conditions; but these can be encapsulated in our embedding of the subsystem into the larger universe (cf. \cite[Appendix D]{DES_gf}).



  
 \subsubsection{Upshot and promisory note}\label{subsec:promise}
 
 This Section has one upshot and one promisory note, about general considerations on subsystems and symmetries.

The first is to suggest a different treatment of asymptotic boundaries, that maintains invariance under symmetries of boundary states. Though this was long ago partially achieved for null asymptotic infinity (see \citep{AshtekarStreubel} and \cite[p. 52]{AshtekarNapoli}), it has also been  developed in the case of gauge symmetries, for Yang-Mills theory for spatial slices in \citep{RielloSoft}, where the spatial subsystem is extended asymptotically. 

This resolution is at the crux of my disagreement with \cite[p. 11]{Wallace2019b}, who   endorses a pared-down version of symmetries on  subsystems. That is because he takes subsystems  as sufficiently isolated   to warrant an asymptotic-like treatment, and asymptotic conditions often pare down symmetries, as we have just discussed.    
I maintain that there is a good notion of subsystem recursivity for  subsystems---namely, downward consistency---that does not mimic the asymptotic ideal of  perfect isolation. Conversely, there are asymptotic treatments of Yang-Mills theory that do not require an anchor state at the boundary, paring down symmetries. I thus conclude that   a treatment of  subsystems  in gauge theories that respects the downward consistency of symmetries is conceptually and technically justified.
 In particular, this implies that   conventions about the representation of the state do not come to us  `anchored' at the boundary: we have just as much free choice there as elsewhere. 

The promisory note is about counting possibilities and representational conventions, and it will be elaborated on in Chapter \ref{ch:subsystems}. It addresses a pressing question. Namely, although we will see in the course of  this thesis many uses of representational conventions, as a conceptually clarificatory tool, a puzzle remains: why is it that for most applications of a theory within the classical domain we can do without specifying the convention we are using?

The answer is that, from within a single convention, for a single possibility, the particular choice made cannot be compared with any other. 

Thus we can limit the domain in which the explicit use of representational conventions is necessary, as follows. In the study of a single physical possibility---describing features of a given solution of the equations of motion for a single system, for example---a representational convention may be left as implicit. Nothing physically important turns on which representational convention was used, though some conventions may be more convenient than others. 

On the other hand,  if we are to compare different physical possibilities, or different subsystems, we must ensure the comparison is made under a fixed representational convention. Thus such conventions can be become important in questions about assessing when symmetry transformations applied to subsystems are observable or not. \\

  In sum, even if it is not always inevitable, the use of representational conventions in gauge theory is extremely useful. Moreover, it is not only useful but \emph{necessary} when dealing with subsystems and counting possibilities: as we must to assess the observability of symmetry transformations, which we will do in Chapter \ref{ch:subsystems}.


\part{Same-Diff? Conceptual similarities and differences between gauge theory and general relativity}\label{part:I}

Philosophers of physics generally accept as the leading idea of a gauge theory---or as the main connotation of the phrase `gauge theory’---that it involves a formalism that uses more variables than there are physical degrees of freedom in the system described; and thereby more variables that one strictly speaking needs to use. Hence the common soubriquets: `descriptive redundancy’, `surplus structure’, and more controversially, `descriptive fluff’ (e.g. \cite{Earman_gmatters, Earman2004}). 

Although the main idea and connotation of descriptive redundancy is undoubtedly correct---and endorsed by countless presentations in the physics literature---some celebrated philosophers, such as \citet{Healey_book} and \citet{Earman_gmatters} among others, have gone beyond this connotation, and defended a stronger,  \emph{eliminativist view} that  gauge symmetry must be  eliminated, so that any two  models of a theory represent distinct physical possibilities, on pain of radical indeterminism. For them,  the connotation of `fluff' is that it can have no purpose.

But radical indeterminism also threatens theories such as general relativity, embodying diffeomorphism symmetry; a threat revealed by the famous hole argument. In that context, the most convincing---and popular---way to defuse the threat is called \textit{sophisticated substantivalism}. It is \emph{not} eliminativist: it is a form of structuralism, often related to the metaphysical doctrine of \emph{anti-haecceitism}, which takes spacetime points to have no metaphysically robust identity across possibilities. According to this doctrine, points can \textit{only} acquire identity  through their complex web of properties and  relations, as encoded in fields.\footnote{See \citep{Pooley_rel}  for a thorough exposition.} 

A similar resolution is available for gauge symmetry, in the form of `anti-quidditism'; but it is there much less popular.\footnote{ However, recently the position has garnered support, starting with \cite{Dewar2017} and followed by \cite{ReadMartens, Jacobs_thesis, Jacobs_Inv}.} Indeed, in the case of gauge symmetry, attempting to \textit{eliminate} the symmetry-related models is considered a more viable alternative. But is this alternative really more justified in the case of gauge symmetry? If so, why? 
In this Part we will deal with this question, in various forms.


 Chapters \ref{ch:sd_1} and \ref{ch:sd_2}   contrast diffeomorphisms and gauge transformations in various  respects. First, in Chapter \ref{ch:sd_1}, I will proceed at a more formal, high-altitude level and I concentrate on the similarities. In   Chapter \ref{ch:sd_2}, I proceed in more technical, low-altitude detail, and I concentrate on possible differences.

Thus in Chapter \ref{ch:sd_1}, we provide the necessary background material for a comparison between the symmetries of general relativity and Yang-Mills theory. This Chapter gives all the mathematical background of gauge theory and general relativity that we will require in this thesis. The conclusion is that, from a formal point of view, both theories are best understood structurally. This Chapter is based on \cite{Samediff_1}.

  Chapter \ref{ch:sd_2}   compares the symmetries of general relativity to those of Yang-Mills theory along several different axes. It will strengthen several analogies and disarm the disanalogies advocated by \cite{Healey_book}.  In particular, I argue that  the following topics are closely analogous  in gauge theory and general relativity: the Aharonov-Bohm effect; conservation of charges associated to symmetries;  locality of symmetry-invariant quantities, and the initial value problem. So  none of these  topics  provide  grounds for drawing a substantial distinction between gauge  and diffeomorphism symmetry. But the Chapter also homes in on one relevant distinction: whether the symmetry changes pointwise the dynamical properties of a given field. This distinction characterizes gauge symmetry related states---but not generic diffeomorphism-related states---as being `pointwise dynamically indiscernible'. This Chapter is based on \cite{Samediff_2}.


\chapter{Same-Diff? I: Conceptual  similarities}\label{ch:sd_1}

\begin{quote}Same-diff [noun]: \textit{
an oxymoron, used to describe something as being the same as something else. Often used as an excuse for being wrong.} (Urban dictionary). \\

Diff: \textit{A common abbreviation for ``diffeomorphism''. E.g. Diff$(M)$ is the group of diffeomorphisms of the (differentiable) manifold $M$.} 
\end{quote}

\section{Introduction and roadmap for this Chapter}
This is the first of two chapters analysing the similarities and distinctions between the gauge symmetries of Yang-Mills theory and the spacetime diffeomorphisms of general relativity. The first will analyse more formal aspects while the second will analyse more detailed aspects  of this comparison. 

My argument requires brief expositions of symmetries, for both  Yang-Mills theory and  general relativity: the theories that best represent the importance of gauge and diffeomorphism symmetry, respectively. I undertake this analysis in Section \ref{sec:attitudes}   for general relativity and  in   Section \ref{sec:PFB} for Yang-Mills theory. Since there are many good references for the foundations of spacetime physics (e.g. \cite{Earman_world, Maudlin_book}), I will concentrate on developing the conceptual foundations of Yang-Mills theories, such as the theory of principal fiber bundles; and so Section \ref{sec:PFB} is much longer and complete than Section \ref{sec:attitudes}. 

But both Sections \ref{sec:attitudes} and \ref{sec:PFB} close with the interpretation of symmetries that I endorse: \textit{sophisticated substantivalism}. I take this to be  a  \textit{structural} interpretation of the theories:  anti-haecceitist for general relativity diffeomorphisms and  anti-quiddistic for the gauge symmetries of Yang-Mills theory. This jargon can be quickly summarized: haecceitism is the doctrine that objects have an intrinsic identity (or `thisness': {\em haecceitas}); haecceitistic possibilities involve individuals being ``swapped'' or ``exchanged'' without any qualitative difference; and  quidditistic possibilities involve properties  being ``swapped'' or ``exchanged'' without any qualitative difference. Anti-haecceitists about
spacetime points thus deny that there are possible  worlds that instantiate the same distribution of qualitative
properties and relations over spacetime points, yet differ only over which
spacetime points play which qualitative roles. Similarly, the anti-quidditist will insist that there are no two possible worlds
 that instantiate the same nomological structure, and yet differ only over which properties play which nomological roles.  \citep{Black_quidditism} is a standard example of the anti-quidditist position, while \citep{LewisRamsey} is an example of the quidditist one.\footnote{
 An example may help visualise these concepts. For anti-hacceitism, picture a connected graph, in which  the vertices do not have an identity beyond their connectivity, or at least no such identity playing a nomological role. So a permutation of the vertices yields a duplicate of the original graph. It is important to note here that although a point's intrinsic identity may have no nomological role,  they are not easily expunged from our representation, for they are required in order to describe the graph's connectivity. An example for anti-quidditism is similarly straightforward: e.g. construe the edges as being  dyadic relations of the vertices. Again, permutation of the edges will not alter connectivity and so will ``give the same graph again''.) \label{ftnt:graph}} 

As stated, sophistication is a metaphysical thesis: a symmetry that reveals an underlying invariant structure, which is what has ontic significance. And indeed, the relation between symmetries and structure is familiar: the more symmetries there are, the less structure remains invariant under their action; and the fewer symmetries there are, the more structure that remains invariant (cf. \cite{Barrett2018_sym} for a more thorough discussion about this relationship). But  does any symmetry, even one that is arbitrarily defined, reveal an ontologically significant  underlying structure? 

This question  is contentious (cf. e.g. \cite{Dewar2017, ReadMartens, Jacobs_thesis}). In Section \ref{sec:soph_cheap} I will further explicate it, and design my own criterion to single out those symmetries that should be interpreted as revealing ontologically significant underlying structure. That criterion is whether the symmetry in question can be construed as simply a  change of notation in the formalism: that is, whether active symmetries have passive counterparts. If they do, we can  reveal, in each chart, the common structure of the symmetry-related models indirectly, as expressible quantities that are coordinate-invariant. We  find that both gauge transformations and diffeomorphisms, in suitable formalisms, satisfy this criterion. 


 Thus this chapter will adjudicate whether we can find salient differences between the symmetries of Yang-Mills and general relativity at a broad, formal, or high-altitude, level.  The verdict will be that we cannot: both theories find a natural expression with a structural, or `sophisticated' attitude towards symmetries.

 In Section \ref{sec:conclusions_soph}, I conclude Chapter \ref{ch:sd_1} with a brief overview of what we have achieved.

\section{Diffeomorphisms in general relativity}\label{sec:attitudes}

This Section will be briefer than the following one, on gauge symmetry, since the interpretation of redundancy in general relativity is less controversial than  in gauge theory. Nonetheless, I would like to give it a non-standard treatment, that gives due attention to  the  definition of smooth structure through charts and atlases. 

In Section  \ref{sec:diff_sym}, I will introduce the isomorphisms and symmetries that occur in general relativity. In  Section  \ref{sec:active_passiveGR} I will define the smooth structure of the manifold through atlases. This perspective on smooth manifolds will then help us understand the origin and significance of the symmetries discussed in Section  \ref{sec:diff_sym}. 


\subsection{Isomorphism and symmetries in general relativity, briefly introduced}\label{sec:diff_sym}
In brief, I will take general relativity in the metric formalism, where the most general models of the theory, sometimes labeled \emph{kinematically possible models} (KPMs) (so as to avoid confusion with those models that satisfy the equations of motion, which are labeled \emph{dynamically possible models} (DPMs)), are given by the tuples: $\langle M, g_{ab}, \nabla, \psi\rangle$. Here $M$ is a smooth manifold,   $g_{ab}$ is a Lorentzian metric (a $(0,2)$-rank tensor with signature $(-,+,+,+)$);  $\nabla$ is a covariant derivative operator, and $\psi$ represents some distribution of matter and radiation. I will assume $\nabla$ is the the unique Levi-Civita one, i.e. obeying $\nabla g_{ab}=0$. I will call the space of these KPMs $\mathcal{M}$, and, if we simplify to fixing $M$ and consider the theory in vacuo, i.e. setting $\psi=0$, then $\mathcal{M}=\mathrm{Lor}(M)$, the space of Lorentzian metrics over $M$.\footnote{Indices $a, b, c,$ etc are taken to be \emph{abstract} (cf. \cite[Ch. 2.4]{Wald_book} for an explanation), i.e. only denote the rank of the tensor, but no coordinate basis. I will denote coordinate indices by Greek letters: $\mu, \nu$, etc. }

In terms of the category-theoretic language (introduced in Chapter \ref{ch:syms}: see footnote \ref{ftnt:category}): the category of smooth manifolds has as objects the smooth manifolds, and  diffeomorphisms as the isomorphisms; diffeomorphisms are those maps that preserve the smooth global structure of manifolds.

The matter and gravitational fields are maps from points of the manifold to some other value space; we will look at this definition in detail when we discuss vector bundles in Section  \ref{subsec:PFB_ex}. The dependence of the fields on spacetime points implies that an action by a diffeomorphism on this base set will lift to an action on the fields. 
We can represent such an action of the diffeomorphisms  of $M$ on $(g_{ab},  \psi)$,   by the pull-backs, $(f^*g_{ab}, f^*\psi)$. It is also useful to represent the local, infinitesimal action of diffeomorphisms. Namely, for a one-parameter family of diffeomorphisms $f_t\in \Diff(M)$,  such that $f_0=\mbox{Id}$, we write the flow of $f_t$ at $t=0$ as the vector field $X^a$. Then, infinitesimally we obtain: 
\be\label{eq:Lie_g}\left.\frac{d}{dt}\right|_{t=0} f^*_tg_{ab}\equiv\mathcal{L}_Xg_{ab}=\nabla_{(a}X_{b)}, 
\ee
where $\mathcal{L}_X$ denotes the Lie derivative along $X^a$.\footnote{For a map $f:M\rightarrow N$, for  $\eta$ a one-form on $N$,  $f^*\eta$ is a one-form acting on $v\in T_x M$ as $f^*\eta(v):=\eta (Tf(v))$, where $Tf:TM\rightarrow TN$ is also called the \emph{push-forward} of the map (taking tangents to curves in $M$ to the tangents to the images of those curves under $f$), and is sometimes denoted by a $f_*$. For a scalar function $\rho$ on $N$, and $x\in M$,  $f^*\rho(x)=\rho(f(x))$.  Since, when $f\in \Diff(M)$,  maps and their inverses  are both smooth, we can mostly ignore the distinction between push-forward and  pull-back.\label{ftnt:push}  }

What are the `natural' isomorphisms of the composite objects $(M, g_{ab}, \psi)$? 
Standard mathematical practice takes isomorphisms in this category to be just those induced by the diffeomorphisms of the base set $M$, and in Section  \ref{par:active_passive}, we will give a brief argument for this; for now, we accept it.    Then, \emph{in vacuo}, 
 two models $\langle M, g_{ab}\rangle$ and $\langle M, \tilde g_{ab}\rangle$ are isomorphic if and only if there is a diffeomorphism of $M$, $f\in$Diff$(M)$, such that $f^*g_{ab}=\tilde g_{ab}$. If matter and radiation fields are included, an isomorphism would require the same map  to similarly relate their distributions in the two models.

Thus we have described the isomorphisms of this space of KPMs. Spacetime physical theories usually assume that  isomorphisms are symmetries of the theory, in the sense that a large, salient set of quantities, and their values, will be physically represented equally well by any isomorphism-related model. Indeed, if one model satisfies the Einstein equations, an isomorphic model will also satisfy them. We can actually do a bit better than that, in the spirit of Definition \ref{def:inf_sym} presented in Chapter  \ref{ch:syms}. We endow $\mathcal{M}$ with a (infinite-dimensional) manifold-like structure of its own, and define an action functional on this space: $S:\mathcal{M}\rightarrow \RR$. In vacuum, this is the Einstein-Hilbert action functional:
\be \label{eq:EH_action}
S[g]:=\int_M\d^4 x\,\sqrt{g}\,R,
\ee
where $R$ is the Ricci-scalar function (a function of second-order partial derivatives of the metric), and $\d^4 x\,\sqrt{g}$ is the infinitesimal volume-element of $M$.
 We then extremize $S[g]$ for given boundary conditions, e.g. in vacuum, and for a fixed manifold $M$, so that elements of $\mathcal{M}$ differ only by their metrics.  Then, omitting  indices, assuming from the extremization requirement that $S[g+\delta g]-S[g]=0$ for all directions $\delta g\in T_{g}\mathcal{M}$, the equations of motion emerge as conditions on the `base metric' $g$, and certain vector fields on $\mathcal{M}$ leave $S$ invariant,  e.g. in vacuum $S[g+\widehat{\delta g}(g)]-S[g]=0$, for all $g$, where $\widehat{\delta g}:g\rightarrow T_g\mathcal{M}$ is a smooth vector field on this infinite-dimensional field space, $\mathcal{M}$. With another set of minimal assumptions \citep{Lee:1990nz}, these vector fields can be identified as the flow---the infinitesimal versions---of the maps $(g_{ab}, \psi)\rightarrow (f^*g_{ab}, f*\psi)$. Namely, these directions are given by $\mathcal{L}_Xg_{ab}$ of \eqref{eq:Lie_g}, and they generate the isomorphisms induced by the diffeomorphisms of $M$. 
 
 Alternatively, we can extract the same set of symmetries from the \emph{initial value problem} for general relativity (we will have more to say about this in Section \ref{subsec:IVP}; see   \cite[Secs. 7.5-6]{Landsman_GR} for a sketch of the (complicated) proof, and \cite[Ch. VI]{choquet2008} for more details). The broad  idea is that that the equations of motion of the theory are not hyperbolic, and so as they stand lack uniqueness given initial values; moreover, the initial values are not arbitrary, but must also satisfy constraints.  To extract a proper hyperbolic equation, we require the solution to respect a representational convention, as in Section \ref{sec:rep_conv}. That is, solutions that lie along a gauge-fixing section \emph{are} unique, and different solutions of the same initial values have the same projection under \eqref{eq:h_proj} to this gauge-fixing section, and thus they are related by an isomorphism of the theory. 

 Therefore, in vacuo, we will say that $\langle M, g_{ab}\rangle$ and $\langle M, \tilde g_{ab}\rangle$ are both isomorphic and symmetry-related  iff there is an $f\in \Diff(M)$, such that $\tilde g_{ab}=f^*g_{ab}$. We write this as: 
\be\label{eq:equivalence}
\langle M, g_{ab}\rangle\sim \langle M, f^*g_{ab}\rangle.
\ee
Thus we identify the symmetry group as $\G:=\Diff(M)$, which acts on the space of Lorentzian metrics over $M$, namely, $\mathcal{M}=\mathrm{Lor}(M)$.

Another point to note: diffeomorphisms act transitively \textit{on} $M$; any point can be carried to any other point. This  means that there is no non-trivial orbit for $\Diff(M)$ defined as a subset of $M$. Of course, $\Diff(M)$ does \emph{not} act transitively on $\mathrm{Lor}(M)$: there we can easily identify the orbits of $\G$ by \eqref{eq:equivalence}, and one orbit does not cover the entire space of models. 

 This means that diffeomorphisms and gauge-symmetries are indiscernible at the level of entire models. As we will see in Chapter \ref{ch:sd_2}, to discern them we must zoom in on their action on the base set, or what we will call the \emph{pointwise} action of the symmetries.

This summarizes the general  view on symmetries for general relativity.

But why relate diffeomorphisms with the dynamics of the spacetime metric, as opposed to that of any other field on spacetime? In the following two paragraphs we provide a provisional answer to this question (we will complement that answer only in \S \ref{subsec:IVP}).

\paragraph{Why   invoke general relativity when discussing diffeomorphisms?}\label{subsec:diffeos_gr}

Here I would like to make a broader point, tangential to the topic of sophistication, about this and the following Chapter's focus on general relativity. For the other theories, based on fields other than the metric, would also, by the arguments of Section  \ref{par:active_passive} (see last paragraph) carry diffeomorphism-invariant structure. And similarly,  action functionals for e.g. a scalar field $\psi$, would be preserved by diffeomorphisms, and we would find that  two distributions, $\psi$ and $f^*\psi$ would jointly either satisfy or fail to satisfy the equations of motion, i.e. they would be related by a dynamical symmetry.  So the reader would be right to ask: why link the interpretation of a spacetime manifold to the symmetries of Lor$(M)$ and not to some other $\F$? Or, equivalently,  why the focus on general relativity when discussing spacetime diffeomorphisms?  

There is a historical reason and a physical reason for this focus. The historical reason is that general covariance and the equivalence principle played a very important role in Einstein's  discovery of the theory. The physical reason is that given some mild assumptions about physical theories written in terms of diffeomorphism-invariant actions over spacetime, every matter field  must couple to a metric. That is, at every spacetime point, all fields in the action functional will have some coupling to a metric. 
Thus the  transformation properties under diffeomorphisms  of each factor in the action  are dictated by their coupling to the metric and the transformation properties of the metric.\footnote{The fact that all  fields seem to couple to a single metric is, I think, a vindication of the geometric, as opposed to Brown's competing dynamical, understanding of the metric field. See \cite{Brown_book} for more about this debate. There are many ways to see the universality property of the metric vis \`a vis covariance, in particular using  the action functional formalism (cf. Section \ref{sec:diff_sym}). All fields employed in the standard model have some tensorial or spinorial weight, and one uses the metric (or e.g. tetrads) to build 4-dimensional scalar densities from them. For instance, a dynamically non-trivial theory will contain terms that are non-linear in the given tensor field (i.e. the field appears at least quadratically). Since the metric is used to contract indices, it will necessarily couple to  these fields, and the covariance of the metric will determine the covariance of the fields. More generally, the integral involved in the action functional contains the metric in its measure: in a chart, it is the volume element $\d ^4 x \sqrt{g}$ that is appropriately invariant under coordinate transformations. Note however, that this relationship says nothing about the signature of the metric: the same argument could be made using a Riemannian (and not a Lorentzian) signature. Conversely, using this covariance property of tensor fields, one can derive the universality of the metric as  a close cousin of the equivalence principle,  from the assumption of Lorentz invariance in perturbative quantum field theory \citep{Weinberg_equiv}.) }

 As we will discuss in  \ref{sec:syms_consts}, this association between general relativity and diffeomorphisms is most apparent in the Hamiltonian formalism. In that formalism,  the `generators' of the symmetries arise automatically from the dynamics of geometry.

\subsection{Sophistication for diffeomorphisms}\label{sec:active_passiveGR}\label{par:active_passive}

In this Section I will first define the smooth structure of $M$ through charts and atlases, in what I will call a \emph{chart-nominalist} interpretation in \S \ref{subsec:chart_nom}. The interpretation is nominalist in the sense  that the charts are not understood as surveying some pre-existing abstract structure: they \emph{induce} the structure. This interpretation will help us understand the reason why the symmetries of general relativity should at least contain the (induced action of the) diffeomorphisms. 
 And in  \S \ref{subsec:soph_anti} I provide a brief summary and defense of sophisticated substantivalism: a structuralist position that maintains commitment to a `thin' notion of existence of spacetime points so that they have no intrinsic identity across models, and by so doing regards symmetry-related models as representing the same physical possibility. In Section \ref{subsec:soph_stab} I provide reasons to regard the plethora of isomorphic models as conceptually harmless.

\subsubsection{Chart nominalism}\label{subsec:chart_nom}

I here define charts as bijective maps from subsets $U$ of $M$ (whose union covers $M$), to $\RR^n$, that have  smooth transition functions wherever they overlap. That is, given $\phi_1, \phi_2:U\rightarrow \RR^n$, so $U$ is the intersection of the domains of $\phi_1, \phi_2$, then $\phi_2\circ \phi_1^{-1}$ is a smooth bijective function from a subset $\phi_1(U)$ of $\RR^n$ to $\phi_2(U)$.\footnote{The very notion of smoothness invokes the use of charts: k-smoothness is defined, for a function $f:M\rightarrow \RR$,  as differentiability up to $k$-th order of the representative functions of $f$ on each chart $\phi$, namely, as $k$-th differentiability of $\tilde f:=f\circ\phi^{-1}:U\rightarrow \RR$. } Any such complete collection of charts  is called \emph{an atlas} for $M$, and any two compatible atlases---whose transition functions are smooth and with smooth inverses---are equivalent. The smooth structure of the manifold is defined as the equivalence class of atlases; or equivalently, as the maximal atlas, including all compatible charts. A maximal atlas can be taken simply to  \emph{define} the smooth and topological structure of the manifold. In particular, one does not need to remain faithful to some prior   topological or smooth structure of $M$: the topology, as well as the
differentiable structure, are bequeathed to $M$  by the
charts of a maximal atlas.\footnote{The set of all domains of
charts in the atlas forms a topological base for the manifold: it is closed under finite intersections and arbitrary unions, and its union is the
whole manifold. With respect to this topology   all charts are  homeomorphisms, by construction. Cf. \citep[p. 22-23]{Lang_book} for a textbook definition of smooth structure in this manner,  and \cite{Wallace_coords} for a conceptual treatment. }  Now  I will call this understanding of $M$ \emph{chart-nominalism}.

Any chart that is dragged by a diffeomorphism gives another chart. So, for $f\in \Diff(M)$ and a given tensor field $\mathbf{T}:=T^{a_1, \cdots, a_k}_{b_1, \cdots, b_l}$, we obtain a transformed field $\tilde{\mathbf{T}}:=f(\mathbf{T})$. Suppose that, under a chart $\phi_1:U_1\rightarrow \RR^n$, the components of $\mathbf{T}$ at a point that lies in $\phi_1$'s domain are given by $T^{\mu_1, \cdots,\mu_k}_{\nu_1,\cdots,  \nu_l}$. Then, there will be a second, compatible  chart, $\phi_2:U_2\rightarrow \RR^n$, for which the components of  the transformed field, $\tilde{\mathbf{T}}$, are \emph{also given} by the \emph{untilded}  $T^{\mu_1, \cdots, \mu_k}_{\nu_1, \cdots, \nu_l}$. The relation between $\phi_1$ and $\phi_2$ is, of course, just  $\phi_1=\phi_2\circ f$, where $U_2=f(U_1)$. Thus, given the joint description of $\mathbf{T}$ by the charts of a given atlas  for $M$, there will be a second atlas for which the \emph{different} tensor, $\tilde{\mathbf{T}}=f(\mathbf{T})$, has that same description. In equations: $\phi_1(\mathbf{T})=\phi_2(\tilde{\mathbf{T}})$, and similarly for every chart of the first atlas.  In  words, the images (i.e. the values of components) of the transformed tensor under the new charts  are the same as the images of the untransformed tensor under the old charts. The fact that the domains of these charts will differ seems inconsequential, since, in the chart-nominalist interpretation, the manifold structure (topological, smooth, etc) is \emph{defined} by the charts.
The active transformation therefore amounts to a change of (a non-maximal) atlas. And any diffeomorphism will leave a maximal atlas completely invariant; a good thing, since otherwise we would not be able to identify `smooth structure' with `maximal atlas'.

We can address the question posed in Section  \ref{sec:diff_sym} (after \eqref{eq:Lie_g}), namely, why it is safe to assume that the isomorphisms of the  n-tuple $(M, g_{ab}, \psi)$ is not a smaller set than  the isomorphisms of $M$. The reason is that any stronger notion of isomorphism would imply, through the   construction through atlases and the passive-active correspondence above, that  the composite objects would not be fully covariant under  all possible (i.e. smooth) coordinate transformations, thus allowing only a subset of all the coordinate systems that are compatible. This subset would perhaps correspond to another mathematical structure on $M$, which would then be sufficient to serve as the base set for the theory in question (see \cite{Wallace_coords} for a more thorough analysis of this idea). Conversely, it is natural to require that the isomorphisms of the fields and quantities over the base points of $M$---labeled dependent and independent variables respectively---are not a larger set than the isomorphisms of $M$. This is essentially the content of two principles about symmetries proposed by \citet[p. 45-47]{Earman_world}: jointly, the two principles require that the dynamical symmetries should  be just those induced by isomorphisms of the base set. The idea finds a natural extension to gauge theory in the principal fiber bundle formalism; and it will be further developed in terms of the active-passive correspondence in Section \ref{subsec:active_passiveYM}. 


\subsubsection{Sophistication and anti-haecceitism}\label{subsec:soph_anti}

Following a nomenclature suggested in \cite[p. 220]{Belot2003} and adopted by  \cite{Moller}  (see also \cite{ReadMartens}),\footnote{\cite[p. 220]{Belot2003} talks about finding a `perspicuous formulation' of the symmetry-related models, \cite{Moller} also talks about perspicuous characterization.} we find the following chronogeometric interpretation of the models of general relativity to be \textit{metaphysically perspicuous}. This interpretation  takes diffeomorphism-invariant mathematical quantities to represent physically significant quantities, understood as quantities about coincidences of material point-particles, elapsed proper times along a particle worldline, etc. For example, if one makes a journey from one planet to another, all empirically measurable quantities  about the trip will be represented as diffeomorphism-invariant functions. These include: the time elapsed along the journey, whether the spaceship is intrinsically accelerating or not as it passes some asteroid,  all operations involved in signaling with particles or light pulses, etc.  
And I understand `perspicuity'  essentially as `intelligibility' of the mathematical structure; the characterization does not yet satisfy any formal criteria. In Section \ref{sec:soph_cheap}, we will try to provide such criteria.

This interpretation is compatible with  \emph{anti-haecceitism}, since one can  understand points of the manifold `thinly', as devoid of an intrinsic identity across  possibilities; we can individuate points only as places in a structure. 
   In that sense, all spacetime distributions of the metric that are related by a diffeomorphism are taken as representing the very same state of affairs, since they only differ as to which regions, or points, of the manifold support which pattern of the field; distributions related by a diffeomorphisms are qualitatively identical (cf. both \cite{Pooley_routledge, Pooley_Read} for details).

The position labeled \emph{sophisticated substantivalism}, introduced in the context of the hole argument,    is anti-haecceitist in this way, and illustrates   structuralism for general relativity, since it awards physical significance only to any quantity that is   diffeomorphism-invariant.
It does so by endorsing \cite{EarmanNorton1987}'s condition of  Leibniz equivalence---that isomorphic models represent the same physical possibility---while still allowing isomorphic models to `peacefully co-exist', without reduction.\footnote{ The label `sophisticated' was used with a negative connotation in \cite{belot_earman_1999} (see also \cite{belot_earman_2001}), to characterize `sophisticated substantivalists' as those who sought to simultaneously keep an ontological commitment to spacetime points while rejecting their context-independent, or primitive, identity across possibilities. Later on, the label was stripped of its pejorative connotation and accepted by many philosophers. \cite{Dewar2017} extended the adjective ``sophisticated'' to other symmetries: where it is understood as allowing commitment to structure without reduction or elimination of the isomorphic copies. More recently,  \citet[p.959]{Belot50} labelled this type of position: ``straight-up cheap anti-haecceitism", which he characterized as adopting ``a qualitative counterpart theory and
[...denying] the existence of worlds that are qualitative duplicates of one another. Then
one can maintain that there is only one possible world corresponding to a whole
family of mathematical spacetimes related to one another by generalized shifts.
The threat of indeterminism vanishes [...]"}

 In sum, in general relativity the  interpretation of the diffeomorphism-invariant structure  is easy to state in words:  it is chronogeometric; it is about how distant the spacetime points stand in relation to each other in a network. 
The sophisticationist identifies the structural content of the theory   with the set of  symmetry-invariant quantities, and she takes these, and only these quantities to denote, or to have ontic significance; nonetheless, she is permissive about non-unique representations of this structure.

 I take these considerations to be helpful for the interpretation of diffeomorphism symmetry. But they do not give more general formal criteria, for arbitrary theories, about when exactly we should accept structural interpretations that are sophisticated in this way. And in particular, I don't find the discussions in the standard textbooks of general relativity illuminating on this issue.\footnote{Here is \citet[p. 438]{Wald_book}:``If a theory describes Nature in terms of a spacetime manifold $M$ and tensor fields, $T$, then if $f:M\rightarrow N$ is a diffeomorphism, the solutions $(M, T)$ and $(N, f^*T)$ have physically identical properties. Any physically meaningful statement about $(M, T)$ will hold with equal validity for $(N, f^*T)$. On the other hand, if $(M, T)$ and $(N, T')$ are not related by a diffeomorphism, \emph{and if the tensor fields $T$ represent measureable quantities}, then $(N, T')$ will be physically distinguishable from $(M, T)$." (my emphasis). } I will try to provide some of these criteria  in Section  \ref{sec:soph_cheap}. There we will find, in the case of Yang-Mills theories, that  the same considerations apply, mutatis mutandis, with quidditism in place of haecceitism and properties in place of objects (or points). 
 



\section{Gauge transformations in Yang-Mills  theories}\label{sec:PFB}
This Section will explore details of symmetries in gauge theories: more especifically, of Yang-Mills theories.

Speaking metaphysically, the  previous Section \ref{sec:attitudes} construed the symmetries of general relativity as isomorphisms of a natural geometric structure. And there is a possible  misgiving that the symmetries of gauge theory are less natural, and thus have a less natural structural interpretation than those of general relativity. 

I believe that the concern is indeed justified in the  case of gauge transformations in the gauge-potential formalism for electromagnetism, which we discuss in Section  \ref{sec:crude_A}. But that formalism is not the last word in the theoretical development of Yang-Mills theories. In  Section  \ref{sec:PFB_mot} I motivate the need for  a more complete,  geometric understanding of what the fields and gauge symmetries of modern physics are about.  
 Sections \ref{sec:PFB_formalism} through \ref{sec:Atiyah} present in more detail the   mathematical formalism that we will need going forward. These are the three  most mathematically detailed Sections in the thesis.\footnote{And I have pondered whether to relocate this Section as an appendix, but in the end thought this would leave too wide a gap in the main text. Nonetheless, if the Section is too mathematically involved or too dry, it can be skipped and referred to as needed.} Section \ref{sec:PFB_formalism} introduces the mathematical definition and details of principal fiber bundles and develops the active passive correspondence for the isomorphisms of gauge theory; and Section \ref{sec:Atiyah} describes the intrinsic definition of Yang-Mills fields over spacetime, alternatively labeled as sections of the bundle of connections or sections of the bundle of connections. 


\subsection{Symmetries need not be isomorphisms: an example from gauge theory}\label{sec:crude_A}

In electromagnetism, the basic dynamical variable is the electromagnetic field tensor, $F_{ab}$. Upon choosing a spacetime split into spatial and time directions, the components of the electromagnetic tensor become the familiar electric and magnetic fields: $F_{i0}=E_i$, and $F_{ij}\epsilon_i^{jk}=B_i$  (where we used the three-dimensional totally-antisymmetric tensor, $\epsilon$, or the spatial Hodge star, to obtain a 1-form). 

The Maxwell equations in Minkowski spacetime are written, in a coordinate basis, in terms of $F_{\mu\nu}$, as:
\be\label{eq:EM}\partial^\mu F_{\mu\nu}=j_\nu, \quad \text{and}\quad \partial_{[\mu}F_{\nu\kappa]}=0, \ee
where $\pp_\mu$ are the coordinate derivatives,  $j$ is the current, and square brackets denote anti-symmetrization of indices. The second equation of \eqref{eq:EM} is called `the Bianchi identity', and it is read as a  constraint on the field tensor. A geometric explanation for this constraint is that $F_{\mu\nu}=\partial_{[\mu}A_{\nu]}$, or, in exterior calculus notation, $\d \mathbf{A}=\mathbf{F}$, where $A_\mu$ is called \textit{the gauge-potential}. At least locally, this relation follows from the Poincar\'e lemma. 

 Gauge-potentials for electromagnetism are locally just smooth one-forms on the manifold, and the natural notion of isomorphism here is just the one inherited from differential geometry: again, pull-backs by  diffeomorphisms. That is, the KPMs of the theory  are given by  $\langle M, \mathbf{A}\rangle$, where $\mathbf{A}=A_\mu \d x^\mu$, i.e. the potentials are  sections of the cotangent bundle---real-valued one-forms over each topologically trivial patch---on the manifold $M$.  Since they are differential forms, we could rehearse the argument of Section  \ref{sec:diff_sym} and conclude that  the isomorphisms of the space of models are again pull-back by diffeomorphisms. 
 
 But the dynamics of the theory are another matter.  
 The equations of motion of this theory---now assuming \emph{in vacuo}, i.e. $j=0$, for simplicity---are: 
 \be\label{eq:eom_A} \partial^\mu\partial_\nu A_\mu-\partial^\mu\partial_\mu A_\nu=0.
 \ee
 These equations are obtained from the action functional:
  \be\label{eq:action_A}S[A]:=\int_M \partial_{[\mu}A_{\nu]}\partial^{[\mu}A^{\nu]}=\int_M *\mathbf{F}\wedge \mathbf{F},
 \ee
 where $*$ is the Hodge-star operator (which takes an argument differential form to its (ortho)complement) and $\wedge$ is the exterior (wedge) product between forms. 
 If we then follow the definition of symmetries given in Chapter   \ref{ch:syms}, we arrive at the standard gauge transformations.\footnote{See also Section \ref{sec:diff_sym} for a more thorough account of how we would go about defining the symmetries also in this case.} 
 
  Namely, it is easy to see that, since the partial derivatives commute, $\d^2=0$ or $\pp_{[\mu}\pp_{\nu]}\phi=0$, for any scalar $\phi$, the transformations that preserve the value of \eqref{eq:action_A} for any $\mathbf{A}$ (and  that take \textit{any} solution of \eqref{eq:eom_A} to another solution) consist  in  adding the gradient of a smooth function to the gauge-potential one-form: $\mathbf{A}\rightarrow \mathbf{A}+\d \xi$, for $\xi\in C^\infty(M)$, and where $\d$ is the exterior derivative.\footnote{This conclusion could be reached following essentially the same procedure advocated in Section \ref{sec:diff_sym}. Note that the symmetries involve only differential geometric operations---such as exterior differentiation---and thus composition with diffeomorphisms is well-defined. Indeed, the two operations  commute, since the exterior derivative commutes with the pull-back: for $f\in$Diff$(M)$,  the object and arrow $(\mathbf{A}, \xi)$ gets mapped to $(f^*\mathbf{A}, f^*\xi)$. \label{ftnt:comm_iso}} The dynamical symmetries are therefore `larger' than those expected from the geometric properties of the fields.
  
  But as we will see in the next Section, there is a formulation of gauge theory that articulates its symmetries in a more `organic' fashion.

\subsection{Fiber bundles as the mathematical representation of fields and symmetries}\label{sec:PFB_mot}

The modern mathematical formalism of gauge theories relies on the theory of principal  and associated fibre bundles. We will not give a comprehensive account here (cf. e.g. \citep{kobayashivol1}), but only introduce the necessary ideas. 

Our  intuitive idea of a field over space is something like temperature. A temperature field can be written as a map from space to the real numbers, $T:M\rightarrow \RR$. Being told that there are fields that have a more complicated `internal structure' than temperature---for instance, vector fields that over each point of spacetime can point in different directions---we may want to generalize the  scalar map above to  $\rho:M\rightarrow F$, a map from spacetime to some internal vector space $F$. 

For tensor bundles, made up of tensor products of tangent and cotangent vectors, $F$ is ``soldered'' onto spacetime, $M$.\footnote{For instance, we can identify elements of the tangent bundle with tangent vectors of curves on the base manifold. In more detail, supposing  the internal vector space $F$ has the dimension of $M$, a \emph{soldering form} gives an isomorphism between each $T_xM$ and $F$, in a smooth way. }
 But the fields employed in modern theoretical physics  live in more general vector bundles, $F$, which are not thus soldered to spacetime. Generically,  those fields have many components at each point, which are not associated to spacetime directions. 
 
 The worry might arise that to examine the symmetry structure of a certain gauge group we would have to examine its action for each matter field separately: how it acts on electrons, on neutrinos, on quarks, etc; and these actions could, in principle,  differ in their general features. But nature is kind: the symmetry group acts similarly, though perhaps with different representations on the various matter  fields, meaning that the parallel transport of internal quantities is compatible for all the fields. This `coincidence' is conveniently described if we encode the symmetries through the formalism of \emph{principal fiber bundles} (PFBs): they contain the essential symmetry structure of each type of interaction---e.g. electromagnetic---independently of the individual matter fields that are susceptible to this interaction.

The first Subsection below, \S \ref{subsec:PFB_idea}, will present the main idea of principal fiber bundles. The aim of this section is to convince the reader through non-mathematical arguments that a principal fiber bundle admits a structural interpretation of its relevant quantities, to the same extent that the metric admits a structural interpretation of  its relevant quantites. 
Next, in  \S  \ref{subsec:PFB_ex}, I will show, through a more familiar example, how is it that principal bundles can orchestrate the interaction of a single given force with all the various matter   fields.

\subsubsection{Principal fiber bundles: the main idea}\label{subsec:PFB_idea}

 States of different species of matter are represented in (as sections of) different vector bundles: one vector bundle per field. A principal fiber bundle `orchestrates' the symmetry properties of all these matter  fields. As articulated convincingly by \cite{Weatherall2016_YMGR}:  even if vector bundles represent possible local states of matter, the \emph{connection} of a principal bundle orchestrates the symmetry properties of all the fields that interact through some given force.  Charged scalar fields, electron fields, quark fields, etc., all interact electromagnetically; and indeed they respond to the same electromagnetic fields (mutatis mutandis, for other interactions, e.g. replacing  `electromagnetism' by the `strong force'). This means that the covariant
derivative operators on the vector bundles in which these fields are valued  have the same
parallel transport and curvature properties.   Such  universality is mathematically enforced because these vector bundles are associated to  the same connection on that principal bundle, and this means they have their covariant derivative operators defined by that connection. 

We can thus, with a clear conscience, focus our efforts on understanding symmetry as it is mathematically manifested in a principal fiber bundle formalism.  
And the main idea underlying the physical significance of this symmetry structure  is  perhaps best  summarized in the original paper by  \citet{YangMills}: 
\begin{quote} The conservation of isotopic spin is identical with the requirement of invariance of all interactions under
isotopic spin rotation. This means that when electromagnetic interactions can be neglected, as we shall hereafter assume to be the case, the orientation of the
isotopic spin is of no physical significance. The differentiation between a neutron and a proton is then a
purely arbitrary process. As usually conceived, however,
this arbitrariness is subject to the following limitation:
once one chooses what to call a proton, what a neutron,
at one space-time point, one is then not free to make any
choices at other space-time points. \end{quote}
What is a proton and what is a neutron at a given point is essentially a \textit{relational} or, more broadly, a \textit{structural} property in $P$.\footnote{ Of course this example, which originally motivated Yang and Mills, applies only in the context of the (approximate) isospin symmetry. Otherwise, the electric charge tells protons and neutron apart in an intrinsic manner.} 

The only physically relevant information seems to be  sameness across different points of spacetime: thus, once  we label a given particle as e.g. a proton at one point of spacetime, the structure of the bundle specifies what would also count as a proton at a neighbouring spacetime point. 
These constraints are imposed by a \textit{connection-form}: the main geometric structure of the bundle. 
A connection-form $\omega$ allows us to define which points of neighbouring fibres can be taken as equivalent to an arbitrary starting-off point in an initial fibre. 

 In this framework, \textit{curvature}  acquires meaning as non-holonomicity. Let $p$ be a given point in the bundle; and take its projection onto spacetime, $x$ to be the starting point of  two spacetime curves that  later reconverge to another spacetime point, $y$. These two curves have a unique type of `lift' to curves in the bundle passing through $p$, called a \emph{horizontal} lift: such lifts  represent parallel transport. Even though the projected paths in $M$ close-off at  $y$,   the end-points of their horizontal lifts will in general  differ. It is this disagreement that  carries physical consequences. That is, the bundle encodes structural, or relational,  properties, that arise from comparisons: and  which, at least infinitesimally,  are captured by certain function(al)s of the connection, namely,  the curvature. And yet, globally, or non-infinitesimally, these comparisons may still carry information that is not captured by the curvature.

 
\subsubsection{PFBs from tangent spaces}\label{subsec:PFB_ex}

To gather intuition about principal fiber bundles (PFBs)  as the `organizers' of symmetry principles, it is worthwhile to introduce them in the context of the familiar tangent vector fields on $M$. 

I begin with the main idea of a fibre bundle and then consider the tangent bundle. The main idea of fiber bundles is that they are spaces that locally look like a product, i.e. a fiber `bundle'. So the many fields of nature would be represented as maps that take each point of spacetime (or space)  into its respective value space, or fiber. 

We denote fiber bundles by $E$; they are smooth manifolds that admit the action of a surjective projection $\pi:E\rightarrow M$ so that locally $E$ is of the form $\pi^{-1}(U)\simeq U\times F$, for $U\subset M$ and $F$ is some `fiber': a  space that `inhabits' each point of $M$ and in which the fields take their values. 

 But the decomposition $\pi^{-1}(U)\simeq U\times F$ is not unique, and will depend on what is called `a trivialization' of the bundle, which is basically a coordinate system that makes the local product structure explicit. Thus, in principle there is no unique identification of an element of $F$ at a point $x\in M$ with an element of $F$ at a point $y\in M$. In principle, there is no identification of a vector, or even of a scalar quantity, like temperature, as possessed at different points of spacetime.

So, to be explicit: $F$ is some space where we can have quantities in spacetime take their value; for instance,  a scalar field could take values in $\RR$ or $\bb C$, whereas a more complicated field such as a  vector  field or a spinor field, could take values in $\RR^4, \bb C^4$, etc. A choice of \textit{section} of the bundle  represents fields taking values in $F$: e.g. a spinor field, or a quark field, etc, which are all vector bundles, in that $F$ is a vector space. A field-configuration for $E$ is called (confusingly, see Section  \ref{sec:PFB_formalism} and footnote \ref{sec:section_triv}) \textit{a section}, and it is a map $\kappa: M\rightarrow E$ such that $\pi\circ\kappa=\mathrm{Id}_M$. Sections replace the functions $\tilde\kappa:M\rightarrow F$, that we would employ if the fields that physics uses had a fixed, or  ``absolute''---i.e. spacetime independent---value space. We denote smooth sections like this by $\kappa\in C^\infty(E)$.

 A useful example of a vector bundle is  the tangent bundle, $TM$. A  smooth tangent vector field is  a smooth assignment of  elements of $TM$ over $M$,  denoted  $X\in C^\infty(TM)$, with $\pi:TM\rightarrow, M$, mapping $X\in T_xM\rightarrow x\in M$. The tangent bundle   $TM$ \emph{locally}  has the form of a product space, $U\times F$, with $F\simeq \RR^4$.  But even if $TM$ were globally trivializable, so that a product structure could be found for its totality,  this would  not mean we could identify an element $v\in \RR^4$ at different points of $M$. Differential geometry teaches us to attach a vector space to each point of $M$ and to have vectors at different points objectively related only according to some definition of parallel transport along paths in $M$.

This example is also useful to articulate what we mean by a principal fiber bundle that `orchestrates the parallel transport' of the other fields. Here the principal bundle that orchestrates parallel transport  of tangent vectors (and tensor bundles in general)  can be taken to  be the bundle of linear frames of $TM$, called `the frame bundle' (where `frame' means `basis of the tangent space $T_xM$'), written $L(TM)$.  The fibre over each point of the base space $M$ consists of all of the linear frames of the tangent space there, i.e. all choices $\{\mathbf{e}_I(x)\}_{I=1, \cdots 4}\in L(TM)$, of sets of spanning and linearly independent vectors (here the index $I$ enumerates the basis elements).\footnote{Depending on the theory, we will take different subsets of the linear frames, and of the corresponding structure group. For instance, for general relativity, we take the structure group as $O(4)$ (or $SO(3,1)$)  acting on the orthonormal bases.}  

So each point $p\in P$ of the frame bundle above a point $x\in M$ (i.e. such that $x=\pi(p)$) is just a  basis for the tangent space $T_xM$; and there is a one-to-one map between the group $GL(\RR^4)$ and the fibre: we can use the group to go from any frame  to any other (at that same point), but there is no basis that canonically corresponds to the identity element of the group.  This example illustrates a feature of principal fiber bundles that distinguishes them from vector bundles: in the former, the fibers are isomorphic to some Lie group $G$; and there is no ``zero'' or identity element on each fibre, as there is in a vector bundle. 

\begin{figure}[h]
  \centering
  \includegraphics[width= 0.5\textwidth]{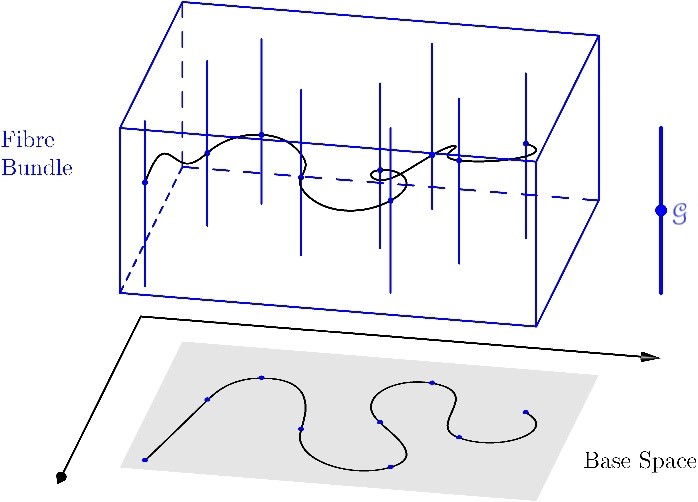}
  \caption{
A principal bundle over spacetime, with $G$ as a structural group. $[\gamma]$ is a curve on spacetime, that is horizontally lifted to $\gamma$, in $P$.
  }\label{fig:PFB}
\end{figure}

 If we imagine the orbits of the group, or the fibers, as being in the vertical direction, as in Figure \ref{fig:PFB},  directions transversal to the fiber will connect frames over neighbouring points of $M$.  We thus dub as \textit{horizontal} those directions by which a connection identifies---or `links'  and takes  as identical---frames on neighbouring fibers.\footnote{
In general relativity, we could take this to be  a torsion-free connection-form on $P$ by $\d \mathbf{e}^I=\omega_J^I \mathbf{e}^J$, where $\omega$ here satisfies the expected equations \eqref{eq:omega_defs} (and we used the one-forms  algebraically dual to the vector basis: $\mathbf{e}^J(\mathbf{e}_I)=\delta^J_I$). This equation translates to one using the covariant derivative $\nabla$ as: $\nabla \mathbf{e}_I=\omega^J_I \mathbf{e}_J$.\label{ftnt:gr_viel} } That is: to link fibres, we need to postulate more structure: a connection.

To see how these horizontal directions encode parallel transport of vectors, we need to return to the tangent bundle $TM$, from the frame bundle, $L(TM)$. We proceed as follows:  take a point of $TM$, i.e. a vector at a given point $x\in M$, $X_x\in F$ as an element of the fiber $F=T_xM\simeq\RR^4$, where the ordered quadruplet are the components of $X_x$ according to a frame, $\{\mathbf{e}_I(x)\}\in L(TM)$. So, we write $X_x=a^I \mathbf{e}_I\in T_xM$ as the ordered quadruplet $(a^1, \cdots, a^4)\in \RR^4$. Of course, if we rotate the frame by an element of the group in question, i.e. $GL(\RR^4)$, say by a matrix $g^{IJ}=\rho(g)$, where $\rho:G\rightarrow GL(\RR^4)$ is the matrix representative of the abstract group, then, as long as we undo that rotation on the components, we obtain the same vector, in the original frame. That is, $a^K g_{KL}^{-1} g^{LI}e_I=a^I e_I$. Thus, if we write a doublet $(p,v)$ as, respectively, the frame and the components, we want to identify $(gp, vg^{-1})$ (where we have simplified the notation for the action of the group to be just juxtaposition). This is a standard construction of an \emph{associated bundle}, denoted by $TM\simeq L(TM)\times_\rho \RR^4$. 

Once we have constructed associated bundles in this way, parallel transport, for any vector bundle comes naturally from a notion of horizontality in the principal bundle. To find the parallel transport of the vector $X_x$ along $Y_x$, we:
\\\noindent (i) choose one  frame $p$ at $x$, and find the corresponding---parallel transported---frames as one moves horizontally along (a direction $\tilde Y_p$ that projects to) $Y_x$,
\\\noindent  (ii) write out the component of $X$ at $x$ in that frame. If, along $Y$, $X$  were equal to its parallel transport, these components would remain numerically constant, since the frame is assumed to be the same, or parallel transported, i.e. identified across points of spacetime. So we can
\\\noindent  (iii) compare the parallel transported components of $X$ with the actual components of $X$; their non-constancy corresponds to the failure of $X$ to be parallel transported, and to the non-vanishing covariant derivative of $X$.  In this way a covariant derivative is just the standard derivative of the components in the horizontal---or parallel transported---frame. This is, in words, the description of the covariant derivative of $X$ along $Y$ at $x\in M$. 

The picture is useful in that it applies to any vector bundle on which the structure group $G$ in question acts. For instance, in the standard model of particle physics, the fundamental forces are associated to Lie groups, and each field that interacts via such  a force lives in a vector bundle that admits an action of the corresponding group. Thus for a given vector bundle with typical fiber $F$, we have a linear representation of the Lie group in question, $G$, $\rho:G\rightarrow GL(F)$, and we can take the principal connection---the notion of horizontality in the PFB with structure group $G$---to induce a notion of parallel transport in the bundle $E$ with fiber $F$. Indeed, we can take the same procedure as above, building a linear frame for $F$ at each point; parallel-transport then encodes an appropriate $G$-covariant way to identify vector values along paths in the base space $M$.

\subsection{Principal fibre bundles: interpreting the formalism}\label{sec:PFB_formalism}

 We have introduced the the main idea of a principal fiber bundle in Section  \ref{subsec:PFB_idea}, and  its function, i.e. as determining parallel transport, in Section  \ref{subsec:PFB_ex}. Now we give the formal definitions. In Section \ref{subsec:general_cons} I will briefly introduce the general formalism for the principal bundles, including the idea of connection forms and gauge potentials. Section \ref{subsec:active_passiveYM} is parallel to Section \ref{subsec:chart_nom}: there I will discuss the relationship between active and passive gauge transformations.

 \subsubsection{The general construction}\label{subsec:general_cons}
  A principal fibre bundle is, in short, just a manifold where some group acts. In detail: it is a smooth manifold $P$ that admits a smooth free action of a  (path-connected, semi-simple) Lie group, $G$: i.e.  there is a map $G\times P\rightarrow P$ with $(g,p)\mapsto g\cdot p$ for some left action $\cdot$ and such that for each $p\in{P}$, the isotropy group is the identity (i.e. $G_p:=\{g\in{G} ~|~ g\cdot p=p\}=\{e\}$). 
Naturally, we construct a projection  $\pi:P\rightarrow{M}$ onto equivalence classes, given by  $p\sim{q}\Leftrightarrow{p=g\cdot{q}}$ for some $g\in{G}$. That is: the base space $M$ is the orbit space of $P$, $M=P/G$, with the quotient topology, i.e. it is characterized by an open and continuous $\pi:P\rightarrow M$. By definition, $G$ acts transitively on each fibre, i.e. orbit. The automorphism group of $P$---those transformations that preserve the  structures---are \emph{fiber-preserving} diffeomorphisms $\tau:P\rightarrow P$, i.e. such that $ \tau(g\cdot p) =g\cdot \tau(p)$. Purely internal, or gauge transformations can be identified as those for which $\pi\circ\tau\circ\pi^{-1}=\mathrm{Id}_M$; that is,  as purely `vertical' automorphisms of the bundle; (the orbits are usually drawn going up the page, as in Figure \ref{fig:PFB}, hence `vertical').

\paragraph{-- The Ehresmann connection-form.}
On $P$, we consider an Ehresmann connection $\omega$, which is a 1-form on $P$ valued in the Lie algebra $\mathfrak{g}$ of $G$ that satisfies appropriate compatibility properties with respect to the fibre structure and the group action of $G$ on $P$. We will first see how such a $\mathfrak{g}$-valued 1-form on $P$ selects a ``vertical'' subspace of the tangent space $T_pP$ at $p\in P$, which ``points in the direction of the fiber'', and how it selects a ``horizontal'' subspace---which gives the notion of parallel transport linking nearby fibres, which we introduced in Section  \ref{subsec:PFB_idea}.

Given an element $\xi$ of the Lie-algebra $\mathfrak{g}$, we define the vertical space $V_p$ at a point $p\in P$, as the linear span of vectors of the form 
\be\label{eq:fund_vec}v_{\xi}(p):=\frac{d}{dt}{}_{|t=0}(\exp(t\xi)\cdot p), \quad \text{for}\quad \xi\in \mathfrak{g}.\ee
 And then the conditions on $\omega$ are:
\be\label{eq:omega_defs}
\omega(v_\xi)=\xi
\qquad\text{and}\qquad
{L_g}^*\omega=g^{-1}\omega g,
\ee
  where ${L_g}^*\omega_p(v)=\omega_{g\cdot p}({L_g}_* v)$ and where ${L_g}_*$ is the push-forward of the tangent space for the left-action $g:P\rightarrow P$. Thus, we can only characterize the action of $\omega$ on vector \textit{fields} on $P$, i.e. on sections of the vector bundle $TP$,  say $\zeta\in C^\infty(TP)$, if they are \textit{left-invariant}, i.e. if $\zeta_{g\cdot p}={L_g}_*\zeta_p$. Such vector fields generate the automorphisms of $P$.
  
   At each orbit, we obtain the  infinitesimal transformation: 
  \be\label{eq:Lie_omega}
   \mathcal{L}_{v_\xi}\omega=[\xi, \omega].\ee
But if the vector field $\zeta$ as above  is the generator of a vertical automorphism, we obtain, instead of \eqref{eq:Lie_omega}, 
\be\label{eq:Lie_omega_full}
   \mathcal{L}_\zeta\omega=[\omega(\zeta), \omega]+\d_{\text{\tiny{P}}}\zeta,\ee
where $\d_{\text{\tiny{P}}}$ is here the exterior derivative on the smooth manifold $P$.\footnote{We could also write this non-infinitesimally, in the more traditional notation: $\tau^*\omega=\Psi\omega\Psi^{-1}+\Psi^{-1}\d_{\text{\tiny{P}}}\Psi$ where $\tau(p)=\Psi(p)\cdot p$, as introduced below.\label{ftnt:13} }

 A choice of connection is equivalent to a choice of covariant `horizontal' complements to the vertical spaces, i.e. $H_p\oplus V_p=T_pP$, with $H$ compatible with the group action.   That is, since $\omega$ is $\mathfrak{g}$-valued and gives an isomorphism between  $V_p$ and $\mathfrak{g}$, the first condition of \eqref{eq:omega_defs} means that: i) the kernel $\mathsf{Ker}(\omega_p)=H_p$, and ii)  since $V_p=\mathsf{Ker}(\pi_*)$, $H_p$ will be 1-1 projected by $\pi_*$ onto the tangent space $T_{\pi(p)}M$. Thus the vectors spanning $\mathsf{Ker}(\omega_p)$ are the so-called \textit{horizontal} vectors in the bundle, and each represents a unique `horizontal lift' at $p$ of a direction at $T_{\pi(p)}M$. This condition also requires that, much like the metric, the connection form is nowhere vanishing. The second condition of \eqref{eq:omega_defs} guarantees that the notion of horizontality covaries with the choice of representative of the fiber (e.g. the choice of frame in the frame bundle example above), that is: a vector $v\in T_pP$ is horizontal iff ${L_g}_* v\in T_{g\cdot p}P$ is horizontal. 

Therefore, in terms of the bundle of linear frames (cf. Section \ref{subsec:PFB_ex} above), we could translate the two conditions of \eqref{eq:omega_defs} as saying that: 1) for each direction on spacetime, there will be a unique way to parallel propagate a given linear frame in that direction, and 2) there is no difference between applying a change of frame before or after parallel propagation: changes of frame commute with propagation. Note moreover, that the isomorphisms of $\omega$ are defined just as the transformations induced from the isomorphims of the underlying structure: the (fiber-preserving) diffeomorphisms of the  bundle. Thus we conform to a straightforward extension of \citet[p. 45-47]{Earman_world}'s principles, briefly  described in the last paragraph of Section \ref{par:active_passive} (see also \cite{Jacobs_thesis, Guy_ghosts}). We will pick this up, in relation to the standard definition of gauge transformations as symmetries of the gauge potential, in Section \ref{subsec:active_passiveYM}, once we have defined the relationship between gauge potentials and (Ehresmann) connection-forms.
  

 \paragraph{-- The  curvature of the connection.}
The forces and interactions ``communicated'' to all the vector bundles by $\omega$ are encoded by  the curvature of $\omega$, a Lie-algebra-valued 2-form on $P$:
   \be\label{eq:curv_PFB} \Omega=\d_{\text{\tiny{P}}} \omega+\omega\wedge_{\text{\tiny{P}}} \omega,
   \ee 
   where $\wedge_{\text{\tiny{P}}}$ is the exterior product on $\Lambda(P)$; it gives anti-symmetrized tensor products of differential forms.  
  Using the decomposition of the tangent space $H_p\oplus V_p=T_pP$, we have associated orthogonal projectors, $\hat H_p$ and $\hat V_p$, with $\hat H:p\mapsto \hat H_p:T_pP\rightarrow H_p$, we can rewrite the curvature   \eqref{eq:curv_PFB}, using \eqref{eq:fund_vec}, as: 
  \be\label{eq:Omega_com} v_{\Omega(\bullet, \bullet)}=\hat V([\hat H(\bullet), \hat H(\bullet)]_{\mathsf{TP}}),
  \ee
  where $\bullet$ is used as the open slot of a differential form, and the square brackets here denotes the commutator of vector fields on $P$. The intuitive idea is as before: one goes around an infinitesimal horizontal parallelogram and finds a certain displacement along the orbit. 
  
  The analogue of \eqref{eq:Lie_omega_full} for the curvature is:\footnote{We could also write this analogue non-infinitesimally, in the more traditional notation: $\tau^*\Omega=\Psi\Omega\Psi^{-1}$ where $\tau(p)=\Psi(p)\cdot p$ (cf. the previous footnote, \ref{ftnt:13}).\label{ftnt:14} } 
  \be\label{eq:Lie_Omega_full}
   \mathcal{L}_\zeta\Omega=[\omega(\zeta), \Omega]. 
   \ee
In other words, the curvature is fully left-invariant: there is no inhomogeneous term in its transformation. This is the crucial property that will, in Section  \ref{sec:KK}, distinguish the gauge symmetries from the diffeomorphisms, yielding the distinction that I labelled $\Delta$ in Chapter  \ref{ch:syms}.

\subsubsection{Active and passive correspondence}\label{subsec:active_passiveYM}
As with the definition of a manifold using an atlas (cf. Section  \ref{sec:active_passiveGR}), here too, the intrinsic construction of bundles above ``hides under the hood'' the explicit formulation via local trivializations. Namely,  we use local trivializations  and  conditions on the transition functions between charts to define the  bundle structure. This Section is parallel to  Section \ref{par:active_passive}; both provide the background for  a straightforward correspondence between active and passive transformations in their respective contexts: for  diffeomorphisms in Section \ref{sec:gr_re}, and for gauge transformations in Section \ref{sec:soph_antiq}.


\paragraph{-- Local sections}\label{sec:local_section}
Locally over $M$,  it is possible to choose a smooth embedding of the group identity into  the fibres of $P$. The maps ${\sigma}$ are called {\it local sections} of $P$.\footnote{ It is somewhat confusing that a \textit{section of a vector bundle} is an entirely different object: it is a vector field. So, for instance two different choices of the electron field are two different sections of its vector bundle, and thus are not counted as `equivalent' in the way that two sections of a principal bundle are. \label{sec:section_triv}} 

 That is, for $U\subset M$, there is a map ${\sigma}: U\rightarrow P$ such that $P$ is locally of the form $U\times G$. Namely, $\sigma$ induces a diffeomorphism $U\times G\simeq \pi^{-1}(U)$, given by $\bar \sigma:U\times G\rightarrow P$, such that:
\be\label{eq:sbar}\bar \sigma: (x, g)\mapsto g\cdot {\sigma}(x), \quad\text{whose inverse is}\quad \bar \sigma^{-1}: p\mapsto (\pi(p), g_{\sigma}(p)^{-1})\ee 
where $g_\sigma:\pi^{-1}(U)\rightarrow G$ gives   $g_{\sigma}(p)$ as the unique group element taking $p$ to the local section, i.e. $g_\sigma(p)$ is the group element such that $g_{\sigma}(p)\cdot p={\sigma}(\pi(p))$.\footnote{The precise form of $g_{\sigma}$ will of course depend on ${\sigma}$. } Thus we have a condition: 
\be\label{eq:g_equi}g_\sigma(g\cdot p)=g_\sigma(p)g^{-1}.\ee
 Call this \emph{equivariance} of $g_\sigma$ between the given action of $G$ on $P$ and $G$'s action on itself by conjugation. We saw an identical condition in Section \ref{sec:rep_conv}, in Equation \eqref{eq:cov_g}, about the equivariance of representational conventions. 
And indeed,  we used the same notation, $\sigma$, for either a representational convention and for a section of a principal bundle. The reason is simple: they are intimately related. Namely, in field theory, a representational convention can be seen as a state-dependent choice of section.

In the same vein, can also define local sections in two alternative ways. As maps  $\sigma:U \to P$ such that $\pi\circ \sigma = \mathrm{id}$; or even  without reference to the base manifold (i.e. spacetime): as submanifolds of $P$ that intersect each orbit in an open set only once (and thereby transversally), as we did in Section \ref{sec:rep_conv} (see footnotes \ref{ftnt:stab} and  \ref{ftnt:slice}).  This last definition is more useful when we have no direct parametrization of $P/G$, and must resort to  quantities that are intrinsic to $P$, as was the case with the space of models $\F$ and our lack of intrinsic parametrization of $[\F]$.

\paragraph{-- Transition functions}

A transition between the trivializing diffeomorphisms $\bar \sigma$ and $\bar \sigma'$ takes an $(x, g)$ in the domain of $\bar \sigma$ to an element in $U\times G$ in the domain of $\bar \sigma'$ by first taking $(x,g)\mapsto p=g\cdot \sigma(x)$ and then using the inverse $p\mapsto (\pi(p), g_{\sigma'}(p)^{-1})$. Using \eqref{eq:g_equi} this gives:
\be\label{eq:diffeo_gs} (x, g)\mapsto \left(\pi(g\cdot \sigma(x)), g_{\sigma'}(g\cdot \sigma(x))^{-1}\right)=\left(x,(g_{\sigma'}(\sigma(x))g^{-1})^{-1}\right)=\left(x, gg_{\sigma'}(\sigma(x))^{-1}\right).\ee This transformation  is a diffeomorphism of $U\times G$. It acts as the identity on $U$ and,  at each $x$,  as the right action of $g_{\sigma'}(\sigma(x))^{-1}$ on $G$.  We will call the map 
\be\label{eq:transition} g_{\sigma'}\circ \sigma=:\mathfrak{t}_{\sigma \sigma'}\,\,\text{the transition function between $\sigma$ and $\sigma'$}.
\ee
This reflects the transition between the representational conventions given in \eqref{eq:transition_h}. 

More generally, given an atlas of charts $U_\alpha\subset M$, and local sections $\sigma^\alpha$, their patching requires us to consider transition functions
\be
\t_{\alpha\beta}\equiv \t_{\beta\alpha}^{-1} : U_\alpha\cap U_\beta\to G.
\ee

These transformation properties translate between choices of local sections across overlapping charts, and must satisfy the cocycle conditions (compatibility over threefold overlaps $U_{\alpha\beta\gamma}=U_\alpha\cap U_\beta \cap U_\gamma$):
\be
\text{on $U_{\alpha\beta\gamma}$: \quad}\t_{\gamma\beta}\t_{\beta\alpha} = \t_{\gamma\alpha}.
\label{eq:cocycle}
\ee

\paragraph{--  Gauge transformations }
Vertical automorphisms $\tau$ can be represented  with a group-valued function on $P$,  $\Psi:P\rightarrow G$, where $\tau(p)=\Psi(p)\cdot p$ with $\Psi(g\cdot p)=g\Psi(p)g^{-1}$, which is $\Psi$'s equivariance condition.
  
 Then any vertical  automorphism $\tau$ induces a diffeomorphism of $U\times G$, as follows. Let $\tau(p):=\Psi(p)\cdot p$, as above.  Then, for a section ${\sigma}$ and  a general $p=\bar \sigma (x,g)\in  \pi^{-1}(U)$,  
using \eqref{eq:sbar} gives: 
 \be  \tau\circ \bar \sigma: (x,g)\mapsto \tau(g\cdot \sigma(x))=\Psi(g\cdot \sigma(x))\cdot(g\cdot {\sigma}(x))=(\Psi(g\cdot \sigma(x))g)\cdot {\sigma}(x).
 \label{eq:taup}\ee
 As expected, the vertical automorphism $\tau$ just takes $\sigma$ to a different section, $\sigma':=\Psi(s)\cdot \sigma$.
 Moreover, since $\bar \sigma^{-1}(g\cdot \sigma(x))=(x, g)$, we obtain that $\bar \sigma^{-1}\tau\circ \bar \sigma$ is a `coordinate transformation', or diffeomorphism of $U\times G$, in analogy to \eqref{eq:diffeo_gs}:
  \be\label{eq:auto_gt}\bar \sigma^{-1}\tau\circ \bar \sigma: (x, g)\mapsto (x, \Psi(g\cdot \sigma(x))g)=(x, g\Psi({\sigma}(x))),\footnote{ Since the map $\psi_{\sigma}:U\times G\rightarrow G$ given by $(x,g)\mapsto g\Psi({\sigma}(x))$ is smooth (since ${\sigma}$ and $\Psi$ are), and, for fixed $x$, the $\psi_{{\sigma}(x)}:G\rightarrow G$ given by  $g\mapsto g\Psi({\sigma}(x))$ is clearly a diffeomorphism of $G$ (since it is just the action of $G$ on the element $\Psi({\sigma}(x))$). The inverse is of course just $(x,g)\mapsto (x, g\Psi({\sigma}(x))^{-1})$, which enjoys the same properties.}\ee
  where we used the equivariance property of $\Psi$. And so the vertical automorphism only acts on the group part of the product $U\times G$, with $g_s:=\Psi\circ {\sigma}:U\rightarrow G$.

We call $g_s\in \G$  \emph{gauge transformations}; these are the local, passive counterparts of the active $\Psi:P\rightarrow G$, described above (and, to be defined, they require a trivialization).\footnote{The set of all $g_\sigma$'s on a  given patch defines $\G:=\{g(x), \,\,x\in U\}$, which inherits from $G$ the structure of an (infinite-dimensional) Lie-group, by pointwise extension of the group multiplication of $G$ over $U$.}  

 
 \paragraph{-- The gauge  and curvature potentials.}
Given  local sections $s$ on each chart domain $U$, i.e. maps $s:U \to P$ such that $\pi\circ s = \mathrm{id}$, we define a local spacetime representative $\mathbf{A}$ of $\omega$,  as the pullback of the connection, $\mathbf{A}^\sigma:=\sigma^*\omega \in \Lambda^1(U_\alpha, \mathfrak{g})$; (here  $\sigma$ is \textit{not} a spacetime index; we momentarily keep it in the notation as a reminder of the reliance on a choice of section).\footnote{Note that $\mathbf{A}$ only captures the content of $\omega$ in directions that lie along the section $\sigma$. The vertical component of $\omega$---which is dynamically inert, as per the first equation of \eqref{eq:omega_defs}---can be seen (in a suitable interpretation of differential forms, cf. \cite{Bonora1983}) as the BRST ghosts. This interpretation   geometrically encodes gauge transformations through the BRST differential \cite{Thierry-MiegJMP}. Although interesting in its own right, we will not explore this topic here. See \cite{GomesStudies, GomesRiello2016} for more about the relationship between ghosts and the gluing of regions. \label{ftnt:ghosts}} We will expand on the significance of these sections in Section \ref{subsec:active_passiveYM} below. 

In a basis for a given chart on $U\subset M$, we write: $\mathbf{A}=A_\mu^I \,\d x^\mu \tau_I,\,\, \tau_I\in \mathfrak{g}$ is a Lie-algebra basis,  and $\, A_\mu^I \in C^\infty(U)$.\footnote{Clearly, $I$ are Lie-algebra indices and $\mu$ are spacetime indices. We take $\{\d x\otimes \tau\}$ to stand in for the frame  discussed in Section  \ref{subsec:PFB_ex}, as the basis for a vector bundle $T^*U\otimes \mathfrak{g}$.} As in \eqref{eq:auto_gt}, vertical automorphisms are represented as gauge transformations, which, infinitesimally, for a Lie-algebra valued function $\xi^a\in C^\infty(U, \mathfrak{g})$,  act as 
\be\label{eq:gauge_trans}\delta_\xi A_\mu^I= \pp_\mu\xi^I+[ A_\mu, \xi]^I=\D_\mu\xi^I,
\ee
where $\D_\mu(\bullet)=\pp_\mu(\bullet)+[A_\mu, \bullet]$, the gauge-covariant derivative, is defined to act on Lie-algebra valued functions.

 Since the exterior derivative  and the pullback operation commute, we  also have, from \eqref{eq:curv_PFB} for the spacetime representative of the curvature:  
 \be\label{eq:curv} \mathbf{F}^\sigma :=\sigma^*\Omega=\d {\mathbf{A}^\sigma}+{\mathbf{A}^\sigma}\wedge {\mathbf{A}^\sigma}
   \ee 
where now  $\d$ and $\wedge$ are the familiar exterior derivative and products in  $\Lambda(M)$. But, unlike the gauge potential (cf. \eqref{eq:gauge_trans}), the curvature transforms homogeneously under a gauge transformation:
\be\label{eq:curv_trans}\delta_\xi F_{\mu\nu}^I= [ F_{\mu\nu}, \xi]^I.
\ee
 Later, in Section  \ref{sec:Atiyah}, we will see how the bundle of connections enables us to define global spacetime representatives of $\omega$ (and $\Omega$) in coordinate-independent ways.

    \paragraph{-- Gauge transformations v. Transition functions}

Transition functions look similar to  gauge transformations, and indeed act very similarly on the gauge potentials. These similarities reflect the fact that, on the overlap $U_{\alpha\beta}$, both $\mathbf{A}_\alpha$ and $\mathbf{A}_\beta$ descend from the same $\omega$ through different choice of sections.

Take $\alpha, \beta$ to label  not just the open set in a section's domain, but a choice of section on each domain. So suppose there is a unique choice of section per chart domain. On the overlaps $U_{\alpha\beta}=U_\alpha\cap U_\beta$ the transition functions relate the $\mathbf{A}_\alpha$'s: 
\be\label{eq:transition_A}
\text{on $U_{\alpha\beta}$: \quad} \mathbf{A}_\beta = \t_{\alpha\beta}^{-1} \mathbf{A}_\alpha \t_{\alpha\beta} + \t_{\alpha\beta}^{-1} \d \t_{\alpha\beta},
\ee

A collection of gauge transformations $g_\alpha : U_\alpha \to G$ act on the respective $\mathbf{A}_\alpha$ {\it and}  $\t_{\alpha\beta}$'s as follows:
\be
\begin{dcases}
\mathbf{A}_\alpha \stackrel{g}{\mapsto} \mathbf{A}_\alpha^g = g_\alpha^{-1} \mathbf{A}_\alpha g_\alpha + g_\alpha^{-1} \d g_\alpha & \text{on $U_\alpha$}\\
\t_{\beta\alpha} \stackrel{g}{\mapsto}  \t_{\beta\alpha}^{g} = g_\beta^{-1}\t_{\beta\alpha}g_\alpha & \text{on $U_{\alpha\beta}$}
\end{dcases}
\label{eq:gaugetransf}
\ee
from which one derives using \eqref{eq:curv}:
\be
\mathbf{F}_\alpha \stackrel{g}{\mapsto} \mathbf{F}_\alpha^g = g_\alpha^{-1} \mathbf{F}_\alpha g_\alpha  \quad \text{on $U_\alpha$}.
\label{eq:gaugeF}
\ee


We reiterate that the introduction of transition functions is generally necessary because, global sections do not exist unless the bundle is trivial, i.e. unless $P= M \times G$ {\it globally} not just locally. In the trivial case, and only in the trivial case, all transition functions can be trivialized to be the identity, i.e. $\t_{\beta\alpha} = g_\beta g_\alpha^{-1}$ for some choices of $g_\alpha$'s.  Only {\it then}, equation \eqref{eq:transition_A} is trivialized and the collection of $\mathbf{A}_\alpha$'s yields a global gauge potential 1-form $A$.

But this construction leaves a remaining puzzle:  once we fix the typical fiber over $M$, we can characterize sections of an associated bundle in an abstract, geometric or frame-invariant manner, as we characterize standard vector fields over $M$.   So what happens when we apply this characterization to the typical fibers of the gauge potential? 
 

\subsection{The bundle of connections}\label{sec:Atiyah}

In Section \ref{subsec:Atiyah} I introduce the bundle of connections and their sections, also known as connections of the Atiyah-Lie bundle, as a global, spacetime representative of the connection-form. In Section \ref{subsec:omega_unity} I provide what I judge to be a perspicuous physical interpretation of the formalism.\footnote{The bundle of connections appeared almost simultaneously in  \cite{AtiyahLie} and  \cite{Kobayaschi_bundle}. It is often  referred to as the \emph{Atiyah-Lie bundle}. See also \cite[Ch. 17.4]{Kolar_book}. To avoid confusion, it is better to refer to a section of the bundle of connections, which is itself a generalization  of a connection to what are known as Lie algebroids (see \cite{mackenzie_2005}), as an Atiyah-Lie connection.} 

\subsubsection{Sections of the bundle of connections}\label{subsec:Atiyah}
At first sight, we face one difficulty: $\mathbf{A}$ is an object that mixes tensorial indices with internal indices. The natural principal bundle for the tensorial part, as discussed above (see \cite[Sec. 3]{Weatherall2016_YMGR}), would be a sub-bundle of the  frame bundle $L(TM)$. The internal part, corresponding to $\mathfrak{g}$, would require a sub-bundle of $L(P\times_\rho\mathfrak{g})$.\footnote{Here $\rho=\mathrm{Ad}:G\rightarrow GL(\mathfrak{g})$,   where $\mathrm{Ad}_g v=g^{-1}v g$ is the natural, adjoint action of $G$ on $\mathfrak{g}$, appearing in \eqref{eq:omega_defs} and \eqref{eq:gauge_trans}).} 

To work this out, one would need to \emph{splice} bundles of these different characters together (see e.g. \cite[Ch. 7.1]{Bleecker}). Although it is possible to construct the bundle in this way, it would involve the introduction of yet more formalism.  But there is an alternative way, that leads to the same answer (see the Proposition in \cite[Ch. 17.5]{Kolar_book}, for ther equivalence) 



 Parallel transport is determined by horizontal directions in the bundle, as we saw in Section  \ref{sec:PFB}, and  we know that the horizontal bundle $H\subset TP$, is left-invariant (see text preceding equation \eqref{eq:Lie_omega}).  So, if we know what parallel transport is at $p$, we know what it is at $g\cdot p$. By getting rid of this redundancy, we can find a global spacetime representation of the connection $\omega$. 
To do that, we first note that there is a 1-1 relation between (Ehresmann) connection-forms and \textit{left-invariant} sections of $TP$ (see \cite[Ch. 4]{kobayashivol1}). 

Left-invariant vector fields are not unconstrained sections of the vector bundle $TP$, i.e. $C^\infty(TP)$. But they \emph{are} unconstrained sections  of $TP/G$, the so-called \textit{bundle of connections} (see e.g. \cite[Sec. 3.2]{Ciambelli}; \cite[p.9]{LeonZajac}; \cite[p.60]{sardanashvily2009fibre}; \cite[Ch. 17.4]{Kolar_book} and \citep[Ch. 7]{Jacobs_thesis}). In other words, the difference between sections of $TP$ and $TP/G$  is that, while both can be seen as sections over $TM$ (with $\pi_*$ the projection), the latter---$TP/G$---is more constrained, since it can only encode left-equivariant objects defined on the first, $TP$.

The main idea in the construction of this bundle  is to take the projection map $\pi_*:TP\rightarrow TM$, and make it `forget' at which point or ``height'' of the orbit   it was applied.  The formalism  represents parallel transport of internal quantities for the directions in spacetime, rather than for directions in the bundle $P$.  
Thus $TP/G$ is most naturally a vector bundle over $TM$ rather than over $M$ or  $P$. But since $TM$ is itself a bundle over $M$, $TP/G$ can  also be construed as a bundle over $M$.

To define the fiber of $TP/G$, recall that a point in $TP$ is
locally described by $(p, v_p)$ with $v_p \in T_pP$, and the group $G$ acts (freely and transitively)  as
$(p, v_p)\mapsto (g\cdot p, {L_g}_*(v_p))$, which is the relation by which we define the left-invariant vector fields. Thus $TP/G$ is defined by identifying
\be(p, v_p) \sim (g\cdot p, {L_g}_*(v_p)), \quad \text{for all}\quad g\in G.\ee
Since locally (i.e. given some trivialization of the tangent  bundle) for $x=\pi(p)$ and $\xi\in \mathfrak{g}$, we can represent $p=(x, g):=g\cdot \sigma(x)$ and $v_p=(X_{x}, \xi):=\xi+\sigma_*(X_x)$ we have, locally, $(p, v_p)=(x,g, X_x, \xi)$. 
If we take the quotient, we obtain that the elements of the new vector bundle will be locally of the form $(x, X_x, \xi)$.  


Given a point on $M$, and a tangent direction on $M$, and a local trivialization of the bundle, an element of the vector bundle $T^*P/G$  spits out a Lie-algebra element. Thus, as in the standard manner of obtaining $\mathbf{A}^\sigma$ from $\omega$, here we also locally recover, in a trivialization, that the representative of the connection, call it $\Gamma$,  is the $\mathfrak{g}$-valued 1-form on $M$; 
$\Gamma$ is global, but in a local trivialization, it would be represented by $A_\mu^I$, where, the indices refer to a Lie-algebra and a tangent bundle basis.  So $\Gamma$ stands to the abstract tensor $g_{ab}$ as $A_\mu^I$ stands to  $g_{\mu\nu}$.
 The values of $\Gamma$ according to different trivializations are  related by the transformation \eqref{eq:gauge_trans}, just as the values of $g_{\mu\nu}$ are related by coordinate transformations. These are correlates of  the passive transformations we saw in Sections \ref{par:active_passive} and \ref{sec:PFB_formalism}.  Thus we find, as announced in the introduction to this Section, the appropriate analogy, comparing a section of $TP/G$ with a global vector field, $\mathbf{X}$, which we can write locally with coordinates, $X^\mu\pp_\mu$, where $\mathbf{A}^\sigma$ stands in analogy to the components $X^\mu$. Thus, the sections of the bundle $TP/G$ will be frame-invariant, and therefore, invariant under passive gauge transformations. \\

 We can sum up  as follows: a section of $T^*P/G$ should be seen as the global, coordinate-independent generalization of $\mathbf{A}^\sigma$; the advantage of a section of $T^*P/G$ over  the standard gauge potential is that it is globally defined and it is independent of internal coordinates (coordinates for the Lie algebra, and tangent bundle); and the advantage over the connection-form is that it is a section of a vector bundle with $T^*M$ as its base space. The disadvantage is that it is highly abstract. Nonetheless, this formulation allows a strong analogy between the basic kinematical variables of the gauge theory and the metric, in a coordinate-independent manner.

\subsubsection{The unifying power of  the bundle of connections}\label{subsec:omega_unity}
To finish  Section \ref{sec:Atiyah}, let us briefly focus once  again on the geometrical meaning of $\omega$. The unifying power of the principal connection is that it defines compatible parallel transport for any field/particle that interacts with the force associated to $G$, even for the as-of-yet undiscovered forces and groups. 

We can think of $\Gamma$, the section of the vector bundle $T^*P/G$, as one more physical field on spacetime. Since it is a section of a certain vector bundle, upon introducing coordinates (or frames) it admits changes of bases with which it is described, and these can be construed as passive gauge transformations. But just as the connection $\omega$ is invariant with respect to these passive transformations, so will be $\Gamma$. Nonetheless, it remains variant under the active transformations. This is analogous to how geometric objects on differentiable manifold are interpreted as invariant with respect to passive coordinate transformations, but are not invariant under active transformations. 

In physical terms, we can associate \emph{fundamental forces} to structure groups. We associate each structure group to a field $\Gamma$, as above. Then any field that interacts with this force will couple to the appropriate $\Gamma$. As we move from one point of spacetime to another, it is this coupling that will provide a standard of constancy for the field with respect to that interaction. $\Gamma$ represents a structure: that of covariant differentiation, or parallel transport, of the internal quantities that are sensitive to the given force.

\section{Formal criteria for sophistication }\label{sec:soph_cheap}

In Section \ref{subsec:soph_anti} we discussed `metaphysically perspicuous'  interpretations of structure, that were taken to render sophisticated substantivalism  plausible. Here I will discuss more formal criteria for a mathematical structure to support a sophisticated interpretation, in particular for gauge theories. But I should warn the reader that this Section is more speculative, and  also more metaphysical, than most of the thesis. 

In Section  \ref{par:cheap} I will briefly introduce the debates about extending  sophistication from general relativity to other theories. Although something like sophistication has been suggested for gauge theories since its inception in terms of principal bundles (in \cite{YangMills}), only recently has the extension of the position from the context of general relativity  been more thoroughly conceptually analysed.  It was in  \cite{Dewar2017} that questions such as `When can sophistication be applied?', and `to what interpretative advantage?' started being seriously considered.  

 In Section  \ref{sec:skeptic} I reassess how a straightforward version of sophistication fares in the case of general relativity, and compare it to the treatment of  Section  \ref{subsec:soph_anti}. I point out that this straightforward formal criterion for a `metaphysically perspicuous' interpretation of the symmetry-related models---namely, that their core ontology be based on a definition of a symmetry-invariant structure---is either  too strict or too abstract to fully convince the skeptic about structuralism. By thus  reassessing the case of general relativity, I then present a resolution in terms of a correspondence between passive and active symmetry transformations, in Section  \ref{sec:gr_re}. In Section  \ref{sec:soph_antiq} I apply this resolution to the case of Yang-Mills theory, and provide the corresponding metaphysically perspicuous interpretation of the theory.

\subsection{Sophistication on the cheap?}\label{par:cheap}


For general relativity, the symmetries of the theory, given in equation \eqref{eq:equivalence}, are induced by the action of the diffeomorphisms on the spacetime manifold. In vacuum, these symmetries coincide with the isomorphisms of a well-understood mathematical structure:  (semi-)Riemannian geometry (cf. \citep{Oneill}). This coincidence,  perhaps aided by  our everyday acquaintance with space and time, helps us to accept the accompanying anti-haecceitist, structural  interpretation of diffeomorphism-invariant quantities as being ``metaphysically perspicuous'', as elaborated in Section  \ref{subsec:soph_anti}. 

Now, should we also accept the anti-quidditist, structural interpretation of gauge-invariant quantities as being similarly metaphysically perspicuous? 
\cite{Dewar2017}, in a paper that kicked off  considerable recent literature, suggests we should. 

 To clarify the extension of sophistication to a wider class of theories beyond general relativity, \citet[p. 502]{Dewar2017} contrasts it with reduction, or what I in Chapter \ref{ch:syms} called  `eliminativism':  eliminativism  advocates
altering the syntax (i.e. the variables) of the theory, or more precisely, excising the structure that could distinguish the  isomorphic models. On the other hand, according to Dewar, whenever a theory has symmetries, we could extend  sophistication to cases where those symmetries do not correspond to any of the better known examples of mathematical structure, by altering the semantics   of that theory, such that
``we obtain [the new semantics] by taking the old objects,
and declaring, by fiat, that the symmetry transformations are now going to ``count''
as isomorphisms''. He calls this extension the  `external' method \cite[p. 502]{Dewar2017}.

 We are here in the vicinity of two related philosophical debates about symmetry. The first is about whether symmetry-related models of a given theory should invariably be regarded as being physically equivalent even in the absence of  a clear, or metaphysically perspicuous, understanding of the common ontology of the models. \cite{Moller} labels the `yes' answer---symmetry-related models are physically equivalent---as the \emph{interpretational approach} to symmetries;  and he contrasts it with  the \emph{motivational approach}, which requires a characterisation of the
common ontology of symmetry-related models before acknowledging physical equivalence: which is the answer he endorses.  

But of course, opting  between those approaches  turns on just what we can accept as a characterization of the common ontology. This brings us to the second, related debate, about whether sophistication can be attained too easily. For sophistication by brute force, or `external' sophistication, as advocated by \cite{Dewar2017}, may not satisfy the motivationalist.\footnote{Here is \citet[p. 502]{Dewar2017}:\begin{quote}
Is there a way to precisify what is meant?
Here is one way to do so. Rather than trying to define the objects of the new
semantics ‘internally’, as mathematical structures of such-and-such a kind (paradigmatically,
as sets equipped with certain relations or operations), we instead
define them ‘externally’: as mathematical structures of a given kind, but with
certain operations stipulated to be homomorphisms (even if they’re not ‘really’
homomorphisms of the given kind). For example, one way to define vector spaces
is to define them as sets equipped with operations of addition and scalar multiplication,
obeying appropriate axioms. This is the internal method. The alternative
is to define them as spaces of the form $\RR^k$, with the further feature that linear
transformations are declared to be homomorphisms—and in particular, that invertible
linear transformations are isomorphisms. This is the external method.\end{quote}  \label{ftnt:dewar_structure}}  As argued by \cite{ReadMartens}: if there is no `metaphysically perspicuous package'' accompanying the understanding of the symmetry-related models, `sophistication' seems to be gotten on the cheap.  To borrow some of their  words \citep[p. 9]{ReadMartens}: ``the traditional sophisticationist methodology must be applied in order to regard those models---interpreted as being isomorphic---as in fact representing the same physical states of affairs.''\footnote{They in fact advise this even for the internal method.}

And while both parties accept that finding a reduced theory---in which the basic variables of the theory no longer admit a non-trivial action of the symmetries---would provide sufficient explication to satisfy the motivationalist, neither gauge theory nor general relativity can be formulated in this manner; at least not without significant explanatory and pragmatic deficits (cf. Chapters \ref{ch:sd_2} and Section \ref{subsec:soph_anti}; I will return to this in Section \ref{subsec:skeptic_why}). Thus demanding reduction is too strict, and  we must keep searching for an answer to the second debate---about when sophistication sufficiently clarifies underlying structure---in order to decide our answer to the first debate: about (what amounts to) a metaphysically perspicuous package for the symmetry-related models.

\citet[p. 9]{ReadMartens} endorse \citet[p. 502]{Dewar2017}'s `internal'  sophistication:  equivalence can be endorsed in cases where the symmetries  coincide with isomorphisms of some familiar structure.
But, as indicated above (cf. footnote \ref{ftnt:dewar_structure}), it is difficult to prevent any given example of  external sophistication from being construed as internal: given  a space of models and a symmetry transformation acting on this space, one could apparently just announce that  structure is defined implicitly, as `whatever is left invariant' by the action of the symmetries.  In so doing, the symmetries become isomorphisms of the implicitly defined structure. So, we have found a new notion of structure that `internalizes' the symmetry. Thus being `internal' or `external' relies on whether the structure is `familiar' or not: a vague and conservative criterion that may well be taken to not illuminate the question. 

In this vein, both \citet{Jacobs_thesis} and \cite{ReadMartens} argue that an implicit definition, as in Dewar's external method, would not sufficiently explicate the common ontology of the symmetry-related models.
Here it is useful to follow \cite{Klein_erlangen}'s  distinction, taken as a starting point for the Erlangen programme; that is reflected in \cite[p.502]{Dewar2017}'s jargon  of  `external vs internal', or, in the words of \citet[Ch.  4.1]{Jacobs_thesis}:\\
\noindent \textit{The symmetry-first approach (external)}: finding a structure  implicitly  as `what stays constant across the symmetry-related models';\\
\noindent \textit{The structure-first approach (internal)}: finding the symmetry-related models as those that possess the same structure. 

 As Jacobs puts it: \begin{quote}
Structure-first sophistication consists of: the
stipulation of a set of relations over the theory's sub-domains, such that the bijections
which induce dynamical symmetries of the theory's models leave these relations
invariant. If we agree that an interpretation of a theory provides a metaphysically
perspicuous picture if it tells us plainly and clearly which entities the theory posits
(ontology) and what the fundamental relations between these entities are (ideology),
then structure-first sophistication is perspicuous in this sense.
\cite[p. 62]{Jacobs_thesis} \end{quote}
On \cite{Jacobs_thesis}'s favoured structure-first (or internal) approach, the aim is to give an `intrinsic' characterisation of
a structure in terms of relations defined over its domain, such that this structure
is invariant under the dynamical symmetries of the theory. This broad idea works out beautifully for the example that he focuses on (scaling of masses in Newtonian gravity). In that case, one successfully characterizes the mathematical structure of the theory first, and then deduces the isomorphisms that preserve that structure. At least in that example, it seems that sophistication is apparently \emph{not} condemned as cheap.

 In sum, the common idea of \cite{Dewar2017, ReadMartens, Jacobs_thesis} is that we can escape the accusation of cheapness in two steps: (1) insisting symmetries  coincide with isomorphisms of some structure that is given `internally', and (2) first defining the structure, and then finding the symmetries that preserve it.

\subsection{Obstacles to Sophistication: the case of general relativity reassessed}\label{sec:skeptic}\label{subsec:soph_stab}

In this Section, I play Devil's advocate to my remarks of Section \ref{subsec:soph_anti}, which defended the abstract characterization of structure furnished by sophistication. The skeptic that I am impersonating is not satisfied that a `perspicuous representation of the common ontology of two isomorphic models of general relativity  is just ``fields on the manifold, where the latter is interpreted anti-haecceitisically''. He may be motivated either by worries regarding structuralism as a foundation of mathematics, or simply by a hankering for a more concrete or intelligible characterization of structure: a hankering often shared by physicists working in quantum gravity. In Section \ref{subsec:skeptic_why} I lay out the skeptic's main arguments. Then in Section \ref{subsec:drag} I defend the skeptic from a counter-argument that I dub `the drag-along'. This counter-argument says that structure is easily conceived of as invariant under the isomorphisms of the theory under an appropriate mathematical construal of those isomorphisms. In    Section \ref{subsec:my} I announce the broad lines of my own position regarding the drag-along response.

\subsubsection{The skeptic's arguments}\label{subsec:skeptic_why}
Even in the supposedly clear case  of general relativity, the skeptic argues, the precise identification of the structure that is invariant remains too abstract, and so opaque. For the main idea in defining structure through symmetry-invariance is that the models of the theory are structured sets, $\mathcal{D}=(D, R)$, consisting of a base (unstructured) set $D$ and relations among the elements of this set, which we can here give just one label, $R$. 
 For example, in general relativity, a model ``consists of a base manifold $M$ over which
we have defined some (geometrical) structure in the form of the tensor fields'' \cite[p.60]{Jacobs_thesis}.  But what exactly are the relations that stay invariant under the symmetries, which are given by the (pullback of) active diffeomorphisms (cf. Section  \ref{par:active_passive})? Tensor fields certainly do not remain invariant.  A statement such as ``the metric tensor is $g_{ab}$'' is not symmetry-invariant; it can be true in one  model and false in a  symmetry-related one. The distances between the points of $M$ seen as an unstructured base set are also not invariant, since the distance between $x$ and $y$ according to $g_{ab}$ is not the same distance as according to $f^*g_{ab}$. The same reasoning would of course apply to angles, Riemann curvature scalars, etc.

 Straightforwardly understood as a property of structured sets, the structure that remains invariant under isomorphisms is characterized as the abstract set of diffeomorphism-invariant quantities: just as the symmetry-first, or external, approach---not the structure-first, or internal, approach---would suggest.  
  And setting aside the drag-along view of those isomorphisms---which I will consider in Section \ref{subsec:drag}---most diffeomorphism-invariant quantities are integrals of scalar densities over the manifold; or, alternatively, we could characterize the full invariant content of a model by using a vast disjunction, or by relying on existential quantification over points and regions of the manifold: e.g. ``there is a two-dimensional surface, of area $X$, bounded by two geodesics, of length, $L$, and it intersects with another such surface at a point'', or ``for all points $x$, there exists a unique point $y$ whose geodesic distance from $x$ is greater than all other points".  This last sentence is true of, say, the two-sphere, where it describes anti-podal points, and it is false of the plane. In particular all of these sentences are invariant under isometries.\footnote{In the example of the connected graph of footnote \ref{ftnt:graph},  the analogue would be to characterize properties of the entire graph, e.g.: there are 4 points and 10 edges, or, there is a single vertex that is connected by $n$ edges, etc. There, as here, we can functionally characterize points, lines, etc, in a symmetry-invariant manner. }
 And indeed, this vast disjunction  is essentially how we characterized the metaphysically perspicuous interpretation of the general relativisitic, diffeomorphism-invariant structure, in Section \ref{subsec:soph_anti}.\footnote{Quantities about coincidences of material point-particles, elapsed proper times along a particle worldline,  all operations involved in signaling with particles or light pulses, etc. }
 
 I have two comments about this approach: one metaphysical and one mathematical. The metaphysical one is based on \cite{Dasgupta_bare} and the mathematical one is based on \cite{Mundy1992} (and related work: \cite{Weatherall_hole, Shulman_hole, BradleyWeatherall_hole}).
 
 \cite{Dasgupta_bare} describes this approach elegantly, using the notion of \emph{grounding}. For our purposes, grounding is just an explanation that need not be causal (World War II was grounded on a great number of people mobilizing and fighting in Europe; but it was caused by   political factors). According to Dasgupta, what we have been thus far calling sophisticated substantivalism (which he calls `thin substantivalism') is just a claim that, according to the theory  facts are grounded \emph{qualitatively}, not \emph{individualistically}.  And he gives a formal definition of `qualitative' in terms of predicate logic: these are just facts that can be expressed with identity, but without  constants, such as $(\exists x), Fx$, but not $Fa$. Facts that are expressed using  constants (i.e. spacetime points) are grounded  individualistically, and belief in them corresponds to what he labels `thick substantivalism'.\footnote{See \cite{Adams1979} for an early defense of `primitive thisness' and a distinction towards primitive trans-world identity: a distinction we are here ignoring. } 
 
  Dasgupta takes  the metaphysician in this context of general relativity to be saddled with two tasks (p. 147, ibid): ``(1) to clearly articulate what the underlying qualitative facts are like,
and (2) show that they are sufficient to explain (in the metaphysical sense)
individualistic facts about the manifold.'' With task (1) in mind, he proposes (p. 149, ibid) a language ``PL of predicate logic
with identity but no constants, in which every predicate expresses a qualitative
property or relation.  Every fact that can be expressed in PL is a purely
qualitative fact.'' So this language is suited to expressing certain facts, such as  that expressed with the 2-place sentence ``someone loves someone'', or $(\exists x, y) (x$ loves $y)$. So we can take every sentence in this language as qualitatively grounded.  The idea is fundamentally \emph{holistic} (p. 151, ibid): each world is described all at once by a complete sentence in this language (with a possibly infinite number of connectives).
 
 This leads me to the mathematical comment. The type of qualitative description sought by Dasgupta is also the subject of \cite{Mundy1992}, who seeks an axiomatization of Riemannian geometry. He characterizes the obstacle to qualitative description as follows (p. 516, ibid):
 \begin{quote}
 Coordinate representation
creates an illusion of ``rigid designation" of individual points, but by formal semantics
all terms in a theoretical language are built from non-logical primitives, hence designate different objects under different interpretations. Modal metaphysics fosters the
same illusion, via ``transworld identity" between elements of different models.
 \end{quote}
 And he contrasts the `rigid designation' of coordinate representations (corresponding to Dasgupta's ``individualistic'') and the intrinsic geometric designation of objects of a given theory. So he continues (p. 517, Ibid):
 \begin{quote}
[...] standard differential geometry still requires postulation of one or
more manifold coordinate systems for the space, and hence is not ``coordinate-free",
despite its invariance under manifold recoordinatization.
This contrasts with intrinsic geometric axiomatizations, using truly coordinate-free
primitives such as the affine betweenness relation $B(p,q,r)$ and the segment congruence relation $C(p,q,r,s)$[...]. Euclidean geometry is expressible intrinsically through axioms $T_E$ in a language $L_E$ over these two primitives.
Equivalence with the coordinate formulations is shown by a representation theorem:
that each model of $T_E$ has Cartesian coordinates, unique up to orthogonal transformations, representing $B$ and $C$ by the standard coordinate formulas.
\end{quote}

  Mundy's axiomatization of Riemannian geometry follows in the tradition of many others (e.g. \cite{Robb1936}). Most notably, in their seminal work, \cite{EPS} set out to build differential topology, conformal and Lorentzian structure from basic postulates concerning the paths of light and massive particles (see also \cite{LinnemannRead} for a pedagogical introduction). Accordingly, Mundy defines $T_R$ (Semi-Riemannian geometry) as above, in which $C$ refers to metric behavior of clocks and rods, and $B$ refers to geodesic motion. 
  The upshot is that Mundy’s formalization
admits no sentence which would be made true by $\langle M, g\rangle$, but not by $\langle M, f^*g\rangle$ (or vice versa).\footnote{  
  The idea that the appropriate, `intrinsic' mathematical formalism cannot express the difference between isomorphic models is a form of radical structuralism (that is also present in e.g. \cite{Shulman_hole}'s use of homotopy type theory, or in certain uses of category theory). We will reject it in Section \ref{subsec:my}.}

But the skeptic  is not satisfied.  ``Very well, you can compactly describe the type of structure that we are dealing with, by using an appropriate axiomatization. But a non-enumerable list of characteristics hardly clarifies the \emph{specific} symmetry-invariant structure at hand'', he contends.  

The skeptic not only wants an internal definition of the \emph{type} of  structure (cf. footnote \ref{ftnt:dewar_structure}) but also a concise characterization of the particular structure of each model.  The paradigmatically successful example is that of the vector spaces of footnote \ref{ftnt:dewar_structure}, where the dimension of the space (and specifying the field of scalars) fully characterizes the entire invariant structure. But it is not the finiteness of the characterization that makes it successful:  a scalar field in a fixed background, whose state is located in an infinite-dimensional space of possibilities, such as a temperature or density, could be considered an equally transparent characterization of the ontology. And while, in the case of Riemannian manifolds,  axiomatic definitions of the \emph{sort} of structure may also exist,  it is the characterization of the \emph{structure of each model that is opaque}, says the skeptic.

 To buttress these judgements, the skeptic 
points out the  difficulties  in finding an \emph{explicit}  basis of  symmetry-invariant quantities for the diffeomorphisms (as mentioned in Section \ref{sec:method}), since these difficulties are major  hurdles to all approaches to quantum gravity.\footnote{See e.g. \cite{Thiemann_2003, Rovelli_book, Donnelly_Giddings} for general arguments surrounding this difficult issue of `gravitational observables', and \cite{Harlow_JT} for an explicit basis in the simplified context of a two-dimensional gravitational theory called Jackiw-Teitelboim gravity.} 

Going into more detail
: far from being a debate of merely  metaphysical interest, the skeptic contends, the  opacity of structure  has implications for the practitioner of quantum gravity, as most thoroughly described in \cite{belot_earman_1999, belot_earman_2001}. They write:
\begin{quote}Far from dismissing the hole argument as a simple-minded
mistake which is irrelevant to understanding general relativity, many
physicists see it as providing crucial insight into the physical content
of general relativity. \cite[p. 169]{belot_earman_1999} \end{quote}
And here is Isham,  on the difficulty of explicitly characterizing structure:
\begin{quote}
The diffeomorphism group moves points around. Invariance under such an
active group of transformations robs the individual points of $M$ of any
fundamental ontological significance . . . [the argument] is closely related to the question of what
constitutes an observable in general relativity---a surprisingly contentious
issue that has generated much debate over the years. \cite[p. 170]{Isham_POT}\end{quote} 

\subsubsection{The drag-along view and its limitations}\label{subsec:drag}

\paragraph{The drag-along}

As we saw in the previous Section, an   axiomatization is one way to realize this idea, but it cannot articulate the structure of each model. Meanwhile, an infinite disjunction and conjunction of qualitative facts is a way to characterize the structure of each model, but it is opaque.


But there is another popular way to resist the idea that the distance, relations, etc. between points varies across models, thus rendering all tensors trivially invariant under isomorphisms. Namely, we can try to `wear our anti-haecceitism on our sleeves' so to speak, and thereby pre-empt the skeptic's demand for clarification. 

This is accomplished by construing diffeomorphisms in terms of the `\emph{dragging-along}' of properties and relations.  
Here is how   \cite[p. 12]{Stachel_Iftime_short} understand the doctrine:
   \begin{quote}
[...] the points of the manifold are not individuated independently of
the metric field. This means that space-time points have no inherent chronogeometrical
or inertio-gravitational properties or relations that do not depend
on the presence of the metric tensor field. This implies that when we drag-along
the metric, we actually drag-along the physically individuating properties and
relations of the points. Thus, the pull-back metric does not differ physically from
the original one. It follows that the entire equivalence class of diffeomorphically-
related solutions to Einstein’s empty space-time field equations corresponds
to one inertio-gravitational field.
Put in other words, [...points] 
 lack haecceity as individualized points of that space-time (``events'') unless and until a particular metric field is specified.
\end{quote}
Thus the proposal is that if we are given an isomorphism that sends the fields at $x$ to the fields at $y$, that is,  $f$ sends the properties at point $x$ in one model to a point $y$ in another model, where $f(x) = y$, then we should ``rebrand'' $y$ in the codomain model as ``really being'' $x$;  or ``replace $y$ with $x$''. It is this proposal  that \cite{GomesButterfield_hole} call {\em the drag-along response} to the hole argument.

 Or more simply put: we have two  metrics  on $M$ that are isomorphic: $g_{ab}$ and $\tilde g_{ab}=f^*g_{ab}$. If we are to compare what they say about points or regions of $M$, in principle we can use any  diffeomorphism  $h\in \Diff(M)$---comparing  a value over the point in the domain with a value over the point in the image.\footnote{The reader may here be worried that the two tensors belong to different vector spaces and so cannot be compared. The idea requires that we compare scalars formed from each tensor: so when we use $h$, we compare, for any two vectors $v, w\in T_xM$, the value of $g_{ab}v^aw^b$ with the value of $\tilde g_{ab}Tf(v)^aTf(v)^b$, where $Tf$ is also often called the push-forward of $f$ (cf. footnote \ref{ftnt:push}). } But clearly, the argument goes, we  should use the diffeomorphism that gives rise to the isometry, namely $f$, so that the  {tensor} $g_{ab}(x)$ should be compared with the   tensor $\tilde g_{ab}(f(x))$. And indeed, by the definition of $\tilde g_{ab}$, the two tensors seem  identical using this standard of comparison. Thus I agree that isometry gives  the only appropriate standard of comparison of isomorphic models  that brings the physical content of spacetime  regions and points   into coincidence. This argument provides a way to understand any tensor on $M$ as trivially diffeomorphism-invariant, and it realizes anti-haecceitism in a very concrete manner.\footnote{This is how we could construe the drag-along as an explicitly anti-haecceitist doctrine: since there is no non-qualitative identification of points across models, the best we can do is conceive of $\tilde g_{ab}$ and $g_{ab}$ as inhabiting  diffeomorphic (but not identical) manifolds: $M$ and $\tilde M$. The natural standard of comparison is a diffeomorphism  $h:M\rightarrow \tilde M$, that sets $h$ as the isometry, i.e. such that $h^*g_{ab}=\tilde g_{ab}$, or, in the previous nomenclature, $h=f$. In the jargon of possible worlds \citep{Butterfield_hole}, this is a perfectly natural counterpart relation between the spacetime points of $M$ and $\tilde M$. But it is not the only possible counterpart relation. Moreover, our chart-nominalist construction of Section \ref{subsec:chart_nom} gives us a single $M$, construed as a smooth structure. Thus,  for what follows, not  much is gained, and much more would have to be explained,   by using different base sets, $M$ and $\tilde M$, and for this section,  we can restrict to the case in which $\tilde M=M$, without loss of generality. \label{ftnt:tildeM}}   
 
The advocate for drag-along can go further:  saying that, compared  using any other diffeomorphism, the tensors  will differ:  for   $h\neq f^{-1}$, we (generically) have $g_{ab}(x)\neq (h^*\tilde g_{ab})(x)$, and that these are physical differences.
 I agree that these are,  \emph{pointwise}, physical differences. For example,  compared using the identity on $M$, at the point $x$, $g_{ab}$ may be flat  whereas $\tilde g_{ab}$ is not. (We will make this argument more precise below, in Equation \eqref{eq:drag}).\footnote{This is in line with \cite{Weatherall_hole}:\begin{quote}
When we say that $\langle M, g_{ab}\rangle$ and $\langle M, \tilde g_{ab}\rangle$ are isometric spacetimes,
and thus that they have all of the same invariant, observable
structure, we are comparing them relative to  [the isometry...] if one only considers
[the isometry], no disagreement arises regarding the value of the metric at
any given point, since for any point $x\in M$, $g_{ab}(x) = \tilde g_{ab}(f (x))$
by construction. \cite[p. 336]{Weatherall_hole}
\end{quote} }   

Let us more precisely verify the statements of the two previous paragraphs.  Let us call   $h\in \Diff(M)$ the base set map used for pointwise comparison of $g_{ab}$ and $\tilde g_{ab}$. Above $h$ was set either equal to the (inverse of the) map $f$---that gives rise to the isometry, $f^*$---or to the identity, $\mathsf{Id}$; here I am generalizing it to any diffeomorphism.  It is clear that if we demand that 
\be g_{ab}(x)=(h^*\tilde g_{ab})(x)=((f\circ h)^*g_{ab})(x),\quad \forall x\in M,
\ee
then $f\circ h$ is an isometry of $g_{ab}$. Assuming $g_{ab}$ is generic, it has only the identity as a (trivial) isometry, and so $f=h^{-1}$. 

This a strict condition, asking for a base set comparison that matches the metric point by point. We could ask for a less strict condition, about whether  the physical content of region $U$ according to $g_{ab}$ is identical to that of region $\tilde U$ (diffeomorphic to $U$, e.t. setting $\tilde U:=h(U)$) according to a different metric $\tilde g_{ab}$. 
\be\label{eq:local_equiv} g_{ab}{}_{|U}\sim_{_U} \tilde g_{ab}{}_{|\tilde U}\quad \text{iff}\,\, \exists\, r_U\in \Diff(U;\tilde U)\,\, \text{such that}\,\,\forall \tilde x\in \tilde U,\,\exists x\in U,\,\, g_{ab}(x)=(r_U^*(\tilde g_{ab}(\tilde x))),\ee
i.e. such that a diffeomorphism $r_U$ from $U$ to $\tilde U$ will bring tensors $g_{ab}$ and $\tilde g_{ab}$ over each point in the two regions $U$ and $\tilde U$ into coincidence. 

For \eqref{eq:local_equiv} to be satisfied, $\tilde x=r_U(x)$ is a necessary condition.  
When   $\tilde g_{ab}=f^*g_{ab}$, the last equality on \eqref{eq:local_equiv} becomes:  
\be g_{ab}(x)=((f\circ r_U)^* g_{ab})(x), \quad \forall x\in U.\ee
 Thus if the metric is generic, the equivalence condition is satisfied iff $f$, with the domain restricted to $\tilde U$,  results in  a diffeomorphism to $U$, i.e. $f_{|\tilde U}\in \Diff(\tilde U; U)$, since then we would find some $r_U$ to compensate it (and of course,  if this is to hold for every open set $U$, then we get again  $f=h^{-1}$). This is exactly what we expected.  
 Moreover, setting $\tilde U=f^{-1}(U)$,  we trivially obtain $ g_{ab}{}_{|U}\sim_{_U} \tilde g_{ab}{}_{|\tilde U}$. Moreover, if  $\tilde U=U$, i.e.  $h=\mathsf{Id}$, as long as  $f_{|U}\not\in \Diff(U)$, we also automatically obtain that $g_{ab}{}_{|U}\not\sim_{_U}\tilde g_{ab}{}_{|U}$. So we have proven, for $U\subset M$ and $\tilde g_{ab}=f^*g_{ab}$: 
 \be \label{eq:drag}\text{For}\,\,f_{|U}\not\in \Diff(U):\,\, g_{ab}{}_{|U}\not\sim_{_U}\tilde g_{ab}{}_{|U};  \quad\text{while, for any}\,\, f\in \Diff(M)\quad g_{ab}{}_{|U}\sim_{_U} \tilde g_{ab}{}_{|f(U)}.\ee 

 In words: all that the relations of \eqref{eq:drag} say is what we had stated loosely before:  that this particular choice of $g_{ab}$ and $\tilde g_{ab}$ will, if $f_{|U}\not\in \Diff(U)$, associate different physical, or diffeomorphism-invariant, properties to $U$---whatever they are---but will always associate the same  physical properties to   $U$ according to $g_{ab}$ and   $f(U)$ according to $\tilde g_{ab}$.  In the same spirit, if $g_{ab}$ and $\tilde g_{ab}$ were such that $f_{|U}\in \Diff(U)$,  then we would still  have $g_{ab}$ and $\tilde g_{ab}$ associating the same physical properties to $U$, even if $f\neq \mathsf{Id}$.

But why should individual regions and points have any importance? If we were to take not a subregion, but the entire manifold, i.e. $U=M$, then necessarily $\tilde U=h(U)=M$, and $f_{|U}\in \Diff(U;\tilde U)$ even for the identity base set mapping $h=\mathsf{Id}$. Thus,  for $U=M$, unlike \eqref{eq:drag}, we obtain that $g_{ab}{}_{|U}\sim_{_U} \tilde g_{ab}{}_{|U}$ for any $h,f \in \Diff(M)$.

This distinction allows me to argue  not only that the drag-along understanding of anti-haecceitism is not mathematically mandatory, but also that it is {\em limited}.

\paragraph{The drag-along: not mandatory.}

As we just saw, in many ways,  \cite{Weatherall_hole} is right: given isometric models $\langle M, g_{ab}\rangle, \langle M, \tilde g_{ab}\rangle$, and a diffeomorphism  of $M$  used to compare the metrics pointwise (as a map between the base sets of the two models, (cf. footnote \ref{ftnt:tildeM}), and if we  want local physical quantities on the domain and co-domain of the map to match, say on every open set $U\subset M$,  then  the drag-along \emph{is mandatory} for a physical interpretation of the theory, as \cite{Weatherall_hole} claims. Moreover, this resolution seems to evade the shortcomings of the purely axiomatic approach, since it retains the standard mathematical object of the theory, $g_{ab}$, as the fundamental variable.

However, the reason the drag-along, though useful and enlightening, is not \emph{mandatory}, is that we need not assign physical meaning \emph{pointwise}, i.e. to each point of spacetime, in whichever way points are understood. 

Some results in the quantum gravity  literature strengthen this conclusion: for instance, in the Hamiltonian formalism, to be studied in Section \ref{sec:syms_consts} and Chapter \ref{ch:Coulomb}, \citet{Torre_local} has shown that,  for the spatially closed universe,  there are no local symmetry-invariant quantities (i.e. quantities which can be written as integrals of the canonical variables and a finite number of their derivatives over the Cauchy surface). Diffeomorphism-invariant functions are usually non-local: this is a notorious problem for quantum gravity (see \cite{Thiemann_2003, Rovelli_book, Donnelly_Giddings}). \citet[p. 170]{Isham_POT} continues the quote from  the end of Section \ref{subsec:skeptic_why}:
\begin{quote} In the present
context, the natural objects [that costitute `observables' in general relativity] are $\Diff(M)$-invariant spacetime integrals [...]
Thus the `observables' of quantum gravity are intrinsically non-local.\end{quote}
Or, as described more recently in \cite[p. 3]{Harlow_JT}:
\begin{quote}
...in gravity any local observable by itself will not be diffeomorphism-invariant, so we must dress it [...creating] for it a gravitational field that obeys
the constraint equations of gravity. In practice such observables are usually constructed
by a ``relational” approach: rather than saying we study an observable at some fixed
coordinate location, we instead define its location relative to some other features of the
state.\footnote{As we will see in Chapter \ref{ch:sd_2} (cf. Section \ref{subsec:sims}) and Chapter \ref{ch:Coulomb}, all theories with elliptic initial value constraints imply similar sorts of non-locality. This non-locality will be quantified to a certain extent in Chapter \ref{ch:subsystems}.}
\end{quote} 
The non-local character of invariant quantities should be no surprise  for the anti-haecceitist who, like \cite{Dasgupta_bare} (cf. Section \ref{subsec:skeptic_why}), relies on qualitative descriptions of points. For such descriptions usually rely on existentially quantifying over all the points on a manifold, and this quantification `morally' replaces an integral sign. So, the drag-along approach is not mandatory for the same reason that the axiomatic approach fails to compactly describe the structural content of each model.

For us, the important point here is that, \emph{as a whole}, i.e. over the entirety of $M$, $g_{ab}$ and $\tilde g_{ab}$ are indiscernible: they agree about  all of the same diffeomorphism-invariant facts, however we choose to relate their base sets. 
Thus, for distinguishing physical possibilities of the entire model, the drag-along, i.e. setting $h=f$, although  immensely useful,  is not mandatory. \\

\paragraph{The drag-along: limitations.}

For practical purposes, the main limitation of the drag-along is that general relativity, and our other spacetime theories, in some circumstances use means of identifying points \emph{other than} by drag-along. And they need to do so, on pain of trivialising important constructions: even elementary ones like the Lie derivative;  or more complex ones, like the   space of asymptotically  flat models. I will now explain this danger of trivialization for these two concrete examples.

 More precisely, about the Lie derivative:  for $f_t$ the flow of a vector field $X^a$, the Lie derivative is usually defined using both the isometry induced by $f_t$ and the identity map for the base set $h=\mathsf{Id}$: we drag the tensors $f_t^*g_{ab}$ over the fixed base set $M$ (writing out the Lie derivative definition of \eqref{eq:Lie_g} of Section \ref{sec:diff_sym}):
\be\label{eq:Lie}
{\mathcal{L}}_{{X}}{g}_{ab}(x):=\lim_{t\rightarrow0}\frac{1}{t}({g}_{ab}(x)-f_t^*{g}_{ab}(x)).
\ee
But if we instead use the drag-along to compare the metrics in the definition of the Lie derivative, we obtain: 
\be\label{eq:Lie_drag}
{\mathcal{L}}_{{X}}{g_{ab}}(x)=\lim_{t\rightarrow0}\frac{1}{t}({g_{ab}}(x)-f_t^*{g_{ab}}(f_t(x))\equiv 0 \; !
\ee Though a Lie derivative could be defined algebraically---as being a derivation satisfying certain axioms, such as commutation with the exterior derivative and with the contraction (or interior product)---this is not how it is mostly used. Indeed, by  restricting ourselves to such an algebraic definition we would lose the straightforward  relation between the Lie derivative and its flow, given in \eqref{eq:Lie_g}, since that relation requires us to understand the pull-back along a diffeomorphism. And it goes without saying that understood in this way, in connection with the diffeomorphisms, the Lie derivative is very useful in the study of geometry: besides being used to study how geometric objects `flow' along given directions---e.g. how the metric can flow to itself and thus possess Killing fields---it allows us to obtain local conservation laws from Noether's second theorem, as we will see in Chapter \ref{ch:sd_2}. 

The problem  for drag-along arising from  asymptotically flat spacetimes is that to treat such spacetimes we need to employ a background structure that is fixed. (This was briefly described in Section \ref{sec:subsystems} (see in particular footnote \ref{ftnt:Belot_ADM}).) Thus two models  $(M, g_{ab})$ and $(M, \tilde g_{ab})$ both have to be flat on  \emph{the same region}, for which we need a notion of identity of the base set---which we called $h=\mathsf{Id}$ above---that remains invariant across possibilities.  And so, if the two models are to represent the same physical possibility, they must be related by an isometry  that \emph{is the identity on that  (asymptotic) region}. The remaining isomorphisms that preserve flatness but are not the identity in the fixed region  are taken to relate different physical possibilities. In sum,  we must `pare down' the isomorphisms of the theory to that subset that preserves flatness in a region, as a non-dynamical structure; and even once this is done, some of these isomorphisms---those that differ from the identity map on those regions---will be taken to relate different physical possibilities.   

Thus this  example represents a limitation of the drag-along way of comparing isometric models.   As \cite[Sec. 4.4]{Belot50} emphasizes, here there is some tension between the standard view that isometric models must be `gauge equivalent' and the way we model asymptotically flat systems: 
 \begin{quote}The truth of the matter is this, I believe: while relativists do often speak as if
solutions of general relativity are gauge equivalent if and only if isometric, they
drop this way of speaking when asymptotic boundary conditions (like asymptotic
flatness at spatial infinity) are in view. [...] The crucial point is that general relativity has sectors in
which gauge equivalence and isometry coincide [...] and sectors in which gauge equivalence is a more
discerning relation than isometry (in which there are isometric solutions that are
capable of representing distinct possibilities).
 \end{quote} 

In the same way, these limitations  militate against the radical structuralist view of \cite{Mundy1992, Shulman_hole, Stachel_Iftime_short}.


\subsubsection{My own view}\label{subsec:my}

I should make it clear that I agree with \cite{Mundy1992, Weatherall_hole, Shulman_hole, BradleyWeatherall_hole} in many respects: mathematically,  the agreed-upon notion of isomorphisms for pseudo-Riemannian manifolds is isometry, and for that reason, isometric manifolds are, for what concerns any substantive mathematical   theorem (e.g. that you could find in a textbook, for instance, \cite{Oneill}), the same. As  \citet[p. 6]{BradleyWeatherall_hole} put it: ``mathematicians generally intend to attribute to mathematical
objects only structure that is preserved by the relevant notion of isomorphism''.

Moreover, I believe the drag-along is an important tool for assessing whether two universes   represent the same physical possibility. But I do not think that   this tool is mathematically mandated: that requirement would leave us bereft of  other tools, that are necessary for other applications.




  Sophistication, on the other hand, merely requires  that isomorphic models model the same physical universe. Thus this position allows us to keep the set-theoretic resources to at least articulate examples like that of the Lie derivative and asymptotic flatness. 
  
  Having said that, I also believe that, ideally, in line with the \emph{motivational} approach to symmetries \citep{Moller},  the existence of examples which do not sit well with the given notion of structure  motivate us to either (i) find a new internal, axiomatic structure which has the resources to treat the examples; or (ii)  reconceive the examples as representing some previously neglected background structure---e.g. by conceiving of asymptotic flatness as  introducing implicitly some set of observers---which, when described with due care, \emph{would} sit well with the original notion of structure. 
  
  Option (i) is definitely more difficult;\footnote{Perhaps a case can be made that this was achieved by Penrose in defining spacetimes that admit of conformal compactification: this implements a new notion of isomorphism, which matches the notion of symmetry, and according to which asymptotic conditions are invariant. But I have not discussed this type of boundary condition here, and  will not dwell on this very rich example.} but I think that for the examples above (the Lie derivative and asymptotic flatness) option  (ii) is up to the task. Asymptotic flatness, as mentioned, should be accommodated into the standard notion of structure by being reconceived in terms of subsystems and observers (see also Section \ref{sec:subsystems} and Chapter \ref{ch:subsystems}); the Lie derivative, as we will see in Section \ref{subsec:rep_hole}, can be reconceived as the effect of an infinitesimal change in the representational convention.

 \subsection{Responding to the skeptic using chart-nominalism in general relativity}\label{sec:gr_re}
  Here is my attempt to assuage the skeptic's worries about an overly long and heterogeneous description of the diffeomorphism-invariant structure, without endorsing the drag-along response I have argued against. I will use local trivialisations, or charts, in two related ways.  The first argument uses charts to deflate the ontic significance of symmetry-related models by glossing it as notational; and the second uses existential quantifiers and charts  to provide a relatively concise definition of symmetry-invariant structure.

In the chart-nominalist spirit (cf. Section \ref{subsec:chart_nom}), an important---and  often neglected---conceptual point about interpreting the symmetries of general relativity is that they have a well-understood correspondence to  passive transformations. One might well think we have already provided such a correspondence: in Section \ref{par:active_passive}, we saw how an active diffeomorphism would correspond to a `reshuffling' of the charts within a given maximal atlas. However, this type of reshuffling is not what  is usually understood as a passive diffeomorphism (although I believe one could construe such a change  as being entirely notational): the reshuffling may not be expressible using a single (non-maximal) atlas.   

The idea to be pursued in this Section is slightly different:  if we understand the intrinsic diffeomorphisms of each chart as mere notational change, and the manifold is built by patching charts together, we can also put a passive gloss on a subset of the active diffeomorphisms (the isomorphisms of the category of smooth manifolds) that corresponds to the intrinsic diffeomorphisms of the charts.    And this understanding should help clarify the symmetry-invariant structure, since  passive transformations are agreed by all parties to only bear on pragmatic concerns about representational conventions (cf. Section \ref{sec:rep_conv}); they are not  construed to have any special significance for ontology. 

Thus in Section \ref{subsec:passive_active_GR},  I will  exhibit this correspondence between the isomorphisms of the theory and a passive, or notational, change that the theory allows. Then, in Section \ref{subsec:chart_inv} I will take this correspondence to deflate, even in the eyes of the skeptic, the ontic significance of the multiplicity of symmetry-related models, since they can construed as notational variants of each other. Then, in Section \ref{subsec:rep_inv} using the ideas of \emph{representational conventions} (cf.  Section \ref{sec:rep_conv}),  I will reconceive these notational variants as  intra-theoretic choices of relations and quantities with which we  describe the physical, i.e. symmetry-invariant, features of the given system.  Finally, in Section \ref{subsec:rep_hole}, I employ representational conventions to reconceive Lie derivatives and show that this use does not ressurect the threat of indeterminism.   

\subsubsection{The passive-active correspondence of diffeomorphisms}\label{subsec:passive_active_GR}
A passive diffeomorphism is  a smooth function from  $\RR^n$ to itself, with smooth inverse, which I will write as $\bar f\in \Diff(\RR^n)$, that is interpreted as `translating' between two charts $\phi_1$ and $\phi_2$ of an atlas; so that  $\phi_1=\bar f\circ \phi_2$. Here we construe $\bar f$ passively, as a pure notational change:  when the domains of two arbitrary charts $\phi_1, \phi_2$  overlap, we have a transition function between the charts that is a diffeomorphism between subsets of $\RR^n$. From the above,  $\bar f:=\phi_1\circ \phi_2^{-1}:\phi_2(U_1\cap U_2)\subset\RR^n\rightarrow \phi_1(U_1\cap U_2)\subset \RR^n$.  This transformation simply does not act on quantities on $M$: we interpret it as only changing their description.\footnote{ When the domains overlap, one could see an active diffeomorphism as a right action of the diffeomorphisms on the charts, $\phi_2=\phi_1\circ f$, whereas the passive diffeomorphism above would correspond to a left action of the diffeomorphisms on the charts  $\phi_2=\bar f\circ \phi_1$.} 

But clearly, reconceiving $\bar f$ actively,  we can uniquely `associate' it  to \emph{a  local} diffeomorphism  on $M$, the domain manifold. More explicitly, as in Section \ref{subsec:chart_nom}: given the charts,  we can reconstruct an active diffeomorphism relating a tensor ${\mathbf{T}}$ to a tensor $\tilde{\mathbf{T}}$ on a patch, from the transition function $\bar f$: namely,  by going down to $\RR^n$ by the chart, applying $\bar f$, and then going up from $\RR^n$ by the same chart:
\be\label{eq:pull_atlas} (\phi^{-1}_1\circ\phi_2) (\mathbf{T})=f(\mathbf{T})=: \tilde{\mathbf{T}} \quad\text{where}\quad f:=\phi_2^{-1}\circ \bar f^{-1}\circ\phi_2\in \Diff(M).\ee 

Most authors would be wary of identifying active and passive transformations in such an explicit fashion: for one thing, they will point out, active diffeomorphisms act globally, whereas passive transformations act on each chart. 

But first of all,  it is undeniable that quantities $Q$ such that $\phi_2(Q)$  are invariant under $\bar f$, i.e. such that 
\be\label{eq:pass_inv}\bar f(\phi_2(Q))=\phi_2(Q),\ee will be  invariant under $f$, i.e.:
\be f(Q)=\phi_2^{-1}\circ \bar f^{-1}\circ\phi_2(Q)=\phi_2^{-1}\phi_2(Q)=Q,\ee
since \eqref{eq:pass_inv} also implies that $\bar f^{-1}(\phi_2(Q))=\phi_2(Q)$. Second, any active infinitesimal diffeomorphism, represented by the vector field $X^a$, will, by the same construction that gave rise to \eqref{eq:pull_atlas}, correspond to some infinitesimal diffeomorphism of each chart in the given atlas. 

Therefore, there is a 1-1 correspondence between  quantities that are invariant under the diffeomorphisms that are connected to the identity and those quantities that are invariant under   coordinate transformations---or notational changes---that are connected to the identity.  
Intuitively, this relation is mathematically rather simple, since we know that local patches of $M$ are locally diffeomorphic  to $\RR^n$, and we can therefore naturally move diffeomorphisms from one space to the other.\footnote{ For illustration, take the change from spherical to cylindrical coordinates: $$ r=\sqrt{\rho^2+z^2}; \,\, \theta=\arctan\left(\tfrac{z}{\rho}\right); \,\, \varphi=\varphi, $$
where $r$ and $\rho$ are, respectively, cylindrical and spherical radius, $\varphi$ is the azimuth angle, $z$ is the cylindrical  height, and $\theta$ is the elevation angle. Passively, we take these coordinates to refer to the same points of $\RR^3$. But we can also construe this diffeomorphism actively, for $\RR^3$ comes endowed with some background structure of its own. So the map above takes a given ordered triple, seen as an element of the product $\RR^3$,  to a different ordered triple: just plug in values of $(\rho,  \varphi, z)$ and find where they go as $(r(\rho,z), \varphi, \theta(z,\rho))$. Under this interpretation, a given $(a,b,c)\in \RR^3$ is being \emph{actively} mapped to $(\sqrt{a^2+c^2}, b, \arctan\left(\tfrac{c}{a}\right))\in \RR^3$.  }   Nonetheless, if we interpret passive transformations differently, as better understood,  then the relation provides a powerful interpretational tool for the invariant quantities.

So we have two ways of associating active and passive transformations. The first way says that an active diffeomorphisms takes quantities as described by one chart in an atlas to the same description but under a different, compatible atlas. (Since the smooth structure is understood as a mere postulation of the maximal atlas, which is preserved by this re-shuffling,  this just means that an active diffeomorphism preserves the smooth structure.)  The second is more useful, since it does not require us to explicitly change the atlas: it says that,   using \eqref{eq:pull_atlas},  passive diffeomorphisms---diffeomorphisms of $\RR^n$ on the image of the charts---recover some of the active diffeomorphisms.  In particular, we get a 1-1 correspondence between  the active and passive infinitesimal diffeomorphisms (as defined in Definition \ref{def:inf_sym} and  used in \eqref{eq:Lie_g}), and therefore, by integrating in time, we get a 1-1 correspondence between invariants under coordinate transformations and invariants under diffeomorphisms (both connected to the identity).\footnote{Thus if we want to gloss symmetry just as notational, we have strong reason to restrict considerations to  symmetry groups that are connected to the identity, as discussed in Chapter \ref{ch:syms}. But I should note that the mismatch between the full group of diffeomorphisms and the subgroup that is connected to the identity is stark: according to one suitable topology---called `weak Whitney topology'---any open set of $\Diff(M)$  containing the identity also contains elements that are not connected to the identity. Another reason to restrict to the symmetry groups that are connected to the identity, as we will discuss in \S \ref{sec:charges}, is that in the Hamiltonian framework, symmetries are generated by the symplectic flow of constraint functions, or momentum maps, and are thus always connected to the identity. Indeed, I take these latter facts to justify the physicist's focus  on invariance with respect to coordinate transformations, as opposed to the more abstract invariance under active diffeomorphisms. \label{ftnt:con_id}}

\subsubsection{Using the active-passive correspondence to shed light on structure} \label{subsec:chart_inv}
    If we are allowed to use existential quantifiers to describe invariant structure, using charts we can find a complete description of \emph{local} geometric facts that is much less disjunctive than those suggested in the second paragraph of Section \ref{subsec:skeptic_why} (that listed geometric invariants such as: ``for all points $x$, there is a unique point $y$ such that...''). For example, the definition below of metric structure quantifies existentially over charts, and is  relatively concise:
     \begin{defi}[Chart-based definition of metric structure]\label{def:chart}
     The manifold has a chart in which the local metric is given by the matrix of real functions $g_{\mu\nu}$.\end{defi}
And we can unpack this characterization in more invariant (but more vague) terms along the lines of: there is a coordinate-grid---a set of directions that cover an area and are infinitesimally linearly independent, in the sense of linear algebra---on which lengths and angles are encoded by  $g_{\mu\nu}$. Though I will not attempt to make such a characterization  more precise here, it is clear that   we could clothe it in a suitably invariant language;  (ultimately, we could employ the entire framework of either \cite{Mundy1992} or \cite{EPS}, discussed in Section  \ref{subsec:skeptic_why}, to characterize this chart).

Definition \ref{def:chart}  completely describes local physical facts in a diffeomorphism-invariant, geometrically interpretable,  manner. In other words,  the definition  has a straightforward geometrical interpretation and its truth-value can be assessed in \emph{each and every} symmetry-related model of the theory, and models that are symmetry-related (or isomorphic) will agree on that truth-value. Indeed, since  physicists usually describe a local geometry by just stipulating some $g_{\mu\nu}$, this fits rather nicely with our theme of closing the gap between their practice and debates within the philosophy of physics.\footnote{ 
   Using abstract tensors and existential quantifiers, we could also say:
``in the set of symmetry related models, there is one model whose metric tensor is $g_{ab}$.'' This property is almost trivially symmetry-invariant. But unlike the statement involving charts, whose quantifier employs only the differentiable structure of the base set $M$ and is therefore applicable---true or false---in each model, the  property described in this footnote requires existential quantification over models, and thus cannot be located within  each model; it is only a property of the entire set of symmetry related models;  it refers to that entire set.  
  
   Compared with the infinite disjunctive list of geometric invariants of Section \ref{subsec:skeptic_why}, another possible  shortcoming of the description using charts is that it intuitively seems less empirically accessible than each item in the list. But cashing out this misgiving requires some footwork (as we will see in Chapter \ref{ch:sd_2}, Section \ref{sec:antiheal}). }

Indeed, it could even be used to efface the difference between symmetry and isomorphism in asymptotically flat spacetimes, as described by \cite{Belot50} (see Section \ref{sec:subsystems}). The property that `a chart exists such that the (coordinate-dependent characterization) of asymptotically flat obtains' (cf. footnote \ref{ftnt:Belot_ADM} in Section \ref{sec:subsystems}), is  invariant.\footnote {In the coordinate-dependent definition of asymptotically flat spacetimes (cf. footnote \ref{ftnt:Belot_ADM}),  we have fixed the notation in a certain manner so that a certain type of property becomes manifest. Namely, among the  coordinates there must be one that functions like a radial distance. This distance is so that, as we go radially outward, spacetime becomes flatter and flatter. Clearly, this restriction pares  down the number of choices of coordinates  that make the asymptotically flat property of $g_{\mu\nu}$ manifest. Nonetheless, each  isomometric class of metrics either has or does not have such a chart. }  

Thus I propose that by using existential generalizations over charges, we can give a concise and informative description of the geometry. 
Of course, any model that admits a  chart with the given description (e.g. $g_{\mu\nu}$) will necessarily admit other descriptions. But these are the notational choices described in the first part of my response to the skeptic (cf. Section \ref{subsec:passive_active_GR}):  they are related through passive transformations which  can be understood in terms of changes of conventions. 
 
\subsubsection{Using representational conventions to shed light on structure} \label{subsec:rep_inv}
While for classical general relativity, Definition \ref{def:chart} reflects how in practice theorists specify spacetimes  locally,  neither quantum gravity  nor gauge theorists try to build a basis for the space of isomorphism equivalence classes (cf. last paragraph of Section \ref{sec:syms_tech}), $[\F]$ in this manner. 
 One  problem is that, \emph{for a given} $g_{\mu\nu}$,  Definition \ref{def:chart} picks out just the  meagre set of equivalence classes to which  $g_{\mu\nu}$ belongs.\footnote{Cf. footnote \ref{ftnt:stab2} for the formal definition of `meagre'.} It is very far from giving us a complete handle on representation of all possible invariant properties of all the models. 

Therefore this resolution may still  not assuage the skeptic's worry in Section \ref{subsec:skeptic_why}, about the difficulties  of finding an \emph{explicit}  basis of  symmetry-invariant quantities. 
But an answer to  this  worry is not conceptually far off. As described  in Section \ref{sec:rep_conv}, the perspicuous characterization of the symmetry-invariant basis we seek is given by a representational convention: a smooth injection from the space of equivalence classes of models under the isomorphism to the space of models: $\sigma:[\F]\rightarrow \F$, $[g_{ab}]\mapsto \sigma([g_{ab}])\in \F$. But again, since we have rejected reduction, or eliminativism, we cannot describe $[g_{ab}]\in[\F]$ intrinsically, and thus we replace $\sigma$ by an equivalent projection operator $h_\sigma$. These projections, seen as maps on $\F$  can be explicitly written down in terms of charts; and they are symmetry-invariant, complete characterizations of each and every model.

For instance, in the case of Lagrangian general relativity, it is common to have $\sigma$ correspond to De Donder gauge (see Section \ref{subsec:gf}), which fixes coordinates so that the densitized metric is divergence-free. Using that Section's notation, $\mathcal F$, for the functional whose value fixes a gauge (or section), we have in this example: 
\be \mathcal{F}(g)=\pp_\mu(g^{\mu\nu}\sqrt{g})=0.\ee
 Geometrically, De Donder gauge anchors our coordinate systems on spacetime waves (solutions of the relativistic wave equation, written as $\square x^\mu=0$). For the  Hamiltonian formalism of general relativity (cf. Section \ref{subsec:sims} and Chapter \ref{ch:Coulomb}),  another common choice is to physically anchor time coordinates using `CMC gauge' (see footnote \ref{ftnt:CMC}): this choice adjusts clocks and simultaneity surfaces so that simultaneous observers measure the same local expansion of the universe. In this case, $\mathcal{F}(g_{ij}, \pi^{ij})=g^{ij}\pi_{ij}=$const.\footnote{  To take some other recent examples, the relational observables of \cite{Lima_proj, Frob_proj} can be seen as projections onto the gauge-fixing surface of a non-linear generalization of de Donder gauge, if one looks just at the invariant metric perturbation. The idea is to construct  relational observables using `field-dependent coordinates'  (see e.g. Eq. (2.22)). Similarly, light-cone coordinates and `geometric clocks' (see e.g. \cite{Giesel_geom_clocks}) anchor the representation, or the coordinates, to the state. The `dressed' diffeomorphism-invariant observables of \cite{Donnelly_Giddings} are relativized to a state since the quantities are anchored on non-null spacetime curves. For a general relationship between different types of gauge symmetry and non-locality, see \cite{Elements_gauge}.   For the constraints on finding local observables in the Hamiltonian context,  see also \cite{Torre_local}. \label{ftnt:nonlocal}}

Using  representational conventions (and the notation $\cal F$ for gauge-fixing) we could,  instead of Definition \ref{def:chart}, say:
  \begin{defi}[Gauge-fixing as structure]\label{def:donder}
     The local structure is given by the values of the map $h:\F\rightarrow \F$, where $h(g_{\mu\nu})$ satisfies some extra condition, $\mathcal{F}(h)=0$, as described in Section \ref{sec:rep_conv}.\end{defi}
By \eqref{eq:gauge-fixing}  and \eqref{eq:h_proj_F}, this is an invariant characterization of the local structure, in that, for any isomorphic model $\tilde g_{ab}$, $h(\tilde g)=h(g)$. Indeed, many, if not all  of the examples of `relational gravitational observables' used by physicists employ just such  projections and representational conventions. These are the type of `relational dressings' mentioned in the \citet[p.3]{Harlow_JT} quote of Section \ref{subsec:drag}: ``rather than saying we study an observable at some fixed
coordinate location, we instead define its location relative to some other features of the
state.'' Definition \ref{def:donder}  also provides good examples of `functional roles for points', as we will see explicitly in the next Section.

Nonetheless, there is a multitude of choices for these projections, which correspond to different ways of characterizing  the geometry. Each such characterization  corresponds to one choice of properties and relations that will be used to specify a  coordinate system. So there are many ways of expressing the invariant structure with properties and relations. In this way, maps between representational conventions, or different such choices, as in \eqref{eq:transition_h},  recover the full set of passive diffeomorphisms, just as they recover the set of passive, state-independent, gauge transformations in \eqref{eq:transition} (and \eqref{eq:transition_A}). 

\subsubsection{Using representational conventions to shed light on the Lie derivative and indeterminism} \label{subsec:rep_hole}

In Section \ref{subsec:drag}, we saw that, given some model,  the drag-along response avoids the indeterminism threatened by the multitude of isomorphic models. But the response did not sit well with standard mathematical tools, such as the Lie derivative, which require some relative shift between isomorphic metrics with respect to a fixed ``underlying'' set of  spacetime points.  But now   the Lie derivative could be  construed as a formalization of the effect of infinitesimal changes between representational conventions. Let us see how this goes, and what such changes imply for indeterminism.

 Taking this opportunity to use yet a different convention, more directly related to the use of the Lie derivative: we will take the convention  associated to a \emph{Kretschmann-Komar  coordinate system} \cite{Kretschmann_inv, Komar_inv}. The idea is to take the four scalar functions $R^{(\mu)}(g_{ab}(x)),\,\, \mu=1, \cdots 4$, formed by certain real scalar functions of the Riemman tensor.\footnote{\cite{Komar_inv} finds these real scalars through an eigenvalue problem: 
 $$(R_{{a}{b}{c}{d}}-(g_{{a}{c}}g_{{b}{d}}+g_{{a}{d}}g_{{b}{c}}))V^{{c}{d}}=0,$$
 where $V^{{c}{d}}$ is an anti-symmetric tensor. } The use of an index in parenthesis emphasizes that this is just a list, and not a vector field.  For generic spacetimes (i.e. excluding Pirani's type II and III spaces of pure
radiation, in addition to excluding symmetric type I
spacetimes), these scalars are functionally independent. 

In such spacetimes, we \emph{define} coordinates  (omitting the dependence on the metric on the left-hand-side and the index's parenthesis):
\be\label{eq:Komar_coords} \bar x^\mu:=R^\mu(g_{ab}(x)).\ee
In a less coordinate-centric language, the idea here is  to fix the identity of spacetime points through a certain choice of their qualitative properties, quantifying existentially over the points of $M$. That is, we define point $\bar x(g_{ab})$ through an inverse relation, in which $g_{ab}$ is the variable argument (as in the above: `$\bar x(g)$ is the point in which a given list of curvature scalars $R^{(\mu)}(g)$, $\mu=1, \cdots, 4$ takes a specific list of values, $(a_1, \cdots, a_4)$'). This inverse relation specifying points in terms of their qualitative properties is in a sense holistic (since it   quantifies over points in $M$), and yet it furnishes a specific qualitative  identity  \emph{for spacetime points} across isomorphic models; (and indeed, it can also furnish a qualitative identity  relation across non-isomorphic models, cf.  \citep{GomesButterfield_counter}).\footnote{ \citet[p. 468]{Curiel2018} construes qualitative identity of points similarly: \begin{quote}
Once one has the identification of spacetime
points with equivalence classes of values of scalar fields, one can as easily say
that the points are the objects with primitive ontological significance, and the
physical systems are defined by the values of fields at those points, those values
being attributes of their associated points only per accidens.
\end{quote} But he does not construe diffeomorphisms as intra-theoretic changes of convention about the choices of scalar fields, as we will.  And though such a definition may very well qualitatively identify point $y$ in model $\langle M, g_{ab}\rangle$   with point $y'$ in another model, $\langle M, \tilde g_{ab}\rangle$, an argument that is not aligned with the main argument of \cite[Sec. 3]{Curiel2018} (that there is no way to identify points belonging to different spacetimes), this is not our focus in this argument. }

Now, given some metric tensor $g^{\kappa\gamma}$ in coordinates $x^\kappa$ (cf. footnote \ref{ftnt:abstract}), we can compute the metric in the new, $\bar x^\mu$ coordinate system as:
\be \bar g^{\mu\nu}=\frac{\pp R^\mu}{\pp x^\kappa}\frac{\pp R^\nu}{\pp x^\gamma}g^{\kappa\gamma}=0.
\ee
But this is just a family of 10  scalar functions indexed by $\mu$ and $\nu$. In the words of \cite[p. 1183]{Komar_inv}:\begin{quote} 
[it is] component by component a well defined scalar constructed from the metric tensor and its derivatives. If
we consider two metric tensor fields and ask whether
they represent the same physical situation, differing
perhaps by being viewed in different coordinate systems,
we now have a ready criterion for determining the
answer. Clearly, at corresponding points in any identifiation of the two spaces, the values of all scalars must
agree if the spaces are to be equivalent. We are therefore
compelled to identify points in the two spaces which
have the same ``intrinsic" coordinates [defined by \eqref{eq:Komar_coords}]. Furthermore at these corresponding points it is
necessary that the ten scalars [i.e. $ \bar g^{\mu\nu}$]
have the same values in the two spaces. 
Thus we find that the functional form of the 10 scalars
 [i.e. $ \bar g^{\mu\nu}$] as functions of the four scalars [i.e. $R^\mu$]: (a) is uniquely
determined by the metric space independently of any
choice of coordinate system, and furthermore (b)
uniquely characterizes the space.\end{quote}

Much as  we did here, Komar sees an alternative rendition of \eqref{eq:Komar_coords} as a gauge-fixing, or coordinate condition. Instead of defining new coordinates through \eqref{eq:Komar_coords}, we write out $g_{ab}$ in coordinates and, unlike \eqref{eq:Komar_coords}, have those same coordinates appear on the left and right hand side  of the equation, namely: 
\be\label{eq:point_gf} x^\mu=R^{(\mu)}(g_{\kappa\gamma}(x)), \quad\text{or equivalently}\quad \pp_\nu R^{(\mu)}(g_{\kappa\gamma}(x))=\delta_\nu^\mu,\ee  an equation that is to be solved for a choice of coordinats $x^\mu$, seen  as a function of the metric.
But also like us,  \citet[p.1186]{Komar_inv} highlights the difference between (what we would call) a projection operator and the idea that we are somehow `breaking [gauge] covariance':
\begin{quote}
The usual
argument, that employing coordinate conditions may
destroy the general covariance [...] does not apply in this case. For considering \eqref{eq:point_gf}
as a coordinate condition [as opposed to the definition of 10 scalar functions of the metric] is only a heuristic device to
make it easier to visualize how to manipulate the quantities with which we are dealing. In point of fact,
we know how to interpret these quantities as true
observables. (An apt analog in electromagnetic theory
may clarify this point of view. The transverse components
of the vector potential may be considered as
the gauge-invariant true observables; or they may be
considered as the components of the vector potential
in a particular gauge, namely the radiation gauge. )
\end{quote}
This is the difference between:  restricting the models to belong to a given  gauge-fixing section---e.g. `the radiation gauge'---and projection to that section---`the gauge-invariant observables'. We will flesh out precisely this electromagnetic analogy  in Sections \ref{sec:antiheal} and Chapter \ref{ch:Coulomb}.

Finally, we could easily have chosen a different set of scalars $R^{(\mu)}$ provided we preserved their functional independence. For example, we could have multiplied $R^{(1)}$ by two and taken the log of $R^{(2)}$. Or we could take different linear combinations of $R^{(\mu)}$; and  so on. Indeed, if we take $R^{(\mu)}$ as coordinates of $\RR^4$, we could find an alternate set by applying any diffeomorphism (seen as a recombination of the original functions of the metric). Suppose we call two alternative sets of basis  in \eqref{eq:Komar_coords} $\bar x^{\mu}$ and $\bar y^{\mu}$.  Now  within the  $\bar x^{\mu}$ choice, we can compare the point labeled by e.g. $(1,2,3,4)_{\bar x}$   to the point $(1,2,3,4)_{\bar y}$.\footnote{
That is, as translated to the $\bar x^{\mu}$ choice, we can explicitly write the comparison of the point labeled by e.g. $(1,2,3,4)_{\bar x}$   to the point $(1,2,3,4)_{\bar y}$; leaving the $\bar x^{\mu}$ subscript implicit:
\be (1,2,3,4)-(\bar x^1((1,2,3,4)_{\bar y}), \bar x^2((1,2,3,4)_{\bar y}),\bar x^3((1,2,3,4)_{\bar y}),\bar x^4((1,2,3,4)_{\bar y})).\ee} This is the type of comparison, that, when pushed to the infinitesimal limit, generates the Lie derivative, which is here seen  as an infinitesimal version of a diffeomorphism that is in its turn seen as a change of representational conventions.

More straightforwardly: we can just track how quantities change from the point identified by e.g. $(1,2,3,4)_{\bar x}$ to an infinitesimally nearby point $(1+\delta \bar x^1,2,3,4)_{\bar x}$: this generates the Lie derivative of those quantities along $\frac{\partial}{\partial \bar x^1}$. 
 Importantly, note that, in this formulation, the Lie derivative does not require some non-qualitative identity of spacetime points: it only requires a continuous parametrization of spacetime points by their qualitative properties.\footnote{
Incidentally, if $g_{ab}$ admits non-trivial automorphisms---or, in the standard nomenclature, admits Killing vector fields---then the $\bar x^{\mu}$ will not be a coordinate chart in the usual sense. For instance, if the metric is homogenous along some direction, we could have e.g.  $\bar x^1$  admitting a single value. In this case, we would automatically find that Lie derivatives along $\bar x^1$ vanish, for any function. Of course, if we want to have functions which are not homogeneous along that direction, and whose Lie derivative would not vanish, we would have to include more fields in the model, apart from the metric, and these could also be used to define physical coordinates $\bar x^{\mu}$. So the idea hangs together. Nonetheless, there are  definitely diffeomorphism invariant quantities definable in a (vacuum) homogeneous model that cannot be described using physical coordinates: for instance, topological invariants. }

Here the reader may be getting uncomfortable: with all this talk about re-conceiving the diffeomorphisms, have we let physical indeterminism back in? No, we have not. Within a single representational convention, invariance under diffeomorphisms is guaranteed, as  remarked by Komar in the quote above and shown in Section \ref{sec:rep_conv} (see in particular Equations \eqref{eq:proj_h}, \eqref{eq:cov_g}, and \eqref{eq:h_proj_F}, and remarks following \eqref{eq:h_proj_F}). Nonetheless,  we may still focus our attention on different  complete sets of invariant quantities to describe the evolution of states.  These different descriptions won't be numerically identical, nor should we expect them to be: they are meant to measure  different things. But since they are complete---in the sense of uniquely characterizing the solution---they will be inter-translatable. The example of Section \ref{subsec:isos_descs}, about translations in a theory of non-relativistic point-particles in Euclidean space gives a simple and visualizable case. We could choose differently among the   particles for one  to serve as the center of coordinates. For each choice, the description is relational and invariant, and though the descriptions according to different particles are not numerically identical, they are inter-translatable. Ultimately, the idea is that we can take a structural view of the theory, \emph{a l\`a} \cite{Weatherall_hole}, that is indifferent to any `bare' identity of spacetime points; and yet recover the isomorphisms of the theory merely as different choices of convention. 
\\


 This ends my defense of a thesis that  many others have also defended, and that is common practice within theoretical physics: that the  structure in general relativity that is invariant under both passive and active (infinitesimal) diffeomorphisms is, in words, chronogeometrical. 
 
 We can sum up this defense as follows: in the pursuit of a conceptually transparent understanding of the invariant structure that is common to all the infinitesimal-symmetry-related models, once we have an active-passive correspondence for the symmetries, we can set aside invariance under active transformations---the actual isomorphisms of the theory---and focus solely on invariance under passive coordinate transformations. For these different  choices can be understood in a purely pragmatic, or notational sense, as intra-theoretic choices of representational conventions.  \\

  To finish the substantial project of Section \ref{sec:PFB}, I will now make similar remarks about the symmetries of Yang-Mills theory, extending the conclusions above to the Yang-Mills case.

\subsection{Sophistication as anti-quidditism in Yang-Mills theory}\label{sec:soph_antiq}
   As briefly discussed in  \ref{subsec:active_passiveYM},   gauge theory (in the principal fiber bundle formulation) also has a correspondence between active and passive symmetry transformations. Each quantity that is  invariant under passive transformations will correspond to one that is invariant under the active ones. Therefore, we can apply reasons  very similar to those of Section \ref{sec:gr_re}, to gauge theories.
   
 Contrary to $\omega$ (and $\Gamma$) and $\Omega$, the spacetime representatives ${\mathbf{A}^\sigma}$ and $\mathbf{F}^\sigma$ are defined over  charts of the spacetime $M$, rather than over the bundle $P$. 
In other words, although $\omega$ and $\Gamma$ are globally defined on $P$, the ${\mathbf{A}^\sigma}$ are only defined on the respective patches $U_\alpha$ of ${M}$ through the choice of a local section $s$. At a fixed $\omega$ or $\Gamma$,  different choices of section give different ${\mathbf{A}^\sigma}$; the difference between the gauge potentials are solely due to different choices of trivialization, i.e. they are passive. 

In more detail,  for any function $f:\pi^{-1}(U)\rightarrow \RR$, if we represent it in a trivialization, i.e. such that $ f_\sigma:U\times G \rightarrow \RR$ (with $f(p)=f_\sigma(x, g)$, where $p=g\cdot \sigma(x)$) we will find  that, for any automorphism (fiber-preserving diffeomorphism of $P$) $\tau$, the function $ f\circ\tau=:\tilde f$ has,  under a different trivialization  $s'$,  the same coordinate representation as the original function had under $\sigma$, i.e. $f_\sigma=\tilde f_{\sigma'}$. This is the correspondence between active and passive transformations on each trivialization patch, that we saw  in Section  \ref{subsec:passive_active_GR} for diffeomorphisms of spacetime.  Then, to emphasize, maps between representational conventions, as in \eqref{eq:transition_h},  recover the  set of passive, state-independent, gauge transformations in \eqref{eq:transition} (and \eqref{eq:transition_A}).
 
  In this respect it is elucidating to construe passive transformations as changes of bases (cf. the related footnote \ref{ftnt:frame_conv} at the end of Section \ref{subsec:isos_descs}). 

\subsubsection{Passive gauge transformations as changes of bases.}\label{subsec:internal_basis}

In the  example of Section \ref{subsec:PFB_ex} that we used as a motivation,  the principal bundle $P$ was originally identified with $L(TM)$ and the vector bundle $E$ was identified with $TM$, and we took $G=GL(\RR^4)$. In other words, $P$ was the space of frames of the tangent bundle, with no preferred frame and in which the group acted transitively on frames over each point $x\in M$. This elucidating example can be naturally extended to a more general setting, and to a general understanding of principal fiber bundles, as discussed in \citep{Weatherall2016_YMGR}. 

 Given some real (for simplicity) vector space $F$ and structure group $G$ and $\rho:G\rightarrow GL(F)$, and $P$ a $G$-principal bundle over $M$,  we can find the associated vector bundle  over $M$, which is denoted $E:=P\times_\rho F$. Conversely, the frame bundle for a given vector bundle $E$, $L(E)$ (formed by the bases of $E_x$ for each $x\in M$) is a principal bundle $P'$ with structure group  $GL(F)$. But we can form another principal bundle $P$, as a sub-bundle of $P'$ as follows. Since $P'\simeq L(E)=L(P\times_\rho F)$ we can see $P$ as a sub-bundle of $L(P\times_\rho F)$ corresponding  to a subset of frames of $L(P\times_\rho F)$ related by $\rho(G)$.\footnote{ Of course, this raises a puzzle: if the principal bundle is construed as just a bundle of linear frames, how can we justify  the restriction of $\rho(G)$ to a subset of the most general group of transformations between frames? As discussed by \cite[Sec. 4]{Weatherall2016_YMGR},  the restriction corresponds to the preservation of some added structure to $F$. In other words, when $F$ is not just a bare vector space, but e.g. a normed vector space, we would like changes of basis to preserve this structure, e.g. the orthonormality of the basis vectors, and this restricts the bundle of linear frames to the appropriate sub-bundle. To see this, define $P\times_\rho F$ as the equivalence class for the doublet $(p, v)\in P\times F$  with  $(p, v)\sim (g\cdot p, \rho(g^{-1})v)$. Suppose  that $F$ is a Riemannian vector space, with metric $\langle\cdot,\cdot\rangle$. We can  induce a metric in $P_F=P\times_GF$  defining, for any  $p$ and $v,v'\in F$:
$\langle[p,v],[p,v']\rangle:=\langle v,v'\rangle$.
To be well-defined, we must have:
$$\langle[p,v],[p,v']\rangle=\langle[g\cdot p,\rho(g^{-1})v],[g\cdot p,\rho(g^{-1})v]\rangle=\langle \rho(g^{-1})v,\rho(g^{-1})v'\rangle:$$
which is true only if the action of the group on $F$ is orthogonal with respect to the metric. This corresponds to $G=O(n)$; similarly, $SO(n)$ adds an orientation to $F$. Similarly, $G=U(n)$ corresponds to a complex vector space structure and a Hermitean inner product; $G=SU(n)$ adds an orientation (see \cite[p. 2403]{Weatherall2016_YMGR}). The moral is that the added structure on $F$ induces an added structure on  the associated vector bundle only if the  transformation group preserves that added structure. }

 As emphasized by \citet[p. 2401]{Weatherall2016_YMGR}, the conceptual advantage of this construal of $P$ is that, assuming the action of $\rho:G\rightarrow GL(F)$ is faithful,  we can intepret \emph{passive} gauge tranformations as just point-dependent changes of bases of the value space $F$ (i.e. the allowed changes of frames for $E$). In other words, a section $s:U\rightarrow P$ may be understood as a frame field for a certain vector bundle ($P\times_\rho F$), and changes of section may be understood as the allowed change of basis at each point. \citet[p. 2404]{Weatherall2016_YMGR} writes: 
 \begin{quote}
 We are thus led
to a picture on which we represent matter by sections of certain vector bundles (with additional structure), and the principal bundles of Yang–Mills theory represent various
possible bases for those vector bundles. 
 \end{quote}
 But I think Weatherall skates over an important distinction between passive and active gauge transformations when he continues:\footnote{Having said that, there are hints further along the paper that  \citet[Sec. 5]{Weatherall2016_YMGR} distinguishes the two types of invariance, and highlights the lack of invariance of the connection-form under the active transformations.}
 \begin{quote}[...]
 these considerations lead to a deflationary attitude towards
notions related to ``gauge'': a choice of gauge is just a choice of frame field relative
to which some geometrically invariant objects [...] may be represented, analogously to how geometrical objects may be
represented in local coordinates.\end{quote}
As I remarked in Section  \ref{sec:gr_re}, invariance under coordinate change can only play this deflationary role once an active-passive correspondence for the symmetries of the theory is established, as it was above.\\


Here too, this property---that the active isomorphisms are locally equivalent to a passive transformation---gives a gloss of `notational redundancy' to the  symmetry in question, 
 a type of redundancy  most authors agree to be well understood. 
Indeed, invariance under different coordinate representations is usually \emph{equated} with `physical status'. Thus, to  take two examples at random:
\citet[p. 82]{Nozick} explains: ``Once we possess the covariant representation under which the
equations stay the same for all coordinate systems, the quantities in the (covariant) equations are the real and objective quantities.''  And, in the introduction to his magisterial book,  \citet[p. vii]{Dirac_QM} writes: 
``The important things in the world appear as the invariants
[...] of these [coordinate] transformations.''

  To a certain extent, this verdict vindicates  some previous philosophical comments about  the  connection (for example, see Maudlin in \cite[p.6]{Synopsis_gauge}).
\\

We have therefore found a perspicuously sophisticated view of the symmetry-related models. For gauge theories the KPMs  are given by $\langle P, \omega\rangle$, which is both isomorphic and symmetry-related to $\langle P, \omega'\rangle$ iff there is a fiber-preserving diffeomorphism $\tau\in$Diff$(P)$, such that $\omega=\tau^*\omega'$. In this formalism, just like in general relativity,  the infinitesimal symmetries of the action match the  isomorphisms of the underlying structured base set, $P$, and these isomorphisms can be given a passive construal, which allows a perspicuous interpretation of the structure that is common to all the symmetry-related models.

Thus we can articulate the ontic commitments of gauge theory that ensue from this attitude as follows. Each possible world, or physical possibility for the force-field---each particular choice of structure---is given by  one way in which internal quantities are parallel transported over spacetime. 
The structural  interpretation of Yang-Mills theory is conceptually similar to the  structural interpretation of general relativity (discussed in Section \ref{par:active_passive} and again in Sections  \ref{sec:skeptic} and \ref{sec:gr_re}). For general relativity, each possible world, or physical possibility---each particular choice of structure---is given by one way in which spacetime points are chronogeometrically related or distributed. For Yang-Mills, the structure  represented by $\omega$ refers to the parallel transport of internal quantities. In even fewer words: general relativity is about the external geometry, whereas Yang-Mills theory is about the internal geometry.\\



\section{Summing up }\label{sec:conclusions_soph}

The question driving this Chapter has been whether we should endorse   structuralism for spacetime diffeomorphisms, but not for gauge symmetries. As briefly mentioned in Section  \ref{sec:syms_tech}, there is an established jargon in modern metaphysics for the two structuralist construals in play here. The structuralist construal of points  is often called {\em anti-haecceitism}. And the structuralist construal of properties that I recommend is often called {\em anti-quidditism}. 
In these terms, the question driving this Chapter was whether anti-haecceitism was right for spacetime symmetries, but some variant of reduction should be preferred to anti-quidditism about gauge symmetries.

Sections \ref{sec:attitudes} and \ref{sec:PFB} rejected this claim at a formal level, and showed that the interpretation of symmetry-related models in both types of theories finds a home within the sophistication outlook.  In these Sections, I rejected the idea that accepting redundancy is accepting defeat; that it is a price we must pay. For the redundancy of gauge theory can be conceived structurally, in as perspicuous a manner as in general relativity. Gauge theory describes an `internal' geometry just as general relativity describes an `external' geometry. 

But in both cases the skeptic about implicit characterizations of structure may remain unconvinced, and in  Sections  \ref{subsec:skeptic_why} and \ref{subsec:drag} I fortified the skeptic's reasons as much as I could. But in Sections \ref{sec:gr_re} and  \ref{sec:soph_antiq} I attempted to   mollify the skeptic by glossing redundancy as notational, as it is often done in the practice of physics.  Thus I employed an active-passive correspondence to interpret the common structure of the symmetry-related models. In this respect, among other things, I showed that  existentially quantified coordinate-based descriptions of the models  are symmetry-invariant and easy to interpret, and that the infinitesimal symmetries of the theories are in one-one correspondence with passive transformations. Lastly, using the formalism of representational conventions of Section \ref{sec:rep_conv}, I gave a practical gloss to these passive transformations: they relate different choices of relations with which to describe the physics.

Of course I could not conclusively show that the structural representation is equally recommended in both cases; I can only show that it is equally conceptually transparent in both cases. In the next Chapter I  will assess  some authors' concrete attempts to draw distinctions between general relativity and gauge theory, seeking to strengthen \emph{the license} for sophistication of gauge symmetries to \emph{a recommendation}. So my final advocacy for this recommendation will be completed in Chapter \ref{ch:sd_2}.    


\chapter{Same-Diff II: concrete comparisons}\label{ch:sd_2}

This Chapter compares the symmetries of general relativity to those of Yang-Mills theory along several different axes.   The only non-trivial distinction that survives this thorough analysis is parallel to a well-known distinction between Abelian and non-Abelian theories. That  distinction runs as follows. 

In electromagnetism---an Abelian theory---the field-strength tensor is a \textit{gauge-invariant} variable; but in the non-Abelian theory  the field-strength is gauge-\textit{variant}. In the standard formulation of general relativity, the Riemann curvature tensor, which is usually taken as the  variable dynamically analogous to the electromagnetic field-tensor,  likewise \emph{varies} under diffeomorphisms. For example, it varies under the infinitesimal version of the transformation, and this variation is   given very simply by the Lie derivative of the tensor.\footnote{Beware: that a quantity such as the Riemann curvature varies under diffeomorphisms tends to be forgotten in the hole argument literature's emphasis on dragging along metric fields. This forgetfulnees bears on the philosophical morals of the hole argument \cite{GomesButterfield_hole}. }

However,  there is a more advanced mathematical formalism for non-Abelian Yang-Mills theory in which its field-strength tensor is \emph{invariant} (viz. the bundle of connections, summarized in Section \ref{sec:Atiyah}). 
Such a mathematical formulation is available for Yang-Mills theories, but \emph{not} for general relativity. That is because while both the Riemann curvature and non-Abelian Yang-Mills curvature are gauge-variant, in the \emph{usual} formalisms their gauge variance is conceptually distinct. Under an isomorphism of non-Abelian Yang-Mills theory, the curvature tensor still transforms in a prescribed, algebraic manner (i.e. without any derivative); while in general relativity the Riemann curvature transforms with a derivative: as mentioned above, it transforms like any other tensor, through a Lie derivative. This conceptual distinctness is reflected in the more mathematically sophisticated formalism for gauge theory---the bundle of connections---in which the non-Abelian field strength is  fully gauge-invariant, just like it is in the Abelian case.

We also need to allow for the fact that---as the Aharonov-Bohm effect shows---there are degrees of freedom of the gauge potential that are `non-locally possessed', as \citet[p.192]{Healey_book} puts it. This means that  the Yang-Mills curvature can only represent the locally possessed degrees of freedom of the fields. 

So here  is my proposal for  the main difference between the symmetries of Yang-Mills theory and  of general relativity:\\\indent
$\Delta$: Yang-Mills theory, but not general relativity, \emph{admits a formalism in which the local, dynamical content of the theory is fully  invariant under the appropriate symmetry transformations.} 

As just explained: for \emph{non}-Abelian Yang-Mills one has to advance to the bundle of connections to obtain such a formalism. I will call this distinction, $\Delta$, for `distinction', or `difference'. 

Can $\Delta$ be cashed out in terms of a reduced formalism? That is: does the distinction support the folklore that for gauge theories we should adopt such a formalism, i.e. be eliminativist? In this Chapter \ref{ch:sd_2}, as in the previous Chapter \ref{ch:sd_1}, I will argue  that `No': a structuralist interpretation  is equally recommended, and equally conceptually transparent, for both general relativity and Yang-Mills theory. In short, Yang-Mills theory no less than general relativity should resist eliminativism while embracing structuralism.

\section{Introduction and roadmap for this Chapter}
The following section, Section \ref{sec:KK}, deals with the attempt to distinguish gauge and diffeomorphism in terms of the labels: `external' and `internal'; and this will lead me to my promised real difference, labelled $\Delta$ above.

I begin by considering one obvious reason to distinguish gauge symmetries from spacetime diffeomorphisms: that the former acts `internally'---shuffling around properties at each spacetime point---whereas the former acts `externally': shuffling around the spacetime points themselves. But before we attribute too much significance to this distinction, we need to be sure it does not just spring from our more everyday acquaintance with the  `medium-sized dry goods' of spacetime---where diffeomorphisms act---than with the `internal spaces',  where the gauge-transformations act. And I argue that we \emph{cannot} be sure of this. For think of how most macroscopic bodies are electrically neutral, so that electromagnetic forces are not easy to perceive; and the other non-gravitational forces are confined to subatomic length scales. But this difference between the forces described by gauge theories and by gravitational physics does not necessarily provide a significant distinction between diffeomorphisms and gauge symmetries. In other words, the obvious distinction between external and internal symmetries may not be a fundamental one; and indeed, it has been  challenged by `bundle substantivalists', such as \citep[Ch. 6.3]{Arntzenius_book}. 
To better probe the everyday distinction between external and internal, I will briefly summarize the interpretation of gauge theories via the  empirically equivalent Kaluza-Klein formalism. The formalism employs only external, or spacetime directions, and by so doing casts doubt on whether we can really thus distinguish external and internal transformations. 


But one conceptual distinction between gauge theories and general relativity  survives  in the Kaluza-Klein formalism, and it is present also in the principal fiber bundle formalism. For, even in the Kaluza-Klein framework, internal directions are `background' structures: they do not respond to the distribution of matter and energy.  This rigidity corresponds to the distinction between the pointwise actions of gauge symmetries and diffeomorphisms on the dynamical quantities of Yang-Mills and general relativity respectively, that I called $\Delta$  above. That is:  the gauge   and Riemann curvature tensors transform, pointwise, in qualitatively different ways (even if both are in their own sense, covariant). The gauge curvature transforms homogeneously and the Riemann curvature, generically, does not.

The bundle of connections, introduced in Section \ref{sec:PFB_formalism}, allows us to polish this distinction.
In this formalism, the curvature of the gauge potential is fully gauge-invariant, in the non-Abelian as well as  in the Abelian  or electromagnetic case (even though in both cases the gauge potential is still gauge variant). In this formalism,  the curvature, or field-strength tensor, exhausts the local gauge-invariant degrees of freedom, in the same way that the familiar electromagnetic field-strength tensor does, in the Abelian version of Yang-Mills theory.  The contrast then is that  the Riemann curvature tensor in general relativity is not pointwise invariant if we drag it along directions that generate the infinitesimal spacetime diffeomorphisms;\footnote{Even if the models of the theory as a whole may have the same physical (structural) content cf. Section \ref{subsec:drag} and \cite{GomesButterfield_hole}.} but the gauge field strength is invariant if we drag it along directions that generate the infinitesimal gauge symmetries. 


Therefore, taking a cue from the Atiyah-Lie formalism, one could say that gauge transformations are isomorphisms of the theory that relate states that are \emph{dynamically locally  indiscernible}. 
In contrast, spacetime diffeomorphisms are those isomorphisms of the theory  that implement  \textit{locally discernible} changes in the dynamical quantities. The distinction applies whether we take a bundle substantivalist view, like Kaluza and Klein, or not. 
That is,  we identify the curvature, or field strength tensor, with the local dynamical part of the gauge field; and these quantities are invariant under  the action of gauge transformations, but they are not invariant under the action of diffeomorphisms.   This is $\Delta$. \\

This is a rather abstract, conceptual difference between the symmetries of the two theories. The remaining Sections of this Chapter will focus on more concrete comparisons.


I will  focus my analysis on such comparisons between diffeomorphisms and gauge symmetries  that have not been given much attention in the literature thus far. This will helpfully   narrow the scope of the Chapter. But even in this limited scope, aiming for  completeness requires some artificial separation of the topics to be discussed. That is also unavoidable: both gauge and general relativity are rich, multi-faceted theories, and may be compared along  different axes.

 In Section  \ref{sec:common}, I begin by criticizing the more obvious, and more concrete, attempts to conceptually distinguish diffeomorphisms in general relativity from gauge transformations in Yang-Mills theory. This Section considers a disparate array of attempts and requires a few mental ``gear-changes''. More specifically, I will discuss:
\\\indent (1) the representation of each type of  symmetry in the Hamiltonian, or Dirac, analysis of constraints,
\\\indent (2) the relation between the symmetries and charge conservation;
\\\indent (3) the comparison between the Aharonov-Bohm effect in gauge theory and general relativity. 

Regarding (1) there is a distinction to be drawn in the context of Dirac constraint analysis, between, on one side, certain diffeomorphisms, and, on the other, the remaining diffeomorphisms and gauge symmetries. Namely, the diffeomorphisms whose generators act by shuffling points in time are of a different type than both those shuffling points in space and gauge transformations---which are of the same type. I take this discrepancy to be part and parcel of the infamous Problem of Time (cf. \cite{Kuchar_Time}), and to lie outside the scope of this Chapter (which is already quite broad!). So, to sum up about (1): the spatial diffeomorphisms and the gauge symmetries are found to be similar in every important way.

Regarding (2), the main question is whether there are quasi-local symmetry-invariant conserved charges that are associated to the symmetries. The answer is that there are in electromagnetism,  but there aren't  (generically) in either general relativity  or in non-Abelian Yang-Mills theories. Therefore the difference  here is really between Abelian and non-Abelian symmetries (namely, those that generate commuting or non-commuting transformations, respectively), and not between diffeomorphisms and gauge symmetries. 

Regarding (3), the obvious suggested difference is that in general relativity a ``phase difference'' is acquired infinitesimally along each of the trajectories, while in the Aharonov-Bohm effect the acquired phase is strictly a global concept: one can only relate it to closed loops, i.e. one cannot pinpoint `where' in the trajectory the phase is acquired. I will show that this interpretation involves a simultaneous \emph{mis}construal of phase acquisition by a charged particle and of the gravitational analogue for the Aharonov-Bohm effect.   

Next, in  Section \ref{sec:Healey}, I tackle a less well-known attempt to distinguish the types of symmetries, formulated by \cite{Healey_book}. 
Healey's  argument is  functionalist, in the sense of Lewis' idea of uniquely specifying an item as the occupant or realizer of a functional role \citep{Lewis_func, Lewis_defT}. He says  that we can distinguish individual representatives of the metric---e.g. one particular representative of the metric tensor field amongst all of the isomorphic copies---by stipulating extra conditions on it.  And he  argues that one cannot similarly distinguish amongst  gauge-equivalent representatives of the same physical possibility. 
I will reconstruct Healey's argument for such ``gauge-exceptionalism'' and show that it is misguided: it only arises if we unduly, and arbitrarily, restrict the  tools we take to be available  for the functional definition.

In the last section, Section \ref{sec:conclusions_sd2}, I will conclude this Chapter.

\section{Distinguishing gauge and diffeomorphisms symmetries}\label{sec:KK}

 The reader may be wondering when I will address the `elephant in the room': "You can argue all day about the similarities of the two types of theory, but \emph{surely} there is a fundamental distinction between symmetries acting internally---as gauge symmetries do---and those acting `externally', as spacetime diffeomorphisms do".   I do agree that it is difficult to countenance the \textit{absence} of such a fundamental distinction between  internal and external dimensions of the universe.                
  But  if we had a more intimate acquaintance with the value space of these  fields, for instance, if isotopic spins were macroscopically detectable, would we perhaps reach the same sort of intuition about the corresponding internal directions of a principal fiber bundle that we have for spacetime directions? 

To probe our intuitions about these questions, in Section  \ref{sec:KK_intro} I will introduce a formalism that unifies the gauge transformations and the diffeomorphisms. The unification geometrizes different forces as effects of curvature in  Riemannian geometry, and it is obtained by enlarging spacetime in a particular manner.

In Section  \ref{sec:pos_argument} I will show that internal and external symmetries   are nonetheless qualitatively different. 

In the principal bundle formalism of Section \ref{sec:PFB_formalism}, the reason can be seen as follows: even though vertical automorphisms are diffeomorphisms, they act more like the isometries of a given metric, or like the change of basis for tangent vector fields, than like generic diffeomorphisms. This reflects both our proposal for a distinction, $\Delta$, that first appeared in the preamble to this Chapter, and the differences pointed to in Section \ref{subsec:general_cons} (text after Equation  \eqref{eq:Lie_Omega_full}); and it will remain the most significant difference between the two types of symmetry to the end of the chapter. 


\subsection{Finding  common ground between diffeomorphisms and gauge transformations: The Kaluza-Klein framework}\label{sec:KK_intro}\label{sec:sim_KK}

As we saw in Section \ref{sec:Atiyah} the bundle of connections formulation of the gauge connection brings the metric formulation of general relativity and the formulation of gauge forces into close proximity. 

In this Section we will expunge the distinction between internal and external  by introducing  a Kaluza-Klein formulation of gauge theory. This formulation geometrizes gauge interactions at the expense of adding extra dimensions to spacetime; it thus effaces the distinction between internal and external directions. 



First, it is important to distinguish a position that  takes internal dimensions to have ontic (and not just notational) status, such as \cite[p. 185]{Arntzenius_book}'s bundle substantivalism,  from one that goes beyond this, and  also `geometrizes' these internal dimensions, such as the Kaluza-Klein framework \citep{KaluzaKlein} (a project that involves several issues that lie beyond the scope of what I am concerned with here). 

The main idea of the Kaluza-Klein framework is remarkably simple  (cf.  \cite[Ch. 9]{Bleecker}): use an inner product $\kappa$ on the Lie-algebra $\mathfrak{g}$,\footnote{This is usually called the \textit{Killing form}: in simple matrix representations of the Lie algebra, it is just the trace of the matrix product. } the metric $g_{ab}$ on $M$, and the connection-form $\omega$ on $P$, to induce a Lorentzian inner product on $P$ by ``just summing the external and internal co-fibrations'', which we can then treat through the tools of general relativity/Riemannian geometry. The induced metric is given by: 
\be\label{eq:KK_metric} \eta(\bullet, \bullet)=g_{ab}(\pi_*(\bullet), \pi_*(\bullet))+\kappa(\omega(\bullet), \omega(\bullet)), 
\ee
or, more economically: $\eta=\pi^*g_{ab}+\kappa\circ \omega$. 

We could now compute the Ricci scalar for this metric, and the corresponding Einstein-Hilbert action. Upon variation of this action, we find \textit{both} the Einstein equations for $g_{ab}$ and the Yang-Mills equations for $\omega$ as the extremum condition. Another surprising feature  of the formalism is extracted from  geodesic motion:  since vertical directions are Killing directions of the metric (due to the covariance of $\pi$ and $\omega$),\footnote{Cf. the discussion at the start of Section \ref{subsec:KK_diff}. } vertical velocities are conserved during geodesic motion.\footnote{This is just an application of a basic theorem of (pseudo)-Riemannian geometry: the angle between a Killing direction and a geodesic remains constant.  } Identifying a particle's charge with its vertical velocity (i.e along the orbit of the group) then guarantees  conservation of charge. Moreover, upon projection of the geodesic onto the base space, we get a dynamical trajectory that correctly captures the deviation from geodesic motion by  the Lorentz force  on the particle due to the curvature, $\mathbf{F}$. A mighty formalism indeed!\footnote{This discussion briefly summarized what was achieved in the Kaluza-Klein formalism, circa 1920-1940. The main ideas of geometrizing electromagnetism appeared in \cite{KaluzaKlein}, but that work left out the weak and strong nuclear forces. Klein extended the formalism in 1938, in a paper  presented at, and published in the proceedings  of, a Conference  on New Theories in Physics held
at Kazimierz (Poland) in 1938. The paper is reproduced in \citep{Klein1938}. See \cite[Ch. 3 and Ch. 6]{Dawning_book} for a historical account. It should be said that the formalism has an immense  scope of application: from string theory to the relational-absolutist debate, where it has been recently been used to show how the relationist can deal with  rotations \citep{GomesGryb_KK}. }

\subsection{The real difference between gauge transformations and spacetime diffeomorphisms}\label{sec:pos_argument}

This Section establishes  the difference, $\Delta$, announced in the preamble to this Chapter, between gauge and spacetime diffeomorphism symmetries. 
 
All symmetries will leave the equations of motion, or the action, or the symplectic form, or whatever structure that is \textit{dynamically} relevant, invariant, according to our definition of symmetries, in Chapter  \ref{ch:syms}. There is no distinction between symmetries to be made at that coarse dynamical level. 
But this leaves open the possibility that  translating the representation of the fields along vertical directions is qualitatively and quantitatively different than translating them non-vertically: and the idea will give the real difference $\Delta$.  Translations along vertical directions correspond to isomorphisms  of the theory that have a homogeneous effect on the dynamical quantities, like the curvature. The automorphisms that correspond to diffeomorphisms are those that fundamentally change the properties of the dynamical variables. 

In  Subsection \ref{subsec:KK_diff},  I functionally characterize the differences between internal and external dimensions, and between gauge transformations and spacetime diffeomorphisms, in the Kaluza-Klein framework. 

In  Subsection \ref{subsec:Atiyah_diff},  we use the Kaluza-Klein discussion as inspiration to characterize the distinction more broadly. We  compare the diffeomorphisms with the gauge transformations, in a coordinate-free, manner, without charts, by using a proxy for the bundle of connections.

\subsubsection{The difference between internal and external dimensions, from the viewpoint of Kaluza-Klein}\label{subsec:KK_diff}

We start with the Kaluza-Klein picture. Here it is easy to  single out the vertical directions: they are the Killing directions of the Kaluza-Klein metric, $\eta$ in \eqref{eq:KK_metric}. In other words, from the second property of \eqref{eq:omega_defs} and since $\kappa$ is taken to be invariant under the (adjoint) action of the Lie group, and $L_g^*\omega=g^{-1}\omega g$, we obtain: $\kappa(\omega({L_g}_*\bullet), \omega({L_g}_*\bullet))=\kappa(\omega(\bullet), \omega(\bullet))$; and since $\pi_*$ is itself invariant under the action of the group: $g_{ab}(\pi_*({L_g}_*\bullet), \pi_*({L_g}_*\bullet))=g_{ab}(\pi_*(\bullet), \pi_*(\bullet))$.\footnote{Thus, infinitesimally, for the vertical metric, these fundamental directions obey: 
$$\mathcal{L}_\zeta(\kappa(\omega, \omega))=2\kappa(\mathcal{L}_\zeta\omega, \omega)=2\kappa([\zeta,\omega], \omega)=0
,$$
by the symmetry  properties of  $\kappa$ (i.e. of the trace). \label{ftnt:comm} } 

For some direction $\zeta$ in $P$ that is \emph{not} along a gauge orbit, neither the first nor the second term in the metric \eqref{eq:KK_metric} will be preserved. And even if the spacetime metric $g_{ab}$ is preserved along some (projected) direction, i.e. such that $\mathcal{L}_{\pi_*\zeta}g_{ab}=0$, it could still be the case that $\mathcal{L}_{\zeta}\kappa(\omega, \omega)\neq 0$.
 Equality obtains iff the gauge curvature \textit{also} vanishes along $\zeta$.\footnote{Namely, constancy requires that  $\mathcal{L}_{\zeta}\omega_p= [\omega_p(\zeta), \omega]$ (see footnote \ref{ftnt:comm}). At a point,  $\zeta$ can be decomposed into a vertical  and a horizontal direction; if it is vertical, we recover the discussion above; and if it is horizontal, we obtain, using the Cartan `magic formula', $\mathcal{L}_{\zeta}\omega=\Omega(\zeta, \bullet)$. But this last term need not be zero, and thus will not match $[\omega_p(\zeta), \omega]$, which is identically zero for horizontal directions.}   And thus the Kaluza-Klein metric will only  be preserved  non-vertically  if  the model is entirely `featureless' along those directions.
 
And so, demanding some modal robustness from our definition---and thereby ignoring such highly homogeneous exceptions---we characterize gauge transformations to be those diffeomorphisms that are generated by Killing fields of the Kaluza-Klein metric. All Kaluza-Klein spaces will have these directions. And  ``being a Killing direction'' is a description that we could, with some allowance for vagueness, characterize as `filling a functional role': it specifies a diffeomorphism-invariant property,  `tracking' the same directions in all diffeomorphism-related models.\footnote{ In contrast, note that had we simply stipulated the identity of, or named, these directions (cf. footnote \ref{ftnt:identity}), they would not remain Killing under a diffeomorphism, for the metric field would `slide over' the curves under a pull-back. The above definition does not care `which curves' fill the above role, for it identifies the curves by the role. } 
 
 In sum, in the Kaluza-Klein framework, the gauge directions are singled out by their rigidity: unlike the fields along other directions, they are constrained to be of a certain form. That is, this split by functional roles corresponds to two sets of directions in $P$: those with a rigid structure---the vertical---and those with  richer, more complex structure: the non-vertical. 


But there is still an uncomfortable feature of the Kaluza-Klein metric, and of this definition of gauge transformations. General vertical vector fields (and not just fundamental vector fields),  those that change from fiber to fiber, are not Killing vector fields. Should these not count as gauge transformations? Surely they should. We will now move on to a more general definition of the distinction.

Essentially, the idea is that  given the value of the connection on $p$, we know what its value on $g\cdot p$ must be: if we write $L_g{}^*\omega=...$ (i.e. the pull-back of the connection by the left group action) we know what will appear on the right hand side, viz. $g^{-1}\omega g$. In contrast, note we do not know how to fill in the analogous equation, $f^*g_{ab}=...$, unless, that is, $f$ is an isometry of $g_{ab}$, in which case $f^*g_{ab}=g_{ab}$. 


\subsubsection{The difference, from the viewpoint of the principal bundle}\label{subsec:Atiyah_diff}

 At the end of Section  \ref{sec:Atiyah}, we had found that the section of the bundle of connections possessed many of the same properties as the metric: it could be written globally, on spacetime, in a coordinate-free manner. Moreover, were we to write down the field explicitly with a trivialization, or coordinate choice, the objects of both theories would be appropriately covariant under coordinate choice. Thus, for instance, if the gauge or the Riemann curvature vanishes at a point, it will vanish in every coordinate system that covers that point. But this is a statement about a passive transformation. 
Let us consider general  active transformations---the automorphisms of the structure---to see how the disanalogies come up.
   
 Two isomorphic connections $\omega$ and $\omega':=\tau^*\omega$---which correspond uniquely to the global sections of $TP/G$,   $\Gamma, \Gamma'$---may be associated to different horizontal directions at a given point $p\in P$.  There is an inhomogeneous term that enters the transformation of $\omega$ under a generic vertical automorphism (see equation \ref{eq:Lie_omega_full} or footnote \ref{ftnt:13}), and thus a horizontal vector for $\omega$ is not necessarily horizontal for $\omega'$. 
 
 Nonetheless, $\omega$ and $\omega'$ still encode a quantity, the  \textit{curvature} $\Omega$, whose transformations has no inhomogeneous term, since it is appropriately equivariant  (see equation \eqref{eq:Lie_Omega_full}). 
And upon quotienting $TP$ to $TP/G$ (as a vector bundle over $TM$ or $M$) by using the bundle of connections, all right equivariant functions (like curvature), become gauge-\emph{invariant}. 

But the same cannot be said of the Riemann curvature: the vanishing of the gauge curvature at some point of the underlying set (be it $P$ or $M$) is physical, that of the Riemann curvature is \emph{not}. Of course, the Riemann curvature is covariant, and so `zero'  is a coordinate-independent value \emph{at a fixed point of $M$}, but  diffeomorphisms will shift the point, and thus the value (cf. Section  \ref{par:active_passive}). This does not imply the Riemann curvature has no structural interpretation: it does, in the same sense that the metric has a chronogeometric interpretation (cf. Section  \ref{sec:skeptic}).

Thus, we can locate or specify vertical automorphisms, among all the fiber-preserving diffeomorphisms of $P$, by their roles:  as those that transform the curvature homogeneously, or, equivalently, that leave the field-strength in  the bundle of connections  \emph{invariant}.  Such transformations locally preserve the dynamical variables, and this  characterization is independent of our underlying attitudes towards symmetry-related models: even if we take a substantivalist view of $P$ \citep[p. 185]{Arntzenius_book}, we can identify gauge transformations as those that relate indiscernible local, dynamical,  states. This is the distinction, $\Delta$, that we have been seeking. Namely, as stated at the preamble of this Chapter: Yang-Mills, but not general relativity, \emph{admits a formalism in which the local, dynamical content of the theory is fully  invariant under the appropriate symmetry transformations.}

In $\Delta$ the word `local' flags the pointwise validity of the distinction, and is crucial. For globally, two isomorphic models are indiscernible by definition, so spacetime diffeomorphisms also relate globally indiscernible states. To be concrete, suppose we define $\tilde g_{ab}=f^*g_{ab}$. Then, at \emph{a given point} $x\in M$, $\tilde g_{ab}$ and $g_{ab}$ may disagree about many things, including the Riemann curvature. But there is no sense in which, globally, $g_{ab}$ and $\tilde g_{ab}$ disagree about the state of the Universe (see also \cite{GomesButterfield_hole}). As presaged in the last paragraph of Section  \ref{sec:diff_sym}, any  distinction to be found between the two kinds of symmetry had to be a local one; and it is. 

And in $\Delta$, the word `dynamical' refers to the use of the curvature, and not to the use of the dynamics (through the equations of motion, action functional or Hamiltonian). Indeed, this characterization is independent of the dynamics of the theory.  And here we see the use of the bundle of connections: In the standard formulation of the Abelian theory in a principal bundle, the curvature  exhausts all the \emph{local} gauge-invariant degrees of freedom; indeed the curvature itself can be seen to merely stand for the \emph{local} gauge-invariant degrees of freedom of the theory. In the non-Abelian theory, things get much more complicated in the standard picture: the curvature is not invariant, and we must take traces of products of the curvature tensor to convey local (i.e. pointwise) gauge-invariant functions. And the theory does not have infinitely many physical degrees of freedom per spacetime point, and so we must choose a basis among all these invariant functions: a difficult task. The bundle of connections cuts this Gordian knot by having a notion of field-strength that is invariant at each point, and which exhausts the number of local physical degrees of freedom of the theory, just like  the field-strength does in the Abelian theory.\footnote{ 
Of course, in neither the Abelian nor the non-Abelian theory, do the local gauge-invariant degrees of freedom exhaust \emph{the totality} of physical degrees of freedom: there are global physical facts about parallel transport, facts that are not encoded in the curvature. (And  the connection is not gauge-invariant in the Atiyah-Lie bundle, or, equivalently, it is not homogeneously equivariant in the principal bundle).}

To further clarify the meaning of $\Delta$, note that the definition applies to all models of the theory: i.e. it is modally robust. 
   And indeed, for gauge theories, gauge transformations as vertical automorphisms of a principal bundle provide directions of local dynamical indiscernibility  for \textit{all models} of the theory. That is: irrespective of the particular state  and at all points  of the underlying domain.
And of course, generically, the non-vertical  generators of the automorphisms of $P$ fail to meet such strict criteria, and are thus associated to shifts of the base manifold; that is, they generate spacetime diffeomorphisms. Under these more general shifts, local dynamical quantities transform non-trivially.\footnote{
Subsets of diffeomorphisms could  be put into a tighter comparison with gauge symmetries if we have $\mathcal{L}_XR^a_{bcd}=0$, where $X$ are vector fields (the infinitesimal generators of diffeomorphisms) and $R^a_{bcd}$ is the curvature of the metric. It can be shown that $\mathcal{L}_XR^a_{bcd}=0$ occurs when $X$ is a Killing field. It follows from realizing that $\mathcal{L}_XF(g_{ab})_{|x}=\int \, \d y \frac{\delta F(x)}{\delta g_{ab}(y)}\delta g_{ab}(y)$ where $\delta g_{ab}= \mathcal{L}_X g_{ab}$.  There are more complicated proofs that use the Bianchi identities and the algebraic properties of the Riemann tensor. 
But there are two problems with this: i) metrics with Killing directions form a meager set in the space of metrics (under any reasonable topology) and ii) the Killing vector fields are also a meager subset of the space of all vector fields (seen as generators of (small) diffeomorphisms). 

 }

Thus, the functional distinction $\Delta$   is modally robust, in the sense that it is present for generic models of general relativity and Yang-Mills theory. 
 The gauge transformations really \textit{do} map between local quantities that are less discernible than those mapped by generic diffeomorphisms. In the Atiyah-Lie formalism, we can polish this distinction, since we find  symmetries that do not change the \textit{local} dynamical variables of the theory (e.g. the curvature of the Atiyah-Lie connection), irrespective of spacetime point or state of the field. This concludes the characterization of  distinction $\Delta$, first proposed in the preamble to this Chapter (and subsuming the differences pointed to at the end of Section  \ref{subsec:general_cons}).

\section{Three natural comparisons assessed}\label{sec:common}
Chapter \ref{ch:sd_1} focussed on  formal differences between the symmetry structures of general relativity and Yang-Mills theory. This Section will focus on three more concrete comparisons between diffeomorphisms and gauge symmetries that will  naturally  come to mind for anyone thinking  about the issue. 

My argument requires brief expositions of symmetries, for both  Yang-Mills theory and  general relativity.   In Section  \ref{sec:syms} I undertake this brief exposition.

Section \ref{sec:syms_consts} compares the symmetries in  the constrained Hamiltonian formalism. The symmetries are shown to be similar in certain important respects; in particular, in their relationship to a (non-signalling) type of non-locality. But here we also find the seed of the robust difference between the symmetries found in \ref{ch:sd_1}: a conceptual difference in the transformation properties of the canonical variables in the two theories.

The second type of comparison, investigated in Section  \ref{sec:charges}, is about the relation between symmetries and charges. It is often remarked that diffeomorphisms are unlike other symmetries of nature, since we cannot obtain from them a physical, conserved charge. That is: generically, we cannot associate to a diffeomorphism symmetry  a physical quantity in a spacetime region whose change in time is solely due to a flux through the region's spatial boundaries. While I agree that such an obstruction to the relation between symmetries and charges exists for diffeomorphisms, I will show that they also exist for any \emph{non}-Abelian Yang-Mills theory. 

The third comparison, investigated in Section  \ref{sec:AB},  is about  the Aharonov-Bohm effect, in gauge theory. There are a few arguments here. One is that the Aharonov-Bohm effect presents  a \emph{sui generis} type of physical under-determination. Resolving this under-determination, it has been argued, requires an eliminativist approach to the theory, unlike the under-determination posed by the `hole argument' in general relativity. I will consider gravitational analogues to dispel most of the distinctions that have been attributed to the Aharonov-Bohm effect.

\subsection{The symmetry generators (constraints) and their locality}\label{sec:syms_consts}

Gauge symmetries are often said to be a necessary evil: it is said that if we are  to obtain an unconstrained, local description of the physics, we need to include redundancy. Let us investigate the meaning of this folklore.

First, since the formalism is slightly unfamiliar to philosophers of physics, in Appendix \ref{sec:Ham} we  quickly review, in the simple case of a non-relativistic mechanical system, how constraints arise in Hamiltonian mechanics, and their broad relation to symmetries and under-determination; (for a more complete treatment, see \cite[Ch. 1]{HenneauxTeitelboim}, and for  philosophical introductions, \cite{Wallace_LagSym} and \cite[Section 2]{GomesButterfield_electro}). This material will laso be used in Chapter \ref{ch:Coulomb}. Here we briefly summarize. 

In the Lagrangian formulation of mechanics, one is given a \emph{Lagrangian} $ L(q^\alpha(t), \dot q^\alpha(t))$, and obtains, from the least action
principle $\delta  S=0$, the Euler--Lagrange equations:
\be \label{equ:Euler-Lagrange}\frac{d}{dt}\frac{\pp  L}{\pp \dot q^\alpha}=\frac{\pp  L}{\pp  q^\alpha}
.\ee

If we use the chain rule for the $\frac{d}{dt}$ derivative, we get from \eqref{equ:Euler-Lagrange}:
\be\label{equ:accel_lagrange} \ddot q^{\beta}\frac{\pp ^2 L}{\pp \dot q^{\beta}\pp \dot q^\alpha}+\dot q^{\beta}\frac{\pp ^2 L}{\pp  q^{\beta}\pp \dot q^\alpha} =\frac{\pp  L}{\pp  q^\alpha}
.\ee
From this equation it becomes clear that the accelerations are uniquely determined by the positions and velocities if and only if the matrix $M_{\alpha\beta}:=\frac{\pp ^2 L}{\pp \dot q^{\beta}\pp \dot q^\alpha} $ is invertible. If it isn't, our system possesses some kind of redundancy in its description. As we know, this is indicative of {\it gauge symmetries}.  

Here, we seek a representation of symmetries in the Hamiltonian formalism. Thus, one replaces $\frac{\pp  L}{\pp  \dot q^\alpha}=:p_\alpha$, and the question  whether the accelerations are determined by the positions and velocities is translated to the question whether the momenta $p_\alpha$  are invertible, as functions of the velocities. If $M_{\alpha \beta}=\frac{\pp  p_\beta}{\pp  \dot q^\alpha}$ is not invertible, there are constraints among the $p_\beta$, which we write as $\Phi^I(p_\alpha, q_\beta)=0$, where $I$ parametrizes the  constraints.\footnote{Here we assume that the rank of $M_{\alpha\beta}$ is constant, and that the ensuing constraints obey regularity conditions. See \cite[Ch. 1.1.2]{HenneauxTeitelboim}. Also note that the index $I$ here should not be confused with a Lie algebra index, $I$, as we have used it before.  } 

When these constraints are conserved by the equations of motion and are compatible amongst themselves---in the jargon of the Dirac algorithm: are first-class---they correspond, by (the converse of) Noether's theorem, to symmetries of the system. Indeed, some of the great advantages of the Hamiltonian formalism are: (i) that the symmetries are not postulated, but, given a few extra assumptions, algorithmically identified; and (ii) the relation between constraints and symmetries is very straightforward. Namely, the symmetries act on any quantity through the Poisson bracket between that quantity and the symmetry's corresponding generator, which is just the constraint. 

In a bit more detail: as scalar functions on phase space  the constraints $\Phi^I$ (for each $I$), have  (differential geometric) gradients, $\d \Phi^I$, which are in one-one correspondence with  vector fields $X_{\Phi^I} =: X_{I}$ due to the symplectic structure of phase space. Namely, given the symplectic form $\omega$ (a closed, non-degenerate two-form on phase space), we can define vector fields from  one-forms (and vice-versa). Applying this definition to  $\d \Phi^I$ we obtain the associated vector fields $X_I$ defined by $\omega(X_I, \bullet)=\d \Phi^I$.\footnote{The relation to Poisson brackets is given by:  $\{f, h\} = \d f (X_h) = \omega(X_f, X_h),$
for $f, h\in C^\infty(\mathcal{P})$. (Here, as usual,  $\d f(X)$ is the contraction between 1-forms and vectors; and $\d f(X)$ is equal to $X(f)$ i.e. the directional derivative of a scalar function $f$ along $X$). \label{ftnt:gens} } The key idea is that, just as the flow specified by the Hamiltonian function conserves energy, these vector fields associated to $\Phi^I$ are tangential to, and so preserve, the intersection of all  the constraint- and energy-surfaces.  That is, in a less geometric (and maybe more familiar) language: they not only commute with the Hamiltonian and conserve energy, but also conserve the charges associated with the constraints. From the more geometrical viewpoint, it is easy to check that, on the constraint surface, $\omega$ is degenerate, and the $X_I$ form its (integrable)  kernel. This is the origin of gauge symmetry in the Hamiltonian formalism (cf. \cite{earman_ode} for a more comprehensive philosophical treatment). 
The same framework and ideas will be required again in Chapter \ref{ch:Coulomb}.\\

In Section \ref{subsec:IVP}, now in  possession of the appropriate tools  of the Hamiltonian formalism and its relation to symmetries and constraints, we revisit the topic that was the focus of  the first subsection of Section \ref{par:active_passive} and provide a complementary answer to why gravity, or rather, geometrodynamics, was the right setting to discuss diffeomorphism symmetry. 

In Section \ref{subsec:sims}, we will see how the canonical symmetries of both Yang-Mills theory and general relativity are alike. Namely, both emerge from elliptic differential constraints. These constraints are responsible for a (mild) form of non-locality, in the sense that the allowed values of the fields at a point depend on the values of the field a finite distance away. 

In Section \ref{subsec:diffs}, we will focus on the symmetries that are generated by the constraints. Here we will see a qualitative difference between the actions of these symmetries on the canonical momenta of the two theories.

\subsubsection{Initial value problem and indeterminism}\label{subsec:IVP}

Back in Section \ref{par:active_passive}, we briefly discussed the reasons why general relativity, above any other theory, is associated with diffeomorphism symmetry. Since other dynamical fields could just as well exhibit the symmetry, this preffered association provided a (minor) mystery. And the mystery deepens if we think of symmetries as generated by constraints. To see this, take as an example  the Hamiltonian treament of  Klein-Gordon fields: this treatment does  \emph{not}  usually mention initial value constraints, and so does not mention any symmetry.  

Let us be more explicit. Take a Klein-Gordon scalar field of mass $m$, whose equations of motion are:
\be \nabla^a\nabla_a\psi -m^2\psi=0.
\ee
This is one equation for one variable, and it is a hyperbolic equation, which has a well-posed initial value problem (IVP) (cf. \cite[Theorem 10.1.2]{Wald_book}). That is, given smooth initial data on a surface $\Sigma$, there is a unique solution throughout spacetime.\footnote{Moreover, the solutions depend continuously on the initial data, but I won't be concerned with this facet of the initial value problem.} Accordingly, a Hamiltonian treatment identifies no local constraints among the dynamical equations. 

Now, compare that with the  relativistic Maxwell equations in vacuum, given (in coordinates) in \eqref{eq:eom_A}
\be\label{eq:gen_eom_A} \nabla^a\nabla_b A_a-\nabla^a\nabla_a A_b=0. 
 \ee
Though at first sight this appears to be similarly well-posed---it has four equations for four variables---that is not the case. As we mentioned above, this is an underdetermined system for $A_a$---for instance, in any coordinate system, $A_0$ can be set to whatever function we like, and thus it is not determined by its initial values. The issue is of course familiar from Chapter \ref{ch:sd_1}:  a verdict of indeterminism would be premature, since it would ignore the fact that different solutions are related by isomorphisms, which are generated by the elliptic initial value constraints.

In more detail, the IVP of this system is both under- and overdetermined. It is overdetermined because \emph{not all} initial values satisfy the constraints, and it is underdetermined because for each valid initial value, \emph{many} solutions can be found. In both cases, one finds unique solutions satisfying some gauge-fixing, or within a representational convention. That is: the solutions are unique because, when the gauge conditions are added, the equations of motion of the theory become hyperbolic (and thus have a standard well-posed IVP). Thus two solutions must have the same projection under $h_\sigma$ in \eqref{eq:proj_h} (such as \eqref{eq:h_proj}, below); therefore they must differ only by the action of some isomorphism.\footnote{ See \cite[Secs. 7.5-6]{Landsman_GR} for a sketch of the (complicated) proof, with more extensive references.}

Incidentally, this construal of symmetries as arising from indeterminism (see also \cite{earman_ode, Wallace_LagSym} for similar arguments) also dispels the idea that diffeomorphism symmetry is a `trivial variable redefinition'. For instance, it flatly denies \cite[p. 454]{Curiel2018}'s argument that general symplectomorphisms and diffeomorphisms are on the same footing: 
\begin{quote}
Hamiltonian mechanics has a similar arbitrariness: one is free to choose any
symplectomorphism between the space of states and the cotangent bundle of
configuration space, that is, one may choose, up to symplectomorphism, any
presentation of phase space (or, in more traditional terms, any complete set of
canonical coordinates), without changing the family of solutions the possible
Hamiltonians determine (Curiel, [2014]). One is not driven to investigate the
ontic status of points in phase space merely because one is free to choose any
symplectomorphism in its presentation. Indeed, one can run an argument
analogous to the hole argument here, substituting ‘phase space’ for ‘spacetime
manifold’, ‘symplectomorphism’ for ‘diffeomorphism’, and ‘symplectic struc
ture’
for ‘metric’. Does that show anything of intrinsic physical or metaphys
ical
significance? No serious person would argue so.\end{quote}

But if we identify initial value constraints as the origin of symmetries in physics, how do we see the diffeomorphism symmetry of the Klein-Gordon equation, which has an unconstrained initial value problem? 
The resolution is to note that the initial value problem invokes the metric to define what is spatial and what is temporal (and what is null). Usually, when we consider the intial value problem of the Klein-Gordon field, we have a fixed background foliation of spacetime. But applying a diffeomorphism to spacetime changes that foliation, and so changes our 3+1 decomposition of the Klein-Gordon field as well. Nonetheless, we would like the different solutions that merely correspond to different foliations to not be counted as physically distinct.
Now, even if we ignore backreaction and take a single background  {spacetime}, different foliations  change the decomposition of the spacetime metric and thereby   the initial values of $\psi$. So this provides a connection between transformations of the metric (and its conjugate momentum) and transformations of the initial values of $\psi$, that should not affect the physical state of the world. But does that mean we must treat every field within a general relativistic framework? No. 

 In the Hamiltonian treatment, even for a single background spacetime such as Minkowski, to describe diffeomorphisms we also require different \emph{spatial} metrics and their conjugate momenta: we cannot leave these fixed, as we did the spacetime metric. Thus,  in the 3+1 context, treating diffeomorphism symmetry requires geometrodynamics, i.e. the inclusion of a phase space for the spatial metric and its conjugate, even if our focus is on the dynamics of the Klein-Gordon field. And in this case, each leaf of any given foliation must obey geometric relations between the spatial metric and its conjugate momenta: these are the Gauss-Codazzi relations (see e.g. \cite[Ch. 4]{Oneill}).  Independently of the dynamics that we later invoke for the spacetime metric, the Gauss-Codazzi relations become the constraints of the initial value formulation. 
Finally,  in  extending these constraints to encompass other sources of energy-momentum, we are  assured that the generated symmetries will also apply to other fields, such as $\psi$ (see \citep{Teitelboim1973}). 

\subsubsection{The similarities between canonical spatial diffeomorphisms and gauge transformations }\label{subsec:sims}
In the case of general relativity, the configuration variables are spatial, i.e. Riemannian metrics, $g_{ij}$ (with $i,j,k$ denoting spatial indices), and their conjugate momenta are denoted by $\pi^{ij}$.\footnote{ Geometrically, given a spacelike foliation of a globally hyperbolic spacetime by leaves which are surfaces of simultaneity, we have $M\simeq \Sigma\times \RR$, where $\Sigma$ is a three-dimensional manifold, and $g_{ij}$ corresponds to the pull-back of the spacetime metric $g_{\mu\nu}$ onto its leaves, $\Sigma$, and  the momenta are essentially the extrinsic curvature of the leaves. In the Yang-Mills case, $A^I_i$ corresponds to the pull-back of $A_\mu^I$, and the conjugate momenta, $E^I_i$, is essentially the curvature contrasted with the normal to the foliation, $F^I_{\mu\nu} n^\mu$.\label{ftnt:Sigma}} There are two sets of constraints that emerge: one is associated to ``refoliations'' of spacetime, namely, redefinitions of the surfaces of simultaneity, and the other is associated to diffeomorphisms that map each leaf to itself. Here, I will set aside the generator of refoliations, due to several interpretational difficulties, including whether it really generates a bona-fide symmetry or not (the infamous `Problem of Time', cf. \cite{Isham_POT, Kuchar_Time}).\footnote{ But I have two cursory remarks. The first is that,  unlike the spatial diffeomorphisms, which arise without further restrictions or qualifications from the Einstein-Hilbert Lagrangian, one can only establish a map between the spacetime refoliations and a Hamiltonian constraint if one restricts the domain of this map to spacetimes that satisfy (some of) the equations of motion \citep{Lee:1990nz}. This seems like an important fact, but it also seems to concern the nature of time and symmetry, not the nature of diffeomorphisms: thus supporting my here setting aside the refoliations. The second remark is that, in the case of general relativity, determining whether the constraint surfaces and their intersections actually form a regular submanifold of the phase space of general relativity is a  non-trivial matter. As far as I understand the current state of the problem, this can only be shown in what is dubbed ``CMC-gauge''. That is,  the intersection of the constraint surfaces with $g_{ij}\pi^{ij}=$const. has been proved to form a regular submanifold; (cf. \cite[p. 172]{fischermarsden}).)\label{ftnt:CMC}}

The constraint called the momentum constraint (which can also be seem as the symplectic generator of spatial diffeomorphisms), is: 
\be\label{eq:mom_ctraint} \bar\nabla_i\pi^{ij}=p^j,\ee
where $\bar\nabla$ is the Levi-Civita covariant derivative intrinsic to the surface, associated to $g_{ij}$, and $p^i$ is the momentum density of the fields that source the gravitational field. This is a set of elliptic equations, meaning they have more spatial than time derivatives. 

Applying the same procedure to Yang-Mills theory, we obtain a constraint of a very similar form to \eqref{eq:mom_ctraint}, the Gauss constraint: 
\be\label{eq:Gauss}\D^i E_i^I=\rho^I ,
\ee
where $\rho^I$ is the (Lie-algebra valued) current density and $\D_i$ is the spatial version of the gauge-covariant derivative of \eqref{eq:gauge_trans}. Equation \eqref{eq:Gauss} is of precisely the same character as the momentum constraint in general relativity \eqref{eq:mom_ctraint}; in particular, both are elliptic equations.

In practice, ellipticity means that boundary value problems require only the boundary configuration  of the field, i.e. they  do not also require the field's rate of change at the boundary. The solution of these equations exists \emph{on each slice}; and so the solution  does not correspond to the propagation of a field, as would a solution to a hyperbolic equation.

Indeed, elliptic equations, are, to a certain extent, non-local. For instance, in electromagnetism,  to determine the allowed values of e.g. the electric field $E^i$ at a point $x$,   we need to know the value of the electric field \emph{on the boundary of a small region surrounding} $x$, and we need to know the distribution of charges in this region, and not just at $x$. 
  Note that this goes beyond  the well-established denial of \emph{pointillisme} \cite{ButterfieldPoint}, for it does \emph{not} amount  just to the requirement that the value of a quantity at $x$ depends on other values, \emph{infinitesimally} distant from $x$.\footnote{For instance, a \emph{smooth} vector field would violate \emph{pointillisme} in this milder sense and yet, not  being subject to any elliptic constraint, would be independently specifiable at $x$ and at the boundary of a spatial region surrounding $x$.}

Even if the topology of space is simple, we have a mild form of non-locality: the Gauss constraint implies that by simultaneously measuring the electric field flux on all of a large  surface surrounding a charge distribution, and integrating, we can ascertain the total amount of charge inside the sphere {\em at the given instant}. In its quantum version, the non-locality implies the total Hilbert space of  possible states is not factorizable. 
Indeed, this type of holism, or non-locality is a well-known issue for all theories with elliptic initial value problems: e.g. Yang-Mills theory and general relativity. 

Importantly, note that the dynamics of fields that do not present any symmetries, such as those of Klein-Gordon field, examined in Section \ref{subsec:IVP}, do not introduce such elliptic constraints: they are just hyperbolic.  

The idea that gauge symmetries are the price we pay for local representations can then be characterized as follows. A Lagrangian, such as the Yang-Mills Lagrangian, that employs gauge symmetries, ensures that  the ensuing theory will possess a Gauss-type local conservation law (as we will discuss in Chapter \ref{ch:Coulomb} (see \cite[Ch 7]{Strocchi_book} and  \cite[Section 3]{Strocchi_phil}). As the above argument illustrates, such local, differential conservation laws are tantamount  to the  elliptic differential constraints on the dynamical variables and to the (mild) degree of non-locality mentioned above. That establishes the link from symmetry to conservation laws to  non-locality. 

There is a more direct route, if we restrict ontological status to gauge-invariant quantities. Then  a localized charge, by its lonesome self, is \emph{not} gauge-invariant. To make it so, we need to `dress' it with its Coulomb field, which extends out to infinity, making the composite thing non-local. Here is how \citet[p.3]{Harlow_JT} describes the situation:
\begin{quote}
There is an analogous situation [to gravity] in electromagnetism: fields which carry
electric charge are unphysical unless they are ``dressed'' with Wilson lines attaching
them either to other fields with the opposite charge or to the boundary of spacetime. [...] One way to think about dressed observables in electromagnetism is that they create both a charged particle and its associated Coulomb field, which ensures that the
resulting configuration obeys the Gauss constraint.
\end{quote}

In sum: certain constraints are non-local in the sense that we are not free to simultaneously specify the values of corresponding physical quantities  completely independently at spatially distant points. 
 In the Hamiltonian theory, the non-locality can be, in a well-defined sense, put into direct correspondence with the gauge symmetry in question (as we will see in Chapter \ref{ch:Coulomb}). More broadly, the appearance of elliptic differential constraints among the equations of motion of the theory are associated not only to symmetry generators in the Hamiltonian formalism, but also, via the initial value problem, to an indeterminism about the future evolution of initial data: an indeterminism that precisely matches the isomorphisms of the theory.  

\subsubsection{The difference between canonical spatial diffeomorphisms and gauge transformations}\label{subsec:diffs}
By imposing constraints one can recover the symmetries. That is, both \eqref{eq:mom_ctraint}  and \eqref{eq:Gauss} represent an infinite set of constraints: one per spatial point. A linear combination of these constraints is thus given by an integral where the constraints are multiplied, or `smeared', by an appropriate coefficient function.  And such linear combinations generate, through the action of the Poisson bracket (cf. footnote \ref{ftnt:gens} and preceding text), infinitesimal diffeomorphisms and gauge transformations, respectively.

In other words, given the foliation of spacetime into $\Sigma \times \RR$ (cf. footnote \ref{ftnt:Sigma}), let $\xi^I\in C^\infty(\Sigma, \mathfrak{g})$ be a smooth Lie-algebra valued function and $X^i\in \frak{X}(\Sigma)$ be a smooth spatial vector field.  Then  with $\xi$ and $X$ giving the smearing, we obtain that the action of the vacuum constraints on the (doublet forming the) canonical pair ($(g, p)$ and $(A, E)$, respectively)  is:\footnote{We will denote the canonical Poisson brackets as $\{\cdot,\cdot\}$,
which, when applied to the fundamental phase space conjugate
variables of general relativity, $(g_{ij}(x),\pi^{ij}(y))$, yields: 
$
\{g_{ij}(x),\pi^{k\ell}(x')\}=\delta_{i}^{(k}\delta_{j}^{\ell)}\delta(x, x').
$; when applied to the  fundamental phase space conjugate variables of Yang-Mills theory, yields: $
\{A_{i}^I(x),E^{j}_J(x')\}=\delta_{i}^{j}\delta_{I}^{J}\delta(x, x')
$. It should be noted that we use the standard inner product on the Lie algebra: in this basis $\delta^{IJ}$, which thus does not distinguish between lower or upper indices. Equation \eqref{eq:gts_ham} is derived essentially using integration by parts and the cyclic trace identity. In Chapter \ref{ch:Coulomb}, we will derive these equations at a more leisurely pace (cf. Equation \eqref{eq:E_gt}). }
 \be\left\{ (g_{ij}(x), \pi^{ij}(x)),\int \d x'\, X^\ell \bar\nabla^k\pi_{\ell k}(x')\right\}=(\mathcal{L}_X g_{ij}(x), \mathcal{L}_X \pi^{ij}(x))  \; ,\ee
and 
 \be\label{eq:gts_ham} \left\{ (A^I_i(x), E^{iI}(x)),\int \d x'\, \xi_J\D^j E_j^J(x')\right\}=(\D_i \xi^I(x), [E^i, \xi]^I(x))  \;,\ee
 as we would expect from \eqref{eq:gauge_trans}. 
 
 But there is a difference between the two transformation laws of the canonical variables, that is important. The difference is that the transformation of the momenta, $E$, of the Yang-Mills potential is algebraic, or non-derivative,  in a way that the transformation of the gravitational momenta, $\pi$, is not. For instance, if the electric field vanishes at $x$, so will its flow under the symmetry generator; but a vanishing gravitational momentum at $x$ implies no such vanishing for its flow under the symmetry. This algebraic property of the electric field is merely the Hamiltonian version of the equivariance property of the gauge curvature; and it constitutes the main difference  that in Chapter \ref{ch:sd_2} I    advocated between the two sets of transformations (where the difference is dubbed $\Delta$).

\subsection{Quasi-local charges}\label{sec:charges}

The second comparison that I would like to address concerns the relation between symmetry and conserved charges. It is generally accepted that diffeomorphism invariance   implies that  the stress-energy tensor of matter
derived from a Lagrangian must be covariantly divergence-free; I agree.\footnote{More generally, one can show that local symmetries imply, through Noether's second theorem, that the equations of motion for the field will obey a conservation law corresponding to the conservation of charges; cf. Chapter \ref{ch:Noether}.}  What is more contentious is whether this counts as  a ``local conservation" of stress-energy and, if not,  whether this would distinguish diffeomorphisms from other symmetries of nature, such as gauge symmetries of Yang-Mills theories.  The folklore is that it does \emph{not} count as a local conservation law and thus that diffeomorphisms thereby differ from from gauge symmetries. 

We can here take \cite{Curiel_priv} to express the folklore:\footnote{I only single out Curiel since he puts the issue in his usual clear-thinking style, and not because I believe he has an unusually wrong opinion (I do not). As I said: he only here expresses an  opinion that is widespread even among the thoughtful, well-informed, specialists.}
\begin{quote}Killing fields are symmetries of individual solutions, but that, I think, is all one can say about [the] symmetries of the
theory [i.e. general relativity].  The theory does not seem to me to have any symmetries in the
standard sense of the term as it is used elsewhere in physics.  (That, inter
alia, is why I think that trying to cast it as a Yang-Mills theory is
doomed to failure.) [...] In the other cases where we apply Noether's second theorem and
talk about symmetries and conserved quantities, those conserved
quantities are conserved in the very strong sense that one can write
down local continuity equations that
can be integrated to yield global conservation laws.  One cannot do any
of this based on the fact that a tensor in a generic curved spacetime is
covariantly divergence-free.  So something that looks and acts a little
bit like a symmetry yields something that looks and acts a little bit
like a conservation law.  I find the differences more striking than the
similarities. 
\end{quote}
The role of this Section will be essentially to unpack and criticize this quotation; and, by so doing, to efface the distinction Curiel draws, at least between general relativity and \emph{non}-Abelian Yang-Mills theory. 

First, let us find a relation between local symmetries and conservation, \emph{\`a la} Noether's second theorem. Suppose that the action for a given theory in spacetime takes the following form:
\be
\label{eq:GR}
\int S= \d^4x \,(\mathcal{L}_g+\mathcal{L}_m);
\ee
where I will assume that all of the contributions of the matter fields to the Lagrangian are confined to the second component, $\mathcal{L}_m$, of the total Lagrangian density, and that both terms are individually diffeomorphism-invariant. I will also assume that the energy-momentum tensor for the matter fields is given by: 
\be \label{eq:eom_matter}
T^{ab}=\frac{\delta \mathcal{L}_m}{\delta {g_{ab}}}.
\ee
There is an   important physical feature of the gravitational field underlying equation \eqref{eq:eom_matter}: that the metric is nowhere vanishing and that it couples to every field. These two properties ensure that from the invariance under diffeomorphism of the Lagrangian and the transformation properties of the metric, we are able to deduce local conservation laws for the matter fields. 

Then,  assuming that infinitesimal diffeomorphisms generate the symmetries of the theory, the corresponding infinitesimal variation of the metric, from \eqref{eq:Lie_g} is 
 $\widehat{\delta g_{ab}}:= \mathcal{L}_{\mathbf{X}}{g_{ab}}=\nabla_{(a} X_{b)}$, for $X^a$ a vector field. Thus,  discounting boundary terms, we obtain, from a variation of the action functional \eqref{eq:GR}, setting $G^{ab}=\frac{\delta \mathcal{L}_g}{\delta {g_{ab}}}$:
\be\label{eq:inf_sym_gr} 0=\widehat{\delta  S}=\int \d^4 x((G^{ab}+T^{ab})\mathcal{L}_{\mathbf{X}}{g_{ab}}),
\ee
and after integration by parts, $\nabla^a T_{ab}=0$, even without imposing the equations of motion, i.e. the Einstein equations $G_{ab}=T_{ab}$.\footnote{As mentioned in Section \ref{subsec:drag}, in Equation \eqref{eq:Lie_drag}, here it becomes apparent that a `drag-along' construal of the Lie derivative would preempt this derivation of conservation. } The problem Curiel alludes to is that these \emph{local} conservation laws  cannot  be integrated to yield global conservation laws. 

For in order to obtain a symmetry-invariant quantity from the local conservation law, we must contract $\nabla^a T_{ab}$ with another auxiliary vector field $X$---in order to obtain a scalar---and integrate by parts.  Namely:
\be\label{eq:killing_int}\int\d^4 x\, \sqrt{g} X^b\nabla^a T_{ab}=-\int\d^4 x\, \sqrt{g} \nabla^{(a}  X^{b)}T_{ab}+\oint \sqrt{h}\,\d^3 x\, n^a X^b T_{ab}=0.
\ee
(here $\oint\sqrt{h}$ is the integral density at the boundary, which is a closed manifold).
Now supposing that $\nabla^{(a}  X^{b)}=0$, we get a physical quantity in a spacetime region whose change in time is solely due to a flux through the region's spatial boundaries. For example, if the integration region extends spatially to infinity, where we assume $T_{ab}=0$, then the second equation  says that the spatial integral of $X^aT_{a0}$ is conserved. That is, it takes the same value on an initial and on a final Cauchy surface, ${\Sigma_1}$ and ${\Sigma_2}$, respectively:
\be\int_{\Sigma_1}  \sqrt{h}\,\d^3 x\,  X^a T_{a0}=\int_{\Sigma_2}  \sqrt{h}\,\d^3 x\,  X^a T_{a0}.\ee
 Vector fields for which $\nabla^{(a}  X^{b)}=0$ are called \textit{Killing fields}. For example: from the above argument, for a  time-like Killing vector field $\pp_t$, one gets conservation of energy. 

But if $\nabla^{(a}  X^{b)}\neq 0$, the volume integral on the rhs of \eqref{eq:killing_int} does not vanish, and there is no such conservation law. 

And Curiel is also correct that, in electromagnetism, Noether's theorem guarantees that the symmetry gives rise to a  density whose  rate of change  in time  is solely due to a flux through the region's spatial boundaries. That is, for electromagnetism, the local conservation laws \emph{can} be integrated without obstruction. Namely, from $\pp^a F_{ab}=j_b$ (equation \eqref{eq:EM}, for the Abelian case), and taking the divergence we obtain 
$\pp^a j_a=0$. Integrating this equation, 
\be \int \pp^a j_a=\oint n^aj_a=0.\ee
Again, assuming that the current density vanishes at spatial infinity we obtain that the spatial integral of $j_0:=\rho$ is conserved in time. 

Here, notice the conservation law needs no special condition like the auxiliary vector field obeying $\nabla^{(a}X^{b)}=0$. 
So it would seem that at least Abelian Yang-Mills theory has a very different relation between symmetries and conserved charges.

 However, the \emph{non}-Abelian theory behaves, in this respect, exactly like general relativity. Namely, the local covariant conservation law is:
 \be\label{eq:Noether_covJ}\D_a J^a_I=0
 \ee
We can integrate $\D_a J^a_I$ against any Lie algebra-valued scalar $\xi\in C^\infty(M, \mathfrak{g})$: 
\be \label{eq:quasi_charge}
\int  (\D_{a} J^{a}_I\xi^I)=\oint n_{a} \,J^{a}_I \xi^I-\int J^{a}_I \D_{a} \xi^I=0.
\ee
But this is  a \emph{bona fide} regional conservation law only if the $\xi^I$'s are such that $\D_a \xi^I=0$; for otherwise the integral will not generically reduce to a boundary flux. As with general relativity, generic configurations have no such `stabilizers' (cf. footnotes \ref{ftnt:stab} and \ref{ftnt:stab2}): generically there are no solutions to $\D_a \xi^I=0$, just as there are no solutions to the Killing equations, $\nabla^{(a}  X^{b)}= 0$. In other words, to obtain physically significant conservation laws from the local conservation laws guaranteed by Noether's second theorem,  $X^a$ and $\xi^I$ must have some physical meaning.\footnote{Indeed, only symmetries that are related to stabilizers deserve the label `global symmetries', in the sense of having associated conserved charges.  \cite{BradingBrown_Noether} describe this as follows: ``we cannot follow the procedure used for global gauge
symmetries in the case of local gauge symmetries to form gauge-independent currents. A current that is dependent on the gauge transformation parameter is not satisfactory ---in particular,
such gauge-dependent quantities are not observable.'' One could, in fact, obtain a current $j^a_I$ that satisfies $\pp_a j^a_I=0$, as opposed to $\D_{a} J^{a}_I=0$. This is easy to do: just subtract from the covariantly conserved current the inhomogeneous term from the left hand side of the equations of motion, namely, $j^a_I:=J^a_I-[F^{ab}, A_b]_I$. The problem with using this current is that the conservation laws obtained will be gauge-dependent \cite{GomesNoether}. Here there is also a parallel to be drawn  with general relativity. 
In general relativity, one can use the same procedure as above to find $j_\mu^I\sim \tau_{\mu\nu}$, the so-called Landau-Lifschitz pseudo-tensor  (see e.g.: \cite[Ch. 15.3]{WeinbergQFT2},  \cite[Ch. 11]{LandauLifschitzVol2} and \citep{Seb_Noether} for a comprehensive philosophical and historical analysis of this point).   } 

In sum, both general relativity and \emph{non}-Abelian Yang-Mills theories lack a generic, symmetry-invariant definition of constancy. In both types of theory, it is only certain  configurations that admit the non-trivial automorphisms that implicitly define such `covariantly constant generators of transformations'. 

This establishes some caveats to our usual understanding of the Noether conserved currents. Such caveats are acknowledged in the physics literature, but their more radical consequences  are usually left unsaid. Ultimately, the very concept of a non-Abelian regionally conserved charge, like the concept of regionally conserved energy-momentum in general relativity,  only makes sense over `uniform'  backgrounds---in terms of $\D_a\xi^I=0$ and $\nabla_{(a}  X_{b)}= 0$.

\subsection{The Aharonov-Bohm effect}\label{sec:AB}

In this Section, I will describe why the Aharonov-Bohm effect cannot be used to draw important distinctions between the symmetries of general relativity and of Yang-Mills theory. In the first subsection I  will describe the effect, in the second subsection  I will discuss the gravitational analog of the effect, and whether the analogy still leaves room for a salient distinction between the symmetries of the two theories. This is mostly a technical discussion, focused on the comparison with gravity: I leave further philosophical analysis of the effect to Section \ref{decide!}, in Chapter \ref{ch:Coulomb}.
\subsubsection{The Aharonov-Bohm effect}\label{sec:AB_intro}
To investigate the physical significance of the gauge potential, Aharonov and Bohm proposed an electron interference experiment, in which a beam is split into two branches which go around a solenoid and are brought back together to form an interference pattern.\footnote{\citeauthor{aharonovbohm1959}'s work was  conducted independently of the work by \citet{ehrenberg1949refractive} who proposed the same experiment with a different framing in a work that did not receive much attention at the time. According to \citet{hiley2013ABeffect}, the effect was discovered ``at least three times before Aharonov and Bohm's paper''; with the first being a talk by Walter Franz, which described a similar experiment in a talk in 1939.} This solenoid is
perfectly shielded, so that no electron can penetrate inside and detect the magnetic field directly.\footnote{ Recently, \citet{ShechAB} and \citet{Earman2019} have challenged the idealisations associated with the Aharonov-Bohm effect, and \citet{DoughertyAB} has defended them.}  

The experiment involves two different set-ups---solenoid on or off---which produce two different interference patterns.  As
the magnetic flux in the solenoid changes, the interference fringes shift. And yet,  in both  set-ups, the   field-strength (i.e. the magnetic field)  along the possible paths of the charged particles is zero. So, the general outline of the experiment is: (a) the observable phenomena change when the current in the solenoid changes; and (b) the electrons that produce the phenomena are shielded from entering the region of non-zero magnetic fields; so (c) if we rule out unmediated action-at-a-distance,  whatever  physical difference  accounts for the change  must be located outside the solenoid. 

Thus, to explain the different patterns, one must either conjecture a non-local action of the field-strength upon the particles, or regard  the gauge potential as carrying ontic significance. Taking this second stance, the Aharonov-Bohm effect shows that the gauge potential cuts finer \emph{physical} distinctions than the field-strength tensor can distinguish. How much finer? 

We can simplify our treatment and imagine an electrostatic situation,  considering only the spatial configuration of the fields. In this case we identify the purely spatial component of the field-strength tensor with the magnetic field. Supposing the electron takes the paths $\gamma_1$ and $\gamma_2$ around the solenoid, we can infer from the
amount of the shift  that there is a field-dependent contribution to
the relative phase of electron paths that pass to the left and to the right of the solenoid, given by:\footnote{In units for which $e/\hbar c$=1. }
\be\label{eq:phase_AB} e^{i\Delta}=\exp{\left(i\oint_{\gamma_1\circ\gamma_2}\mathbf{A}\right)}. 
\ee
A gauge transformation  $\mathbf{A}\rightarrow \mathbf{A}+\d\phi$ will not affect \eqref{eq:phase_AB}, since the difference is the integral of an exact form---$\d \phi$---over a manifold without boundary, $\gamma_1\circ\gamma_2\simeq S^1$, and so must vanish. Thus the phase difference $\Delta$ cares only about the gauge-equivalence class of $\mathbf A$.\footnote{This means the phase  cares only about the principal connection $\omega$, not about how we represent it on spacetime. And, since the magnetic field vanishes outside the solenoid in both situations, the connection $\omega$ is different in the two situations, although the curvature of that
connection is the same, viz. zero.}

 To find out more precisely what physical information the equivalence classes of the gauge potential carry that goes beyond that encoded by the curvature,\footnote{In electromagnetism, the curvature encodes all the \emph{local} gauge-invariant degrees of freedom of the potential. In the non-Abelian theory,  traces of products of the curvature encode the \emph{local} gauge-invariant degrees of freedom. \label{ftnt:local_curv}} we  suppose that the underlying spatial manifold has a non-trivial topology, in the sense of a non-trivial de Rham cohomology $H^1(M):=\mathrm{Ker}\,\d^1/\mathrm{Im}\, \d^0\neq 0$, where $\d^1$ is the exterior derivative operator   acting on the space of 1-forms, and $\d^0$ is that same operator acting on smooth functions (or 0-forms). Then there are distinct equivalence classes $[\mathbf{A}^1]\neq [\mathbf{A}^2]$ that can nonetheless correspond to the same electric and magnetic field. More precisely, there are potentials $\mathbf{A}^1, \mathbf{A}^2$ such that:  $\mathbf{A}^1=\mathbf{A}^2+\mathbf{C}$ where $\d \mathbf{A}^1=:\mathbf{F}^1=\mathbf{F}^2:=\d \mathbf{A}^2$, and so $\d \mathbf{C}=0$, and yet  $\mathbf{C}\neq \d \phi$ (for any $\phi\in C^\infty(M)$). This implies $\mathbf{A}^1$ and $\mathbf{A}^2$ are not related by a gauge-transformation and so  are not in the same gauge-equivalence class. Their \emph{local} physical, or gauge-invariant content, represented by $\mathbf{F}$, matches, and yet they differ globally, or in their \emph{global} gauge-invariant content. (See \citep[Sec 4]{Belot1998} for a more thorough philosophical analysis of this paragraph's discussion).

In other words, $[\mathbf{A}]$ carries a local physical component---expressed in the magnetic field, or in the spatial part of the field-strength tensor, $\mathbf{F}$---and a non-local one: expressed in the cohomological content of $\mathbf{C}$. So, we take the field-strength tensor to capture the local, gauge-invariant, dynamical content of the gauge potential. But it doesn't exhaust the non-local  physical content of the gauge potential. 

 The Aharonov-Bohm effect  confirms  that the distinction between the  equivalence classes, $[\mathbf{A}^1]\neq [\mathbf{A}^2]$, that relies only on the non-local part,  is empirically significant: thus implying that the electric and magnetic fields, and the field-strength tensor as well, are not the  \emph{sole} bearers of ontic significance.\footnote{Philosophers have recently focused on questions about the locality and reality of the gauge potential \cite[cf.][]{healey1997ab,Belot1998, Maudlin_response,healey1999ab,nounou2003ab,Mattingly_gauge,Healey_book,lyre2009ab,Synopsis_gauge,Myrvold2010,Wallace_deflating} and \citet{Mulder_AB}.}

\subsubsection{Gravitational analogies and disanalogies}\label{sec:grav_analog}

In light of the Aharonov-Bohm effect, do we need to recalibrate our attitudes towards gauge symmetry? First, it is clear that the effect causes no trouble for the non-eliminativist, `sophisticated approach to symmetry-related models', discussed at length in  Chapter \ref{ch:sd_1}. While the sophisticated approach is not eliminativist, it still awards physical significance to those and only those quantities that are gauge-invariant; and the phase is such a quantity.

 Nonetheless, the effect has spurred eliminativists about gauge: in particular, those that endorse the holonomy interpretation of gauge theory.
The holonomy formalism, as we will discuss further in Section  \ref{sec:elim_elim}, takes as basic variables complex-valued loops on the spacetime manifold. There are several explanatory deficits of the holonomy interpretation (cf. footnote \ref{ftnt:tr_hol}), which I will leave for discussion in Section  \ref{sec:conclusions}. For now, I will just question whether the  Aharonov-Bohm effect is truly a distinctive feature of gauge theory as opposed to general relativity. If it is not, it should serve equally well---or equally poorly!---as a motivation for eliminativism in general relativity. 

There are by now several treatments of the analogues of the Aharonov-Bohm effect within general relativity (cf. \cite{Dowker1967, Anandan1977, Ford_1981}).
All hands agree that non-local effects of gravitational curvature can arise already at the classical level; (this is another distinction that I deem only peripheral to the topic of this Chapter; we will touch on it in Section \ref{decide!}, where we will discuss some broader morals of the Aharonov-Bohm effect). The treatment here is closest in spirit to both \citep{Dowker1967, Ford_1981}. But unlike those papers, I will not exhibit  solutions of the Einstein equations that can incorporate the essential features of the set-up; and unlike \citep{Anandan1977} I am also not interested in the experimental set-up required to verify this effect. The morals that I will draw are also similar in spirit to those of \cite[Section 5]{Weatherall2016_YMGR}. 

 As a first approximation we would like to find, in general relativity, two physically distinct situations in which the curvature remains zero in the entire region declared `accessible' to the system under investigation. 

The geometric curvature is defined analogously to  the gauge curvature: parallel propagate a vector around a loop and check whether it comes back to the original or has been rotated; the curvature measures this rotation, for an infinitesimal loop.  In a similar two-dimensional setting, the following two situations are closely analogous to the two distinct situations of the Aharonov-Bohm effect, i.e. solenoid on or off:
\\\indent (i) the parallel propagation of a vector along $\gamma_1$ and $\gamma_2$, in Euclidean or Minkowski space; and 
\\\indent (ii) before parallel propagating the vectors along the two curves, pick out a point between the two curves and `cut out' a  wedge from the spacetime, encompassing an angle $\theta$, and then stitch spacetime back together along the edges of the wedge. 

This second situation creates a cone, with a singular curvature at its apex, whose value depends on $\theta$.\footnote{ The same type of curvature defect could be obtained by a `cosmic string', cf. \citep{Ford_1981}.} In the first, but not the second situation, the vector will come back to itself, unrotated. In the second situation, there will be a relative rotation, depending on $\theta$. 

The singular curvature between the paths will affect the interference properties of a coherent beam of
particles such as neutrons, or  indeed, of any system whose state has a vector component, e.g. an axis of rotation of a gyroscope.  Thus, for example, in situations like (ii), neutrons which traverse a region of
space where the curvature is identically zero are  nonetheless capable of detecting the
effects of curvature in a far away region of space-time, i.e. at the conical singularities.

Lastly, we can also impute to the gravitational case a cohomological understanding of the non-local effect. For  there is a straightforward extension of the usual de Rham cohomology to flat vector bundles (see e.g. \cite[Ch. 5]{voisin}),  and thus we can, much like in the electromagnetic case,  attribute the difference of the two gravitational situations to different spin connections  whose associated curvature is identically zero in the acessible regions, but which have different cohomological contributions. So, as far as symmetries go, the analogy with general relativity seems again very tight.

 One often repeated objection here is that   for each choice of gauge potential, the profile of the phase gained along the trajectory will look different  \citep[Section 6]{Healey2004}, \citep[Ch. 2]{Healey_book}. Thus there can be no physically significant local accrual of a phase. 
 
    And yes, it is true that there is a type of gravitational  Aharonov-Bohm effect, essentially based on proper time, that is incrementally accrued. This is the type of effect that one would obtain for the phase difference of a spinless particle. But this type of effect is less interesting, as it does not require closed loops: it only requires the selection of particular points along each trajectory (cf. \citep[Section 6]{Healey2004}).\footnote{This caveat may be also related to the importance of quantum physics in  the original Aharonov-Bohm proposal: in the gauge potential case,  we require a closed loop to observe the effect, open trajectories will not do. That is  because there is a single charge that is in a \emph{quantum} superposition along the two trajectories. Otherwise, we could make sense of a phase difference in a gauge-invariant way for open paths as follows: given two charges, with associated wavefunctions $\psi_1(x_1), \psi_2(x_2)$, respectively at points $x_1\in \gamma_1$ and $x_2\in \gamma_2$, we can compare the phases of the two in a gauge-invariant way by transporting along the segment of $\gamma:=\gamma_1\circ\gamma_2$ that connects them. Namely, we transport the phase of $\psi_1(x_1) $ along $\gamma$ to $x_2$ and compare it with the phase of $\psi_2$; this is an invariant (since $\psi_1(x_1)\exp{(i\int_{x_1}^{x_2} A)}\psi_2^{-1}(x_2)$ is gauge-invariant). \label{ftnt:gi_hol} } An analogy between the gravitational and the gauge Aharonov-Bohm effects requires the use of vectors (or sections of vector bundles) in each case.

  And once we have restricted the analogy, the argument about local accrual cuts both ways. For we  should note that the dependence on a choice of gauge only occurs if we use a local spacetime representative $\mathbf{A}$ of the global principal connection $\omega$. That is, if we use only the connection to parallel transport the phase, there will be no dependence on a choice of gauge.
   Nonetheless, using $\omega$ won't help `localize' the accrual of relative rotation:   if the particle's phase  is being parallel transported, its phase is just constant:  parallel transport after all is what defines a constancy of phase across the value spaces over different fibers. Similarly, the direction of a spacetime vector being parallel transported along a curve is either  just constant or ill-defined, in the sense of being dependent on the choice of  coordinates used to describe the rotation. 
 
   So, my point is that in both cases the effect is holistic: although the total effect   is measurable, there simply is no fact of the matter as to how this effect comes about as the result of small, locally accrued differences. There are just facts about parallel transport that are evinced only globally, or rather, only when the paths reconverge.

    A second objection is based on \citep{Anandan_1993} and is highlighted by  \citet[Secs. 6-7]{Healey2004} as \emph{the} main difference between the gravitational and the gauge potential-based Aharonov-Bohm effect.  The objection focuses on one sense in which  the vector rotation can be construed as locally accrued. Namely, since tangent vectors are `soldered onto' spacetime,  the  angle between the parallel transported vector and the tangent to the curve \emph{is} locally accrued. Thus, for example, if the vector was just tangent to one of the trajectories, the accrued difference would be a function of the total intrinsic acceleration (and thus related to the  difference in total elapsed proper times).\footnote{ So I take the  angle between the parallel transported vector and the tangent to the curve to be locally accrued even if that angle vanishes. In this point I depart from \citet[Section 5]{Weatherall2016_YMGR}'s kindred criticism (of Healey's argument for a disanalogy). For Weatherall argues  that a reference of constancy is only meaningful if the particle paths are geodesic, which would constrain the angle to vanish throughout motion. Thus, he says, there would be no local accrual.  But there is a difference between  local accrual not being well-defined and a vanishing accrual. Moreover,  I also believe one could have a meaningful notion of constancy even for non-geodesics, e.g. for an accelerating rocket.} 
    
     But this second objection cuts ice only in the simplified two-dimensional treatment I gave above, and is discarded when more detail is added. That is,  to assess the relative rotation of  the spin of a particle such as that of the neutron, i.e. to assess the relative rotation of polarization vectors, we must use Fermi-Walker transport. In other words, we are calculating a type of \emph{Thomas precession}, which is about the rotation of a spatial vector (a 3-vector),  i.e. the rotation in the plane orthogonal to the timelike trajectory of the particle.  Of course, the angle between a polarization 3-vector and the tangent to the curve is also constantly zero; there is no coordinate-independent way to locally  measure the rotation of the polarization vector. To put it differently: comparing the parallel transported 4-vector to the tangent to the trajectory allows us to locally determine the evolution of one degree of freedom of the 4-vector; but this still leaves open how the remaining polarization degrees of freedom---three if the particle is massive, two if it is massless---evolve along the trajectory. For these, all we can do is compare a relative rotation upon the reconvergence of the paths,  and thus the qualitative analysis presented above still holds.
  
  One disanalogy between the gravitational and the gauge Aharonov-Bohm effects remains: about shielding. We can, in the laboratory, easily shield magnetic sources from the paths of the electron. Shielding gravitational sources, however, is not so easy:\footnote{ See \cite{ChruscielBeig_shield, CarlottoSchoen2016} for some interesting shielding results in perturbative and non-perturbative gravity, respectively. } for that, we may need more esoteric gravitational objects, such as cosmic strings. But  I take this feature as not germane to the comparison. For it is of course due to the particular dynamics of the two theories---electromagnetism is not gravity!---and not due to  the character of their symmetries.   
  
 In sum, apart from differences that are due to dynamical features of gravity (such as, the absence  of particles with negative mass, that would allow shielding), there seems to be no conceptual distinction between the Aharonov-Bohm effect and parallel transport around a conical defect in general relativity. In both cases, one cannot, by surveying a neighborhood of the particles' trajectories, infer whether they will experience a relative shift when they are once again reunited. But a mystery arises only  if we assign undue significance to the coordinate choices used to evaluate global rotations. Thus one cannot use the effect to advocate eliminativism for gauge theories but not for general relativity. 

\section{Healey against localized gauge potential properties}\label{sec:Healey}


 In this Section  I  take issue with one argument in Healey’s (otherwise outstanding!) book on the philosophical interpretation of gauge theories. It occurs in his Chapter on classical gauge theories (Chapter 4); and it is directed at an interpretation of the theories that Healey calls `the localized gauge potential properties view’.
 

 Healey takes the `localized gauge potentials properties view' to  postulate that the gauge potential $A$ at a spacetime point $x$ represents a physically real property of, or at, $x$.  This is, in essence, a quidditist viewpoint, according to  which properties and relations have a nature that outstrips their patterns of instantiation in objects and co-instantiation with each other. The viewpoint has been defended, for instance in \cite[Ch. 6]{Arntzenius_book}. And
\cite[Section 4.2]{Healey_book} is devoted to assessing, indeed rebutting, this view. And although I also reject this quidditist view, I will disagree with Healey's arguments, which are  distinctively philosophical.

Healey's main argument  against the localized gauge potential properties view  is that it suffers from a massive or radical under-determination---of a kind familiar in philosophy, especially associated with the labels `multiple realizability’, and `permutation argument’. More problematically, Healey argues that this threat of under-determination is unique to gauge theories.

In broad lines, my reply will be   that the under-determination is avoided by a `structuralist’ construal of the properties in question, that is as tenable as---no more dubious or controversial than---the `structuralist’ construal of spacetime points (that Healey himself endorses). Since I have already argued at length in favor of these views in Chapter \ref{ch:sd_1}, one could reasonably leave it at that. But in making his argument, Healey touches on an interesting topic: that of specifying, amongst the infinitely many physically equivalent representatives, a particular spacetime distribution of the gauge potentials or of the metric. As I understand Healey, he posits that this specification is easy for the metric, but impossible for the gauge potential. And with this alleged contrast, I will disagree.


I will start in Section  \ref{sec:healey_intro} by laying out Healey's argument in more detail and furnishing what I take to be the more interesting challenge that can be read into his argument. Then I will give my answer in Section  \ref{sec:antiheal}. 


\subsection{Healey's argument from functional roles}\label{sec:healey_intro}

To spell out Healey's  argument in more detail, I will now indulge in a bit of  `Healey exegesis'. 
Healey admits that within a theory there may be many terms for unobservable items; but unobservability by itself is not bad news, since we can employ \citet{Lewis_defT, Lewis_func}'s ideas about simultaneously specifying several theoretical  items that are, in some sense, problematic,  by their each uniquely satisfying some description (usually called ``functional role'') that can be formulated in terms of  less problematic items.  
Lewis's ideas allow us to  fix what such theoretical (or, less broadly,   unobservable) 
terms in a theory refer to, without having  a prior interpretation of those terms, by describing how they fit in a pattern of better understood (or, less broadly, observable) items.

In Lewis's framework, functional roles such as the one Healey here discusses  usually involve a binary division of our theory's vocabulary: a  `troublesome' part, that  we denote with a $T$,  and an `okay' part, that  we denote with an $O$, whose members' reference is already fixed; (so that $T$ is less well understood than $O$). Each $T$-term is to be specified by satisfying a certain pattern of relations, whereas the $O$-vocabulary is assumed to be already interpreted.\footnote{  
Here is Lewis describing $O$ and $T$-terms: `` `$T$-term' need not mean `theoretical term', and `$O$-term' need not mean `observational term' [...] `$O$' does not stand for `observational'. [...] They are just any old terms'' \cite[p. 250]{Lewis_func}. 
So despite the letters `T' and `O', Lewis' proposals are not only about the theory-observation distinction. See \cite{ButterfieldGomes_1} for a recent analysis of functionalism as a species of reduction.}
 So the leading idea of functional definition (and here, of identifying particular realizers for the gauge potential by their patterns of instantiation) is to use $O$ and $T$-terms to jointly fix the reference of the $T$-terms. 
 
 But Healey argues that, in a quidditist interpretation of  gauge theories,  this Lewisian strategy is bound to be plagued by under-determination. We will see the details of his arguments in a moment. I will give two responses: one in Section \ref{subsec:straight} that is a straightforward rejection of quidditism, already developed in Chapter \ref{ch:sd_1} and which I deem to be less interesting. But in Section \ref{subsec:interesting} I also want to give another response, that rebuts a more interesting construal of the details of Healey's arguments. This second response will give us a taste of the power and flexibility of \emph{representational conventions}, mentioned at the end of Section \ref{sec:method}. 
 
 \subsubsection{The straightforward challenge}\label{subsec:straight}
I  illustrate this challenge with Healey's chosen example, based on a toy-theory of  coloured quarks \cite[p. 94]{Healey_book}. In a world in which this toy theory is true, particles can  have one of three colours: red, green and blue. Although colours figure in the laws of the toy theory, these laws are invariant under colour-permutation symmetry.  One way to realize this symmetry is to have the particles be dynamically confined in colour-neutral combinations of red, green and blue. Healey thus supposes that in this world it is a law of nature that red, green and blue always  occur coulour-neutrally and that 
there is no further distinction between the colour-carrying particles; once they occur in certain colour-neutral combinations, not even their locations can differ. 

Healey then argues, correctly, that in this set-up the terms `green', `red' and `blue' are referentially
indeterminate. For $x_1$ could stand for any colour, as long as $x_2$ and $x_3$ stand for the remaining two colours. Thus the individual colour terms cannot directly refer.

 This challenge   fits comfortably into the debate we have already encountered in Chapter \ref{ch:sd_1} and in this Section's preamble: namely, about whether to  conceive a property  as having an intrinsic nature independent of its patterns of association with other properties (quidditism), or not (anti-quidditism). And as we saw at length in Chapter \ref{ch:sd_1}, there is an analogous philosophical  debate about objects, rather than properties. 
 Healey himself rehearses this  debate and says (as most authors do) that the best response to under-determination, for someone who believes that spacetime points are objects, is to take an anti-haecceitist view of spacetime points’ individuation.

In that jargon, the structuralist response against under-determination is easy to state: if all the properties of a certain kind are each exhausted by their each filling a certain theoretical role (of course, with different roles for different properties)---in other words: if each property has no further ``intrinsic nature''---then permuting which properties fill which roles make no sense. There is no such permutation. Similarly, in the general-relativistic case, if we specify points by their chronogeometric relations, then permuting which points fill which roles makes no sense.\footnote{See Section \ref{subsec:skeptic_why}.}

In sum, I agree with Healey about the problem: a literal understanding of symmetry-related gauge potential distributions across spacetime as being physically distinct is untenable.  In other words, for general relativity as much as for gauge theory, a literal interpretation of models would give rise to  radical indeterminism. But I disagree about what is the best solution: I advocate sophistication while Healey advocates eliminativism.  
 I find the sophistication response---judging all of these  different distributions to be physically equivalent---simple and convincing, and applicable to both general relativity and Yang-Mills theories.

 After this rather  perfunctory statement, I proceed to the more interesting challenge lurking in Healey's objections. \\

\subsubsection{The interesting challenge}\label{subsec:interesting}
If one endorses quidditism, as David Lewis did, there is still a second line of response to the threat of physical underdetermination: to appeal to patterns of facts   of ``geography'' to break the underdetermination. As described by   \citet{LewisRamsey}:
\begin{quote}
Should we worry about symmetries, for instance the symmetry
between positive and negative charge? No: even if positive and negative
charge were exactly alike in their nomological roles, it would still be true
that negative charge is found in the outlying parts of atoms hereabouts,
and positive charge is found in the central parts. O-language has the
resources to say so,  and we may assume that the postulate mentions whatever
it takes \textit{to break} such symmetries. Thus the theoretical roles of positive
and negative charge are not purely nomological roles; \textit{they are locational
roles as well}.  [my italic]
\end{quote}

Based on his use of Lewis's ideas, I interpret Healey as saying that  one can   functionally specify a  spacetime metric, but cannot specify a  gauge potential. 
More precisely, here is Healey's argument that the functionalist methodology applies so as to single out spacetime metrics, but not to single out gauge potentials:
\begin{quote}
The idea seems to be to secure unique realization of the terms [...] 
in face of the assumed symmetry of the fundamental theory in which they figure by adding one or more sentences [namely, $S$] stating what might be thought of as ``initial conditions'' to the laws of that theory. These sentences would be formulated almost exclusively in what Lewis calls the $O$-language---i.e. the language that is available to us without the benefit of the term-introducing theory $T$. But they would also use one or more of the [symmetry-related]  terms [...] 
to break the symmetry of how these terms figure in  $T$. They would do this \textit{by applying further constraints} [namely, $S$] that must be met by the denotations of these terms in order that $S\&T$ be true. Those constraints would then fix the actual denotation of the [...] 
[symmetry-related terms] in $T$ so that, subject to these further constraints, $T$ is uniquely realized. [...] [But]  \textit{The gauge symmetry of the theory would prevent us from being able to say or otherwise specify which among an infinity of distinct distributions so represented or described is realized in that situation. This is of course, not the case for general relativity.}  \citep[p. 93]{Healey_book} [my italics]
\end{quote}

But I will ask: Why is this ``of course not the case for general relativity''? And why does Healey see a contrast between general relativity and gauge theory? These questions are central for this Chapter.  

For I do not see in this entire passage an attempt to draw a distinction between anti-quidditism for gauge and anti-haecceitism for gravity, per se.   I believe the more interesting interpretation of  this passage is as an attempt to address other questions, about the \emph{use} of the theories. 
I think the interesting question being alluded to here is whether we can use features of the world around us to single out  a unique model of the theory, or a model with unique features. This other question is interesting because, in practice, we \emph{do} select some model over others when we represent a given physical situation, and therefore in using  the theory we must `break the symmetry' between all of the models. Lewis takes this breaking to justify a type of quidditism; but I do not: I think it is solely based on pragmatic concerns, to do with the \emph{use} of theory.

In more detail:   suppose  that  we cannot find a perspicuous interpretation of general relativity  that includes just the symmetry-invariant quantities as part of the basic syntax of the theory,  as argued in Chapter \ref{ch:sd_1}. And suppose we agree, as I have argued in Chapter \ref{ch:sd_1}, that the best we can then do  is to keep all of the symmetry-related representations on a par. Then we are faced with a mystery:  in practice we \emph{do} select particular representatives over others. For instance, in specifying the local metric structure, with the symmetry invariant chart-based Definition \ref{def:chart}, we in effect must choose one form for the metric.   No particular choice is mandatory, but each must be based on something: physical features, indexicals, ostension, etc.

Thus I take the more interesting interpretation of Healey's passage here to be that this `singling out' of particular models is possible for gravity, but not for gauge theory. If this were so---if there was literally nothing on which we could base our choices of gauge potential representative---we would be more motivated to seek out reduction or elimination for  gauge theory than for gravity.\footnote{And indeed, later in the book Healey uses this distinction as a motivation for seeking  a different, symmetry-invariant ontology of gauge theory, based on holonomies, which I  criticized in Section  \ref{sec:elim_elim} see footnote \ref {ftnt:tr_hol}.} \\

\subsection{Refuting the distinction using representational conventions}\label{sec:antiheal}
\emph{Contra} Healey,  I will argue that having some physical ``hook'' with which to choose representatives does \emph{not} imply that we are breaking the symmetry at a  fundamental level. Different choices of representational conventions would be equally capable of representing a given state of affairs; some may just be  more cumbersome than others, or they obscure matters for the purposes
at hand, even while they may shine light on  complementary aspects of that state of affairs. 
 And in this sense,  we can shift our focus to different features of the world, according to our interest, and thereby single out different representative    models---within both general relativity and Yang-Mills theory.  This Section builds on Section \ref{sec:rep_conv}, and it will be elaborated in one direction in Chapter \ref{ch:Coulomb}---where we find explicit physical cues for  a particular choice of convention.

We are now ready  to rebut Healey's argument from functional roles.
As we have seen, the argument brings an interesting question to the fore: under what conditions could we be justified in choosing for the gauge potential one spacetime distribution over another?
 Selecting such a representative  involves  a tension between: (i) a structural construal of physical properties---as  ones that are invariant under the symmetries in question---and (ii) in practice  selecting unique   representative distributions of the gauge potential, among the infinitely many representatives of the same situation. At first sight, these two requirements, (i) and (ii), are inimical, if not contradictory, for (i) implies we can have no physical guidance for accomplishing (ii)! In the jargon of Section \ref{sec:method}: how do we choose representational conventions?\footnote{A reminder: a representational convention is an injective map $\sigma:[\F]\rightarrow \F$, $[\varphi]\mapsto \sigma([\varphi])\in \F$. But again, since we cannot usually represent elements $[\varphi]$ of $[\F]$ intrinsically, we replace $\sigma$ by an equivalent projection operator that takes any element of a given orbit to the range of $\sigma$, i.e.  $h:\F\rightarrow \F$, or more specifically, $h:\F\rightarrow \text{range}(\sigma)$, with $\mathcal{O}_\varphi\subset \F$ being mapped by $h$ into $\sigma([\varphi])$.} 
 

Below I will show that we can construct a particular representative of the gauge potential as fulfilling a given role, and explicitly check that such a notion is invariant under the permutations of properties. In more detail: I will first show,  in Section \ref{subsec:rep_conv_Healey}, that (1) intra-theoretic resources enable us to pick out gauge representatives; and then in Section \ref{subsec:indexicals}, I  show that (2) indexicals make no difference
for the unique specification of the metric. Point (1) will be buttressed in Chapter \ref{ch:Coulomb}, where we give a very specific reason to choose a very specific convention (Coulomb gauge). \\


\subsubsection{Representational conventions to the rescue}\label{subsec:rep_conv_Healey}
 Starting with (1), I will resolve the tension between (i) and (ii) with explicit examples; by,  in Healey's words: `\textit{breaking} the symmetries', by providing `\textit{further constraints}',  such that we fix the  denotation of `a section $\sigma$', or, equivalently, a particular gauge-potential, $\mathbf{A}$, that is, a particular spacetime representation of each value of the principal connection $\omega$.
 
  Note to begin with that, according to Healey's standards, we are justified in including in our $O$-vocabulary all the `locational roles', which  describe contingent, happenstantial facts about  `where and when' specified events happen; and which I will loosely interpret as `referring to spacetime'. Thus I free myself to include in the $O$-vocabulary, and thereby use in the specification of the roles, the differential geometry of spacetime.  

I will first expound the functional roles in the case of the gauge potential, and then draw the analogies with the metric.

In the simple example of electromagnetism, we require the model  to satisfy certain relations among the parts of the field. For example, in our hierarchy of extra-empirical theoretical virtues, we could place Lorentz covariance very highly (cf. \cite{Mulder_AB} and  \cite{Mattingly_gauge} for advocacy of this criterion and choice of gauge) and therefore prefer an explicitly Lorentz-covariant choice of convention:\footnote{This example is merely illustrative, as in the Lorentzian setting this gauge-fixing is not complete and we would require additional imput about the initial state. But in the Hamiltonian setting, Chapter \ref{ch:Coulomb} lays out a closely related, bona-fide example, including in the case of bounded manifolds. \label{ftnt:illus} }
\be\label{eq:Landau} \mathcal{F}(A):=\nabla^a A_a\equiv \nabla^a (\sigma^*\omega)_a =0.\ee
With this choice, the equations of motion for $A_a$, Equation \eqref{eq:gen_eom_A} becomes the  hyperbolic equation (which gives a well-posed IVP: see Section \ref{subsec:IVP}):
\be \square A_a=0,
\ee
where  $\square:=\nabla^a\nabla_a$ is the  d'Alembertian. 

Here the value of the connection $\omega$ is fixed: all we are trying to do is to determine a particular, state-dependent section $\sigma(A)$ (seen as a state-dependent submanifold of the bundle $P$). The only extra constraints that we have imposed in this equation, namely, that the spacetime divergence of the particular representative of $\omega$ vanishes, use only  the $O$-vocabulary that Healey would grant us, as I mentioned, and therefore should qualify as providing `actual denotation' according to his standards. 

We can also explicitly display in $O$-vocabulary the projection of an arbitrary representative of $\omega$ into a representative satisfying \eqref{eq:Landau}. 
Namely, given any representative of any equivalence class, $\mathbf{A}$, we define, as in Equation \eqref{eq:proj_h}: 
  \be\label{eq:h_proj}h(\mathbf{A})_a:=A_a-i\nabla_a(\square^{-2}\nabla^b A_b),\ee
  where $A_a$ is any 1-form and $\square^{-1}$ is a propagator (which would require more input to be fully determined: cf. footnote \ref{ftnt:illus}). Much as in other representations of  gauge-invariant quantities---such as in the holonomy interpretation---fixing the gauge is  non-local   in the following sense: just as $\int A$ requires the value of of $A$ at several points simultaneously as an input, the projected state $h_\sigma(A)$ requires the value of $A$ throughout the region as  input. This is just a reflection of the non-local aspects of gauge-invariant functions  (cf.  \cite[p. 460]{Earman_local}, \cite[Ch. 4.5]{Healey_book},  \citep{Strocchi_phil, GomesStudies} and the appendix of \cite{GomesButterfield_electro}). 

 As required (cf. Section \ref{sec:rep_conv}), the choice is not physically restrictive. Given any $\mathbf{A}$, we can translate it along the fibers, looking for a  representation of the field that satisfies equation \eqref{eq:Landau};  and for any $\mathbf{A}$,  $h(\mathbf{A})_a$ is a potential that is related to $\mathbf{A}$ by a gauge-transformation, namely $g_\sigma(\mathbf{A})=\square^{-2}\nabla^b A_b$. 
Of course we still have the freedom to change the section $\sigma$, but a different section would not satisfy the original  condition that    uniquely specified $\sigma$.

  And it follows from this construction that the determination of the representative is structural, in the sense that, though $h(\mathbf A)_a$ is, like $\mathbf{A}$,  a Lie-algebra valued one-form,  $h(\mathbf{A})_a$ \textit{is a gauge-invariant functional of $\mathbf{A}$}. That is, for a gauge-related $A'_a=A_a+\nabla_a\chi$, we obtain $h(\mathbf{A}')_a=h(\mathbf{A})_a$, and therefore $h(\mathbf{A})_a$ can be construed as a structural property of the field.  Moreover, it is an ``exhaustive property'', in the sense that each equivalence class (or physical world, according to the theory) will project to a  single $h(\mathbf{A})_a$. 

  Within a single world, or physical situation, or equivalence class $[\mathbf{A}]$,  permuting among the infinity of distinct representative  distributions, namely, permuting among the corresponding $\mathbf{A}$'s, makes no difference to $h(\mathbf{A})$. 
The properties  of $h(A)$ are  exhausted by  its filling a certain theoretical role, or pattern of instantiation: reflecting anti-quidditism. 
  
   Thus, the representational convention:  `find a representation of the electromagnetic potential that is divergence-free', \emph{\` a la} \eqref{eq:Landau},  is a gauge-invariant specification. Nonetheless, it successfully pins down a representation for the electromagnetic potential. 
   This construction thus explicitly contradicts the letter of Healey's under-determination argument in the quotation above.\footnote{Note that we are fixing the section, or, passive transformations, for $\omega$, in accord with the lessons from Chapter \ref{ch:sd_1}, where we invoked the active-passive correspondence in order to clarify  symmetry-invariant structure. Similarly in the case of general relativity, we would in practice   abandon the abstract index notation of tensors and employ equations in particular coordinate systems: the representational convention would be in effect fixing the coordinate system in a state-dependent way. But, formally, a representational convention would work equally well for the active transformations. Then we would obtain $h$ of \eqref{eq:h_proj} as a projection map in the space of models, $\mathcal{M}$: all symmetry-related distributions project down to  the same representative that fills one such role. But technically, it is much more convenient to use the active-passive correspondence to articulate the convention as applied to charts, or trivializations. \label{ftnt:abstract}} \\

  One could still ask what conditions could possibly suggest a choice such as  \eqref{eq:Landau}. The answer, hinted at above,  is that different  pragmatic and theoretical virtues can motivate  different choices.  Of course,  the representational convention cannot be empirically mandated, since, by assumption, the symmetries leave all empirical matters invariant, in both the gravitational and the gauge cases. A uniquely specifying functional role such as \eqref{eq:Landau} can, even should, be one of \textit{mere preference} for particular representations of the field; it is at most suggested by being \emph{suitable} for certain types of questions one might want to ask about the system (such as ``does it respect relativistic causality''?). It is  a pragmatically-guided choice of ``coordinates''.  
  

  In the gauge case, different pragmatic criteria are in play. As I said, the choice of \eqref{eq:Landau} is explicitly Lorentz covariant (and has its virtues thoroughly extolled by \cite{Mattingly_gauge}, who argues  that it should be considered as \cite{Maudlin_response}'s ``ONE TRUE GAUGE'').  \cite{Maudlin_ontology} himself endorses a different choice, Coulomb gauge, to be explored in Chapter \ref{ch:Coulomb}. 

The point, as described in Section \ref{sec:rep_conv},  is that any choice that gives rise to a fully gauge-invariant and complete projection operator equally captures the structure of the states. Different choices represent different lenses through which we capture that structure. Thus we may want to highlight the helicity degrees of freedom of the theory, in which case we would use temporal gauge. Or again, we might choose to split the electric field into one component that is purely `electrostatic'---or rather, due solely  to a Coulombic potential---and another that is purely radiative, as we will describe in Chapter \ref{ch:Coulomb}. Such a   split  for the electric field  corresponds symplectically to a Coulomb gauge for the spatial vector potential: the gauge potential splits into a term that is ``pure gauge'' and one that is radiative (or in Coulomb gauge). 

  
Besides, this discussion   applies equally to the metric, as promised in Section \ref{subsec:rep_inv}. To recap: when we gauge-fix the representational conventions for the metric  (say using harmonic gauge), we functionally specify a single local representative of the  invariant structure by the use of some extra condition $S$ (as we did for electromgnetism).

And this flexibility is also explanatory, or at least is able to shed light on important physical features. Just as is easy to explain the Larmor effect by a Lorentz boost between different frames, the ability to choose different conventions, or gauge-fixings, makes it easy  to explain that a given process in quantum electrodynamics involves just two physical polarisation states \textit{and} that it is Lorentz invariant. In both the special relativistic and the gauge scenarios, two different `frames'---for quantum field theory, the temporal and Lorenz gauge---are necessary to easily explain two different aspects of a given phenomenon.
 
 This concludes the first part, (1), of my response to what I called the `interesting challenge': showing that intra-theoretic resources enable us to choose representational conventions and  pick out representatives in all kinds of theories with symmetries. Now on to the matter of indexicals. \\

\subsubsection{Indexicals can also be internal}\label{subsec:indexicals}
 
(2): Here is one possible objection: ``Fine'', you, or Healey, might say, ``we use the same tools to specify local representatives of both types of fields (amongst all of their symmetry-related models). But I can use \emph{indexicals} to specify a metric and I cannot do the same to specify gauge potentials''. 

The idea here is in effect that you could specify the metric along your own worldline. But the idea is misguided. For to specify the entire metric you will still need to specify a linear frame along your worldline, and how do you do that? You must use relations to other objects, fields, or features of the metric field itself (such as anisotropy directions, or even proper time). All of these are just other (types of) conjuncts of the `functional role' specifying the (particular representative of) the metric,  of the same sort that could be appealed to for  the gauge potential.
 
 Besides,   you could do something similar with electromagnetic interactions: this is what we do when we interpret `photons' that propagate and hit our retinas as purely radiative, i.e. as having no polarization along the time axis or along the direction of their motion.\footnote{There is here a slight awkwardness in the physics lingo. The `photon field' is taken as $A_\mu$, but a `photon' is taken to be radiative, usually written  $p^\mu A_\mu=0$.} We could also  indexically fix the gauge by resorting to the charged matter fields. In the spirit of the opening passage of Yang and Mills' original paper \cite{YangMills}, and suppressing our everyday acquaintance with the  `medium-sized dry goods' of spacetime, I could say: ``what is trapped in this tiny (sub-atomic) box is a proton and not a neutron''. And if, even after all this, you still insist that indexicals somehow apply only to  external relations such as relative distance, and not to internal relations such as relative charge, I could only  retort  that this exceptionalism would  beg the question. 
  \\

Thus, under closer scrutiny,  the distinction suggested by \cite{Healey_book} crumbles. 

\section{Summing up }\label{sec:conclusions_sd2}

Although I have focused on general relativity as the best-known  diffeomorphism-invariant theory,   I have kept out of the discussion those idiosyncrasies of the theory that are of a more dynamical nature. 
In Section \ref{sec:pos_argument} I reported the only substantial distinction I found---$\Delta$: Yang-Mills theory, but not general relativity, \emph{admits a formalism in which the local, dynamical content of the theory is fully  invariant under the appropriate symmetry transformations.} Thus, putting these issues aside, in Section  \ref{sec:common} I could find no other salient difference between gauge symmetries and diffeomorphisms for the  three topics considered:  (i) as to the constraints, a difference exists but it is due to the problem of time (Section  \ref{sec:syms_consts}); (ii) as to  the association between symmetries and conserved charges, a difference exists but it is due to  an Abelian vs. non-Abelian nature of the symmetries (Section  \ref{sec:charges}); and  (iii) as to the Aharonov-Bohm effect (Section  \ref{sec:AB}), although there may be practical  difficulties involved in `shielding'  any source of gravitational curvature, I could find no salient difference: the rotation of the electron phase (respectively, vector) along an open path is either zero or has an arbitrary dependence on the section (resp. coordinates). In both cases the shift in rotation can  be invariantly measured only for closed paths, and then it carries non-local information. In Section \ref{sec:Healey}, I addressed Healey's  different distinction than those studied in Section  \ref{sec:common}, and argued that it also stumbled. In Section \ref{sec:pow_rep}, below, I summarize the lessons learned from Section \ref{sec:Healey}'s refutation of that distinction. Then, in Section \ref{sec:elim_elim}, I show that the difference, $\Delta$, that I conceded in Section \ref{sec:pos_argument}, cannot be used to help the eliminativist. And in Section \ref{sec:conc_sd2}, I briefly conclude.

\subsection{The powers of representational conventions}\label{sec:pow_rep}

   Healey articulates the conceptual analogy between the under-determination argument for gauge and the hole argument. Although the analogy is close, mathematically as well as philosophically (for general relativity also can be formulated as a gauge theory, see e.g.  \cite[Part III, Ch. 5]{Baez_book} and \cite[Ch. 8]{Bleecker}) Healey  sees what he considers an important disanalogy with general relativity; and it is this disanalogy that I  rejected  in  section \ref{sec:Healey}. 

  And although Healey's target is a form of quidditism---a target I also shoot down---his arguments attempt to draw a distinction from general relativity that I found unconvincing. For Healey believes that one can use \cite{Lewis_func, Lewis_defT}'s ideas about functionalism  to fix a particular representative of the metric, i.e. a particular local distribution of properties;  but one cannot do the same for the gauge potential. If this were  so, it would indeed point to a salient difference between spacetime diffeomorphisms and gauge transformations: the latter but not the former would leave us no handle with which to fix representation. Thus gauge under-determination would be of a more problematic sort, and eliminativism---the attempt to formulate a theory with a syntax that never uses gauge-variant terms---would become better motivated. But the truth is that in neither theory are there \emph{general}  physical distinctions  that  select one or the other spatial distribution of the field, be it the metric or the gauge potential. This raised a conundrum, since we often select one representation over another: but what warrants such a choice? 
  
  I then resolved this conundrum  by specifying  roles for the gauge potential to fill, thus fixing a representational convention. The key to the resolution is to note that the `functional' role---e.g. highlighting the `radiative' features of the photon field---operates as a projection map in the space of models, $\mathcal{M}$: all symmetry-related distributions project down to  the same representative that fills one such role.   
 But this choice of convention is not  mandated by the physical facts; it is based on  pragmatic, contingent, or user-centric criteria, that operate similarly in both general relativity and gauge theory.

Finally, I want to emphasize the usefulness of `choosing gauges' in the sense of Section  \ref{sec:antiheal}. By choosing a gauge I mean: gauge-invariantly  specifying    the representative of a physical situation, based on explanatory and pragmatic criteria.  I have claimed here, building on Sections \ref{sec:rep_conv} \ref{subsec:rep_inv},   that such choices are no different than choices of coordinate systems that are adapted to physical situations. 
If we had no such choices available, we would likely lose enormous explanatory ability. In the  words of \citet[p. 1]{Tong_gt}:
\begin{quote}
 The [gauge] redundancy allows
us to make manifest the properties of quantum field theories, such as unitarity, locality,
and Lorentz invariance, that we feel are vital for any fundamental theory of physics
but which teeter on the verge of incompatibility. If we try to remove the redundancy
by fixing some specific gauge, some of these properties will be brought into focus,
while others will retreat into murk. By retaining the redundancy, we can flit between
descriptions as is our want, keeping whichever property we most cherish in clear sight.
\end{quote}

\subsection{Denying eliminativism}\label{sec:elim_elim}

One might think that the difference between the two types of symmetry transformations, labeled $\Delta$ (cf. Section \ref{sec:pos_argument})  can be pressed into service against redundancy, i.e. for eliminativism. For $\Delta$ says that non-Abelian Yang-Mills theories can be formulated using a  mathematical structure (the bundle of connections) in which the curvature represents all local degrees of freedom of the theory in a gauge-invariant manner. Thus one may imagine that, by employing just the curvature of the connection in that  formalism, we could at last  get rid of gauge symmetry. In the Abelian case, we need not even resort to the bundle of connections to make this claim, and so, for the purposes of this discussion, we can set it---the bundle of connections---aside. 
 
There are two problems with this eliminativist claim. First, the curvature represents only, and all, those physical degrees of freedom that are localizable. As we saw in Section  \ref{sec:AB}, there are properties pertaining to the parallel transport of internal and external quantities that are non-local and cannot be easily encapsulated by the curvature. Thus, while we may have a good reason to distinguish the gauge symmetries from the diffeomorphisms, we do not have warrant to eliminate the connections, and their gauge-related counterparts, from the formalism.

Second, even in the absence of something like the Aharonov-Bohm effect, the curvature cannot be articulated as a primitive, local  quantity: it must make reference to the gauge potentials. For instance, if we really believe the physical world embodies Lorentz invariance, we would also like our theories to exhibit  \textit{explicit} Lorentz covariance. This is not possible with local, gauge-invariant fields, such as the standard electric and the magnetic fields, even in the Abelian case. In fact, this difficulty, noticeable in   the `moving
magnet and conductor problem' (as in the opening of Einstein's 1905 paper),  led to the creation of special relativity.

Thus a unified, Lorentz covariant formulation of the electric and magnetic field  necessarily  employs $F_{\mu\nu}$. And $F_{\mu\nu}$ is not just an arbitrary 2-form, but one that necessarily satisfies---satisfies independently of any other contingent fact, such as the metric or the  matter distribution---the Bianchi identities. These constraints are enforced by representing the curvature as a function of  the gauge potential.  One could try to represent  the theory without constraints and symmetries, but that is a difficult task, if we are to keep other pragmatic criteria, such as locality. As can be seen clearly in the Hamiltonian formalism (cf. Section  \ref{sec:syms_consts}), constraints  encode symmetries. Thus, in order to find primitive, gauge-invariant quantities, one should also aim to unshackle them from any constraint.  But finding a formalism that is genuinely symmetry-invariant without the imposition of further local constraints is at least highly non-trivial. 

 For instance, in an attempt to excise mention of the gauge potential from the formalism and thereby obtain a purely gauge-invariant formulation, we could try to include another term in the Lagrangian with the use of a Lagrange-multiplier, a term whose equations of motion would recover the Bianchi identity. Formally, this is easy enough to do:
\be\label{eq:action_LM}S[\boldsymbol{\lambda},\mathbf{F}]:= \int F^{ab}F_{ab}+\lambda^{[abc]} \partial_{[c}F_{ab]}.
\ee
But in fact, since this new term is in fact imposing a constraint,  this move implicitly invokes the gauge potential that we are trying to avoid. For note that, in the previous equation, $\lambda^{[abc]}$ can be rewritten as $\lambda^{abc}=\epsilon^{abcd}\lambda_d$  by the Hodge-duality between three-forms and one-forms. Therefore, after integration by parts (we assume, as usual, that $\mathbf{F}$ falls off quickly and there are no boundary contributions to the integral) we can rewrite the Lagrange multiplier term as: $F_{ab}\epsilon^{abcd}\partial_c\lambda_d =\mathbf{F}\wedge *\d \boldsymbol{\lambda}$, where $\boldsymbol{\lambda}$ is a one-form. Re-inserting this identity in  \eqref{eq:action_LM}, the new equations of motion for $\mathbf{F}$ now yield precisely $\mathbf{F}=\d\boldsymbol{\lambda}$. Re-insertion of this identity into the action recovers the standard Maxwell action, without the Lagrange multiplier term,  but with the trivial notational substitution of $\mathbf{A}$ by $\boldsymbol{\lambda}$. 

The general lesson here is that constraints cannot be so easily eliminated.
\footnote{And the non-Abelian theory is even more inimical to a local, gauge-invariant representation: to start with, we cannot even write the equations of motion without the explicit appearance of the gauge potential.} 

Another attempt to get rid of gauge variance (the one favoured by \cite{Healey_book}),  adopts gauge-invariant but \emph{non-local} fundamental variables for the theory. The most common version adopts \emph{a holonomy} formulation: a holonomy basis of gauge-invariant quantities associates to each loop in spacetime a phase, namely,  the one we found in our treatment of the Aharonov-Bohm effect, in equation \eqref{eq:phase_AB}: 
\be\label{eq:hol} hol_\gamma(A) =\exp{i\int_\gamma A}, \quad\text{for each}\quad \gamma:S^1\rightarrow M.\ee  But the holonomy interpretation goes beyond  interpreting  the holonomies as integrals of the gauge potential: it  promotes the ontic status of these quantities, so that they should no longer be thought of as derivative from the gauge potentials (or from the connection form), but as primitive. 

Under this interpretation, the holonomy formalism carries many explanatory deficits in comparison to the formulation of the theory on the bundle. For instance, since the basis of gauge-invariant variables derived from primitive holonomies is vastly overcomplete, it obeys certain constraints. And, as far as I can see,  these composition properties can only be derived by reference to the original, gauge potential variable, $\mathbf{A}$. Without appeal to $\mathbf{A}$, they must be postulated \textit{ab initio}, and can be, at least in  the non-Abelian case, very unnatural.\footnote{More carefully, we can assign a complex number  (matrix element in the non-Abelian case) $hol(C)$ to the oriented embedding of the unit interval: $C:[0,1]\mapsto M$. This makes it easier to see how composition works: if the endpoint of $C_1$ coincides with the starting point of $C_2$, we define the composition $C_1\circ C_2$ as, again, a map from $[0,1]$ into $M$, which takes $[0,1/2]$ to traverse $C_1$ and $[1/2, 1]$ to traverse $C_2$.  The inverse $C^{-1}$ traces out the same curve with the opposite orientation, and therefore $C\circ C^{-1}=C(0)$.
Following this composition law, it is easy to see from \eqref{eq:hol} that 
\be\label{eq:loop_com} hol(C_1\circ C_2)=hol(C_1)hol(C_2),\ee with the right hand side understood as complex multiplication in the Abelian case, and as composition of linear transformations, or  multiplication of matrices, in the non-Abelian case.

For both Abelian and non-Abelian groups, given the above notion of composition, holonomies are conceived of as smooth homomorphisms from the space of loops into a suitable Lie group. One obtains a representation of these abstractly defined holonomies as
holonomies of a connection on a principal fiber bundle with that Lie group as structure group; the collection of such holonomies carries the same amount of information as the gauge-field $A$ (cf. \cite[Sec. 3]{Belot1998} for a philosophical exposition). However, only for an Abelian theory can we cash this relation out in terms of gauge-invariant functionals. That is, while \eqref{eq:hol} is gauge-invariant, the non-Abelian counterpart (with a path-ordered exponential), is not. For non-Abelian theories the gauge-invariant counterparts of \eqref{eq:hol} are Wilson loops, see e.g. \citep{Barrett_hol}, 
$ W(\gamma):=\text{Tr}\, \mathcal{P}\exp{(i\int_\gamma A)}
$,
where one must take the trace of the (path-ordered) exponential of the gauge-potential. It is true that all the gauge-invariant content of the theory can be reconstructed from Wilson loops;  (see also \cite{Weatherall_holonomy}, for a category-theory based derivation of this equivalence). But,  importantly for our purposes, it is no longer true that there is a homomorphism from the composition of loops to the composition of Wilson loops. That is, it is no longer true that the counterpart \eqref{eq:loop_com} holds,  $W(\gamma_1\circ\gamma_2)\neq W(\gamma_1)W(\gamma_2)$.  The general composition constraints---named after Mandelstam---come from generalizations of the Jacobi identity for Lie algebras, and depend on $N$ for SU($N$)-theories; e.g. for $N=2$, they apply to three paths and are: $W(\gamma_1)W(\gamma_2)W(\gamma_3)-\frac12(W(\gamma_1\gamma_2)W(\gamma_3)+W(\gamma_2\gamma_3)W(\gamma_1)+W(\gamma_1\gamma_3)W(\gamma_2))+\frac14(W(\gamma_1\gamma_2\gamma_3) + W(\gamma_1\gamma_3\gamma_2) = 0$.  \label{ftnt:tr_hol}}

This type of explanatory reliance on the theory with more symmetry is in fact a common issue with many ``relationist'' approaches. To give a simple, oft-repeated example: in relational particle dynamics, even if the vastly overcomplete set of  inter-particle separations are somehow taken as primitives, they are not independent. They must obey constraints ---e.g.. the triangle inequality---which can be either posited \emph{ab initio}, or, more naturally, arise from the dimensionality (and geometry) of the (substantival) space in which they are embedded (see \citep[Section 6]{Belot2003}, who also notices the analogy with the holonomies case).\footnote{Indeed, I am not convinced such a relational approach to particle mechanics captures all of the symmetry-invariant content of the theory. For instance, it is not clear how to write more general, symmetry-invariant topological properties of the substantival space---such as dimension---solely in terms of inter-particle separations. } 
In short, the reduced picture is too deeply rooted in the formulation of the non-reduced theory  to have any conceptual transparency on its own.

But ultimately, as we saw in Section \ref{sec:rep_conv}, we employ representational conventions in many of our physical theories. As we will argue more fully in Chapter \ref{ch:subsystems},  in the presence of subsystems,  we \emph{must} keep the ability to alter that convention: as described in  \ref{subsec:isos_descs}, this involves, or recovers, the idea of gauge transformations (cf. \eqref{eq:transition_h}). This is, essentially, the reason \cite{RovelliGauge2013} gives for why we need gauge degrees of freedom (I will elaborate on this in Section \ref{sec:rep_conv_gluing}).


\subsection{Conclusion}\label{sec:conc_sd2}
   The lesson of this Chapter and the previous  one is that  gauge transformations and diffeomorphisms are structurally very similar, with the exception of one robust dissimilarity ($\Delta$). Although we could find, at the end of the day, this conceptual difference between the two, our investigations in these two Chapters have found  no smoking gun that would  validate eliminativism for gauge while endorsing sophistication for diffeomorphisms. 
 
 
  Thus we can  understand the ontological commitments of both theories as structural: one describes chronogeometric relations, in a well-understood sense, and the other describes the parallel transport of all sorts of charges that figure in the standard model, in a well-understood sense.

 Therefore, we conclude that fiber bundle structuralism is a valid, explanatory perspective about the ontology of gauge theories. It suggests a form of anti-quidditism,  as valid and explanatory as anti-haecceitism is for chronogeometric structure. 
 So for us, the point here is of course: if anti-haecceitism is good for spacetime, why not also adopt anti-quidditism about gauge? Or, as they say in England: what is sauce for the goose  is also sauce for the gander.

  \part{What  gauge symmetries can and cannot do}\label{part:II}
  

We will start this Part with Chapter \ref{ch:Noether}, where I will  give one answer to the question of how  mere redundancy can be scientifically fruitful. 
The answer it provides is based on Noether's theorems. 

According to Noether's first theorem, a rigid symmetry explains charge conservation. 
But, fortunately, charge conservation does not exhaust the content of Noether's theorems. For  charges  interact with fields, and therefore charge conservation will have consequences for the dynamics of those fields. This Chapter argues that  the real power of Noether's theorems is to ensure that the details of these interactions will conserve charge; provided, that is, that the theory in question has a sufficiently redundant mathematical representation. In more detail, Noether's first theorem guarantees that theories in  which the symmetries are rigid---in which the redundancy is specified in exactly the same way at all spacetime points---must have certain globally conserved charges;  the electric charge is again a standard example. Alone, this theorem offers  no guidance on how a dynamical field might interact with the charges. Noether's second theorem guarantees that,  once the rigid symmetry is weakened so that we are free to redescribe quantities at one spacetime point independently of their description at another,  the details of that interaction   will conserve charge.

So this Chapter argues that gauge theories illustrate the  immense value that a symmetry, or redundancy of description, has for theory-building,  for it consistently combines different parts of the system: the charges and the fields they interact with.\footnote{ Although it is well-known that Noether's first theorem implies charge conservation, and the second theorem  implies relations between the theory's equations of motion, this particular interpretation---that weakening a rigid symmetry enforces compatibility between charge conservation and the  dynamics of the corresponding fields---was developed in full in \cite{GomesButterfieldRoberts}.}

Chapter \ref{ch:Coulomb} builds on the arguments of Section \ref{sec:antiheal}. There we responded to Healey's worries that choosing particular distributions of the gauge potential over spacetime was distinct from choosing particular distributions of the metric over spacetime. Here we will show precisely, in a specific example, how we can choose such a representational convention.\footnote{We will give a more systematic treatment of representational conventions in Chapter \ref{ch:subsystems}.} 
For that, we work with the Hamiltonian formalism, and exploit symplectic orthogonality. The main idea  is that there is natural split of the electric field into two parts: one  part  is like its electrostatic component, or rather, it is the component that is uniquely and synchronically determined by the contemporaneous distribution of charges. The other part is the remainder, and it corresponds to radiation. It turns out that the gauge potential can be split into two corresponding parts, by employing symplectically orthogonality: one part is symplectically orthogonal to the radiative component of the electric field, and it is `pure gauge'  (it would express a vanishing  magnetic field). The other part satisfies Coulomb gauge, and it is symplectically orthogonal to the Coulombic part of the electric field. 


  \chapter{The Gauge Argument: A Noether reason}\label{ch:Noether}

  \section{Introduction and roadmap for this Chapter}\label{sec:intro_noether}
  In this Section  I will reprise the themes of Part \ref{part:I}  while introducing the project of this Chapter. 
  
  All interpretations of modern gauge theories adopt two core assumptions at their foundation. The first is that gauge symmetry arises when there are more variables in a theory than there are physical degrees of freedom. Hence the well-known soubriquets: gauge is `descriptive redundancy', `surplus structure', and `descriptive fluff'. Correspondingly, considerable effort has been devoted to techniques for eliminating gauge redundancy in order to appropriately interpret gauge theories.\footnote{Cf. \citet{Earman_gmatters, Earman2003_ode, Earman2004b}, \citet{Healey_book} and \citet{RosenstockWeatherall2016c,rosenstockweatherall2018e}. See also \citet{GomesRiello_theta} in response to \citet{Dougherty_CP}.} The second assumption is that a theory with gauge symmetry constitutes the gold standard of a modern physical theory: witness the gauge symmetry invoked in the Standard Model. This leads to a remarkable \emph{puzzle of gauge symmetry:} if interpreting gauge symmetry requires eliminating it, then why is gauge symmetry so ubiquitous? 

Of course, a number of answers---alternatives to simple eliminativist interpretations of gauge---have already been articulated in this thesis, and more await. 
The purpose of this Chapter is to articulate another answer to this question: namely, that gauge symmetry provides a path to building appropriate dynamical theories---and that this rationale invokes  the two theorems of Emmy \citet{noether1918a}.\footnote{For details on the historical development of Noether's theorems see \citet{ks2011noether}. For a modern statement of the first and second theorems, cf. \citet{Olver_book}, Theorems 5.58 (p. 334) and 5.66 (p. 343) respectively. }
 
Noether's first and better-known theorem (commonly  called simply \emph{Noether's theorem}) implies that global (or what we will call {\em rigid}) symmetries of a classical Lagrangian field theory---i.e. symmetries in which the redundancy is specified in exactly the same way at all spacetime points---correspond to charges that are conserved over time, such as energy and angular momentum. For example, the conservation of an electron's charge can be viewed as arising from the (redundant) global phases of the electron's wavefunction. But we will be equally concerned with Noether's second theorem, which is about  local (or what we will call {\em malleable}) gauge symmetries---meaning that the specified redundancy varies between spacetime points. Agreed: this theorem's physical significance is of course already well recognized, including in the philosophical literature (\citet{bradingbrown2000n,BradingBrown_chapter}). In particular, a recent line of work shows how such malleable gauge symmetries encode relationships between spatial or spacetime regions, and thus between parts and wholes in a field theory. We will have more to say about this in Chapter  \ref{ch:subsystems}.

In this Chapter, we will urge that these two theorems give us a further answer to the puzzle, `why gauge?' It is an established, indeed conventional, answer amongst practising physicists. For it is implicit in the well-known \emph{gauge argument} or the \emph{gauge principle} first formulated by Hermann \citet{weyl1929g}. This argument begins with an assumption of local gauge symmetry, and then claims to `derive' the form of the dynamics of quantum theory in a way that exhibits `minimal coupling' to an electromagnetic potential. We claim that this is an instance of a much more general role for gauge, which has not been at all discussed in the philosophical literature: gauge symmetry supports theory construction, in particular by constraining the space of models to those in which charges appropriately couple to forces. Although some philosophers like \citet{BradingBrown_chapter} have pointed out the role of gauge symmetry in theory construction, it is this last coupling of charges to forces that we would like to highlight, which provides the answer to the puzzle of gauge symmetry that we will advocate here.

As experts will be quick to note: the gauge argument in its common textbook form is fraught with difficulties. However, our argument is that these difficulties can be overcome; and indeed that there is a more general gauge argument available for use in the construction of physical theories. We thus proceed in Section \ref{sec:ga-and-woes} to rehearse the usual gauge argument and its woes.

The real limitation of the textbook gauge argument, as we shall see, is that it does not reflect the generality of the kind of argument that physicists typically use. Thus, in Section \ref{sec:ANoether}, I will  present a much more general gauge argument, which I will call the \emph{Noether gauge argument}, in the context of classical Lagrangian field theory. The key to understanding this argument is the combined use of \emph{both} Noether's first \emph{and} second theorem. In the first step, one applies Noether's first theorem to establish the conservation of charge. In the second step, one makes use of the power of Noether's second theorem, to infer specific interpretive information about how these charges couple to gauge fields. We draw out and clarify what that information is, in the presence of various kinds of symmetries that are sometimes referred to as `gauge', in order to illustrate the precise extent to which the gauge argument can be fruitfully used to constrain physical theories. 

\section{The gauge argument and its critics}\label{sec:ga-and-woes}

The textbook gauge argument or gauge principle uses gauge invariance to motivate a quantum theory of electromagnetism. We begin Section \ref{subs:beware} with a brief presentation of this argument as it is usually presented. Classic textbook statements can be found in \citet[\S 6.14]{schutz1980g} \citet[\S 4.2]{gc1989dg}, and \citet[\S 3.3]{ryder1996qft}, among many other places. Then in Section \ref{subs:criticisms} we assess it. The argument has been discussed in the form below by philosophers as well, such as \citet{teller1997m,teller2000gauge}, \citet{brown1998a}, \citet{martin2002g}, and \citet[\S 2]{wallace2009anti}.

\subsection{Beware: Dubious arguments ahead}\label{subs:beware}

We begin by describing a quantum system with the Hilbert space $L^2(\RR^3)$ of wavefunctions, recalling that a unique pure quantum state is represented not by vector, but by a `ray' of vectors related by a complex unit. This implies that the transformation $\psi(x)\mapsto e^{i\theta}\psi(x)$ for some $\theta\in\RR$, referred to as a `global phase' transformation, acts identically on rays, and is in this sense an invariance of the quantum system. But now, the story goes, suppose we replace this with a `local phase' transformation $\psi(x)\mapsto e^{i\phi(x)}\psi(x)$, in which the constant $\theta$ is replaced with a function $\phi:\RR^3\rightarrow\RR$, or indeed with a smooth one-parameter family of such functions $\phi_t(x)$ for each $t\in\RR$. This transformation is `local' in the sense that its values vary smoothly across space and time. The corresponding Hilbert space map $W_\phi:\psi\mapsto e^{i\phi}\psi$ does not act identically on rays. However, one might still wish to postulate that this transformation has no `physical effect' on the system, or is `gauge'. Various motivations for this step are given in the textbooks, often with vague references to general covariance of the kind found in general relativity: which we will return to shortly. But to mimic the standard presentation, we will simply press forward, referring to $W_\phi:\psi\mapsto e^{i\phi}\psi$ as a \emph{local} or \emph{malleable} \emph{gauge transformation}.

The main premise of the argument is to assume that the Schr\"odinger equation must be invariant under this local phase transformation. But, for the free non-relativistic Hamiltonian in the Schr\"odinger (position) representation, this is not the case.\footnote{Obvious variations of the argument exist for relativistic wave equations too \citep[cf.][\S 3.3]{ryder1996qft}.} Writing $\psi_t(x):=e^{-itH}\psi(x)$ with $H = \tfrac{1}{2m}P^2$, one finds that $W_\phi:\psi\mapsto e^{i\phi_t(x)}\psi$ transforms the Schr\"odinger equation to $i\tfrac{d}{dt}\left(e^{i\phi_t(x)}\psi_t(x)\right) = \tfrac{1}{2m}P^2e^{i\phi_t(x)}\psi_t(x)$, which is equivalent\footnote{The LHS is $i\tfrac{d}{dt}e^{i\phi_t(x)}\psi_t(x) = e^{i\phi_t(x)}\left(-\tfrac{d\phi}{dt} + i\tfrac{d}{dt}\right)\psi_t(x)$. For the RHS, use the fact that $e^{-i\phi_t(x)}Pe^{i\phi_t(x)}=P+\nabla\phi_t(x)$, and so $e^{-i\phi_t(x)}P^2e^{i\phi_t(x)} = (e^{-i\phi_t(x)}Pe^{i\phi_t(x)})^2 = (P+\nabla\phi_t(x))^2$. Thus the RHS is $\tfrac{1}{2m}P^2e^{i\phi_t(x)}\psi_t(x) = e^{i\phi_t(x)}\tfrac{1}{2m}(P+\nabla\phi_t(x))^2\psi_t(x)$. Multiplying both sides on the left by $e^{-i\phi_t(x)}$ and rearranging then gives the result.\label{eq1-calc}} to the statement that,
\begin{equation}\label{g-tr-schr}
 i\tfrac{d}{dt}\psi_t(x) = \left(\tfrac{1}{2m}(P+\nabla\phi_t)^2 + \tfrac{d\phi_t}{dt}\right)\psi_t(x).
\end{equation}
Instead of preserving the Schr\"odinger equation, a gauge transformation produces the additional terms $\nabla\phi_t$ and $\tfrac{d\phi_t}{dt}$ in the Hamiltonian.

To correct this situation, the big move of the gauge argument is to introduce a vector  $A = (A_1,A_2,A_3)$ and a scalar $V$, which are assumed to behave under the gauge transformation as,
\begin{align}\label{AV-trans-rules}
  A\mapsto A+\nabla\phi_t, && V\mapsto V-\tfrac{d\phi_t}{dt}.
\end{align}
This has the form of the familiar gauge freedom of the electromagnetic four-potential (see e.g. \eqref{eq:gauge_trans}) that leaves the electromagnetic field unchanged.

To restore invariance of the Schr\"odinger equation under gauge transformations, one thus apparently needs only to assume that the Hamiltonian is not free, but rather given by,
\begin{equation}\label{eq:mc-ham}
  H = \sum_{r=1}^3\tfrac{1}{2m}(P_r-A_r)^2 + V,
\end{equation}
which is known as the \emph{minimally coupled} Hamiltonian. For, replacing the Hamiltonian in the Schr\"odinger equation with this one, we find that the transformation rules for $A$ and $V$ perfectly compensate for the extra terms appearing in Equation \eqref{g-tr-schr}. Thus, gauge invariance of the Schr\"odinger equation is obtained, provided the Hamiltonian contains interaction terms $A$ and $V$ that behave like the 3-vector potential $A$ and scalar potential $V$ for an electromagnetic field.

With an eye towards a modern gauge theory formulated as a vector bundle with a derivative operator, it is even possible to interpret the potentials $A$ and $V$ as associated with a change of derivative operator: writing $\pp_\mu := (\tfrac{d}{dt},\nabla)$ and $A_\mu = (V,A)$, one finds that the procedure above is equivalent to replacing $\pp_\mu$ with,
\begin{equation}\label{eq:cov-der}
  D_\mu := \pp_\mu + iA_\mu = (\tfrac{d}{dt} + iV,\nabla + iA) = (D_t,D).
\end{equation}
This is commonly referred to as a `covariant derivative'. Then, substituting $\tfrac{d}{dt}\mapsto D_t$ and $\nabla\mapsto D$ into the free Schr\"odinger Equation $i\tfrac{d}{dt}\psi = \tfrac{1}{2m}\nabla^2\psi$ and rearranging, we derive the minimally coupled Hamiltonian of Equation \eqref{eq:mc-ham}. Accordingly, this choice of Hamiltonian is sometimes advocated, for example by \citet{Lyre_gauge}, on the basis of a `generalised equivalence principle', according to which electromagnetic interactions with all matter fields ``can be transformed away''.\footnote{This principle arises in particular on a principal fibre bundle formulation of gauge theory; for philosophical appraisals, see \citet{Lyre_gauge}, \citet[\S 5]{Weatherall2016_YMGR}, and \citet[Ch. 6.3]{Healey_book}.} In short, it appears as if minimal electromagnetic coupling has been derived out of nothing: or at least, from an assumption of gauge invariance.
  
\subsection{Criticisms of the gauge argument}\label{subs:criticisms}

That is how the story is usually presented. I agree: it is far from water-tight. The argument begins with a system with a global symmetry, gratuitously generalises it to a local symmetry---which, to emphasise, was not required for mathematical consistency or for empirical adequacy---and then, in order to fix the ensuing non-invariance of the governing equations, proceeds to conjecture a new force of nature. To put it uncharitably: the argument fixes a problem that didn't exist by conjecturing a redundant field, and then turns this game around, claiming to come out successfully by `retrodicting' the existence of electromagnetism. More charitably: the gauge argument suffers from at least two categories of concerns. We will set out each of these three concerns here and in Section \ref{sec:ANoether} present an alternative {\em Noether gauge argument} that answers them entirely.

The first category of concerns is the gauge argument's claim to have derived a dynamics that is specifically electromagnetic in nature. Although a formal set of operators $A_\mu=(V,A)$ have been included in the dynamics, no evidence is given that these operators take the form required for any \emph{specific} electromagnetic potential, or that the coupling to $A_\mu$ will be proportional to a particle's charge $e$, or even that $A_\mu$ is non-zero. And if they could be shown to be non-zero, then as \citet[p.210]{wallace2009anti} rightly asks: ``how do neutral particles fit into the argument?'' A minimally coupled dynamics does not to apply to neutral particles, and yet since the gauge argument never mentioned or assumed anything about charge, it presumably is intended to apply to them.

This concern can be assuaged by scaling back the conclusion of the gauge argument: its aim is not to derive any particular electromagnetic interaction, but rather to \emph{constrain} the dynamics so as to be compatible with gauge invariance. This leaves open the specific character of $A_\mu$, and indeed even the question of whether it is zero. Although not all authors adopt this attitude towards the gauge argument, we advocate it as the preferable attitude, and will develop it in more detail in the subsequent Sections.

A second category of problems arises out of the free-wheeling argumentative style of the gauge argument. For example, it is not a strict deductive derivation of either the electromagnetic potential or the dynamics. At best, the gauge argument appears to show that one \emph{can} adopt a minimally coupled Hamiltonian in order to assure gauge invariance. But this does not ensure that one \emph{must} do so: the door appears to be left open for other dynamics to be gauge invariant, but without taking the minimally coupled form that the gauge argument advocates. As \citet[p.S230]{martin2002g} writes: ``The most I think we can safely say is that the form of the dynamics characteristic of successful physical (gauge) theories is \emph{suggested} through running the gauge argument.''

Another example of free-wheeling argumentation is in the motivation for requiring the local gauge transformations $W_\phi:\psi\mapsto e^{i\phi} \psi$ to be symmetries. Sometimes a preference for this transformation over global phase transformations is dubiously motivated by a desire to avoid superluminal signalling.\footnote{For example, \citet[p.93]{ryder1996qft} writes: ``when we perform a rotation in the internal space of $\phi$ at one point, through an angle $\Lambda$, we must perform the same rotation at all other points at the same time. If we take this physical interpretation seriously, we see that it is impossible to fulfil, since it contradicts the letter and spirit of relativity, according to which there must be a minimum time delay equal to the time of light travel.'' For a detailed critique, see \citet[p.S227]{martin2002g}.} In other cases it is motivated by the coordinate invariance of a spatial coordinate system. But as \citet[p.210]{wallace2009anti} points out, no reason is given as to why we do not similarly consider local transformations of configuration space, momentum space, or any other space, to be symmetries. Nor is there any clear reason why the $U(1)$ symmetry of electromagnetism is chosen as the global symmetry motivating the move to the local symmetry, as opposed (say) the $SU(3)$ symmetry of the strong nuclear force.

Regarding the generalisation of the gauge argument to other global symmetry groups beyond electromagnetism, I wholeheartedly agree with Wallace: one should expect, and indeed we will argue in Section \ref{sec:ANoether}, that an appropriate generalisation of the gauge argument can also be applied to these more general gauge groups.

My approach here speaks to a third category of concerns, that the gauge argument is awkwardly placed as an argument for a quantum theory of electromagnetism. The construction of a covariant derivative operator suggested by the gauge argument is most appropriately carried out not in quantum field theory, but in the classical Yang-Mills theory of principal fibre bundles. Here too I agree with Wallace:
\begin{quote}
  ``In fact, it seems to me that the standard argument feels convincing only because, when using it, we forget what the wavefunction really is. It is not a complex classical field on spacetime, yet the standard argument, in effect, assumes that it is. This in turn suggests that the true home of the gauge argument is not non-relativistic quantum mechanics, but classical field theory.'' \citep[p.211]{wallace2009anti}
\end{quote}
Indeed, it is remarkable that in the presentation of the gauge argument above, the role of the `rigid' or `global' $U(1)$ symmetry is hardly substantial: only the local malleable symmetries play any substantial role in the argument. This is an oddity to be sure, though one that we will correct shortly.

In Section \ref{sec:ANoether}, we will switch perspectives from the \emph{verdammten Quantenspringerei} to the context of classical Lagrangian field theory, and propose a framework that substantially clarifies the roles of rigid gauge symmetries, of malleable gauge symmetries, and of their relationship, which I will call the `Noether gauge argument'.

\section{A Noether Reason for Gauge}\label{sec:ANoether}

\subsection{Overview}

For a more general view of how gauge symmetries constrain the dynamics of a physical theory, I will now, as announced in Section \ref{subs:criticisms},  make a two-step use of the theorems of Emmy \citet{noether1918a}: the first, and then the second. I will refer to this as the \emph{Noether gauge argument}. Agreed: this is by no means a new observation, since practising physicists use this property of gauge frequently!\footnote{A succinct example is \citet{Avery_2016}, who write ``Noether’s second theorem, which constrains the general structure of theories with local symmetry''.} But I believe it is worth highlighting and clarifying exactly the kind of information that can be extracted in various cases, as part of my advocacy that philosophical discussions of gauge should better recognise gauge's significance for theory construction.

The Noether gauge argument proceeds in two steps. First, we choose a rigid gauge symmetry associated with an arbitrary global gauge group, and propose that its action produces a variational symmetry: by Noether's first theorem, this guarantees the presence of a collection of conserved quantities. But  matter fields do not exist in  isolation: they couple to other `force' fields, and possibly to long-range ones. Thus, in the second step, we introduce such a field and apply Noether's second theorem, `loosening' the rigid symmetries to malleable ones; and we show that this provides three  concrete constraints on the dynamics (viz. the vanishing of the three lines in Equation \eqref{eq:vanishing} below). The interpretation of these constraints can be seen on a case-by-case, or sector-by-sector, basis: we will consider their implications for rigid versus malleable symmetries, as well as for $A$-independent versus $A$-dependent ones. Thus in the following Sections we will spell out the consequences of the three constraints for four different sectors of the theory.  In particular, we will find through explicit computation---adopting only a minor additional assumption of non-derivative coupling---that when we couple the matter fields to force fields, gauge-invariance guarantees that the Lagrangian for these fields is massless, and so they constitute long-range interactions.\footnote{The formalism equally applies to spin-2, or gravitational, fields; but, apart from some cursory remarks, we will not discuss these.} The generalised Gauss laws thus are guaranteed to relate the content of the matter current within a region to the flux of the other force fields at distant closed surfaces surrounding such a region.

Disclaimers: first, in the interest of clarity and pedagogy, I will not try to incorporate the full generality of Noether's theorems, which is truly extraordinary but over-complicated for our discussion. In its place, I will make several simplifying assumptions, both about the Lagrangian density and about the action of the gauge group, which are not strictly speaking necessary but which simplify my argument. Second, throughout this discussion, I will follow standard practice and distinguish two equivalence relations for classical fields on a manifold. First, I will write `$=$' to denote ordinary equality between fields, irrespective of the satisfaction of the equations of motion, and refer to this as \emph{strong} or \emph{off-shell} equality. Second, given a fixed Lagrangian, I will write `$\approx$' to denote equality between fields that holds if the Euler-Lagrange equations are satisfied for that Lagrangian, and refer to this as \emph{weak} or \emph{on-shell} equality.\footnote{This common terminology is due to Dirac \citep[cf.][]{HenneauxTeitelboim}.}

 To represent the forces that are sourced by $\varphi$, we take the collection of vector-valued one forms $A_\mu^I$, which take a vector of $M$ at a point of $U$ to $\mathfrak{g}$, with $\mu$ representing the spacetime components of the vector and  $I$ indicating the components in $\mathfrak{g}$.\footnote{What is a `force' and what is `matter' will be further distinguished by their transformation properties under a gauge transformation, in \eqref{eq:gauge_trans}. Matter transforms linearly, whereas forces acquire derivatives of the generator as inhomogeneous terms.} 
 These fields are associated with a dynamics by postulating a preferred real-valued action functional $S(\varphi_i, A_\mu^I)$, whose extremal values are postulated to provide the equations of motion.

We also  assume $G$ has some action (a representation) on $V$, the vector space of local field-values of the matter fields $\varphi$, defining this action pointwise as $g\cdot \varphi(x)=g(x)\cdot\varphi(x)\in V$. Let $t^{ij}_I$  be the $n$-dimensional Hermitean matrix representation on $V$ of $\mathfrak{g}$, i.e. $t:\mathfrak{g}\rightarrow GL(V)$,  where  the $a$ are indices of the Lie algebra space, in the domain of the map, and $i,j$ denote the matrix indices in the image of the map, acting linearly on $V$. Then we take the (malleable) gauge transformations, infinitesimally parametrized by $\xi\in \fG$, to act on our fundamental variables as in \eqref{eq:gauge_trans}, but with the added transformation of the matter field, as:
\be\label{eq:gauge_trans1}
\begin{cases}
\delta_\xi \varphi_i=\xi^I t_I^{ij}\varphi_j=(\xi t\varphi)_i \\
\delta_\xi A^I_\mu=\D_\mu\epsilon^I=\pp_\mu \xi^I+[\xi, A_\mu]^I
\end{cases}.
\ee
where the square brackets are the Lie algebra commutators.  These transformation rules are not as general as they could be, but neither are they arbitrary: they are the first-order terms of the Lie algebra action on the respective vector spaces---in particular, `first-order' in the derivatives of $\xi$ and in powers of $A$ and $\varphi$---and in this sense provide an appropriate approximation of any malleable gauge transformation. We here focus on this special case, equation \ref{eq:gauge_trans1}, only to simplify the presentation of the argument.


Our aim now is to constrain how the matter fields $\varphi$ couple to force fields. Let $\mathcal{L}(\varphi, \pp \varphi, A, \pp A)$ be the Lagrangian defining our action $S(\varphi,A)$, which we assume for simplicity does not depend on higher-order derivatives.\footnote{This can be justified by appeal to Ostrogradsky's theorem; see \citet{swanson2019ostrogradsky} for a philosophical discussion.} Variation along the directions of the gauge transformations above yields (with summation convention on all indices):
\begin{align}\label{eq:vanishing}
  \begin{split}
\left( \frac{\delta \mathcal{L}}{\delta \varphi_i}(t^I\varphi)_i+ \frac{\delta \mathcal{L}}{\delta \pp_\mu\varphi_i}(t^I\pp_\mu\varphi)_i+\big[ \frac{\delta \mathcal{L}}{\delta A_\nu}, A_\nu]^I+\big[\frac{\delta \mathcal{L}}{\delta\pp_\nu A_\mu}, \pp_\mu A_\nu\big]^I\right)\xi_I+\\
\left( \frac{\delta \mathcal{L}}{\delta \pp_\mu\varphi_i}(t^I\varphi)_i+ \frac{\delta \mathcal{L}}{\delta A_\mu^I}+\big[ \frac{\delta \mathcal{L}}{\delta \pp_\nu A_\mu}, A_\nu]^I\right)\pp_\mu\epsilon_I+\\
\frac{\delta \mathcal{L}}{\delta \pp_\nu A^I_\mu}\pp_\mu \pp_\nu \xi^I=0\\
\end{split}
\end{align}
Since the derivatives of $\xi$ are functionally independent, this equation implies that \textit{each line must vanish separately}: the first line is a consequence of rigid symmetries, and the remaining two are of malleable ones. These are the fundamental constraints on the dynamics that we propose to analyse, and the task of the remainder of this Section will be to unpack them.

The requirement that each of these lines vanishes provides a strong constraint on the form of the Lagrangian, and hence on the dynamics. This, I claim, provides the core of the Noether gauge argument. To extract interesting physical information from this constraint, there are four sectors to compare, arising from the use of either rigid or malleable symmetries, and either $A$-independent or $A$-dependent Lagrangians. We treat each sector in turn. 

The results will be: a theory with rigid symmetries can be dynamically non-trivial and complete---i.e. it will not require further constraints---when $A$ does not figure in the Lagrangian. With malleable symmetries and no $A$-dependence, the constraints demand that the dynamics be  trivial, i.e. no kinetic term for the matter field can appear in the Lagrangian. When forces have their own dynamics, that is, when  the Lagrangian is $A$-dependent, a theory with rigid symmetries may be incomplete, and require further constraints to render the dynamics of $A$ compatible with charge conservation; an example will be given. It is only in the last case, where we have malleable symmetries and $A$-dependence, that the equations of motion coupling forces to charges is automatically consistent with the conservation of charges (and so no further constraints are required). Thus we will see the power of malleable symmetries and $A$-dependence together to secure an interacting dynamics that conserves charge. And this will be our Noether gauge argument. 

\subsection{$A$-independent, rigid symmetries}\label{4.3}

First, suppose we are as in the first step of the textbook gauge argument: there is no $A$ in sight, and the symmetry is rigid, so that  $\pp_\mu  \xi^I = 0 = \pp_\mu \pp_\nu  \xi^I$. Then the vanishing of the first line of Equation \eqref{eq:vanishing} reduces to
\begin{equation}\label{eq:first-line}
\frac{\delta \mathcal{L}}{\delta \varphi_i }(t^I\varphi)_i+ \frac{\delta \mathcal{L}}{\delta \pp_\mu\varphi_i}(t^I\pp_\mu\varphi)_i=0.
\end{equation}
But by the Euler-Lagrange equations $\mathsf{E}(\varphi)_i\approx 0$, where $\mathsf{E}(\varphi)_i= \frac{\delta \mathcal{L}}{\delta \varphi_i}-\pp_\mu  \frac{\delta \mathcal{L}}{\delta \pp_\mu\varphi_i}$, we have
\begin{equation}
  \frac{\delta \mathcal{L}}{\delta \varphi_i}\approx\pp_\mu  \frac{\delta \mathcal{L}}{\delta \pp_\mu\varphi_i},
\end{equation}
where we again are using `$\approx$' to denote `on-shell' equality. Applying this to Equation \eqref{eq:first-line} we find that
\be\label{eq:A_rig}
  \pp_\mu (\frac{\delta \mathcal{L}}{\delta \pp_\mu\varphi_i}(t^I\varphi)_i)=\pp^\mu J^I_\mu(\varphi)\approx 0
\ee
where we have defined the matter current as
\be\label{eq:J_phi}
  J^I_\mu(\varphi):=\frac{\delta \mathcal{L}}{\delta \pp_\mu\varphi_i}(t^I\varphi)_i.
\ee
In summary, we have derived what is guaranteed by Noether's first theorem, that the current $J^I_\mu(\varphi)$ is conserved on-shell. Or, turning this around: symmetry requires the Lagrangian to be restricted so that $J^I_\mu(\varphi)$ defined in Equation \eqref{eq:J_phi} is divergenceless. Having constrained the space of theories in this manner, there are no more equations to satisfy: conservation of charge is consistent with the dynamics and no further constraints need to be imposed. 

\subsection{$A$-independent, malleable symmetries}

In the next case, suppose that we allow---in addition to Section \ref{4.3}'s equations---the ones arising from a $\pp\epsilon\neq 0$, while still not allowing for an $A$ in the theory. We get, in addition to equations \eqref{eq:J_phi} and \eqref{eq:A_rig}, from the vanishing of the second line of Equation \eqref{eq:vanishing}:
\be\label{eq:noA_mal}
\frac{\delta \mathcal{L}}{\delta \pp_\mu\varphi_i}(t^I\varphi)_i=J^I_\mu(\varphi)=0. 
\ee
So here the  conserved currents are forced to vanish. Clearly this condition is guaranteed for all field values if $\frac{\delta \mathcal{L}}{\delta \pp_\mu\varphi_i}=0$, which requires a  vanishing kinetic term. A careful analysis of more general cases reveals this is the only generic solution.\footnote{For instance, assume $\frac{\delta \mathcal{L}}{\delta \pp_\mu\varphi_i}$ depends only on $\pp_\mu\varphi_i$, then since $t^I_{ij}\varphi^i$ can take any value, we must have $\frac{\delta \mathcal{L}}{\delta \pp_\mu\varphi_i}=0$. Now, suppose $\frac{\delta \mathcal{L}}{\delta \pp_\mu\varphi_i}$ depends on $\varphi_i$ as well. Since $\varphi_i$ has no spacetime indices to match the $\mu$ of the gradient $\pp_\mu\varphi_i$,   to make a Lagrangian scalar, we  would need the $\varphi_i$ contribution to this term to itself be a scalar, call it $F(\varphi)$. So for example: $\frac{\delta \mathcal{L}}{\delta \pp_\mu\varphi_i}= \pp_\mu\varphi_i(\varphi_j\varphi^j)$, or more generally $\frac{\delta \mathcal{L}}{\delta \pp_\mu\varphi_i}=F'(\pp \varphi)_{i\mu}F(\varphi)$ (where we raise indices with an inner product of $V$); and as in the example $F(\varphi)=\varphi_j\varphi^j=0$ iff $\varphi=0$. But then the same argument as before suffices, since we can still allow $t^I_{ij}\varphi^i$ to take any value in $V$ (for an appropriate,  non-zero value of the scalar formed just from $\varphi$, e.g. the contraction $\varphi_j\varphi^j$).  Or, in other words, for $\varphi\neq 0$,  $\frac{\delta \mathcal{L}}{\delta \pp_\mu\varphi_i}(t^I\varphi)_i=0$ iff $F^{-1}(\varphi)\frac{\delta \mathcal{L}}{\delta \pp_\mu\varphi_i}(t^I\varphi)_i=0$ where $F^{-1}(\varphi)\frac{\delta \mathcal{L}}{\delta \pp_\mu\varphi_i}$ depends only on $\pp_\mu\varphi$; and thus we are back to the first, simple  case.} 

This analysis pinpoints the obstacle appearing in  the textbook gauge argument that we rehearsed in Section \ref{subs:beware}. When the matter field Lagrangian has a non-trivial kinetic term, malleable transformations cannot be variational symmetries. That is: if we impose malleable symmetries without introducing a gauge potential, we cannot consistently also allow a term in the Lagrangian including  $\pp_\mu\varphi_i$. It is to allow such terms and still retain the malleable symmetries that the next two Sections will introduce the gauge potential.

\subsection{$A$-dependent, rigid symmetries}

We first proceed precisely as in the first case,  introducing the $A$ field, but still \textit{keeping the symmetries rigid}. 
Using the equations of motion for $A$ as well as those of $\varphi$, i.e. $\mathsf{E}[A]=0$ as well as  $\mathsf{E}[\varphi]=0$, we get,   in direct analogy to \eqref{eq:A_rig}, a conserved current that is a sum of two currents:\footnote{To be explicit, the $A$-dependent terms that appear in the first line of \eqref{eq:vanishing} are $\big[ \frac{\delta \mathcal{L}}{\delta A_\nu}, A_\nu]^I+\big[\frac{\delta \mathcal{L}}{\delta\pp_\nu A_\mu}, \pp_\mu A_\nu\big]^I\approx \big[\pp_\mu \frac{\delta \mathcal{L}}{\delta\pp_\nu A_\mu}, A_\nu]^I+\big[\frac{\delta \mathcal{L}}{\delta\pp_\nu A_\mu}, \pp_\mu A_\nu\big]^I=\pp_\mu\big[ \frac{\delta \mathcal{L}}{\delta \pp_\nu A_\mu}, A_\nu]^I$. } 
\be\label{eq:conserv} \pp_\mu \left(\frac{\delta \mathcal{L}}{\delta \pp_\mu\varphi_i}(t^I\varphi)_i+\big[ \frac{\delta \mathcal{L}}{\delta \pp_\nu A_\mu}, A_\nu]^I\right)=\pp^\mu (J^I_\mu(\varphi)+\tilde J^I_\mu(A))\approx 0
\ee
and nothing more; there are  no further conditions that the terms of the Lagrangian need to obey.  (Here, the definition of $\tilde J^I_\mu(A))$ is given by \eqref{eq:conserv}.)

So, unlike the previous case, which admitted only a trivial kinetic term for the matter field $\varphi$, this sector will admit many possible dynamics. The problem here is of a different nature: the theories are not sufficiently constrained; the equations of motion do not automatically guarantee conservation of charges.

Let us look at an example of how things can go wrong in this intermediate  sector containing forces but only rigid symmetries, for the simple, Abelian theory. In the Abelian theory, $\tilde J(A)\equiv 0$, since quantities trivially commute. Thus Equation \eqref{eq:conserv} only contains the standard conservation of the matter charges and the symmetries are silent about the relationship between this charge and the dynamics of the forces. 

Consider a kinetic term of the form $\pp_{(\mu}A_{\nu)}\pp^{(\mu}A^{\nu)}$ where round brackets denote symmetrization. So this differs from the standard Maxwell theory kinetic term for the gauge potential: namely, $F_{\mu\nu}F^{\mu\nu}:=\pp_{[\mu}A_{\nu]}\pp^{[\mu}A^{\nu]}$ where square brackets denote \textit{anti}-symmetrization. But the symmetrized version is nonetheless gauge-invariant (under {\em rigid} transformations). Now, the Euler-Lagrange equations for this theory differ only very slightly from the Maxwell-Klein-Gordon equations. The equations of motion for $A$ yield:
\be\label{eq:wrongL}
\pp^\mu (\pp_{(\mu}A_{\nu)})=J_\nu
\ee
 in contrast with the usual $\pp^\mu (\pp_{[\mu}A_{\nu]})=J_\nu$. 
But clearly, unlike the usual case, the divergence of the left hand side does \textit{not} automatically vanish:
\be\label{eq:counter_exA}
\pp^\nu\pp^\mu (\pp_{(\mu}A_{\nu)})=\pp^\mu\pp_\mu\pp^\nu A_\nu=\square\pp^\nu A_\nu\not\equiv0.
\ee
At this point, we would have to go back to the drawing board and introduce more constraints on the theory: this theory does not couple forces to charges in a manner that guarantees charge conservation.

Thus we glimpse my overall thesis: only by introducing malleable gauge symmetries do we restrict interactions between forces and their sources so that they are consistent with the conservation of the matter current. 

Of course, in this example it is easy to see what is the smoking gun: the kinetic term $\pp_{(\mu}A_{\nu)}\pp^{(\mu}A^{\nu)}$ is \textit{not} invariant under malleable transformations. According to the next Section---our fourth sector---requiring this stronger form of invariance will restrict us to the space of consistent interactions.  No tweaking required.

\subsection{$A$-dependent, malleable symmetries}
In this fourth sector, we again obtain \eqref{eq:conserv}, from the vanishing of the first line of \eqref{eq:vanishing}, since nothing changes at that level. But, from the vanishing of the second line in Equation \eqref{eq:vanishing}, we have:
\be
 -\frac{\delta \mathcal{L}}{\delta A_\mu^I}=\frac{\delta \mathcal{L}}{\delta \pp_\mu\varphi_i}(t^I\varphi)_i+\big[ \frac{\delta \mathcal{L}}{\delta \pp_\nu A_\mu}, A_\nu]^I=J^I_\mu(\varphi)+\tilde J^I_\mu(A).
\ee
Once again using the Euler-Lagrange equations for $A$, to substitute the left-hand side, we find that
\be \mathsf{E}(A)^I_\mu=\frac{\delta \mathcal{L}}{\delta A_\mu^I}-\pp_\nu\frac{\delta \mathcal{L}}{\delta \pp_\nu A^I_\mu}\approx 0.
\ee
Defining $\frac{\delta \mathcal{L}}{\delta \pp_\nu A^I_\mu}=:k_{\mu\nu}^I$, we now obtain: 
\be\label{eq:final_conserv} J^I_\mu(\varphi)+\tilde J^I_\mu(A) = -\pp^\mu k_{\mu\nu}^I+  \mathsf{E}(A)^I_\mu\approx -\pp^\mu k_{\mu\nu}^I
\ee
This equation links both the matter and force currents to the dynamics of the force field. 

We already know from the vanishing in the first line of Equation \eqref{eq:vanishing} that the sum of the currents is divergence-free on shell (cf. Equation \ref{eq:conserv}). Thus, taking the divergence on the left hand side of \eqref{eq:final_conserv}, we must have $\pp^\nu \pp^\mu k_{\mu\nu}^I= 0$. Since two derivatives of a scalar field are necessarily symmetric, all we need in order to satisfy conservation is that:
\be
k_{\mu\nu}^I=-k_{\nu\mu}^I\quad \text{or}\quad k_{\mu\nu}^I=k_{[\nu\mu]}^I,
\ee
 which is just what we have from the vanishing of the \emph{third} line of Equation \eqref{eq:vanishing}. Thus, the result of including malleable symmetries, in this simple case, restricts us to consider Lagrangians in which the derivatives of $A^I_\mu$  only enter in anti-symmetrized form: $\pp_{[\mu}A^I_{\nu]}$. This restriction excludes the previous example of equation \eqref{eq:wrongL}. 

More generally, if we try to find a Lagrangian that includes  force fields without obeying the relations obtained from the malleable symmetries, the equations of motion of the force fields and those relating force fields and matter may require further constraints to be compatible with charge conservation, as we saw in the counter-example in the previous section. This is the power of local gauge symmetries: they link charge conservation---taken as empirical fact or on a priori grounds---with the form of the Lagrangian for the force fields. 


\subsection{Masslessness: An invitation}
There is yet more information that can be gleaned from the Noether gauge argument, which is contained in equation \eqref{eq:final_conserv}: upon integration, it yields a boundary term and a volume integral. That is, it gives  a relation between a quantity at a far-away boundary---related to the flux of the force field components $A^I_\mu$--- and the matter content inside this region. In the Abelian case, for the $0$th component of the equations of motion, this just gives the standard Gauss law. But more generally, being detectable at arbitrarily long distances makes the `forces' associated to $k_{\mu\nu}$ long-range.\footnote{This is a classical treatment. Quantum mechanically, non-Abelian theories would suffer from confinement, which lies outside the scope of this discussion.} 

It is common to conclude\footnote{For example, in the context of quantum electrodynamics, compare \citep[p. 343]{WeinbergQFT1}.} on this basis that gauge invariance \emph{forbids the presence of a mass term} for $A_\mu$. However, like the textbook gauge argument, the general form of this argument for masslessness is often heuristic in character. In particular, it assumes that each term in the Lagrangian is independently gauge-invariant. Then it is true that, on its face, a term like  $m^2 A^I_\mu A_I^\mu$ is not invariant under malleable symmetries.\footnote{Note that the Proca action does include a mass term for the photon field, i.e. the gauge potential $A$, but it is only gauge-invariant with $m=0$, in which case it just reduces to the standard Maxwell equations. For $m\neq 0$, one must have, in relation to the Maxwell equations, gauge-breaking, or `gauge-fixing', conditions. For a discussion, see \citet[\S 3-2-3]{ItzyksonZuber}.} 

To show in full generality that masslessness is required would go beyond the scope of this Chapter: we leave it as an exercise to the ambitious reader to explore! However, we can still improve on the standard heuristic argument without much effort in a special case that includes electromagnetism, by enforcing Equation \eqref{eq:vanishing} off-shell, for all models $(\varphi, A)$, and requiring that any mass term $m$ be field-independent and that the only coupling between the matter field and $A$ be just $A_\mu^I J^\mu_I$. 

A term in the Lagrangian of the form $m^2 A^I_\mu A_I^\mu$ would not leave any trace in the first line of Equation \eqref{eq:vanishing}.   But from the second line, again assuming no on-shell constraint between values of the matter and force field, from the second term, we obtain $m^2A_\mu^I$, which can only cancel with something dropping out of $\big[ \frac{\delta \mathcal{L}}{\delta \pp_\nu A_\mu}, A_\nu]^I$. Call this term $\kappa_{\mu\nu}(A, \pp A)$, which is such that $[\kappa_{\mu\nu}, A^\mu]^I\propto A^I_\nu$ for all $A$. Take, in this basis, $A^I_\nu$ to be constant, e.g. $\delta^I_{1}\delta_\nu^{y}$, so that $A=\tilde A:=\tau_1\otimes dy$, where $y$ is a spacetime coordinate function and $\tau_1$ is one element of the Lie-algebra basis.  This implies that the partial derivatives inside $\kappa_{\mu\nu}(\pp A, A)$ vanish, that is, $\kappa_{\mu\nu}(\pp \tilde A, \tilde A)=\kappa_{\mu\nu}(\tilde A)$, and therefore that it is a polynomial of $A$ with no derivatives, i.e. it is a polynomial that contains a single element of the  Lie-algebra basis, $\tau_1$, and the spacetime 1-form basis, $dy$. But this means that the commutator $[\kappa_{\mu\nu}, \tilde A^\mu]^I$ vanishes, and therefore cannot be proportional to $\tilde A$. As a result, in order to maintain off-shell invariance under malleable transformations, a mass term for $A$ cannot be included in the Lagrangian.



\section{Conclusion}\label{sec:conclusion}

I have given a detailed defence of the use of gauge symmetries for theory-building, in the spirit of the textbook gauge argument.

I  defended a  general `Noether gauge argument' in classical field theory. In particular, gauge symmetries of various kinds were fed into the powerful theorems of Emmy Noether, in order to produce precise constraints on the possible dynamics. The result is more than a simple argument that gauge symmetry is useful for theory construction: gauge symmetry constrains how one can consistently combine charges with the fields they interact with. Noether's first theorem of course implies charge conservation; but the second theorem then implies relations between the theory's equations of motion, which amounts to a coupling constraint. In other words, converting a rigid symmetry into a malleable one enforces the compatibility between charge conservation and the dynamics of the corresponding fields: a result which has not yet been stressed by the philosophical literature.

Of course, one may still feel that once our theories---present and future---have been successfully constructed, why not throw gauge symmetries away, and move down to a description in which such redundancies have been eliminated? I would reply that, in addition to the several reasons for gauge identified in this thesis, we find that gauge symmetries provide an explanatory reason---a reason drawing heavily on Noether's two theorems---for the way charges couples to fields.

 \chapter{How to choose a gauge: the case of electromagnetism}\label{ch:Coulomb}
 

 \section{Introduction and roadmap for this Chapter}\label{intro}

Recently,   \cite{Maudlin_ontology} has exhorted not only philosophers, but also modern theoretical physicists, to be clearer about their theories' ontology.  
He writes: `both the glory and the bane of modern physics is its highly mathematical character. This has provided both for the calculation of stunningly precise
predictions and for the endemic unclarity about the physical ontology being postulated'  \citet[p.5]{Maudlin_ontology}. We agree. Today, more than in yesteryear, a theory's mathematical formalism is often interpreted \emph{en bloc}. No special care is taken to specify: which parts represent ontology, `what there is' (and within that: what is basic or fundamental, and what derived or composite); and which parts represent `how it behaves' (which Maudlin (p. 4) calls `nomology': in particular, dynamics); and which parts represent nothing physical, but instead mathematics (which, though unphysical, can of course be invaluable for calculation). 

We also endorse Maudlin's programme to develop presentations of our physical theories that are clear about these distinctions; (though one should of course accept  that what we usually consider to be one theory may admit two such meritorious presentations---no uniqueness claim is required.)\footnote{So this programme is consistent, in particular, with rejecting the logical positivists’ aim of presenting physical theories with a once-for-all division of fact and convention: a rejection we share with e.g. \cite{Putnam_analytic}.}

As a case study of his programme, Maudlin considers a theory that, in the hierarchy of mathematical sophistication of modern physics, sits relatively low, viz. classical electromagnetism. He then applies the results to assess some proposed interpretations of the Aharonov-Bohm effect. 

In this Chapter, we also will consider classical electromagnetism,  and with an overall aim similar to Maudlin's---to clarify interpretative issues. But we construe this as a clarification of representational conventions (cf. Section \ref{sec:method}), and we focus on the Hamiltonian formulation of the theory in Section  \ref{sec:HamEM}. For the reader that is unfamiliar with the  Hamiltonian framework, we expound it   for finite-dimensional systems, rather than for field theories, in Appendix  \ref{sec:Ham}. This exposition includes a comparison with the Lagrangian framework, and the treatment of constraints. Despite our adopting the Hamiltonian framework, some of our conclusions, technical as well as interpretative, will be concordant with Maudlin’s: in particular, about the Coulomb gauge having a special role.

It is often remarked that the Hamiltonian framework’s use of a time parameter, and-or of a `3+1’ split of spacetime, carries the price, for special relativistic theories like electromagnetism, that one loses manifest Lorentz invariance. We of course accept that this is a limitation; though we note that often (including in our discussion below) the Lagrangian framework also uses a time parameter.

But  for interpreting the gauge aspects of electromagnetism, the Hamiltonian framework has two 
  countervailing benefits. 
 First: its pairing of momentum and configurational  degrees of freedom means that we can  pair degrees of freedom of the electric field with degrees of freedom of the vector potential. Second: the Hamiltonian framework illuminates symmetries that are time-dependent of the sort associated (in both it and the Lagrangian framework), with constraints and the failure of determinism.  This illumination comes from the way the Hamiltonian framework algorithmically identifies  subsets of the equations of motion that represent such symmetries   (viz. `the Dirac algorithm'; see Section \ref{sec:syms_consts} as well as Appendix \ref{sec:Ham}). Combining these benefits gives an illuminating, and interpretatively clear,  splitting of the  electric field into its Coulombic and radiative parts. Besides, this splitting gives a special role to the Coulomb gauge,  i.e. div$({\bf A}) = 0$.\footnote{ Maudlin also advocates this gauge, but for very different reasons than us. We discuss the differences in Sections \ref{sec:Gauss} and \ref{concl} (cf. footnote \ref{analo}), but in short: he makes a controversial ontological proposal, while we, less contentiously, see the gauge as ``merely’’ natural, because of its reflecting the splitting of the electric field.}

  So our plan is as follows. Section  \ref{sec:conceptual} addresses some conceptual disputes  surrounding this Chapter's enterprise of choosing a gauge, in particular, in relation to \citet{Maudlin_ontology}. So we address whether the type of non-locality present in gauge theory can be used to pick a gauge in the classical domain of the theory. This requires us to briefly revisit the Aharonov-Bohm effect (cf. Section \ref{sec:AB}) and compare with the type of non-locality studied in \cite{Myrvold2010} (called non-separability). Following \cite[Section 11]{Belot2003}, we also discuss the nomological status of points off of the constraint surface. This helps justify our focus on parametrizations of the constraint surface as the first step in choosing a gauge condition that we judge to be `non-contingent' (in the sense that the form of the equations, not their solutions, are non-contingent).

  In Section \ref{sec:HamEM} we will apply the  ideas discussed in Section \ref{sec:conceptual} (cf. also Section \ref{sec:syms_consts} and   Appendix \ref{sec:Ham}) about the constrained formalism  to electromagnetism: the Gauss constraint emerges from the Dirac algorithm, and it is the generator of gauge symmetries. Finally, we will split the electric field into the part that is uniquely fixed by the Gauss constraint---and so by the instantaneous distribution of charges---and the remainder. We thereby find a symplectically corresponding decomposition of the gauge potential into a part that is pure gauge and a remainder: call them $X$ and $Y$, respectively. If we single out the  scalar degree of freedom of the electric field that  is fixed by the Gauss constraint, by representing it as a gradient, it turns out that, on the other side of the symplectic correspondence,  Coulomb gauge is singled out. More precisely: if we demand that in our split of the gauge potential, the part  $Y$ of the gauge potential that remains after we extract the pure gauge part    $X$, 
  is dynamically independent from the part of the electric field that is fixed by the Gauss constraint,  then this non-gauge part  $Y$  will obey the Coulomb gauge equation.  
  
 In other, somewhat less technical, words: my main idea is as follows. The Coulombic part of the electric field is its electrostatic-like component, which is determined by the instantaneous distribution of charges. The electric field has the gauge potential as its conjugate. And as to the Coulombic part of the electric field, there is a part of the gauge potential that is not at all conjugate to it (i.e. is symplectically orthogonal to it). That part of the potential satisfies the Coulomb gauge condition.\footnote{So the nomenclature is confusing, since the Coulombic part of the electric field corresponds to, i.e. is conjugate to, the {\em other} part of the gauge potential than that which satisfies the Coulomb gauge condition. Thus one might facetiously propose that for clarity, we should: (i) associate the adjective `Coulombic’ with Coulomb’s historical work on electrostatics, and so apply it only to the electrostatic-like component of the electric field; and (ii) as regards the gauge potential,  re-name the gauge condition as the {\em `ortho-Coulomb condition’}.} Besides, this result is worth expounding, since one usually thinks of a choice of gauge as motivated by calculational convenience, often for a specific problem---and so from a general theoretical perspective, as completely arbitrary: whereas this result shows that the choice is related  to a physically natural, and general, splitting of the electric field.

In the short final Section \ref{concl}, we conclude.

\section{Conceptual issues}\label{sec:conceptual}

 Both the Lagrangian and Hamiltonian frameworks show the special status of constraints: they are the generators of symmetries and must be imposed prior to any equation of motion, if there is to be a correspondence between the frameworks. In Section \ref{decide!}, we address whether the non-locality implicit in  the Gauss constraint (see Section \ref{sec:syms_consts}) can be used to meaningfully pick out part of the electric field. This clarification requires us to briefly reassess ideas of non-locality and non-separability of \cite{Myrvold2010}, also in relation to the Aharonov-Bohm effect.  In Section \ref{sec:off}, we give some philosophical remarks about the status of states that are not in the constraint surface.

\subsection{Three morals related to the Aharonov-Bohm effect}\label{decide!}
 In Chapter \ref{ch:sd_1},  I have urged an interpretation of gauge theories---more specifically: the family of Abelian and non-Abelian classical Yang-Mills theories, which includes electromagnetism---that takes a single physical possibility to correspond to an entire gauge-equivalence class.\footnote{\label{analo}This is disanalogous to Maudlin, since in the enterprise of interpreting gauge, he investigates ontology associated with different representatives of each equivalence class. So his reasons for advocating the Coulomb gauge are very different from our reasons for 
 seeing it as special.} In Maudlin's jargon of ontology (`what there is') and nomology (`how it behaves'), this interpretation, explicated in Part \ref{part:I}, in brief,  is that:\\
 \indent  \indent (i) \emph{the ontology} consists of a field over spacetime, that encodes the relations  between charges that interact via a particular type of force (e.g. electromagnetism), i.e. the field encodes  a relation of sameness of charges across  spacetime called `parallel transport'; \\
 \indent  \indent (ii) \emph{ the nomology} describes how this sameness relation is constrained by the distributions of charges in spacetime. \\
 Besides: although this interpretation was articulated in the Lagrangian (covariant) framework, we see no significant obstacle to an appropriate translation to the Hamiltonian framework (though  there are some subtleties  about how  to do this: cf.  the end of Section \ref{sec:Ham_symp} and \citep[Sec. 11]{Belot2003}).
 Thus   one main  aim of this Chapter is to  give such a translation.     

 In either  framework,  this summary, (i) and (ii), is of course a ``high altitude'' view of the theory. Nonetheless,  we  can already see how it accommodates facts that are characterised only non-locally  in terms of their spacetime properties. Thus, as we saw in Chapter \ref{ch:sd_2}, Section \ref{sec:AB}, one can parallel transport an internal quantity around a loop in spacetime, and whether that quantity comes back to its original value or not can depend on facts {\em outside} the vicinity of the loop.
  
  In the standard jargon of electromagnetism, this type of non-locality is associated with  the Aharonov-Bohm effect and the introduction of \emph{gauge} and {\em vector potentials}.  The most important lesson of the effect is   of course that there is physical significance in gauge. This is highlighted already in the opening of the original paper   \citep{aharonovbohm1959}:
\begin{quote}
In classical electrodynamics, the vector and scalar potentials were first introduced as a
convenient mathematical aid for calculating the fields. It is true that in order to obtain a
classical canonical formalism, the potentials are needed. Nevertheless, the fundamental
equations of motion can always be expressed directly in terms of the fields alone.
In the quantum mechanics, however, the canonical formalism is necessary, and as a result,
the potentials cannot be eliminated from the basic equations. Nevertheless, these equations,
as well as the physical quantities, are all gauge invariant; so that it may seem that even in
quantum mechanics, the potentials themselves have no independent significance.
In this paper we shall show that the above conclusions are not correct and that a further
interpretation of the potentials is needed in quantum mechanics. 
\end{quote}

But we would like to stress  two points of clarification about  this quotation.  They  lead to three ``morals’’: 

(1): First: as emphasized in Part \ref{part:I}, although the potentials appear prominently in the theory, we need not interpret as physically different, two potentials that are related by a {gauge transformation}.  Thus the  `independent significance' of the gauge potentials needs to be taken {\em modulo} such transformations. 
 
 
  If one rejects this first point, then instead of articulating an ontology for the equivalence class of gauge potentials, one would try to articulate ontologies for different choices of an element of the class: i.e. for different choices of `gauge-fixing'. As we mentioned in footnote \ref{analo}, this is \cite{Maudlin_ontology}'s approach; (cf. also \cite[Sections 8-9]{Mulder_AB}).  In the end, Maudlin settles on Coulomb gauge as 
   combining a relativistic nomology with an ontology that requires a preferred foliation of spacetime: and is thus friendly to the pilot-wave/Bohmian   interpretations of quantum mechanics that he favours.   \\
  
(2):  Our second  point of clarification was described in Section \ref{sec:AB}: although the  experimental set-ups implemented hitherto for observing the Aharonov-Bohm effect have a quantum component---viz. an effect on the phase of a quantum wave-function---the {\em classical} theory already interprets the situations that give rise to the effect as physically distinct. So even if hitherto no {\em classical} way to experimentally manifest this distinction has been found, there aren't any  no-go theorems vetoing such a classical manifestation: that is, a measurement (ultimately, a pointer-reading) that uses some classical probe to register which of two $\bf A$-configurations, that are physically distinct since not differing merely by a gradient $\d \lambda$, but that determine the same magnetic field, is realized.

\medskip

Both these two points of clarification  (especially of course, the first) are  recognised in the literature.\footnote{The second is  well expressed by  \citet[pp. 532-533, pp. 550-554]{Belot1998}, who also addresses judiciously the question of what an interpretation of classical electromagnetism can tell us about a quantum world. A related question concerns what  we can learn about the world from the sort of idealized descriptions usually given of the Aharonov-Bohm effect: for example, excluding from space the region where the local gauge-invariant quantities such as curvature are distinct, i.e. excluding the solenoid. For recent controversy about such idealizations, cf. \cite{ShechAB, Earman2019, DoughertyAB}.
} 
 But what matters for us is that the second point is related  to   classical electromagnetism's exhibiting  a  type of \emph{non-locality}. This is usually expressed in terms of holonomies; (though a complete treatment without holonomies can be given---see \cite[Sec. 6]{GomesRiello_new}, \cite{Gomes_new}). Thus \cite{Myrvold2010} notices that, for classical electromagnetism {\em in vacuo}, and for a simply connected region: one can use the composition laws of the holonomies to decompose any gauge-invariant function on  this region into gauge-invariant functions of its component---i.e. mutually exclusive, jointly exhaustive---sub-regions. But, Myrvold continues: for a region that is not simply connected, there are certain ``large'' holonomies that cannot be obtained from composition of holonomies confined to the sub-regions. That is, there are certain global gauge-invariant functions that are not \emph{separable}.  Myrvold concludes that this type of non-locality only arises for non-simply connected regions (and is exhibited by the Aharonov-Bohm effect). 
 
At first sight, this conclusion is in tension with our own discussion, in Section \ref{subsec:sims}, of the role in the Hamiltonian framework of the {\em Gauss constraint}, i.e. the equation, in elementary terms, that the divergence of $\bf E$ is equal to the charge density $\rho$. As we saw there,  this equation   involves a sort of {\em non-locality} even in the simply-connected domain. For it implies that by simultaneously measuring the electric field flux on all of a large  surface surrounding a charge distribution, and integrating, we can ascertain the total amount of charge inside the sphere {\em at the given instant}.  So this non-locality is classical, and regardless of whether the volume enclosing the charge distribution is simply connected  (the surface surrounding the charge distribution could be topologically a doughnut, not a sphere)---which is apparently, at odds with Myrvold's conclusion.      
But here we should recall that Myrvold's analysis is restricted to electromagnetism {\em in vacuo}; and fortunately, when we remove the restriction, the tension disappears. That is: when we allow for charges, the same  line of argument  that Myrvold uses also proves that there are gauge-invariant functions that are not separable---even for simply connected regions.  In short: in the presence of charges, we have  non-locality (and non-separability) even for simply-connected regions.  The Appendix \ref{sec:elim_elim} will give more details.
\medskip

We can sum up this discussion of our relation to Maudlin’s enterprise, and of the Aharonov-Bohm effect, in three morals, as follows:\\
\indent \indent (1): Even setting aside Aharonov-Bohm phases, {\em classical} electromagnetism in the Hamiltonian formalism shows a certain kind of (non-signalling) non-locality, namely in the Gauss constraint.  (This was discussed in Section \ref{sec:AB}).\\
\indent \indent (2):   There is no   unique physically preferred split of the field's degrees of freedom into purely local and purely non-local ones. But each such choice of split can be made to correspond to a choice of gauge-fixing. \\
\indent \indent (3): In particular, the Coulomb gauge corresponds in this way to a natural choice of splitting of the electric field; (which, incidentally, buttresses some of \cite{Maudlin_ontology}'s arguments). That is: this gauge follows naturally   from considering the Gauss constraint to single out a `scalar' part  of the electric field that is determined by the instantaneous distribution of charges.

 \subsection{Off the constraint surface}\label{sec:off}
 Our sketch of the Hamiltonian treatment of constraints would be incomplete without some mention of {\em symplectic reduction}: a large and important topic  (briefly mentioned in footnote \ref{ftnt:magic}).  However, it is usually pursued, not  (as in this Chapter) by using the the Legendre transformation being many-one to motivate  restricting one’s attention to the constraint surface $\Gamma$ in phase space (as in this Chapter); but by postulating {\em ab initio} a smooth action of a Lie group $G$ on phase space, and studying the consequences. So the relation to the Lagrangian framework, and to constraints originating from $\frac{\partial p_\beta}{\partial {\dot q}^\alpha}$ being not invertible, tends to be obscured. But the rest of this Chapter will not need an account of symplectic reduction: for which, cf. e.g. \cite{Marsden2007} for a  complete, but concise, exposition, \cite{Butterfield_symp} for a philosophical introduction, and \cite{GomesButterfield_glimpse} for the relation to the Lagrangian framework.

However, symplectic reduction prompts a philosophical topic we want to address. It is about possibility, i.e. about how we should think of the non-actual i.e. unrealized states in the state-space. We have hitherto said nothing about this, since our discussion has  prompted no alteration from how one normally thinks of possibility within the Hamiltonian dynamics of an {\em unconstrained} system. There, one naturally regards which energy hypersurface the system is actually on (or equally: the actual value of any first integral of the motion) as a  matter of initial conditions, mere happenstance. And for all we have said so far, it seems that in general, this attitude applies equally to constrained systems. For in general, it seems that the constraints, labeled $\Phi^I$ in  Section \ref{sec:syms_consts}, could have taken values other than zero  (cf. Appendix \ref{sec:Ham} and Equation \ref{eq:ctraints}): the state could have been {\em off} the actual constraint surface $\Gamma$. But the theory of symplectic reduction reveals a wrinkle: indeed, two wrinkles. \cite[Section 11]{Belot2003}  discusses them, as do  \cite{GomesButterfield_glimpse}; and we sketch them here.

First, in some cases there is reason to deny that the states lying off the constraint surface $\Gamma$ are genuinely possible. 
Belot's (and our) simplest  example is {\em relationism}, {\em \`a la} Leibniz and Mach, about space. The case can be made for a system of $N$ point-particles in Euclidean space. For  this system, while the `absolutist' will take the configuration space $\cal Q$ to consist of all the ways $N$ particles can be placed in ${\RR}^3$, i.e. to be ${\RR}^{3N}$, the relationist will say that two such placements that differ by a spatial translation and-or a rotation should be regarded as one and the same. That is, the relationist advocates a {\em relative configuration space}, whose points are sets of relative distances between the particles. This space can be presented as the quotient of  ${\RR}^{3N}$ by the obvious action of the Euclidean group; (modulo some technicalities about excising unsuitably symmetric points of ${\RR}^{3N}$). And when one works through the details of the constraint formalism, it turns out that on the relationist's view, only states on the relevant constraint surface $\Gamma$ within $T^*{\cal Q}$ are genuinely possible.  Indeed, when the  action of a Lie group $G$ on phase space $T^*{\cal Q}$ is induced by a  group action on configuration space $\cal Q$, we have (if a few more regularity conditions discussed by Belot are satisfied) that:\footnote{Incidentally, such a relationist understanding of symmetries would also disallow many of \cite{Belot_sym}'s `problematic'  examples of symmetries that seem to incur observable changes (see Section \ref{sec:syms_tech}). But in Hamiltonian general relativity (cf. \ref{sec:syms_consts}), the symmetry that is related to `refoliations' of the equal-time surfaces can not be so induced by configurational symmetries. But this exception may not be so bad:  as discussed in Section \ref{subsec:sims} (see footnote \ref{ftnt:CMC}), the imposition of refoliation symmetry is more contentious, as it is related to the infamous `Problem of Time' \cite{Isham_POT, Kuchar_Time}. In  \cite{gomes:aop2018}, it is argued that the maximal group of local configurational symmetries is that of conformal diffeomorphisms: a semi-direct product between (spatial) diffeomorphisms and  conformal transformations. Such a group is intimately related to CMC slicings (cf. footnote \ref{ftnt:CMC}) and is instrumental in the formulation of the relationist geometrodynamical theory known as \emph{shape dynamics} (see \cite{SD_first, SD_linking, Flavio_tutorial}).     }
\be T^*{\cal Q}/G\neq T^*({\cal Q}/G), \quad\text{but}\quad \Gamma/G\simeq T^*({\cal Q}/G).
\ee

Of course, not everyone is a relationist! But also in other cases, there is a similar rationale to `endorse the dynamics intrinsic to $\Gamma$', and reject the states not in $\Gamma$. Thus Belot points out that in some field theory cases, theorizing about states off the constraint surface corresponds to treating charges that source the field in question without being affected by it, i.e. treating external sources. For example, in our case-study of electromagnetism: for non-vacuum, the Gauss constraint becomes div$(E) = \rho$, where $\rho$ is treated as sourcing, but as unaffected by, the electromagnetic field. (Of course, there is a close analogy with Poisson’s equation $\nabla \phi = \rho$ in Newtonian gravity, and its modern descendant, Newton-Cartan gravitation: the mass density $\rho$ sources the potential $\phi$ but does not self-gravitate.) Since this is an idealization, one has reason to reject the states off the constraint surface as not genuinely possible, and to endorse the dynamics intrinsic to $\Gamma$---like the relationist above.

Besides, as Belot goes on to say: faced with this idealization, one should seek theories in which the coupling is ``two-way’’. Indeed, there are such theories (references in his footnote 74); and---what matters for our present topic---in these theories, one again  gets \emph{only one constraint surface}, like in the case of vacuum electromagnetism. That is:  once one augments the phase space so as to describe the charges (including: augmenting the Hamiltonian to describe the two-way coupling), one gets just one constraint surface 
in a higher-dimensional space---{\em not} a family of surfaces in the  original lower-dimensional phase space, indexed by the charge distributions. 
  So again, one has reason to endorse the dynamics intrinsic to the constraint surface.  And again: our main theme about gauge structure is illustrated: viz. that null vector fields on the constraint surface are infinitesimal generators of gauge transformations.

Let us summarise this discussion by quoting Belot. He writes (p. 215):---
\begin{quote}

This [i.e. treating charges as external, i.e. as sourcing the field in question yet without being affected by it] amounts to working off of the constraint surface in order to study the field dynamics in the presence of external sources painted onto spacetime independently
of the behavior of the fields. This is, of course, an ad hoc maneuver—if one wants to
study Yang–Mills with sources honestly, one must introduce matter which not only
acts upon the field but is also acted upon by it. And when one pursues this upright
course, one ends up with a constraint which is a direct analog of the usual Gauss
constraint—the null directions of the constraint surface correspond to the infinitesimal generators of gauge transformations. Under this more fundamental
approach, there is no physical interpretation for points lying off of the constraint
surface—and so we have an excellent reason to prefer an intrinsic reading of the
theory.
\end{quote}

\section{The case of electromagnetism}\label{sec:HamEM}

 We turn to our case-study, classical electromagnetism.
There are many accounts of its symplectic structure in the literature, both physical and philosophical: some of them of course excellent.\footnote{Excellent physics expositions include \cite{JacksonBook}. Excellent philosophical discussions include: \cite{Belot1998, Belot2003, Healey_book}. We particularly recommend \cite{Belot2003} for the Hamiltonian  formulation of vacuum Yang-Mills theories.} But there is a ``core'' of ideas and results, 
that is relatively easy to expound and illuminating, without  having to plumb the depths of the (elegant) symplectic geometry of the theory.  This core is, so far as we know, not articulated in the literature: certainly, it is not stressed. 

So as we announced at the start of Section \ref{intro}, the technical aim of this Chapter is to expound this core, and show that it sheds light on various ideas, both formal and interpretative. More specifically, we will try to shed light on (1) classical non-locality  and (2) the preferred splits of degrees of freedom. (These correspond to the three morals   at the end of Section \ref{intro}.) 

 In Section \ref{sec:toy} we will illustrate the ideas that are used in (2) in a simple toy example. We will see how certain considerations of convenience and simplicity can go a long way to selecting gauge-fixing conditions.  
  In Section \ref{sec:symp_EM} we will apply those ideas to electromagnetism, where they give rise to the Gauss constraint and gauge transformations. In Section \ref{sec:Gauss} we will interpret the Gauss constraint as encoding a type of non-locality, {\em \`a la} (1). It defines a part of the electric field, viz. the Coulombic field, that is determined by the instantaneous distribution of charges. (Adopting \cite{Maudlin_ontology}'s terms for a moment: it is not `fundamental ontology', but `derivative ontology', since derived from the charge distribution.) There is here a strong analogy with the elementary Newtonian gravitational potential $\phi$, 
 which is sourced by the mass distribution via Poisson's equation: and which is often said to be ``not physically real'', or  ``a convenient fiction'', since it has no energy or momentum, but only encodes, via its gradient $\nabla \phi$ the infinite battery of counterfactual conditionals about how test-masses located at the spatial point in question would accelerate. Similarly here: the Coulombic field encodes infinitely many counterfactual conditionals about how test-charges would move---if there were also no other part of the field, i.e. no radiative part, enjoying its own dynamics. 
 
   Using the decomposition of fields implied by this understanding, we will in Section  \ref{sec:radcoul} use the symplectic structure of the theory to find the ``block-diagonal conjugate structure''  of the electric and gauge potential fields, and show how this selects the Coulomb gauge.

\subsection{A toy example of natural coordinate choices on phase space}\label{sec:toy}
To make our aims more vivid, we return to the simple example we gave at the end of Appendix \ref{sec:Lag}.  Consider two identical free particles of mass $m$ on a line, with coordinates $q_1$ and $q_2$.   So the phase space $\cal P$ is 4-dimensional, with  coordinates  $(q_1, q_2, p_1, p_2)$.  The canonical Hamiltonian for the system is
\begin{equation}
 H = \frac{1}{2m}(p_1^2 +p_2^2).
\end{equation}
Now we add a constraint to the system, namely:
\begin{equation}
{\cal M} = p_1 + p_2 = 0 \, ;
\end{equation}
(which is clearly first-class). So the constraint ${\cal M}=0$ requires the total canonical momentum $P:=p_1 + p_2$ to vanish,   and defines a 3-dimensional constraint surface $\Gamma$ in $\cal P$. The total Hamiltonian is ${H}_T  = H + a \, {\cal M} $, where $a$ is an arbitrary function of time. It generates the time evolution
\begin{equation}\label{eq:mom_ex}
\dot q_1 = \frac{p_1}{m} + a 
\,, \qquad 
\dot q_2 = \frac{p_2}{m} + a \,,
 \qquad 
\dot p_1 = \dot p_2 =0 \,.
\end{equation}

Here we started with the variables that initially seemed natural, centered on each of the particles. But a more natural choice for coordinatizing momentum space  (i.e. at each fixed value of $(q_1, q_2)$)  would be $P_+(p_1,p_2):=p_1+p_2$ and $P_-(p_1, p_2)=p_1-p_2$. Now   at each  value of $(q_1, q_2)$, the constraint surface is given by: 
\be 
\{(P_+, P_-)\, |\, {\cal M}(z)=0\}=\{(0, P_-) \, |\, P_-\in \RR\} \, .
\ee

In the $q$-coordinates, the natural conjugate variables to $P_-$ and $P_+$ are $Q_-=q_1-q_2$ and $Q_+=q_1+q_2$, respectively. The first, $Q_-$, is the relative distance between the two particles. It  is gauge-invariant, $\{ q_1 - q_2 , {\cal M} \} = 0$; and its equation of motion contain no arbitrariness,
\begin{equation}
\label{eq:dotQ}
\dot q_1 - \dot q_2 =\dot Q_-= \frac{p_1}{m} -
 \frac{p_2}{m}  =\frac{P_-}{m}\,.
\end{equation}
On the other hand,  from \eqref{eq:mom_ex},
\be\label{eq:gauge_q} \dot Q_+=m \, \dot q_1
+ m \, \dot q_2 = 2m a \ee
is ``pure gauge'', since $a$ is an arbitrary function of time. Note that \eqref{eq:dotQ} and \eqref{eq:gauge_q} are given on the full phase space: of course, the symplectic form on the constraint surface would be degenerate, since $P_+=0$ there.   To sum up: $Q_-$, $Q_+$ and $P_-$ are natural coordinates to parametrize the constraint surface.  

In what follows, we will try to provide a similarly natural decomposition of electromagnetism's gauge potential and electric field. ${\cal M}$ will be the Gauss constraint, which we will take to be naturally parametrized by a Coulomb potential. So this potential  will play the role of $P_+$, and thus the radiative degrees of freedom of the electric field will play the role of $P_-$. As regards the configuration variables: $Q_-$ will be given by a more complicated function of the original configuration variables, which projects it into Coulomb gauge, but it will be likewise gauge-invariant. And finally, $Q_+$ will be the pure gauge part of the gauge potential. 

\subsection{Hamiltonian treatment of electromagnetism}\label{sec:symp_EM}

The Maxwell equations were written in Section \ref{sec:crude_A}. In the presence of charges,  the equations of motion in terms of the gauge potential, \eqref{eq:eom_A}, are written as: 
\be\label{eq:eom_Aj}
 \partial^\mu\partial_\mu A_\nu-\partial^\mu\partial_\nu A_\mu=j_\nu.
 \ee
 These equations (together with the equations for the dynamics of the charges constituting the currents) are obtained from the action functional:
  \be\label{eq:action_Aj}S[A]:=\int \d^4 x\left( \partial_{[\mu}A_{\nu]}\partial^{[\mu}A^{\nu]} +A^\mu j_\mu +\mathcal{L}_{\text{\tiny{matter}}}\right),  
 \ee
 where $\mathcal{L}_{\text\tiny{matter}}$ is the Lagrangian density for the matter fields; (for illustration, one can take this  as the Klein-Gordon Lagrangian or as the Lagrangian for a charged point particle).
 
Now we choose a spacetime split into spatial and time directions,  $M=\Sigma\times \RR$.  (We recall Section \ref{intro}'s admission that this carries the price of losing manifest Lorentz invariance.) We also assume  that the fields have appropriate fall-off conditions at spatial  infinity.

Upon such a spacetime decomposition, the components of the electromagnetic tensor recover the familiar electric and magnetic fields: $F_{i0}=E_i$, and $F_{ij}\epsilon_i^{jk}=B_i$ and $j_0=\rho$  (where we used the three-dimensional totally-antisymmetric tensor, $\epsilon$, or the spatial Hodge star, to obtain a 1-form, and $i, j, k$ are spatial indices, i.e. in $\Sigma$), and the first equation of \eqref{eq:EM} becomes the familiar Maxwell equations.\footnote{  For $\mathcal{L}_{\text{\tiny{matter}}}=mv^a v_a$, for $v^a=(\dot \gamma)^a$ the 4-velocity of a charged particle whose trajectory is $\gamma$, i.e. such that $j^a=e v^a$, we obtain the Lorentz force equation  as the equation of motion for the particle.}

Now we perform a Legendre transform. Then the spatial vector fields $\bf A$ and  $\bf E$ are canonically conjugate. In fact, $\bf A$ is the configurational variable, and $\bf E$ is the momentum; since ${\bf E} = {\pp {\cal L}}/{\pp {\dot{\bf A}}}$. The Poisson bracket is defined, for $F, G$ two functionals of the fields $A$ and $E$; 
\be\label{PoisField}
\{F,G\} := \int d^3y \; (\frac{\delta F}{\delta A_i(y)}\frac{\delta G}{\delta E^i(y)} \, - \, \frac{\delta F}{\delta E^i(y)}\frac{\delta G}{\delta A_i(y)}) \, .
\ee

Now, as described in Section  \ref{sec:Lag} and Section \ref{sec:Ham_symp}, the Lagrangian has symmetries, which translate into constraints in the Hamiltonian formalism. Analogously to \eqref{eq:Ham_tot}, we obtain a total Hamiltonian, written as $H_T=H(A, E, j, \lambda)+{H}_{\text{\tiny{matter}}}$, with: 
\be\label{eq:Ham}
H(A, E, j, \lambda)=\int \d^3 x \left(\|\mathbf E\|^2+\|\mathbf{B}^2\|+\lambda(\mathrm{div}(\mathbf E)-\rho) +A^i j_i\right), 
\ee
where $\partial_{[i}A_{j]}\partial^{[i}A^{j]}=:\|\mathbf{B}^2\|$,  $\mathrm{div}(\mathbf E):=\pp^i E_i$. 

The part of \eqref{eq:Ham} that we want to draw attention to is the term $\lambda(\mathrm{div}(\mathbf E)-\rho) $. For $\lambda$ is a scalar function on the spatial surface $\Sigma$: $\lambda$ is the \emph{Lagrange multiplier}. In Appendix \ref{sec:Lag},  it is just the $a^I$ of \eqref{eq:Ham_tot}, but  now in the field-theoretic context, when $I$ becomes a continuous index. The constraints corresponding to ${\Phi}^I$, are just \eqref{eq:Gauss} in the Abelian case, i.e.:
\be 
\label{eq:Gauss_E}
\mathrm{G}(x):=\mathrm{div}(\mathbf E)(x)-\rho(x)=0, \quad \text{for all}\,\,x\in \Sigma. 
\ee
Accordingly, \eqref{eq:Gauss_E} defines not a single constraint, but an infinite set of them: one per spatial point.    The values of $\lambda$ at the various points $x \in \Sigma$ thus give a particular linear combination of the constraints. Hence $\lambda$ is also called a `smearing' of the constraints; and it  is convenient to define the smeared Gauss constraint: $\mathrm{G}(\lambda):=\int_\Sigma(\mathrm{div}(\mathbf E)-\rho)\lambda$. 

To see that we are in the domain of the previous discussion, namely that the constraints are all first class, we can check that the constraints  commute, and also commute with the Hamiltonian constraint. This is easy to verify, since a given linear combination of   symplectic flows, which we call a \emph{smeared} symplectic flow,  $X_\mathrm{G(\lambda)}$, acts on the canonical variables as: 
\begin{align}
X_{G(\lambda)}({A}_i(x))&:=\{  {A}_i, \mathrm{G}(\lambda)\}:=\int \d^3 y\frac{\delta( \lambda(\mathrm{div}(\mathbf E)-\rho))}{\delta E^i(x)}=\int \d^3 y\lambda(y)\pp_i\delta(x,y)=-\pp_i\lambda(x);\label{eq:G_A}\\
 X_{G(\lambda)}({E}^i(x))&:=\{ {E}_i, \mathrm{G}(\lambda)\}:=-\int \d^3 y \frac{\delta( \lambda(\mathrm{div}(\mathbf E)-\rho))}{\delta A_i(x)}=0\label{eq:E_gt}
\end{align}
To obtain these results, note that in each line the third term, i.e. the integral of a functional derivative, gets just one term from  \ref{PoisField}; and that  the final equation in \eqref{eq:G_A} is obtained by integration by parts removing the derivative of a delta-function.\footnote{The Gauss constraint is Lie-algebra valued, even in the Abelian case; that is necessary for it to generate infinitesimal gauge transformations. It just so happens that here   the Lie group $G$ is $S^1 = U(1)$, and so the Lie algebra is $\mathfrak{g}\simeq \RR$. In the non-Abelian theory,   the Gauss constraint is given in \eqref{eq:Gauss}: $G(x)=\D^a_A E_a-\rho$, where, $\D^a_A {E}_b:=\nabla^a E_b-[E_b, A^a]$. 
 Due to the appearance of $A$ in the Gauss constraint, it acts linearly on ${E}$: which is thus only covariant, and not invariant as it is in the Abelian theory. Similarly, the transformation of $A$ acquires an additional commutator, and so gets promoted to a gauge-covariant derivative of $\lambda$, i.e. $\D_i\lambda$. In more detail, apart from integration by parts described in the main text, we get, due to the cyclic trace identity, $\mathsf{tr}(\lambda [E_a, A^a])=[\lambda, E_a] A^a$, leaving only the commutator  $[\lambda, E_a]$ (or. resp. $[\lambda, A^a]$) after functional variation by $A$ (resp. by $E$) and integration; thus we derive  the right hand side of Equation \eqref{eq:gts_ham} and cash in the promisory note issued there. }

So we see that, as expected,  the constraints act on phase space as the familiar gauge transformations: they preserve the value of the electric field (which is gauge-invariant in electromagnetism) and change the value of the gauge potential by a gradient of a scalar function: $A_i\mapsto A_i-\pp_i \lambda$.
Since the flow of the constraint does not change the Hamiltonian (and preserves the constraint as well), it generates a symmetry of the system.

From the more geometric viewpoint, the infinite-dimensional phase space whose canonical coordinates are the electric field and the gauge potential has a bona-fide (infinite-dimensional) symplectic geometry (e.g. modeled on Banach manifolds, cf. \cite[Chapter 2]{Lang_book} and footnote \ref{ftnt:inf_dim}). Thus we have a symplectic form, which  in vacuum is written as:\footnote{It is easy to extend this to the presence of matter. For example, with a Klein-Gordon field, we would add: $\int \sqrt{g} \;\dd \bar\psi \curlywedge \gamma^0  \dd \psi$. See \cite[Section 3]{GomesRiello_new} for more details on the symplectic geometry of the infinite-dimensional space.}
\be\label{eq:symp_form}
\Omega  = \int \d^3x  \,\delta E^i \curlywedge \delta A_i.
\ee
Here we take $\delta E^i$ and  $\delta A_i$ to be the fundamental one-forms on phase space, and $\curlywedge $ to be their anti-symmetrization (the exterior product of   one-forms in an infinite-dimensional space). Note the possibly confusing nomenclature: here $\Omega$ has nothiing to with the curvature 2-form on the bundle, of \eqref{eq:curv_PFB}. 

Just as  for finite dimension, $\frac{\pp}{\pp x^i}, \frac{\pp}{\pp p_i}$ are the vectors tangent to the $x^i$ and $p_i$-coordinates: so also $\frac{\delta}{\delta E^i(x)}$ and $\frac{\delta}{\delta A^i(x)}$, for each $x\in \Sigma$, are vectors in  the infinite-dimensional  phase space. And just as  for finite dimension, we can find new directions, or vector fields, by the linear sum, 
$\sum_i a^i \frac{\pp}{\pp x^i}$: here also, we can find new directions by linear sums, i.e. by integrating the fundamental directions smeared  with certain coefficients, e.g. 
\be\label{eq:infv} 
\bb v=\int \d ^3 y\, v^i(y)\,\frac{\delta}{\delta E^i(y)}.\footnote{ In electromagnetism, the standard basis, namely $(E_i(x), A^i(x))$, for $x\in \Sigma$,  is in this language just a basis that takes the coefficient functions to be all multiplied by Dirac deltas, e.g. a subset of smearings $v^i_x=E^i(x)$ defined as $v_x^i(y)=E^i \delta(x,y)$, for each $x\in \Sigma$ and $E^i\in \RR$. This is what makes that basis completely local. Here we want to find a particular coordinate system, that is not completely local in this way. But in general, one could use symplectic duality to find the dual of any maximal Lagrangian submanifold in phase space.}
\ee (where we use a double-struck notation to indicate that these are vector fields on the infinite-dimensional phase space of classical electromagnetism).

Moreover, configuration space,  which is the space of smooth gauge potentials $\mathcal{A}=\{\mathbf{A} \in C^\infty(\Sigma, {\RR}^3)\}$ is itself a fibered manifold, with orbits given by the action of the (infinite-dimensional) group of gauge transformations, $\mathcal{G}=\{\lambda\in C^\infty(\Sigma)\}$, namely: $\mu:\mathcal{G}\times \mathcal{A}\rightarrow \mathcal{A};$\,\, $\mu(\lambda, \mathbf{A})=\mathbf{A}-\d\lambda$.  This structure exists also for the non-Abelian Yang-Mills theories. For both cases, by using the isomorphism that $\Omega$ provides between $T\mathcal{A}$ and $T^*\mathcal{A}$, this action can be lifted to phase space in the usual manner (yielding \eqref{eq:E_gt}, or its appropriate non-Abelian generalization,  \eqref{eq:gts_ham}).

\subsection{Interpreting the Gauss constraint}\label{sec:Gauss}

We now combine Sections \ref{decide!}’s and \ref{sec:symp_EM}’s discussions of the Gauss constraint with a comment on a proposal of Maudlin’s. This will pave the way for our main result in Section \ref{sec:radcoul}.

As we discussed in Sections \ref{subsec:sims} and \ref{decide!}: the Gauss constraint (now in the form of \eqref{eq:Gauss_E}) involves a (non-signaling) kind of non-locality. For (by the elementary divergence theorem) the integral of $\bf E$ over a surface enclosing a spatial region determines the total  charge  inside the surface {\em at the given instant}.\footnote{Here, `determines’ can be read as `gives us knowledge of’. For one can imagine measuring $\bf E$ precisely throughout the surface, doing the integral, and inferring the total charge. This is, of course, the intuitive basis for quantum theory’s charge superselection rule. For using this procedure, one can measure the total charge at an arbitrarily large distance from the system; and this suggests that measuring charge is compatible with measuring any other quantity on the system, so that the charge operator  commutes with the operator representing that quantity.} Like the Aharonov-Bohm effect, this type of non-locality is classical; but unlike the Aharonov-Bohm effect, it does not require an underlying topologically non-trivial spatial domain for the probe systems. 

As mentioned in this Section’s preamble, this suggests isolating the part of the electric field that is determined by the instantaneous distribution of charges, and thinking of it as `derivative’ from this distribution. We will shortly pursue that idea. But first we notice that \cite[p.10]{Maudlin_ontology} proposes to `turn this around’. Thus he writes:
\begin{quote}
Let us propose that this equation [i.e. div$({\bf E}) = \rho$] represents not a physical law but an ontological
analysis: electric charges just are the divergences of electric fields. In this way we reduce both the
physical ontology and the nomology, and further gain an explanation of why electric charges cannot
exist without electric fields.
\end{quote}
So Maudlin proposes that the Gauss constraint be read as a definition of $\rho$, the electric charge. (He springboards this proposal from the corresponding one (p. 9) about $\bf B$: that div$({\bf B}) = 0$ is completely equivalent to the non-existence of magnetic charges.)

We submit that---perhaps unfortunately---this proposal does not work. There are various problems. The most obvious is that there are particles with the same electric charge but differing in other ways; (they have different masses, and-or different charges for other interactions). 
 That is: particles have characteristics independent of their   electric charge. (And if one attempted to define the source of each distinct field in a similar eliminative manner, one would then have to explain why all the different sources and sinks coincide in space and time.) Other problems include: (i) since electromagnetism is a linear  theory, must charged sources pass though each other? (ii) how do we explain interactions of the electromagnetic field with (apparent) matter, such as light reflection and refraction?\footnote{Notice incidentally that Maudlin is re-inventing the wheel. It is a creditable re-invention, since the wheel has a venerable design: but for all that, the wheel does not roll. That is, without metaphors: Maudlin’s proposal is the initial idea of the {\em electromagnetic world-view} that aimed to reduce mechanics (and all physics) to electromagnetism. In particular, it took the velocity-dependence of particles' masses to suggest that all mass might be of electromagnetic origin. It was advocated in the early twentieth-century by physicists such as Abraham and Mie; for an introduction cf. \cite[Chapter 8]{Kragh1999}. We should also note the ongoing---yet also very speculative---programme to reduce both matter and radiation to structures in spacetime, sometimes called `super-substantivalism': for which, \cite{MisnerWheeler1957} is a classic, and \cite{Lehmkuhl2018} is a fine philosophical introduction. 
 
Maudlin’s proposal also connects with the discussion at the end of Section \ref{sec:off}. Namely: {\em if} the proposal held good so that charge was indeed fully defined by the electric field, then our objection in that discussion, that treatments of charge as sourcing the field but as unaffected by it are an idealization---which should be replaced by a ``two-way’’ coupling---would fall by the wayside.   
}

But the failure of this proposal does not impugn the more modest idea above: that we should isolate the part of the electric field that is determined by the instantaneous  charge distribution, and think of it as `derivative' from the distributions.  So we think of this part---the Coulombic part---of the field as `the price to pay' for a local formulation of the theory; but it can be replaced by the more fundamental distribution of charges acting at a distance. In other words: there is a certain component of the interaction between charges that does {\em not} take into account a field that has its own dynamics: all this component needs is the present distribution of the charges themselves. The remaining part of the electric field is then interpreted  as `fundamental', in the sense that it has its own dynamics: a dynamics that is not reducible to the dynamics of other components of the theory.

The upshot will be that, since $\mathbf{E}$ and $\mathbf{A}$ are conjugate, and we take part of $\mathbf{A}$ to be `pure-gauge',  the above decomposition of $\mathbf{E}$ can be made to  correspond, in a particular way, to a certain decomposition of $\mathbf{A}$. We will do this in parallel to the toy case of \S \ref{sec:toy}, but here defining the gauge-complementary part of $\mathcal{A}$ to be symplectically orthogonal to a natural parametrization of the Coulombic part of $\mathbf{E}$, and by defining the Coulombic-complementary part of $\mathbf{E}$ to be symplectically orthogonal to the pure gauge part of $\mathbf{A}$.

\subsection{Radiative and Coulombic parts of $\mathbf{E}$ and $\mathbf{A}$}\label{sec:radcoul}

Now we come at last to our application of {\em the Helmholtz theorem}. The theorem states that any vector field $\mathbf{Z}$ on $\RR^3$ can be decomposed (`split') as:
\be \mathbf{Z}=\mathbf{X}+\mathbf{Y}
\ee 
for a unique pair of curl-free ($\mathbf{X}$) and divergence-free ($\mathbf{Y}$) vector fields. So this theorem will be our route to the decomposition of $\mathbf{A}$ and $\mathbf{E}$. It guarantees that we can mathematically ``isolate’’ the divergence-free components and curl-free components of $\bf A$ and of $\bf E$. In physics jargon, the divergence-free component is called `transverse’ (also `radiative’), illustrating that it carries two degrees of freedom; and the curl-free component is called `longitudinal’, for one degree of freedom. 

We stress that---quite apart from this Chapter's aims---although the split of the electric field into radiative and  Coulombic components is mathematically useful,   it is far from being of only mathematical relevance; and from being an arbitrary division of the field's degrees of freedom. The split is physically meaningful. For instance, the question often arises about whether a process `radiates' electromagnetic waves. For example, the famous ``freely-falling electron paradox'' asks this question (see e.g. \cite{Saa_electron} for a pedagogic review). That is because distinguishing the components of the electric field that are wave-like from those which are `Coulombic'  is not straightforward. These are the two characteristics that are expected of any definition of radiation: (i) it is transversal: i.e. it only contains polarizations that are orthogonal to the momenta; (ii) if the charges are confined to a compact region, the Coulombic and radiative parts of the electric field are the coefficients in an inverse radial expansion ($1/R^2$ and $1/R$ terms, respectively).   The radiative $\mathbf{E}^r$ that we will specify below has both these features.

Now, $E_i$ and $A^i$ generate the standard coordinates in phase space. These are coordinates in an infinite-dimensional space, which parametrize six degrees of freedom per spatial point.  But if we want to think of the Gauss law as derivative (as we envisaged in Section \ref{sec:Gauss}), we need to find new coordinates  for phase space (defined in terms of the original coordinates) that instead parametrize the radiative and Coulombic parts of the fields, much like what we did in Section \ref{sec:toy}. The Coulombic coordinate will then be uniquely fixed by the charge density distribution, while the radiative part should parametrize the rest of the field. These coordinates will no longer be local in spacetime, but they are nonetheless very useful, as remarked above. And once we have found these new coordinates, we can use the symplectic form $\Omega$ (cf. \ref{eq:symp_form}) to establish the conjugate decompositions of the gauge potential $\mathbf{A}$.  Indeed, it is easiest to expound this re-definition of coordinates in terms of symplectic geometry, since the symplectic form $\Omega$ is a coordinate-independent object. Moreover, as in Section \ref{sec:toy}, it should be noted that the symplectic form is only non-degenerate in the total phase space. Therefore, to seek a decomposition through symplectic orthogonality, that decomposition should not be restricted to the constraint surface; rather it selects appropriate coordinates in the total phase space, and the constraint surface fixes some of these coordinates. In short: we are seeking adapted coordinates to parametrize the constraint surface.

Thus, on physical grounds, we would like to decompose the electric field as $E_i = E^r_i + E^c_i$ (with $r$ for `radiative’ i.e. divergence-free, and $c$ for Coulombic, i.e. curl-free), where 
\be\label{eq:E_rad} \pp^i  E^r_i\equiv 0; \qquad\text{and}\qquad \pp^i  E^c_i=\rho.
\ee
Both components are dynamically independent (the evolution of one is independent of the evolution of the other); and $E^c_i$ is to be completely fixed by the distribution of the charges. This means, in vacuum ($\rho=0$), that $ E^c_i=0$ and $E_i=E^r_i$.  In short: ${\bf E}^r$ is the component of the electric field that is not due to the distribution of charges (hence the label `radiative’); while ${\bf E}^c$ is the component of the field due to the simultaneous distribution of charges. What is the form of this component?

The Gauss constraint reduces the degrees of freedom in $\bf E$ from three to two. So when we decompose ${\bf E}$, writing: $E_i = E^r_i + E^c_i$,  with $E^c_i$ to be completely fixed by the distribution of the charges, as discussed above,  we conclude that $E^r_i$ has two degrees of freedom; so that $E^c_i$ has one, i.e. it is `secretly' a scalar. 
 Since one of the terms---$E^r$---is divergence-free, and we want the degrees of freedom of the other---$E^c$---to be exhausted by the divergence, i.e. to be `secretly' a scalar, it is convenient to introduce the Helmholtz decomposition: we take $E^c_i$ to be curl-free, so that it is the gradient of a scalar. 
We  thus \emph{choose} to write this  vector quantity in the customary way, i.e. as  $E^c_i=\pp_i\phi$, for some scalar function $\phi$.\footnote{ By the Poincar\'e lemma, in every simply connected domain, any curl-free vector field---which in the language of differential forms would be written as e.g. $\d \mathbf A=0$---is of a pure gradient form. In the field of tensor algebra, one often refers (confusingly) to a `spin-decomposition'. Thus a 2-tensor $T_{ab}$ may have vector and scalar components, of the form $\nabla_a V_b$ and $\nabla_a\nabla_b\phi$, respectively, and a vector $V_a$ may have a scalar component, that is written as $\nabla_a\phi$. The extraction of these components usually is made through  something like the Helmholtz decomposition theorem.  In fact, the theorem is a special case of the Hodge decomposition theorem for n-forms \cite[Chapter 43]{Morita_book}.} Then, the Gauss constraint $\mathrm{G}=0$ fixes $\phi$  in terms of its simultaneous distribution of charges, via the Poisson equation:
\be\label{eq:Poisson} \nabla^2 \phi_c=\rho. 
\ee

Now, the purely Coulombic terms of the electric field have their phase space coordinate axis  generated by the following vectors on phase space; (recall the discussion around eq. \ref{eq:infv}): 
\be \bb E_c:=\int \d^3 x\,\pp_i\phi \frac{\delta}{\delta E^i(x)} , \quad \text{for each} \quad \phi\in C^\infty(\Sigma).\ee  
This coordinate can then be fixed by \eqref{eq:Poisson}; with $\phi\stackrel{!}{=}\phi_c$ the electrostatic-like potential.\footnote{One might worry here that not all vector fields define a coordinate system: their vector field commutator must vanish, for that to be the case.   For finite dimensions, this is not a worry for $\sum_i a^i \frac{\pp}{\pp x^i}$, since the coefficients are constant, i.e. phase space independent.  Happily, the same is true here: for the coefficients of the new directions do not themselves depend on the coordinates.}

And now we can ask what are the subset of degrees of freedom of $\mathbf{A}$ that are symplectically orthogonal to the Coulombic part of the electric field.  Let us call these components $A^r_i$. That is, we seek those vector fields 
\be\label{eq:Ar}
 \bb A^r=\int \d^3 x A^r_i \frac{\delta}{\delta A_i}\ee 
that are symplectically orthogonal to the Coulombic coordinates of the electric field. That is, the $A^r_i$ are specified by requiring:
\be\label{eq:ortho} 
0\stackrel{!}{=}\Omega( \bb A^r, \bb E_c)=\int \d^3 x\, A^r_i \,\pp^i\phi=-\int \d^3 x\,\phi\,\pp^i A^r_i, \quad \text{for all} \quad \phi\in C^\infty(\Sigma) \, ;
\ee
where we applied integration by parts in the   third  equality and assumed that there is no boundary contribution to the integrals. Since $\phi$ is an entirely arbitrary test (or coordinate) function  we get: 
\be\label{eq:Coulomb_gauge}
\pp^i A^r_i=0. 
\ee
Equation \eqref{eq:Coulomb_gauge} is,   of course,  the \emph{Coulomb gauge} for   (the radiative component of) the gauge potential. It is a complete gauge-fixing, i.e. it leaves no more gauge-freedom in the potential. Moreover,  just like in Section \ref{sec:toy}, where $Q_-$ was written in terms of the original configuration variables, here we could write: 
\be\label{eq:rad_proj} h(A)_i=A^r_i:=A_i-\pp_i(\nabla^{-2} \pp^j A_j),\ee
which is also called the \emph{radiative projection}. That is, $h(A)$ is a projection operator onto the gauge-invariant content of $A_i$, as given in Equations   \eqref{eq:proj_h} and \eqref{eq:h_proj}, and discussed in Sections   \ref{sec:rep_conv} and \ref{subsec:rep_conv_Healey}. It is a \emph{representational convention}. And it is clear from \eqref{eq:rad_proj} that for any gauge-transformed $A_i^\lambda:=A_i+\pp_i\lambda$, a straightforward computation will show that $h_i(A^\lambda)=h_i(A)$. That is: $A^r$ is gauge-invariant. 

 Correspondingly, instead of arguing for \eqref{eq:E_rad}  from physical grounds as above, we can see the radiative part of the electric field as being selected   as the component of the electric field that is symplectically orthogonal to the pure gauge part of the gauge potential. Namely, we take the pure gauge part, i.e. the vectors that are along the gauge-orbit, to be given by:
\be\label{eq:Ac} 
\bb A^c:=\int \d^3 x \,\pp_i \lambda \,\frac{\delta}{\delta A_i} , \quad \text{for each} \quad \lambda\in C^\infty(\Sigma).\ee 
Then, in parallel to the calculation \eqref{eq:ortho}, namely:  
\be 
0\stackrel{!}{=}\Omega( \bb A^c, \bb E_r)=\int \d^3 x\, E^r_i \,\pp^i\lambda=-\int \d^3 x\,\lambda\,\pp^i E^r_i, \quad \text{for all} \quad \lambda\in C^\infty(\Sigma) \, ,
\ee
 we find the defining equation for the fields that are symplectically orthogonal to  the pure-gauge part of $\mathbf{A}$: 
\be\label{eq:Er}
\pp^i E^r_i=0.
\ee

  Thus we have found that if we choose to represent the  Coulombic  degree of freedom by the gradient of a scalar, we obtain the Helmholtz decomposition for the electric field: a unique decomposition in terms of divergence-free and curl-free components. And we obtain a similar decomposition for the gauge potential. That is, we obtain:
\begin{align}
\mathbf{A}=\mathbf{A}^r+\d \lambda\quad\text{and} \quad \mathbf{E}=\mathbf{E}^r+\d \phi.
\end{align}

 To sum up:  we have shown (by just an integration by parts, in each case) that:\\
\indent \indent (1) the curl-free (longitudinal/Coulombic) component of $\bf E$ is symplectically orthogonal to the radiative part of the gauge potential; (cf. equations \eqref{eq:Ar} to \eqref{eq:Coulomb_gauge});\\
while on the other hand:\\
\indent \indent (2) the divergence-free (transverse/radiative) component  of $\bf E$ is symplectically orthogonal to the pure gauge part of gauge field; (cf. equations \eqref{eq:Ac} and \eqref{eq:Er}).

 Finally, we note as a corollary to these results, that we can similarly orthogonally decompose the symplectic form \eqref{eq:symp_form} as:
\be\label{eq:symp_form2}
\Omega  = \int \d^3x  \left(\,\delta E_r^i \curlywedge \delta A^r_i+\delta E_c^i \curlywedge \delta A^c_i\right) \, :
\ee
which guarantees that the respective phase space directions are (symplectically) independent in each summand. \\

\noindent This concludes our main results. We end this Section with two comments:---\\
  \indent \indent (i): In the presence of boundaries,  the determination of $E^r$ does not require the further stipulation of boundary conditions, \emph{if} the gauge-freedom is taken as unconstrained at the boundary, as we will discuss in Chapter \ref{ch:subsystems} (see \cite{Gomes_new, DES_gf, GomesStudies}). Namely, we obtain that, at the boundary $E^r_i n^i=0$, where $\mathbf{n}$ is the vector normal to the boundary.\\
    \indent \indent (ii):   What we have just seen is a general feature of the symplectic geometric treatment. Namely: the tangent space to the constraint surface and the gauge orbits are the symplectic orthogonal complements of each other: cf. Lemma 1.2.2 in \cite{Marsden2007}. Thus  in particular, the radiative part of the electric field is singled out just by the symplectic form and the gauge orbits.\footnote{The general idea, at least in the first-class, or coisotropic, case, is remarkably simple; and thus merits a quick sketch. From \eqref{eq:Ham_Gamma}, vectors $v$ that are tangent to the constraint surface obey $\d\Phi^I(v)=0$ for every $I$. Thus from \eqref{symquant}, we obtain  $\omega(X_I, v)=0$.} \\

\section{Conclusion and outlook}\label{concl}

Using the symplectic or Hamiltonian formalism, we have shown how decompositions of the electric field correspond, through symplectic orthogonality, to decompositions of the gauge potential (which correspond to choices of gauge), and vice versa. Gauge choices thus have a very natural interpretation in terms of choices of decomposition of the electric field. 

We have argued  that  for various reasons a natural  decomposition of the electric field takes one part, i.e. component,  to be determined by  the instantaneous charge distribution.
We find that  the part of the electric field that remains---i.e. the radiative part, the part that is not involved in the Gauss law and does not ``care about''   the instantaneous charge distribution---is symplectically \textit{orthogonal} to the pure gauge part of the gauge potential. This orthogonality establishes a firm link between gauge symmetry and locality,  as a relation between the Gauss law and the pure gauge part of the potential. But this relation does not, by itself, suffice to select a gauge-fixing.

On the other hand, \emph{if} we define the Coulombic part of the electric field as the gradient of a Coulombic potential, then the gauge \emph{is} fixed:  the  part of the gauge potential that is   symplectically orthogonal to the gradient of the Coulombic potential   \emph{is}  the gauge potential in Coulomb gauge. 

 Thus in summary, again: our main idea is that the electric field has the gauge potential as  its conjugate, and there is a part of the potential that is not at all conjugate to (i.e. is symplectically orthogonal to) the Coulombic part of the electric field, i.e. the part determined by the instantaneous distribution of
charges. {\em That} part of the potential satisfies the Coulomb gauge condition.   And this result is worth expounding, since it shows that a choice of gauge need not be a matter of calculational convenience for some specific problem or class of problems, but can be related  to a physically natural, and general, splitting of the electric field.

 We also saw how these results prompt a comparison with \cite{Maudlin_ontology}. For Maudlin attempted to shift the Gauss law from, in his terms, the nomology to the derivative ontology: namely, by analysing the charges in terms of the electric field. We have argued that this does not work. 
In any case, it does not help with fixing a gauge.  If instead we analyse the Coulombic part of the electric field away, defining it in terms of the charges by \eqref{eq:Poisson}, then  the fundamental ontology  (to use Maudlin's term) can be made to correspond, as we have seen in  \S \ref{sec:radcoul},  to a configuration space parametrised by  the charges and to a choice of the  Coulomb gauge for the gauge potential. 

But we emphasise that  Coulomb gauge is not {\em mandatory}.  This choice is a good example of what we have been labeling a \emph{representational convention} (cf. Section \ref{sec:method}).  It only corresponds, in a well-defined sense, to a particular decomposition that singles out the Coulombic part of the electric field. Nonetheless, in whichever sense that choice of decomposition of the electric field is natural, Coulomb gauge is also natural. 

\medskip

Finally, we offer an {\em outlook}: we stress that the lessons of this Chapter go through, with minor modifications, to the non-Abelian domain, gravity, and also apply in the presence of boundaries; (see e.g. \cite{GomesStudies, GomesRiello_new}). In brief, each extension requires one important modification. 

First, in the non-Abelian case, the Coulombic split for the gauge potential occurs only at the level of \emph{perturbations}. That is, although we can split vectors on phase space, $\bb X$,  the non-Abelian nature of the theory implies that the split is not integrable (see \cite[Section 5]{GomesRiello_new} and \cite[Section 9]{GomesHopfRiello}). Thus the Coulombic split of a state is always ambiguous.\footnote{In the Abelian case, one can integrate the perturbative split along paths in phase space so as to define a split of any final state; integrability guarantees that the end result is path-independent.}  

The same occurs with the gravitational theory, where the Coulombic-like split  of the momentum constraints in general relativity proceeds very much in the same fashion, but with an added difficulty due to the coupled Hamiltonian constraint. Due to this complication, instead of a radiative, unconstrained degree of freedom that is just transverse (like $\pp_i E^i_\rad=0$), one has  to solve the other initial value constraints---the Hamiltonian constraint---jointly.\footnote{ Here the standard way (cf. \citep{York}),  also has a tracelessness condition, so that the unconstrained momenta degrees of freedom are the transverse-traceless momenta, $\pi_{ab}^{TT}$, obeying both $h^{ab}\pi_{ab}^{TT}=0$ and $\bar\nabla^a\pi_{ab}^{TT}=0$.} It similarly implies a mild form of non-locality.

 In the presence of boundaries, in the Abelian case, the radiative and Coulombic components of the electric field are again defined by symplectic orthogonality with the pure gauge part of the potential, which we deem unconstrained at the boundary (cf. \cite[Prop. 3.3]{GomesRiello_new}). Thus we obtain the following modifications, for a bounded region $R$ bounded by $\pp R$, whose normal is $\mathbf{n}$:\\
Instead of \eqref{eq:Coulomb_gauge}, 
\be\label{eq:rad_proj_bdary}
\begin{dcases}
\pp^i A^r_i=0  & \text{in }R,\\
 n^iA^r_i=0  & \text{at }\pp R.
\end{dcases},
\ee  
and with $A^{c}_i=\pp_i\lambda$ for $\lambda\in C^\infty(\Sigma)$ (as in \eqref{eq:Ac}). \\
Instead of \eqref{eq:Er},
\be
\begin{dcases}
\pp^i E^r_i=0 & \text{in }R,\\
 n^iE^r_i=0  & \text{at }\pp R.
\end{dcases};
\ee 
and instead of \eqref{eq:Poisson},
\be
\begin{dcases}
\nabla^2\phi_c=\rho  & \text{in }R,\\
 n^i\pp_i\phi=f & \text{at }\pp R.
\end{dcases},
\ee
where $f$ is the electric flux through the boundary, $f:=n^i E_i$, and here can be seen as an independent variable. Note that: (i) the pure gauge part of the potential is unmodified (since we do not truncate gauge transformations at the boundary $\pp R$), and (ii) the normal to the electric field at the boundary belongs to the Coulombic part. The radiative part gets no extra degree of freedom at the boundary.\footnote{Thus, the radiative part is local, according to straightforward definitions of locality (cf.  \cite[\S 2, p.5]{Wallace2019b}).  The Coulombic field depends on the distribution of $\rho$ in the region \emph{and} on the boundary flux of the electric field. Thus, even if the charge distribution for two worlds matches inside $R$, the Coulombic field therein may differ, since in each world it will depend on the independent variable that is the electric boundary flux.  }


 \chapter{The direct empirical significance of symmetries}\label{ch:subsystems}
 \section{Introduction}\label{sec:DES_intro}

The debate I want to focus on in this Chapter is about whether gauge symmetries can have a \textit{direct} empirical significance. 
 Of course, all hands agree that symmetries have various important empirical implications. The obvious examples come from the Noether theorems (see Chapter \ref{ch:Noether}); and there are many other such indirect empirical signatures of gauge symmetries.
 On the other hand, as we have seen in Section \ref{sec:unobs},  symmetries as applied to the whole universe are not directly observable: this is the \emph{unobservability thesis} (cf. \cite[p. 10]{Wallace2019}).

  But some familiar symmetries of the whole Universe, such as velocity boosts in classical or relativistic mechanics (Galilean or Lorentz  transformations), have a \textit{direct} empirical significance  \emph{when applied solely to subsystems}.  Thus Galileo's famous thought-experiment about the ship---that a process involving some set of  relevant physical quantities in the cabin below decks proceeds in exactly the same way  whether or not the ship is moving uniformly relative to the shore---shows that subsystem boosts have a direct, albeit relational, empirical significance. For though the inertial state of motion of the ship is undetectable to experimenters confined to the cabin, yet the entire system, composed of ship and sea registers the difference between two such motions, namely in the different relative velocities of the ship to  the  water.\footnote{ 
As often is the case in physics, the characterization of direct empirical significance used here may rely on certain approximations. For surely,  with the right equipment (such as a window),  the person within the cabin \textit{could} discern movement of the ship from within,  and different movements of the ship could create different sorts of eddies and turbulence in the sea. This sort of idealization is ubiquitous in physics, and generally unproblematic: see \cite[p. 4]{Wallace2019a}  for an answer to this type of concern.  \label{ftnt:approx}  }

 So the question arises: {\em Can other symmetries---especially gauge symmetries---have a similar direct empirical significance when applied to subsystems?}

For as we have seen,  gauge symmetries 
are normally taken to encode descriptive redundancy. That is, they arise in a formalism that uses more  variables than there are physical degrees of freedom in the dynamical system described. 

This descriptive redundancy means that the natural answer to our question is `No'. For surely, while a ``freedom to redescribe'' may have some indirect empirical implications as we have seen (e.g. in Chapter \ref{ch:Noether}),  it could not have the content needed for a direct empirical significance, like the one illustrated by Galileo's ship. 
This `No' answer was   developed in detail by  \citet{BradingBrown} in response to  various discussions such as  \cite{Kosso}. They take themselves---I think rightly, in this respect---to be articulating the traditional or orthodox answer. 

The `Yes' answer has been argued for by  \cite{GreavesWallace}, building on \cite{Healey2009}.  I will agree with some aspects of both \cite{BradingBrown}'s and \cite{GreavesWallace}'s analysis of symmetries, later complemented by \cite{Wallace2019, Wallace2019a, Wallace2019b}. But, unlike any of them, I will recast the topic so as to focus on gauge-\textit{invariant} information about regions and the  conventions used to represent this physical information. My own conclusion will be that only a finite subset of gauge-transformations, usually called `global' (but here called `rigid'), can have \emph{direct empirical significance}, or \textit{DES}, as it is known in the literature.

Thus the broad notion of `direct empirical significance'  of a symmetry amounts to the following simple definition:
\begin{defi}\label{def:DES} DES are transformations of the universe possessing the  following two properties (articulated in this way by \citep{BradingBrown}, following \citep{Kosso}): \\
\indent \,(i): \emph{Global Variance}--- the transformation applied to the Universe in one state should lead to an empirically different  state; and  yet\\
\indent  (ii): \emph{Subsystem Invariance}---the transformation should be a symmetry of the subsystem in question (e.g. Galileo's ship), i.e. involve no change in quantities solely about the subsystem. \end{defi}

I will take the concept of DES to imply observability of those symmetries; but I will prefer the use of the label DES as opposed to `observable' since, as mentioned above,  symmetries may possess indirectly observable consequences (such as charge conservation, cf. Chapter \ref{ch:Noether}). Moreover, the label  `DES'  is already	settled in much of the literature.\footnote{Whether gauge symmetries have DES has been discussed in \citep{Kosso, BradingBrown, GreavesWallace, Healey2009, Teh_emp, Friederich2014, Ladyman_DES,  Friederich2017,  GomesStudies, Gomes_new, Teh_abandon, Wallace2019a, Wallace2019b, Seb_ramirez, Chasova}. None of these completely encapsulate my own views, but I have various overlaps of agreement with each. 
}

According to condition (ii) of Definition \ref{def:DES}, the empirical significance is to be witnessed by observers that lie outside the subsystem---it cannot be detected by those confined within it. Therefore, DES combines an inside and an outside perspective and, in this limited sense, acquires an epistemic dimension, or at least one that considers physical information as it is intrinsically accessible within a subsystem. 
 

 
By focussing on  gauge-\textit{invariant} information as accessible within the subsystem and also globally, and by characterizing subsystems as respecting the conditions of Section \ref{sec:subsystems}, in particular, downward consistency, given in Equation \eqref{eq:dwc}, I will identify DES as defined by (i) and (ii) above with a particular type of \emph{failure} of \textit{Global Supervenience on Subsystems} (GSS). 

For now, I define GSS heuristically; the formal Definition \ref{def:reg_ins} is given below, once we have agreed on the nomenclature. 

GSS is upheld---indicating the absence of DES---when the physical facts  intrinsic to each of  the complementary subsystems  need not be augmented by any relational fact in order   to uniquely specify all the physical facts of the whole. In other words,  the intrinsic physical states of those subsystems composing the whole uniquely determine the physical state of the whole. 

GSS fails---indicating the presence of DES---if given just the intrinsic physical states of the subsystems, there are physically distinct possibilities to join these states into some physical state of the joined system. 
That is,  the relation between states of the Universe and states of its subsystems are many-to-one, because there is relevant relational information that cannot be registered intrinsically within each subsystem. In these cases, we will say there is \textit{Global Non-Supervenience on Subsystems} and denote it by $\neg$GSS.
\footnote{
 In the context of gauge systems under study here, a \textit{failure of global supervenience on subsystems}, is close in spirit to Myrvold's \textit{global patchy 
non-separability} \cite{Myrvold2010}, which he articulated for
the holonomy approach to gauge theories. But I refrain from adopting this
nomenclature because (i) I do not focus on holonomies (though see Section \ref{sec:elim_elim} and footnote \ref{ftnt:tr_hol}), and (ii) it does not
apply to finite-dimensional systems like Galileo's ship. \label{ftnt:Myrvold1} }


Technicalities apart, 
  my main claim is that both Galilean boost symmetry for particle systems and gauge symmetry for certain field theories carry Direct Empirical Significance (DES)  through a failure of Global Supervenience on Subsystems ($\neg$GSS). This holism is \textit{empirically significant}, since it registers physical---i.e. gauge-invariant---differences in the entire system and we take such differences to lead to empirically distinguishable universes. Moreover, in some cases the implied under-determination of the physical state of the whole universe by the physical state of its subsystems can be encoded in a \textit{subsystem symmetry} (as seen from the `outside perspective'); in these cases $\neg$GSS is identified with DES.

It is easy to see that $\neg$GSS is a necessary condition for DES, but DES also requires   that the physical underdetermination   be in 1-1 correspondence with a group action on a subsystem. In some cases, this will be true. In other cases,  the variety can only be interpreted as arising from a group action intrinsic to the boundary between the regions. In the latter cases,   $\neg$GSS does not have a natural interpretation in terms of DES.\\

 Of all the treatments of DES for local gauge theories,  \cite{GreavesWallace, Wallace2019,Wallace2019a, Wallace2019b}  bear the most similarities to mine here. They focus on subsystems as given by regions and they identify transformations possessing properties  (i) and (ii)  by first formulating the putative effects of such transformations on the gauge fields in these regions. A more refined treatment that takes into consideration extensions of the symmetries to the measuring apparatus (or subsystems) was developed only in \citep{Wallace2019}; it was applied to particle mechanics in \citep{Wallace2019a} and field theory in \citep{Wallace2019b}. 
  The two main differences between this Chapter and that series of work are that:\\
\indent (1): I will be explicit about the need for, and use of, representational conventions. This first demand  is in line with \cite{Gomes_new}: it is a consequence of the   focus on the physical content of the states---a focus that is necessary in order to assess physical significance.\\
\indent (2): My  treatment of the boundary of subsystems---in particular the relation between non-asymptotic and asymptotic boundaries---is different. I believe we should first understand how gauge symmetry behaves in the non-asymptotic case and then translate that understanding to the asymptotic case; whereas \cite{Wallace2019, Wallace2019b} goes in the other direction.

 So naturally, my conclusions will differ somewhat from the previous literature. 
\section{ DES in gauge theories}

In this Section, I will briefly set up GSS in the context of gauge theories. This will involve, in Section \ref{sec:setting}, a brief recap of our definition of a kinematically isolated subsystem from Section \ref{sec:subsystems}. In Section \ref{sec:subs_DES}, I describe the composition of the physical subsystem states in a bit more detail. In Section \ref{sec:int_bdary}, I describe the obstacles to characterizing the physical subsystem states. These obstacles  are rooted in our attempt to localize physical states that carry a  (mild form of) non-locality---as I described in Sections \ref{subsec:sims} and  \ref{subsec:rep_conv_Healey}.

\subsection{Kinematically isolated subsystem}\label{sec:setting}
 As discussed in Section \ref{sec:subsystems}, we would like to define `kinematically isolated subsystems'  such that: (1) there is always a global state that restricts to any given subsystem state (upwards consistency), and (2)  the symmetries of the whole restrict to the symmetries of the subsystem (downwards consistency of symmetries).

Under this second assumption, the universal symmetries  bequeath  symmetries, through the split, to the subsystems, by mere restriction. Moreover, I will take only a kinematical idea of isolation into account, as discussed in Section \ref{sec:subsystems}. I now elaborate these conditions more precisely. 

\paragraph{Technical conditions}
With Wallace, I will take subsystems to be represented as elements $X$ of  a collection  $\Xi$, so $X\in \Xi$. The collection $\Xi$ is partially ordered by inclusion, and bounded by a minimal and a maximal element---representing the empty set and the entire universe, which I will call $\Sigma$.\footnote{So as to match the usual notation for a  Cauchy surface. } And we define a state space for each $X$, ${\cal M}_X$, with $\F_\Sigma:=\F$, such that the state spaces respect the partial ordering (here $\cal M$ is the space of models of a theory;  from Section \ref{sec:syms_tech}). Namely,  for $X\subset Y$, we define  $\iota_{XY}$ as the inclusion map (or embedding), $\iota_{XY}:X\rightarrow Y$, and, schematically,  $r_{YX}$ as the restriction map: $r_{YX}:Y\rightarrow Y_{|X}$, with  $\iota_{XY}\circ r_{YX}=\mathsf{Id}_X$. 
The idea is that the restriction on the subsystems gets `pulled-back'  to a restriction on the state spaces, which we can here schematically denote: 
\be\label{eq:restriction} \iota^*_{XY}:{\cal M}_Y\rightarrow {\cal M}_X.\ee  
We denote it thus since, in cases of interest in field theory, this ``restriction map'' is really a type of pull-back.\footnote{ For cases of interest our state space will be the space of sections of some vector bundle (cf. Section \ref{sec:PFB}). If $M$ is the base space of a vector bundle $E$, and $\iota:N\rightarrow M$ is an embedding map, then $\iota^*E$ defines a vector bundle over $N$ by pull-back (i.e. the fiber over $x\in N$ is the fiber over $\iota(x)\in M$).}   And also with Wallace, we  assume ``upwards consistency'': given $X\subset Y$, for a given $\varphi_X\in {\cal M}_X$, there exists a $y\in {\cal M}_Y$ such that $r_{XY}(\phi_Y)=\varphi_X$. This means that any subsystem state is compatible with some global state. 
(If we set  $Y=\Sigma$, we can lighten notation by omitting the $Y$.) 

  If we recall equation \eqref{eq:group_action} in Section \ref{sec:syms_tech}, we had  
$$\mu:\G\times \F\rightarrow\F; \quad
(g,\varphi) \mapsto \mu(g, \varphi).$$
And for a fixed $g\in \G$, we write  $\mu_g:\F\rightarrow \F$. 
 In accordance with the arguments of Section \ref{sec:subsystems}, I  will assume that for the entire universe,   both the isomorphisms and symmetries coincide, and are given by the action of $\G$ on $\F$. 
Moreover, I assume there is an understood notion of isomorphism for each $\F_X$, $\mathsf{Iso}_X$, whose membership I denote with a bar, and whose action I denote as 
\be  \mu_X:\mathsf{Iso}_X\times \F\rightarrow\F; \quad
(\bar g_X,\varphi_X) \mapsto \mu_X(\bar g_X, \varphi_X).
\ee 
Given $\mathsf{Iso}_Y$,  $\mathsf{Iso}_X$ satisfies: 
\be\label{eq:dwc_iso} \mathsf{Iso}_X =\{\bar g_X\,\,|\,\,  \mu_{\bar g_X}\circ\iota^*_{XY}=\iota^*_{XY}\circ \mu_{\bar g_Y}, \,\, \text{for}\,\,{\bar g_Y}\in \mathsf{Iso}_Y\}.
\ee
This is just downward consistency for isomorphisms.\footnote{It is easy to see that both isometries in the category of (pseudo)-Riemannian metrics (see Section \ref{sec:diff_sym} and gauge transformations  in the category of connection-forms (see Section \ref{sec:PFB_formalism}) satisfy \eqref{eq:dwc_iso}, for $X, Y$ regular subsets of the spacetime manifold (seen as the space of orbits of a principal bundle in the latter case). I leave it as an exercise for the reader to show whether \eqref{eq:dwc_iso} holds whenever the isomorphisms of the category in question are inherited from an action of the isomorphisms on the base set. } 

Whether the isomorphisms $\mathsf{Iso}_X$  of $\F_X$ match the dynamical symmetries $\G_X$ of  $\F_X$  is a different question: we have to first be able to define an intrinsic dynamics for $X$, which is where Wallace's more strict definition of dynamical isolation comes in handy. It amounts to having the dynamical evolution in $X$ be defined  just by an initial state of $X$, with possibly  some initial boundary condition as the only additional data. But if we allow time-varying boundary conditions, i.e. if we augment  ${\cal M}_X$ to include some restricted form of external data, we can again define dynamics (indexed by  external data), and so also obtain the symmetry group $\G_X$ (indexed by the external data).\footnote{ As we will see in Section \ref{sec:int_bdary}, dealing with this indexing is the major challenge we have to face in order to satisfy the downward consistency of symmetries.  But gauge theories are local field theories with no action at a distance, and thus already have in-built a weak notion of dynamical isolation for disconnected subsystems. One can always find a representational convention in which the evolution equations are hyperbolic and of the same form, but with some unspecified boundary condition. Making this precise is difficult, and will be sketched in Section \ref{sec:sub_rep_conv}, but a full answer requires more mathematical footwork: see \cite{GomesRiello_new, GomesHopfRiello, Riello_new, GomesStudies}.\label{ftnt:local_hyper}} 

 Differently from Wallace, I will further demand that  the restriction $r_{X}$ of \eqref{eq:restriction}  co-varies with the symmetries  of $Y$, if there are any. Namely, if $g_Y$ is any symmetry of ${\cal M}_Y$ then the composition with the restriction is also a symmetry of the subsystem; and all the subsystem symmetries are encompassed in this way.  In other words, I will demand that subsystems 
 satisfy \emph{downward consistency for symmetries,} which is like \eqref{eq:dwc_iso} but for symmetries. The demand is that given $\G_Y$,   $\G_X$  satisfies:\footnote{We can alternatively write this condition  (and similarly, with bars, for \eqref{eq:dwc_iso}) as : \be\begin{cases}\text{for all}\,\,g_X\in \G_X,\,\, \exists g_Y\in \G_Y \,\,\,\text{such that}\,\,\, \iota^*_{XY}\circ \mu_{g_Y}=\mu_{g_X}\circ\iota^*_{XY}\\\text{for all}\,\,g_Y\in \G_Y,\,\, \exists! g_X\in \G_X\,\,\,\text{such that}\,\,\, \iota^*_{XY}\circ \mu_{g_Y}=\mu_{g_X}\circ \iota^*_{XY}.\end{cases}\ee }
\be\label{eq:dwc} \G_X =\{ g_X\,\,|\,\,  \mu_{ g_X}\circ\iota^*_{XY}=\iota^*_{XY}\circ \mu_{ g_Y}, \,\, \text{for}\,\,{ g_Y}\in \G_Y\}.
\ee
%
 Thus for any $g_Y\in \G_Y$ a symmetry of ${\cal M}_Y$ and any $\varphi_Y$ that restricts to a $\varphi_X\in \F_X$ of the subsystem, i.e. $\varphi_X=  \iota^*_{XY}\varphi_Y$, 
\be\label{eq:dwc0}\iota^*_{XY}(\varphi_Y^{g_Y})= \varphi_X^{g_X}, \quad \text{for some}\quad g_X\in \G_X ,\ee 
so they are symmetry-related in ${\cal M}_X$.
  Since the symmetries of the maximal element, i.e. the universe, are isomorphisms, and the restrictions of the isomorphisms are  also isomorphisms, \eqref{eq:dwc_iso} and  \eqref{eq:dwc}  then imply that the symmetries of the subsets are also isomorphisms:   $\G_X=\mathsf{Iso}_X$.



In field theory, the local dynamical equations  governing a subsystem that is demarcated by a boundary---possibly with evolving boundary conditions---are identical to those governing a larger  system of which it is a part. But in order to satisfy \eqref{eq:dwc},  to the extent that boundary conditions evolve, that evolution should not pare down the symmetry group of the subsystem equations of motion (see footnote \ref{ftnt:local_hyper}).

This is a watered-down version of Wallace's subsystem recursivity, focused on symmetry: all we require from our definitions of subsystem is that the  symmetries are recursive in this way. 
It is weaker in the sense that subsystems do not need to  be idealized as infinitely far apart, as they would have to be if one  requires complete dynamical autonomy for arbitarily large periods of time. And just to give an example, it is this weaker notion that would allow us to model the interior and the exterior of black holes as subsystem and environment: a case of great interest.

So, given some notion of dynamical subsystem symmetries $\G_X$, we define the corresponding equivalence classes, i.e. $[\varphi_X]$. I can now  define:
\begin{defi}[GSS]\label{def:reg_ins}
Given a sub-collection $\Xi'\subset \Xi$ such that $\cup_{X\in \Xi'}=\Sigma$, i.e. such that the join of the members of $ \Xi'$ encompass the universe. Then GSS holds iff each complete collection of kinematically compatible  physical subsystem states $[\varphi_X]$ gives rise to  a unique global physical state, $[\varphi]$.
\end{defi}
  Shortly, we will elaborate what are the conditions on compatibility of physical, i.e. symmetry-invariant states.

We can summarize our conditions as conditions on $\cal M$ and on the dynamics: (a) $\F$ will be given by an instantaneous state space; and (b)  the restriction  of the symmetries of the whole to the subsystem are  the dynamical symmetries of the subsystem. 

Note that such a kinematical understanding of subsystem recursivity about symmetries can accommodate our intuitions about, and the familiar examples of, direct empirical significance. Consider, for simplicity, a  Galileo's ship scenario with the shore  taken as the environment, in which the subsystem at $t=0$ is inertial and at a finite distance $d$ from the shore. Now, for a fixed time interval $I$, the boosts must be bounded to be smaller than  $d/I$. 
This constraint on the boosts is usually ignored because we employ certain idealizations, e.g. that the shore is infinitely far away, so that $d\rightarrow \infty$. Here  I prefer a different idealization, in which I take $I$ to be  small. This feature will allow  the kinematical understanding of subsystem recursivity  to avoid some of the fuzziness of dynamical isolation, and yet have the resources to articulate a fruitful construal of DES.

 And thus, as it stands Definition \ref{def:reg_ins} is in line with both the Galileo's ship analogy and with the idea of gauge transformations as mere re-description.  Schematically: if the subsystems are `shore' and `ship', and there are  equivalence relations, $\sim$, and we use square brackets $[\bullet]$ for equivalence classes, applicable to states of  subsystems and of the whole, and  given  the physical (i.e. ``gauge-invariant'') states  $[s_{\text{\tiny shore}}], [s_{\text{\tiny ship}}]$ and $[s_{\text{\tiny shore and ship}}]$, there is a many-to-one relation, encoded by the set $I$:
\be\label{eq:seaship}[s_{\text{\tiny shore and ship}}^{(i)}]=[s_{\text{\tiny shore}}]\cup_{(i)}  [s_{\text{\tiny ship}}],\qquad i\in I=\text{Boosts}\ltimes\text{Euclidean}.
\ee
The idea is just that there are many ways to jointly embed the intrinsic states of the two subsystems into the same universe: the physical states $[\varphi_{\mbox{\tiny shore}}]$ and $[\varphi_{\mbox{\tiny ship}}]$ can be `glued' in a variety of ways.  So here the set $I$ that parameterizes the many-to-one relation is  the (inhomogeneous) Galilean group (which is a semi-direct product ($\ltimes$) of boosts and the group of translations and rotations).\footnote{In fact, \eqref{eq:seaship} holds only up to the first order in velocities. As argued in  \cite[Appendix A]{DES_gf}, here $\neg$GSS can outstrip the (inhomogeneous) Galilean group in the sense that there is a larger variety of possible universes composed of the same intrinsic states. To see this, suppose, for simplicity, that sea and shore are subsystems of an $N$-particle universe (`shore' is composed of particles 1 to $n$ and `ship' of particles $n$ to $N$); and that the $S$-symmetries---here $S$ need not be the action; see Section \ref{sec:syms_tech} and Definition \ref{def:sym}---are such that each physical state is characterized by a relational configuration---the interparticle distances (cf. \citep{GomesGryb_KK} for the intrinsic relational dynamics of  subsystems that also possess angular momentum, and how these embed into a Newtonian universe). Here  there may be a very large number of ways we can embed histories of  $[s_{\text{\tiny shore}}]$ and $[s_{\text{\tiny ship}}]$ into the same system. For instance, as long as the accelerations of each subsystem is the same for all particles, we can have very different relative  accelerations between the two subsystems' centers of mass. Of course, these relative accelerations need to be realized through appropriate embeddings of the subsystems into the ambient system (we should be able to find the embedding as a function of the relative accelerations). This is how this picture of $\neg$GSS articulates Newton's Corollary VI (see \citep{saunders2013}).}
This is described more carefully---including how it satisfies \eqref{eq:dwc}---in \cite[Appendix A]{DES_gf}.

\subsection{The subsystems}\label{sec:subs_DES}
Now turning to field theory, we first carve up the system into two (mutually exclusive, jointly exhaustive)  subsystems whose state spaces we label $\F_+$ and $\F_-$, or $\F_\pm$, for short: (a mnemonic notation to help one think of the subsystems as complements of one another, and intersecting only at a common boundary, like  $0\in \RR$ for the positive and negative reals). When these subsystems are made to correspond to regions, we will name the regions $R_\pm$. These are taken as subsets of the spatial manifold $\Sigma$, i.e. such that $R_+\cup R_-=\Sigma$, and I will moreover assume that the interiors of the regions, $\mathring{R}_\pm$, are disjoint, and that their closures, $\bar R_\pm$, intersect at a \emph{boundary manifold}, $S$, i.e. 
\be \mathring{R}_+\cap \mathring{R}_-=\emptyset, \quad  \bar R_+\cap \bar R_-=S.
\ee

Once downward consistency is respected, given regions $R_\pm$ and the restriction maps $r_\pm: \Sigma\rightarrow \Sigma|_{R_\pm}$, or alternatively  the embedding maps $\iota_\pm: {R_\pm}\rightarrow \Sigma$, we would write  $\F_\pm:=\iota_\pm^*\F$ and $\G_\pm$ as above (see \eqref{eq:dwc}).

We extend the use of the equivalence class notation and of the square brackets: 
 $\varphi_\pm\sim_\pm\varphi'_\pm$ iff $\varphi'_\pm=\varphi_\pm^{g_\pm}$ for some $g_\pm\in \G_\pm$, in which case $\varphi'_\pm\in[\varphi_\pm]$. Note that \emph{no} extra conditions on the gauge transformations at the boundary are imposed.\\ 

We can now translate Definition \ref{def:DES} into the nomenclature of this Section: \\
 \noindent $\bullet$\,\,\textit{Global Variance:}~~~ $[\varphi]\neq [\varphi'] $: the two physical states of the Universe are distinct according to the $\sim$ relation.   \smallskip
 
\noindent $\bullet$\,\,\textit{Subsystem Invariance:}~ $[\varphi_\pm]= [\varphi'_\pm] $: regionally the states are physically indistinguishable according to our $\sim_\pm$ relation; that, is for each ($\pm$)  subsystem, the primed and unprimed states are symmetry-related according to their internal models. \\

I will also say that two subsystem physical states $[\varphi_\pm]\in [\F_\pm]:=\F_\pm/\sim_\pm$ are \emph{compatible} (as per Definition \ref{def:reg_ins}),  iff they jointly descend from a global state, $[\varphi]$. This compatibility, or compossibility, will be made rigorous with representational conventions, in Section \ref{sec:sub_rep_conv}. For now, by introducing some, yet-to-be-defined, compositions of physical states, $\boxplus$ and $\boxplus'$, reflecting the allowed variety of global states countenanced by $\neg$GSS, and writing 
\be\label{eq:phys_var} [\varphi_+]\boxplus[\varphi_-]=:[\varphi]\neq [\varphi']:=[\varphi'_+]\boxplus'[\varphi'_-]=[\varphi_+]\boxplus'[\varphi_-]
\ee
we indicate more clearly that the very concept of GSS needs to be gauge-invariant, i.e. physical. 

Note also that the \textit{subsystem states} are intrinsically identical between the $ [\varphi]$ and the $ [\varphi']$ Universes, i.e. between the leftmost and rightmost hand sides of \eqref{eq:phys_var}. Therefore the difference between the two sides of the equality must lie in the relation between the subsystems; this is signalled by \eqref{eq:phys_var}'s use of  $\boxplus$ as well as $\boxplus'$. Ultimately, $\neg$GSS is possible because   there are different domains for the equivalence relations---subsystem or universe.

 More formally,  
  \be\label{eq:reg_ins} [\varphi_{(i)}]=[\varphi^+]\cup^S_{(i)} [\varphi^-],\quad i\in I\quad\text{with}\quad  [\varphi_{(i)}]\neq  [\varphi_{(i')}]\quad\text{iff}\quad i\neq i',\ee
where we label each legitimate/physically possible composition of the two given regional states to form a physically possible universal state  by $i$, with $i$ belonging to some index set $I$, which can depend on the component states. (This dependence, written as $I([\varphi^\pm])$,  will be henceforth omitted). Here $\cup^S_{(i)}$ represents the $i$-th valid \textit{gluing}, i.e. composition,  of the two gauge-invariant data $[\varphi^\pm]$ along $S$. 

  If $I$ is empty  there is \textit{no} possible gluing, i.e. the regional gauge-invariant states are incompatible and cannot conjoin into a universal physical state (\textit{regional incompatibility}). If $I$ has a single element for all the composable states, the gluing is unique, and then  there is  GSS. If otherwise, i.e. if $I$ has more than one element, the universal physical state is undetermined just by the regional physical states: more information about relations between the subsystems is needed,  and  there is {$\neg$GSS}. 




  In Yang-Mills theory,   $\neg$GSS is associated with DES  only in conjunction with those conditions which are necessary for the existence of conserved global charges, so that $I$ can be put into correspondence with a global symmetry group.\footnote{Namely, the association  will obtain  only for reducible configurations---those which have stabilizing gauge transformation (analogous to non-trivial Killing fields for a spacetime metric)---in which case $I$ is isomorphic to the stabilizer group (cf. footnotes \ref{ftnt:stab} and \ref{ftnt:stab2}), or the group of reducibility parameters in the language of \cite{BarnichBrandt2003} who develop the link between stabilizers and conserved charges in great detail. \label{ftnt:reducible}} This procedure thus establishes a link between  an \text{indirect} consequence of gauge---the conservation of charges---and a `direct' one (DES). This is explored in \cite{GomesNoether}: unfortunately, due to a lack of space, I will not describe this connection here. \\

\subsection{The obstacle: gauge symmetries at the boundaries}\label{sec:int_bdary}
 
But here we must face the  issue described in Section \ref{sec:setting}:  without a firm understanding of what the subsystem symmetries are, we cannot get a grip on DES; for how would we  define the equivalence classes of
equation \eqref{eq:reg_ins}? And there are subtleties in construing  the isomorphisms  of the subsystem states as dynamical symmetries.  In particular, there are subtleties about the symmetry-invariance of  a bounded subsystem's own dynamical structures, such as its intrinsic Hamiltonian, symplectic structure, and variational principles in general. These subtleties can obstruct downward consistency and thus obstruct our assumptions about the subsystem symmetries.   
 
 Until recently, generic subsystems that were  defined by regional restrictions were \emph{not} supplied with gauge-invariant boundary conditions. The technical reason is that, setting aside very stringent physical boundary conditions---such as vanishing field-strength in the case of Yang-Mills theory---the variation of the action functional produces boundary terms that are not gauge-invariant. The  conceptual reason  is that gauge theories manifest a type of non-locality. Thus the global, physical phase space (or the corresponding global physical Hilbert space, in the quantum theory) is not factorizable into the physical phase spaces over the composing regions.\footnote{This type of holism, or non-locality is a well-known issue for theories with elliptic initial value problems: e.g. Yang-Mills theory and general relativity, as described in Section \ref{sec:syms_consts}. For a reference that explores non-factorizability in the context of the holonomy formalism,  see \cite{Buividovich_2008}. For the relation between different symmetries and locality in gauge theory, see \cite{Elements_gauge}, and for the non-locality of gravitational invariants see e.g. \cite{Torre_local, Donnelly_Giddings}; for more recent use of this non-factorizability in the black hole information paradox, see \cite{Jacobson_2019}.  For a discussion of the relation between the factorizability of Hilbert spaces and the augmentation of the phase space with `edge-modes', which we will shortly discuss, see \cite{Teh_abandon, Geiller_edge2020,  carrozza2021}. See also footnote \ref{ftnt:nonlocal}. \label{ftnt:edge_nonlocal}} 
 
 Imposing gauge variant boundary conditions means that the standard manner of specifying the field dynamics of a subsystem would  not be  fully gauge invariant (or rather, would not match dynamical symmetries to isomorphisms, as discussed in Section \ref{sec:subsystems} and in the previous Section). The usual response is to pare down gauge symmetries at the boundary: as in the asymptotic case, not all isomorphisms are to count as symmetries.   In this way,  the boundary conditions and the  boundary contributions to the dynamics remain symmetry-invariant, but only  in this pared down  way (see e.g. \cite{ReggeTeitelboim1974} for the first paper to enforce this approach explicitly, and, e.g.  \cite[Section 2]{Harlow_cov} and \cite[Section 2]{Geiller_edge2020} for  recent  treatments).
 
Thus the  standard approach to non-asymptotic boundaries treats the lack of  invariance of the subsystem similarly to the one of asymptotic boundaries, discussed in Section \ref{sec:subsystems}.  Thereby, it breaks downward consistency, given in Equation \eqref{eq:dwc}: the action of the universal symmetry on the subsystem is \emph{not} a subsystem symmetry. And \citet[p. 11]{Wallace2019b} endorses this view of subsystems,  because he takes subsystems  as sufficiently isolated so as to warrant an asymptotic-like treatment.

 
 But I believe a lot of our intuitions and uses of subsystems in gauge theory are excluded by this definition---hence the recent flurry of papers on subsystems in gauge theory. Recently, the pared down treatment of internal boundaries of subsystems has been called into  question  (cf. e.g. \cite{Teh_abandon, Geiller:2017xad, DonnellyFreidel, GomesRiello_new, Riello_symp, GomesStudies, Geiller_edge2020, carrozza2021} and references therein). New mechanisms, for instance, `edge-modes', have been devised to maintain the gauge invariance of the internal boundary under the symmetries of the entire universe.  
Thus in the next Section I will briefly describe how the careful use of representational conventions in the presence of subystems can also provide a solution to this obstruction to the downward consistency of symmetries.

\section{Representational conventions to the rescue}\label{sec:intro_variety}\label{sec:under_intro}\label{sec:sub_rep_conv}

As argued in Section \ref{sec:rep_conv}, the physical state space, composed of abstract equivalence classes, is mathematically intractable; (see also Section \ref{sec:elim_elim} for failed attempts at an intrinsic parametrization of this space). To get around  that, we introduced representational conventions. Now we must use representational conventions if we are to make sense of GSS through equations like \eqref{eq:phys_var}.

In Section \ref{sec:rep_conv_gluing} I  will describe the role of representational conventions in gluing and  in counting possibilities. In Section \ref{sec:rep_conv_dwc} I will describe how representational conventions are able to resolve the obstacles   listed in Section \ref{sec:int_bdary} to the downward consistency of symmetries, or to the matching of isomorphism and symmetry for subsystems. Finally, in Section \ref{sec:results} I will list the final assessement about DES, for different types of systems and subsystems. 

 \subsection{Representational conventions for gluing}\label{sec:rep_conv_gluing}
It is clear that if we are to compare   different physical possibilities---as we have to do in the quantum theory or in assessing DES for subsystems---we must ensure the comparison is made under a fixed representational convention. 
\citet[p. 18]{Wallace2019} highlights this same point:\begin{quote}
given configurations $(q; q')$ of the systems separately, we have
not been given enough information to describe their joint configuration: that
requires, in addition, a representational convention as to how points in the two
configuration spaces are to be compared. Such a convention is inevitably required
whenever we combine subsystems into a joint system. (In practice, the
convention is often given by a choice of coordinate systems, and/or of reference
frames, in the two subsystems.)
Prior to stipulating any such convention, there is no sense in which $(q, q')$ specifies a different configuration from $(R(g)q, q')$, since $q$ and $R(g)q$ are 
representationally equivalent.\footnote{In our notation: $q\equiv\varphi_+, q'\equiv \varphi_-, R(g)q\equiv\varphi_+^{g_+}$.} Given a choice of representational convention [i.e. in our notation introduced in Section \ref{sec:rep_conv}, $\sigma$],
though, it is clear that applying the symmetry transformation to one system
gives rise to a different total configuration (and that this is true independent of
what the actual representational convention is). So: symmetry-related configurations
can be understood as representing different possible configurations \emph{if we
hold fixed the choice of representational convention.} [my italics] \end{quote}

As  to different physical possibilities, Wallace says (p. 4, ibid): ``it becomes relatively simple to understand  modal questions in more directly empirical terms: is a situation where the symmetry transformation is applied to this system, but not to other systems, the same as or different from the original situation?''. The requirement of a fixed representational convention is paramount for DES, since it discloses whether a symmetry transformation has been applied to a given state (recall Galileo's ship and Definition \ref{def:DES} in Section \ref{sec:DES_intro}).

As I argued in Section \ref{subsec:promise},  in the study of a single physical possibility and over a single region---describing features of a given solution of the equations of motion on a single region, for example---a representational convention may be left as implicit. Nothing physically important turns on which representational convention was used, though some conventions may be more convenient than others. 

But when investigating subsystems and their relation to the entire universe,  more than one representational convention is necessarily at play---one per subsystem. Even if these conventions are defined implicitly by the same $\cal F$ (see Section \ref{subsec:gf}), they have different regional domains and therefore  may be incongruous. Here is a simple example:  center of mass---neither of two subsystem's center of mass can correspond to the joined states' center of mass.  Upon composition, a new center of mass will emerge, and we will have to `readjust' both our previous representational conventions. Thus \citet[Sec. 3]{RovelliGauge2013} illustrates the need for transition functions with a similar  example, which he  interprets as illustrating `the reason' we need gauge degrees of freedom: namely,  to ``couple subsystems''. Here is his example: two  squadrons of spaceships are separated by some distance, and each  subsystem's Lagrangian has a translation symmetry. Each squadron eliminates redundancy by employing new variables, based only on relative distances: they get rid of redundancy altogether, by employing some projection $\mathsf{pr}:\F\rightarrow [\F]$. But when the squadrons come together  they are unable  to articulate the composition of the subsystems, since they lack  transition functions. As we saw in  \eqref{eq:transition_h}, in Section \ref{subsec:isos_descs}, it is important that we leave room for a change of representation: mathematically, this corresponds to using a projection,  $h:\F\rightarrow\F$, as opposed to a reduction. 
 Unfortunately, I lack the space here to give a full account of \cite{RovelliGauge2013} using our jargon;  see \cite{GomesStudies}. Suffice it to say here that Rovelli's example is a particular illustration of the discussion  of Section \ref{subsec:isos_descs}, where I  arrived at gauge transformations from changes of representational conventions.

Just as in the case of a general principal fiber bundle $P$, unless the bundle is trivial and therefore admits a description using a single region,  transition functions, given in \eqref{eq:transition}, are \emph{necessary}. That is,  when different state spaces corresponding to different subsystems are  in play, we must allow for state dependent transition functions, i.e. transition functions between the representational conventions, as given in \eqref{eq:transition_h}. We now make this discussion more explicit.\\

Using   representational conventions for subsystems and universe, $\sigma_\pm$ and $\sigma$, respectively, we define: 
\begin{defi}[Gluing physical states]
Two physical states $[\varphi_\pm]$ are composable, iff, given any representational conventions $\sigma$, and $\sigma_\pm$, and the associated states $h_\pm$ (which we  here abbreviate to exclude their argument), as in \eqref{eq:proj_h},  there exist $g_\pm\in \G_\pm$, seen as  changes of conventions, also called `external' transformations, so that $h_+^{g_+} \cup h_-^{g_-}\in \F$. In other words, physical states are composable if there is a symmetry that will bring their two representatives to join smoothly at the boundary. 
\end{defi}

Given some global representational convention $\sigma$,  we can  rewrite the condition for $\neg$GSS, i.e. \eqref{eq:phys_var}, as there being $g_\pm$ and $g'_\pm$ such that:
\be\label{eq:final_gluing2}h_+^{g_+}\cup h_-^{g_-}=:h\neq h':=h_+^{g'_+}\cup h_-^{g'_-}. \ee
Crucially, \eqref{eq:final_gluing2} employs also the global representational convention, to assess whether there is a physically significant difference of the joined states.

It is important to note, as I mentioned above,  that we cannot assume $g_\pm=\mathrm{Id}$: namely, that either the subsystem or its environment comes to us with a fixed, immutable representational convention, that need not be adapted upon composition. That is because the representational convention of the global state may  have its restrictions to the subsystems fail to satisfy the regional  representional convention; exactly this occurs in the example of the center of mass, above.
In other words, we may have: 
 \be\label{eq:mesh} \iota^*_\pm h\neq h_\pm.
 \ee
 Thus, in order to count global possibilities given just the physical state, or, equivalently, the $h_\pm$, some adjustment between the two states in their regional representational conventions may be allowed or even required; that is, we should allow both $g_\pm\neq\mathrm{Id}$. This is the technical reason we do not have to constrain the symmetries at the boundary to be the identity, as we might want to do in the asymptotic case.

For gauge theory,  the assumption of states as supported on the regions $R_\pm$ and adjacency of the regions fixes the embedding of the subsystems. In this case, we can define composition of \eqref{eq:final_gluing2} by:\footnote{More carefully, we would use any extension  $\tilde h_\pm$ of $h_\pm$ into the ambient manifold $\Sigma$ (i.e. any global models $\tilde h_\pm$ such that $\iota_\pm^*\tilde h_\pm=h_\pm)$ and then multiply it by the Heaviside functions, $\Theta_\pm$. We simplify the notation in the main text. \label{ftnt:heavi}} 
\be\label{eq:gluing_gauge} h_+^{g_+}\Theta_++h_-^{g_-}\Theta_-=:h\neq h':=h_+^{g'_+}\Theta_-+h_-^{g'_-}\Theta_-.
\ee The $\Theta_\pm$ in \eqref{eq:gluing_gauge} are the (Heaviside) characteristic functions of regions $R_\pm$.
  Then the conditions for gluing become simply smoothness conditions:
\be h_+^{g_+}=_{|S}h_-^{g_-}
\ee
where the subscript ${}_{|S}$, restricting the equality to $S$, is understood as also matching derivatives.\footnote{For the standard notion of continuity, i.e. when all we require is the value of $f$ at the boundary, and not also of its derivatives, we employ no vertical bar, i.e: $f=_{S}f'$ iff $f(x)=f'(x)~\forall x\in S$. \label{ftnt:C0}} 

In the case of particle mechanics, composition requires also an embedding of each subsystem into the common ambient space (cf. \cite[Appendix C]{DES_gf} and footnote \ref{ftnt:heavi}).

\subsection{Representational conventions and downward consistency}\label{sec:rep_conv_dwc}

 Now, I can briefly, and at a pedestrian level, address the issues posed by the non-locality of gauge theories for the consistent definitions of subsystems, as mentioned in Sections \ref{sec:int_bdary} and   \ref{sec:subsystems}.  In Section \ref{sec:obstacle} I will give more technical details about the obstacle to matching dynamical symmetries to isomorphisms on subsystems. In \ref{sec:rep_conv_sol} I will describe how representational conventions are able to resolve these issues, so that we have a clear view of exactly in which cases we have  $\neg$GSS and DES. In Section \ref{subsec:hoehn}, I compare the fates of gauge symmetries at the boundaries as described using  representational conventions and as described by \cite{carrozza2021}---a very recent work tying loose ends in the `edge-mode' literature.

\subsubsection{The obstacle}\label{sec:obstacle}
First, I will just schematically introduce the issue, as seen through the Lagrangian formalism. The  Yang-Mills action in vacuum   is: 
\be S(A):=\int_{M\times \RR}   F^I_{\mu\nu}F_I^{\mu\nu}.
\ee
On a bounded submanifold, say, $R\times \RR$, where $R\subset M$ is a spatial submanifold of $M$, a variation of the action yields, after integration by parts: 
\be\label{eq:deltaS} \delta S(A)= -\int_{M\times \RR} \delta A_I^\nu(\D^\mu F^I_{\mu\nu})+\oint_{S\times \RR} s^\mu  F_{\mu\nu} \delta A^\nu,
\ee
where $s^\mu$ is the normal to the hypersurface $S\times \RR$ in ${M\times \RR}$. 
 Now, for the first term of \eqref{eq:deltaS} to vanish for arbitrary variations of the gauge potential it suffices that the gauge potential satisfies the Yang-Mills equations. But the second term vanishes only if either (the normal component of) the  field tensor  vanishes along the boundary or $\delta A^I_\mu$ vanishes at the boundary. The first condition is severely limiting; the second is not a gauge-invariant condition. To see this, suppose $\delta A_\mu^I=0$: as per the arguments of Section \ref{sec:rep_conv_gluing}, a transition function to a a different region would be state-dependent. Therefore the variational $\delta$ would act on that transition function, yielding a term that is `non-gauge-covariant' under variations.\footnote{ It is important here that these are time-like boundaries; for the spacelike initial and final surfaces, one can implement whatever initial conditions one likes. And the boundary term gives rise to the symplectic potential: $\theta=\int E^i\delta A_i$, which defines the symplectic structure of the theory, $\Omega:=\delta\theta$. So the above serves as an illustration.  This is done with much greater care in \cite[Secs. 4-7]{carrozza2021}. Most other work deals with non-covariance in the Hamiltonian or symplectic formalism, as I comment in the main text below (see footnote \ref {ftnt:equiv_Omega} and equation \eqref{eq:Omega_noncov}). But the fact remains that the origin of the problems with boundaries in gauge theories is agreed throughout the literature to be of the form $\delta g$; and the above example serves as an illustration. }

In the symplectic formalism, we witness a similar obstruction.  In brief,  denoting the symplectic 2-form by $\Omega$ (i.e. a closed, non-degenerate 2-form on phase space, discussed in Chapter \ref{ch:Coulomb} and Appendix \ref{sec:Ham}, given for electromagnetism in \eqref{eq:symp_form}): infinitesimal generators of gauge transformations, $\xi\in C^\infty(M, \mathfrak g)$ are usually  characterized by their generating phase space vector fields $\xi^\#$ in the kernel of the symplectic-form, that is, gauge transformations satisfy (see footnote \ref{ftnt:magic} in Appendix \ref{sec:Ham}): 
\be\label{eq:Omega_gauge} \mathbb{i}_{\xi^\#}\Omega\approx 0,\ee
where $\approx$ means the equality holds after we impose the kinematical constraints, or conservation laws. (See \cite[Ch. 1]{HenneauxTeitelboim} and \citep{Butterfield_symp, GomesButterfield_electro} for  philosophical introductions;  and also \cite[(2), p. 968]{Belot50}, who makes the same point.) 
 For Yang-Mills theories, with a  general, non-Abelian algebra $\mathfrak{g}$,  \eqref{eq:Omega_gauge} is always satisfied in the absence of boundaries. But in the presence of boundaries, it is only satisfied if $\xi_{|S}=0$ or if the boundary electric flux vanishes: $\mathsf{f}^I:=F^I_{0i}s_i=E^I_is^i=0$ (where $s^i$ is the normal to the boundary $S$ of $R$ within $\Sigma$). Again, these  are either severely limiting isolation conditions or do not respect downward consistency (Equation \eqref{eq:dwc}).\footnote{In more detail, let $\Omega=\int \mathrm{tr}(\delta A \wedge \delta E)$. Then we obtain: 
 \be \label{eq:Omega_noncov}\mathbb{i}_{\xi^\#}\Omega=\int \mathrm{tr}(\d \xi \dd E+[A, \delta E]+[\delta A, E])=\int \mathrm{tr}(\xi\, \delta(\D_A E))+\oint \mathrm{tr}(\xi \delta \mathsf{f}), 
\ee where $\D_A$ is the gauge-covariant derivative  of \eqref{eq:gauge_trans}.
We can extract two important pieces of  information from this equation: (1) the flow  of gauge transformations is Hamiltonian, i.e. such that for each $\xi$ we have a generating function on phase space, $H_\xi$ such that: $\mathbb{i}_{\xi^\#}\Omega=\delta H_\xi$ iff $\delta \xi=0$ and either $\xi_{|S}=0$ or $\delta \mathsf{f}_{|S}=0$. (But, unless $f=0$,  $f$ is not gauge-invariant in the non-Abelian theory, and therefore we cannot fix $\delta \mathsf{f}=0$ gauge-invariantly);   (2) $\xi^\#$ is in the kernel of the symplectic form iff  $\xi_{|S}=0$ or $\mathsf{f}=0$.  \label{ftnt:equiv_Omega}}

In the next Section we will see how, using representational conventions, one can find gauge-invariant regional dynamical structures labeled by the (equivalence classes of the) fluxes at the boundary; see \citep[Sec. 4]{Riello_symp} and \cite[Sec. 3]{GomesRiello_new}. In a similar fashion, in \cite{carrozza2021} these dynamical structures are obtained by stipulating a representational convention solely at the boundary; they call such conventions  \emph{boundary reference-frames}.
 
\subsubsection{How representational conventions help}\label{sec:rep_conv_sol}


As stressed  
in Section \ref{sec:rep_conv_gluing}, variational principles in the presence of boundaries oblige us to adopt an appropriate use of representational conventions.   And indeed,  \cite{GomesStudies, GomesRiello_new, Riello_new, GomesHopfRiello, GomesRiello2018,  GomesRiello2016} overcome  the obstructions to gauge covariance at the boundary  by---in the language of this thesis---explicitly including the representational convention in the  symplectic structure.

Thus, in our jargon, we consider all variations to be performed within the same representational convention. The dependence on the representational convention then appears explicitly in the variational procedure through the projection operator, $h_\sigma$, given in \eqref{eq:proj_h}, since that operator is state-dependent and will thus be acted upon by variations. By taking into account the phase-space dependence of this projection, the projected  dynamical structures of the subsystem  become suitably gauge-invariant (cf. \citet[Section 3]{GomesRiello_new}, where the representational conventions analogous to those given at the end of Section \ref{concl} were used).

Here is the rough idea  why this works (see especially \cite{GomesRiello2016, GomesStudies}): the problem with the extra term in the variation of the action in \eqref{eq:deltaS} or in the contraction with the symplectic form in  \eqref{eq:Omega_noncov} (in footnote \ref{ftnt:equiv_Omega}), is essentially a lack of gauge-covariance under the variation $\delta$. The analogy between finite and infinite-dimensional geometry (see footnote \ref{ftnt:inf_dim} of Chapter \ref{ch:syms}), relates $\delta$ and  spacetime derivative operators, $\d$. And similarly, there is  an analogy between the infinite-dimensional space of models $\F$ and the principal bundle $P$ (see footnotes  \ref{ftnt:stab} and \ref{ftnt:slice} of Chapter \ref{ch:syms}). In the finite-dimensional case, as discussed in Chapter \ref{ch:Noether}, to get around a lack of gauge-covariance under spacetime-dependent gauge transformations, we introduce the connection $\omega$ on $P$ through minimal coupling, i.e.  so that, schematically, quantities acted on by the \emph{horizontal} derivatives (as discussed in Section \ref{sec:PFB_formalism}) $\d\rightarrow\d_h:=\d-\omega$ (where $h$ stands for `horizontal')  remain covariant under spacetime-dependent gauge transformations.  So  too, in the infinite-dimensional case, we can justify the introduction of a connection-form $\varpi$, such that $\delta\rightarrow \delta_h:=\delta-\varpi$ becomes a fully covariant operator under state dependent changes of convention.\footnote{ The subscript $h$ is not  misleading, since $\varpi$ is associated to a horizontal projection operator on $\F$.  The difference between a horizontal projection operator mentioned here and projection onto a gauge-fixing surface, given in \eqref{eq:h_proj}, is that the former is `distributional': it acts as a projection only within  each $T_\varphi\F$. When such a distribution is integrable, meaning that the associated infinite-dimensional connection form $\varpi$ has no associated curvature, then the horizontal spaces foliate the space of models, and each leaf will correspond to the range of a particular choice of $\sigma$, and each initial value  selects an entire leaf. The infinite-dimensional connection-forms studied by Gomes and Riello generically have curvature; but the curvature vanishes in the Abelian theory, i.e. electromagnetism. One can take $\varpi$ to provide an infinitesimal, or perturbative, covariant notion of the projection operator $h$. \label{ftnt:hh}}

 In a similar spirit,  \cite{carrozza2021} show that the  variational principles from complementary regions decouple from each other iff the regional action is made gauge-invariant under a state-dependent gauge transformation.
 Thus, when we replace $\delta$ by $\delta_h$ in the definition of the symplectic form, $\Omega$, we get a new symplectic form, $\Omega_h$, that loses the unwanted term in \eqref{eq:Omega_noncov} and---in a superselection, or, in the language of \cite{carrozza2021}, post-selection sector (see the Intermezzo below)---decouples the variational principles of the regions.

Superselection sectors are necessary because the newly defined, horizontal, symplectic form does not  capture the entire dynamics. In particular, $\Omega_h$ excludes the terms that contain the  Coulombic part of the electric field.  
 Now we will briefly see how this worry arises and can be resolved. However, the following intermezzo can be skipped if we are solely interested in the matter of DES and GSS. For these matters, we are only interested in how the regional dynamical data can be composed or glued, and, for this purpose, these obstructions to regional dynamical autonomy are largely harmless. \\

\paragraph{Intermezzo: Superselection sectors and the obstruction to regional dynamical autonomy.}
The horizontal symplectic form---a symplectic form whose variations are expressed within a representational convention---$\Omega^h$  is both horizontal and gauge-invariant. As a consequence it can be unambiguously projected down to a 2-form $\Omega^h_\text{\tiny{red}}$ on the reduced,  on-shell phase space $\Gamma/\G$.
Moreover, it is easy to show $\Omega^h$ is closed, and thus, since the  pull-back by the reduction commutes with the variational derivative, its projection $\Omega^h_\text{\tiny{red}}$ is also closed.
However, for $\Omega^h_\text{\tiny{red}}$ to define a {\it symplectic} structure on $\Gamma/\G$, $\Omega^h_\text{\tiny{red}}$ would need to be non-degenerate as well. 
It turns out that, {\it in the presence of boundaries}, this is not the case.

Physically, this is simple to understand: $\Omega^h_\text{\tiny{red}}$ fails to provide a symplectic structure for the {\it Coulombic} dof. The reason there is no such failure in the absence of boundaries is that $E_c$ is fully determined by the matter degrees of freedom, and therefore does not need to independently appear in the symplectic structure. However, in the presence of boundaries, $E_c$ is determined by the charge density $\rho$ {\it as well as} $f$ \eqref{eq:Gauss}. Thus, loosely speaking, what is missing in $\Omega^h_\text{\tiny{red}}$ is a symplectic structure for the fluxes $f$. 

In more detail, in the symplectic formalism for electromagnetism, the choice of representational convention discussed in Chapter \ref{ch:Coulomb} symplectically couples two pairs of degrees of freedom: (i) the `radiative components', i.e. $h_\sigma(A)=A_r$ is coupled to the radiative content of the electric field, $E_r$; and (ii) the Coulombic content of the electric field, $E_c$ is coupled to  the `pure-gauge' part of the gauge potential, i.e. the elements of the form $A=g^{-1} \d g$. To recap: the radiative component of the electric field corresponds, roughly, to  radiation, and it  depends neither on the simultaneous distribution of charges nor on the value of $\mathsf{f}$ at the boundary; whereas the Coulombic component is entirely determined by these two pieces of information (see Section \ref{concl}). (Though we have not in this thesis investigated the non-Abelian counterpart to this choice, it works very similarly, albeit only perturbatively, cf. footnote \ref{ftnt:hh} and \citep[Sec. 3.2]{GomesRiello_new}).
The covariance property of the projection $h$ guarantees that the radiative regional phase space structure is  gauge-invariant, as I described above. But since radiative and Coulombic parts  are symplectically orthogonal, the radiative phase space structure leaves out the `Coulombic' part of phase space, which therefore needs to be added in (the reduced symplectic form will otherwise remain degenerate, and thus pre-symplectic).

We can resolve this problem by using (covariant) superselection sectors (or, in the language of \cite{carrozza2021}, post-selection sectors). In the Abelian case, we can find a suitable symplectic description of the symmetries and of the entire invariant content of the region by fixing the electric flux at the boundary (i.e. so that $\delta\mathsf{f}=0$). These fluxes are the external indices, mentioned in  Section \ref{sec:setting}, that we need to describe the regional dynamics invariantly.
In the Abelian case, this means simply that one foliates the reduced phase space $\Gamma/\G$ by subspaces at fixed value of $f$.  As a result, within each superselection sector, $E_c$ is also completely fixed by the matter dof and we are therefore in a situation similar to that of the case without boundary. Thus, in the Abelian case, although $\Gamma/\G$ is not symplectic, each superselection sector is.

In the non-Abelian case, fixing $\mathsf{f}$ would be tantamount to breaking the gauge symmetry at the boundary. 
Therefore the best one can do is to fix $\mathsf{f}$ up to gauge, i.e. demand that $f$ belongs to the set $[\mathsf{f}] = \{ \mathsf{f}= g^{-1} \mathsf{f}g , \text{for some} g\in\G \}$. Or alternatively, we can fix $f$: within some representational convention (e.g. the non-Abelian analog to the convention described in Chapter \ref{ch:Coulomb}), i.e. fix only the horizontal component of $\mathsf{f}$; or, as  in \cite{carrozza2021}, we can fix $\mathsf{f}$   with respect to a boundary reference frame.
Here, following \cite{GomesRiello_new, Riello_symp}, I will call the restriction of $\F$ to those configurations on-shell of the Gauss constraint with $\mathsf{f}\in[\mathsf{f}]$  a {\it covariant} superselection sector.
%

More technically, the physical content of the Coulombic piece can be arranged into `superselection sectors': i.e. different symplectic spaces attached to each (gauge orbit of the) boundary electric flux, and, in the radiative-Coulombic basis,  these sectors are dynamically decoupled from each other. In the Abelian case, the full gauge-invariant phase space structure of a regional subsystem  is thus  indexed by $\mathsf{f}$. In the general non-Abelian case, we would have to partition the phase spaces $T^*\F_\pm=\cup_{[\mathsf{f}]}T^*\F^{[\mathsf{f}]}_\pm$, where $[\mathsf{f}]$ is the equivalence class of electric fluxes (in the Abelian theory, since $\mathsf{f}$ is gauge-invariant, no square brackets are necessary). And for each value of $[\mathsf{f}]$ there is a well-defined gauge invariant regional symplectic form. Indeed,  we can abstractly define the projected, or reduced symplectic form: $\Omega^{[\mathsf{f}]}_{\text{\tiny red}}$, as $\pi^*\Omega^{[\mathsf{f}]\pm}_{\text{\tiny red}}=\mathfrak{i}_{[\mathsf{f}]\pm}^*\Omega^{\pm}$, where $\mathfrak{i}_{[\mathsf{f}]\pm}$ is the embedding of the Gauss constraint surfaces for $[\mathsf{f}]$ (as in footnote \ref{ftnt:magic}).\footnote{To define a symplectic structure over the (gauge-reduced) covariant superselection sector, one needs to add a symplectic structure for the superselected fluxes $f\in[f]$. 
This can be done in a canonical manner, by realizing that $[f]$ is essentially a (co)adjoint orbit in $\G$ and by resorting to the canonical Kirillov--Konstant--Sourieu (KKS) symplectic structure on coadjoint orbits. 
A properly constructed horizontal variation of the KKS symplectic structure  over the fluxes, $\omega^H_\text{KKS}$, can then be added to $\Omega^h$. The resulting 2-form $\Omega^h_{[f]} = \Omega^h+ \omega^h_\text{KKS}$ is gauge-invariant, closed, and projects to a non-degenerate symplectic structure within a reduced covariant superselection sector.
The resulting reduced symplectic structure is also independent of the choice of representational convention. Nonetheless, the basis in which one describes the physical dof \textit{will} depend on that choice.  It is, after all, a representational convention. 

Mathematically, this procedure is closely related to performing the Marsden--Weinstein symplectic reduction  not on the pre-image of the zero-section of the momentum map, but on the pre-image of a coadjoint orbit of the moment map.}


 See \cite[Section 4]{Riello_symp} and \cite[Sec. 3]{GomesRiello_new} for the derivations of the facts about symplectic reduction (as described in e.g.  \cite{Marsden2007}) mentioned in this intermezzo, and for the relation between the reduced symplectic form and the symplectic form in a representational convention (and to Marsden-Weinstein symplectic reduction in general).

In sum: to explicitly describe the entire regional gauge-invariant dynamics, we would need, in the Abelian case, some further boundary conditions, e.g. the fixed (instantaneous) electric-flux. These conditions are expected for the dynamical decoupling of the regions, and indeed reflected in e.g. \cite{carrozza2021}, where they are called  `post-selected' boundary conditions.

\paragraph{Superselection sectors and gluing.}
Here, for the treatment of DES,  we can sidestep the intricacies of the remaining issue discussed in the previous intermezzo by resorting to a  gluing theorem (see \cite[Section 6.1, Theorem 6.1]{GomesRiello_new}), stating  that the standard regional  representational conventions (e.g. the one used in Chapter \ref{ch:Coulomb}) generically uniquely characterize the solutions of \eqref{eq:gluing_gauge}.\footnote{ For illustration purposes, I display the solution of the gluing problem here, using the representational convention of Chapter \ref{ch:Coulomb} (see Section \ref{concl}),  in the Abelian case: 
$$\ln g_\pm = \zeta_{(\pm)}^{ \pm\Pi}\quad\text{with}\quad \Pi=\Big(\mathcal{R}^{-1}_+  + \mathcal{R}^{-1}_-\Big)^{-1}\left(  (\nabla^2_S)^{-1}\mathrm{div}_S( h_+-h_- )_{S}\right), $$
where the subscript $S$ denotes operators and quantities intrinsic (i.e. pulled-back)  to the interface surface $S$; $\zeta_{(\pm)}^u$ is a harmonic function on (respectively) $R_\pm$ with Neumann boundary condition $ \pp_n \zeta^u_{(\pm)}=u$, and $\mathcal{R}$ is the Dirichlet-to-Neumann operator. For the meaning of these operators, and also the analogous solution for the general non-Abelian Yang-Mills gauge theories, see \cite[Sec. 4]{GomesRiello_new}, and  \cite[Appendix D]{Gomes_new}.}  

In this respect, the important result obtained in the course of the proof of the gluing theorem is that, generically, the Coulombic part of phase space is redundant, \emph{for gluing}. That is,  though  the superselection or indexing of the regional phase space (see previous intermezzo) may be important to describe the regional physical facts autonomously, it becomes redundant once we have at hand both the charged matter content and the radiative/gauge-fixed symplectic pair of each region. Because once we glue the radiative pieces, we can generically reconstruct the Coulombic pieces. Here is \citet[p. 57]{GomesRiello_new}: 
\begin{quote}
Once both regional radiatives are known, even the regional Coulombic components
are completely determined---including the electric 
flux $\mathsf f$ through $S$, which is thus no longer
an independent degree of freedom once the radiative modes are accessible in both regions.
Thus, in this case---when the larger (glued) region $\Sigma$ has no boundary---the regional
radiative modes [generically] encode the totality of the degrees of freedom in the joint system. In particular, the conclusion reached in Section 3.4 from a regional viewpoint that $\mathsf f$
through $S$ must be superselected is a mere artifact of excluding [radiative] observables
in the complement of that region.
The addition of charged matter does not change this conclusion.
\end{quote}

In other words, under the dynamics that takes into account representational conventions, we can specify regional states that are fully symmetry invariant and are not further indexed by boundary conditions: $h_\pm(\varphi_\pm)=h_\pm(\varphi^{g_\pm}_\pm)$ (where we have reinstated $\varphi$ as the explicit argument of $h$). Up to some caveats that we will discuss in Section \ref{sec:results}, we have a theorem that characterizes the gluing or composition solely in terms of those  states, thereby avoiding the obstacle of Section \ref{sec:obstacle} (see also the intermezzo above).


\subsubsection{A different resolution}\label{subsec:hoehn}

 The resolution of the  failure of gauge covariance at the boundary is pursued differently in \cite{DonnellyFreidel} and follow-up papers (see e.g. \cite{Geiller_edge2020} for a more complete list).   These papers   add new degrees of freedom at the boundary with appropriate gauge-variance properties so as to cancel out the unwanted terms. In some circumstances the two approaches are related through a suitable interpretation of the new degrees of freedom as our $g_\sigma$ of \eqref{eq:gauge-fixing}: (see e.g. \cite[Section 5]{Riello_symp}, \cite[Section 5]{ReggeTeitelboim1974} \cite[Section 4]{carrozza2021}, and \cite{Teh_abandon}). But I find the introduction of new degrees of freedom problematic; (see \cite[Sec. 7]{GomesRiello_new}, \cite[5.6]{Riello_symp} and \cite[Sec. 3.2]{GomesStudies} for critiques).

The work reported here is, however, compatible with the more recent \citep{carrozza2021},
which stands out among the rest of the literature on  the fate of gauge symmetry in the presence of timelike boundaries of spacetime and which I now briefly summarize.  Results about the fate of gauge symmetries in the presence of boundaries  often only apply at Cauchy surfaces (there are exceptions, e.g. \cite{Harlow_cov}, whose results are  generalized by \cite{carrozza2021}).\footnote{Though it is usual  to employ the covariant symplectic framework, the most interesting formal results in the literature apply after pull-back to a Cauchy surface.} \cite{carrozza2021} fully extend those treatments to spacetime in an elegant manner: by demanding a decoupling of the variational action principles of  complementary spacetime regions separated by a timelike boundary (as in \eqref{eq:deltaS}). This is done in two steps: (i) restricting state space by physical boundary conditions, such as fixing the electromagnetic field at a boundary, i.e. fixing a superselection, or, in their words, post-selected sector (analogous to our fixing the electric flux); and (ii) adding boundary terms to the action functional so that, within the physically restricted state space, the regional variational principles---such as in \eqref{eq:deltaS}---fully decouple. 

As I have described in this Section, point (i) has a subtlety:  in  gauge theory,  it is a non-trivial matter to fix physical, or gauge-invariant, boundary conditions. To accomplish  that, \cite{carrozza2021} help themselves to  reference frames at the boundary, which they construct  from holonomies---cf. Section  \ref{sec:elim_elim} and footnote \ref{ftnt:tr_hol}---given by a certain family of curves from the boundary to a point at infinity corresponding to the group identity. These reference frames provide representational conventions at the boundary, and relative to these reference frames, we can fix gauge-invariant boundary conditions for the gauge potential (and for the electromagnetic curvature tensor, in the Abelian case). The boundary reference frames provide a split of the \emph{boundary} gauge potential (and the boundary curvature, in the non-Abelian case) into a pure gauge part and a physical part: precisely as the representational convention does on the entire spatial region (and not just at the boundary; cf. \eqref{eq:rad_proj_bdary}). 

In the spirit of this thesis, \cite{carrozza2021} award physical significance only to those  `reference frame reorientations' that can be interpreted as changing the relations between the physical content of complementary regions. And they find essentially the same results as I will describe in the next Section.


\subsection{Results}\label{sec:results}
In this thesis, I have taken different physical relations between the physical  regional states exist  when more than one global physical state can emerge from gluing the same physical states of the complementary regions. This happens iff the fixed boundary states (the post-selected states, in the language of \cite{carrozza2021}) have stabilizers (i.e. obey  \cite[Eq. 308]{carrozza2021}). The different relations can be identified with certain group elements, acting either on the region or  on the boundary. 


Since stabilizers are so important for DES, let me remind the reader of what was discussed in footnotes \ref{ftnt:stab} and \ref{ftnt:stab2} (see also footnote \ref{ftnt:reducible}): stabilizers  represent certain degeneracies within any given  representational convention. They occur for reducible states, i.e. states that have stabilizers (cf.  footnote \ref{ftnt:stab}) and are thus not ``wrinkly enough''---do not have features that vary enough---to completely fix the representation. For example, in an N-particle configuration space, we cannot completely fix orientation for configurations that are collinear: these configurations are stabilized by an action of $S^1$ (a rotation around the axis of collinearity). Stabilizers inevitably produce degeneracies in the representational convention: they foil uniqueness, but for physical reasons. When they exist, stabilizers form a rigid  (or  global), finite-dimensional group of gauge transformations.

 As described in  Section \ref{sec:rep_conv_sol} above, though  boundary conditions on the flux are necessary for describing the regional dynamics gauge-invariantly, they are not necessary for gluing (in a simply-connected region). Namely, in this context, the regional radiative content (which includes the charge density) generically determines the glued content (including the electric flux on each region). But if we restrict the space of models to some subset in which the boundary states---the gauge potential---have stabilizers, then the regional radiative states \emph{fail}  to determine the glued content. So in these cases, due to the presence of stabilizers, gluing partially fails. Moreover,  the stabilizers themselves are state-dependent---they must be transformations that preserve the gauge potential  at the boundary; they are always finite in number, and they correspond to the physical edge-modes  found in \cite{carrozza2021} under `Dirichlet post-selection'. They represent a holistic property of the theory ($\neg GSS$): the global state is not fully determined by the regional states, and the extra information can be encoded by group elements, acting on the regions or on the boundary between them, depending on the case: if they act only on the boundary, we do not interpret $\neg GSS$ as DES, since there is no subsystem symmetry that corresponds to this degeneracy. \\ 

So, finally,  what is the verdict about DES? There is no single answer, but here is a brief summary (see \cite[Sec. 5.1]{DES_gf} for an expanded version). The results below expand and precisify those of \cite{GreavesWallace, Wallace2019a, Wallace2019b}, and, where conditions overlap with   those of \cite{carrozza2021}, their results also match.
 \\
  
  First, assuming a trivial topology for $\Sigma$:
  
  \indent (i) in the Abelian case: there is no failure of GSS---no variety of global physical states given the regional physical states---in the absence of charged matter; nor when $\F$ allows charged matter at the interface between the regions. But for the sector characterized by a space of models in which charged matter is present in the regions but not on $S$, we have a degeneracy in the glued state, i.e. we have $\neg$GSS. Now, as described in Section \ref{sec:DES_intro}, $\neg$GSS is a necessary condition for DES, but DES also requires   that the physical variety   be in 1-1 correspondence with a group action on a subsystem.  In this case, the degeneracy \emph{does} correspond to a rigid shift by elements of $U(1)$ on  one of the subsystems and thus we also have DES.   This sector contains the situation depicted by `t Hooft beam splitter experiment (see \cite[p.110]{thooft}  and  \cite[p. 651]{BradingBrown}). It is also interesting to note that a treatment using holonomies (as in the appendix of \cite{GomesButterfield_electro},  confirms this result.

\indent (ii) In the non-Abelian  case, we start with a caveat: the results obtained by Gomes and Riello are valid only perturbatively (but perturbed around any state of $\F$, i.e. around any background). And here too, we must distinguish a few possibilities.

 First: as in the Abelian case,  if the allowed background states of the regions  have the same set of stabilizers, and if these stabilizers  act non-trivially on the regional states as a whole (e.g. by acting non-trivially on the matter fields), then there will be a physical variety of universal  states; so we once again obtain $\neg$GSS. Moreover, this variety corresponds to the action of the  group of stabilizers on  one of the subsystems and so is analogous to the previous case, so we obtain DES. For example,  for  backgrounds $[A_\pm]$ corresponding to the orbits of $A_\pm=0$, for  $G=SU(N)$, and charged matter within the regions: one copy of $SU(N)$ will have empirical significance.   (But such a condition is generically forbidden: generic states in non-Abelian Yang-Mills theory have only a trivial stabilizer.)

Second:  if the sector is such that states at the interface of the region have stabilizers---meaning that there are non-trivial gauge transformations that act as the identity only on the boundary values of the states---then we also get one physical global state per  boundary stabilizer. Here we have $\neg$GSS, since the subsystem states underdetermine the global state, but there is no interpretation  of this degeneracy as an action of a group on the subsystem states, and so no DES. This case represents what I take to be the \emph{physically relevant} notion of edge modes; (see also \cite{carrozza2021}, for a similar argument).\\

    A comparison of these two cases  with the familiar Aharonov-Bohm phases in the Abelian theory (cf. Section \ref{sec:AB}) is also worthwhile. There,   $\Sigma$ is taken to have a non-trivial topology, and the cohomology class of the gauge potential represents holistic physical information that can nonetheless be represented at the boundary by suitable transition functions. There too: there is a discrepancy between the tensor product of the regional physical  state  spaces and the physical state space of the union of the regions.  The discrepancy represents holistic physical information about the total system that is not contained in the individual subsystems. There too, we get $\neg$GSS, but not DES.



\section{Conclusions}\label{sec:conclusions}

In gauge theories, empirical significance can be obscured by redundancy of representation.  Ultimately, that is why the direct empirical significance (DES), or the observability of symmetries  continues to be a debated question. Nonetheless, the standard treatment of DES is almost silent about fixing representational conventions, with the exception of \citep{Gomes_new} and \cite{Wallace2019}, where the assumption {is} partially flagged, as noted in Section \ref{sec:rep_conv_gluing}, but not fully examined. Here I have paid it due attention.

\citet[p. 11]{Wallace2019b}  endorses a pared-down version of symmetries on  subsystems. That is because he takes subsystems  as sufficiently isolated   to warrant an asymptotic-like treatment, and asymptotic conditions often pare down symmetries, as  discussed in Section \ref{subsec:promise}. I maintain that there is a good notion of subsystem recursivity for  subsystems---namely, downward consistency---that does not mimic the asymptotic ideal of  perfect isolation. 

I have here argued that this notion is perfectly able to mathematically articulate and assess interesting foundational questions, such as that of the direct empirical significance of symmetries for non-asymptotic subsystems.   


As a last remark,  I admit that the asymptotic notion of boundaries  is ubiquitous in physics.  In fact, we even model the solar system in this way. In these cases, isomorphisms do not completely match the symmetries and thus they  acquire some physical meaning. 
 But I must admit I do not fully grasp how to connect, in detail, the observer with the asymptotic anchor on representations: more conceptual analysis is needed.

\chapter*{Closing words}

  This thesis has developed several conceptual questions that emerge for gauge theory. I will close with what I believe is the best justification for why gauge theories earn their living, that is, earn our believing in them. 

Gauge degrees of freedom fill an explanatory gap, have a neat relationist  interpretation, and are thoroughly warranted if we value consilience with other important theoretical structures of physics, such as Hamiltonians, actions, Lorentz invariance, etc.  The redundancy of description in gauge theories  comes down to a freedom to choose representational conventions, and is contiguous with the redundancy of description in other theories.  Demands for the elimination of this redundancy from our theoretical description of nature seems to ignore the criteria by which we interpret theories. In the words of 
\citet[p. 218]{Belot2003}:\begin{quote}
  But this much, I suppose, is uncontentious: judgments about the interest and correctness of interpretations of theories which are (in the strictest sense) false must rest ultimately upon judgments about the extent to which various interpretations of a given theory contribute to, and integrate smoothly with, our understanding of the world. Here the following sorts of considerations play a role: background metaphysical commitments and hopes; judgments about the relative perspicuity of various alternative formulations of the theory that we are interested in, and about the links between variant formulations and competing interpretations; and considerations---operating at the technical, conceptual, and metaphysical levels---that arise when we consider how our theory is related to neighboring theories, both more and less fundamental.
  \end{quote}

 \begin{appendix}

 \chapter{The Hamiltonian framework}\label{sec:Ham}

 One of the most distinctive features of the Hamiltonian framework  is the fact that  the Dirac analysis of constraints provides an algorithm for discovering gauge   symmetries. Once the algorithm succeeds, the equations of motion of a theory are divided into ones that come from constraints and those that we normally think of as generating dynamics.

Since the formalism is slightly unfamiliar to philosophers of physics,  in Section \ref{sec:Lag} we will first review, for  the  case of a non-relativistic mechanical system,   how under-determination of motion i.e. indeterminism can arise, in the {\em Lagrangian} framework. Then we will relate this to the constraints in the Hamiltonian framework that are obtained by a Legendre transformation. Then, in Section \ref{sec:Ham_symp} we will introduce the analysis of symmetry using symplectic geometry.  In these two Subsections, we will, broadly speaking, keep the relation between the Lagrangian and Hamiltonian frameworks simple---though sufficient for  our points about  the case of electromagnetism---by certain restrictions of scope about types of constraint. (These restrictions are summarised in the two paragraphs after equation \ref{eq:Ham_Gamma}.)  Thus we will see that both the Lagrangian and Hamiltonian frameworks show the special status of constraints: they are the generators of symmetries and must be imposed prior to any equation of motion, if there is to be a correspondence between the frameworks.  We end the Section with some philosophical remarks about the status of states that are not in the constraint surface.

\section{The Lagrangian framework and indeterminism }\label{sec:Lag}
In the Lagrangian formulation of mechanics, the focus is on curves in the space of all possible instantaneous configurations of the system. For the mechanics of $N$ point-particles parametrized by $\alpha=1, \cdots, N$, let this configuration space be called $\mathcal{Q}$. A \emph{Lagrangian} is a map from the tangent bundle on configuration space to the reals: $\mathcal{L}:T\mathcal{Q}\rightarrow \RR$. Once integrated along a curve, the Lagrangian yields the \emph{action functional} as a function from curves $\gamma$ in $\mathcal{Q}$ into the reals:
\be S(\gamma):=\int_\gamma \d t\, \mathcal{L}(q^\alpha(t), \dot q^\alpha(t)).\ee 
One then obtains, from the least action
principle $\delta  S=0$, the Euler--Lagrange equations \eqref{equ:Euler-Lagrange}. Using  the chain rule for the $\frac{d}{dt}$ derivative, we get from \eqref{equ:Euler-Lagrange}:
the equations for the accelerations, \eqref{equ:accel_lagrange}, which I reproduce here:
\be \ddot q^{\beta}\frac{\pp ^2 \La}{\pp \dot q^{\beta}\pp \dot q^\alpha}+\dot q^{\beta}\frac{\pp ^2 \La}{\pp  q^{\beta}\pp \dot q^\alpha} =\frac{\pp \La}{\pp  q^\alpha}
.\ee
The accelerations are uniquely determined by the positions and velocities if we can isolate $\ddot q^{\beta}$ in this equation. A necessary and sufficient condition for this  is that the matrix $M_{\alpha\beta}:=\frac{\pp ^2 \La}{\pp \dot q^{\beta}\pp \dot q^\alpha} $ be  invertible. If it is not,   the accelerations are undetermined, so that the motion is under-determined by the initial positions and velocities: there is indeterminism at the level of the $2N$ variables, $q^\alpha(t), \dot q^\alpha(t)$.   Assuming we believe that in physical terms, the motion is indeed determined, this indicates a  redundancy in our description of the system. For a philosophical introduction to this indeterminism and redundancy, from a Lagrangian treatment, compare \cite{Wallace_LagSym}.  

Let us see how this redundancy appears in the Hamiltonian formalism. The idea of the {\em Legendre transformation} is that at any point $q \in {\cal Q}$, the Lagrangian $\cal L$ determines a map $\mathrm{Leg}_q$ from the tangent space $T_q{\cal Q}$ at $q$ to its dual space $T^*_q{\cal Q}$. Intuitively speaking, this is the transition ${\dot q} \mapsto p$. To be precise, $\mathrm{Leg}_q$ is defined by
\be
\mathrm{Leg}_q : w = {\dot q}^{\alpha} \frac{\partial }{\partial q^{\alpha}} \in T_q{\cal Q} \; \mapsto \; \frac{\partial {\cal L}}{\partial {\dot q}^{\alpha}} \in T^*_q{\cal Q}  \; .
\ee         
One easily checks that because the canonical momenta $p_\alpha:=\frac{\pp  \La}{\pp  \dot q^\alpha}$ are 1-forms, this definition is coordinate-independent. An equivalent definition, manifestly coordinate-independent and given for all $q \in {\cal Q}$, is the Legendre transformation, $\mathrm{Leg}:T\mathcal{Q}\rightarrow T^*\mathcal{Q}$, defined by
\be\label{eq:Leg}
\forall q \in {\cal Q}, \, \forall v,w \in T_q\mathcal{Q} \; : \; \;  \mathrm{Leg}(w)(v):=\left.\frac{d}{dt}\right|_{t=0}\mathcal{L}(w + t v).
\ee
(Here we take $w,v$ to encode the identity of the base-point $q$, so as to simplify notation, writing $\mathrm{Leg}(w)$ rather than $\mathrm{Leg}(q,w)$ etc.) That is: $\mathrm{Leg}(w)(v)$ is the derivative of $\cal L$ at $w$, along the fibre $T_q{\cal Q}$ of the fibre bundle $T{\cal Q}$, in the direction $v$. Thus $\mathrm{Leg}$ is also called the {\em fibre derivative}.

Given $\cal L$, we define its energy function $E:T{\cal Q} \rightarrow {\RR}$ by 
\be\label{Lenerg}
\forall v \in T{\cal Q}, \; E(v) := \mathrm{Leg}(v)(v) \, – \, {\cal L}(v) \; ;
\ee
or in coordinates, 
\be\label{Lenerg2}
E(q^{\alpha}, {\dot q}^{\alpha}) := \frac{\partial {\cal L}}{\partial {\dot q}^{\alpha}}{\dot q}^{\alpha} \, - \, {\cal L}( q^{\alpha}, {\dot q}^{\alpha}).
\ee
Then one shows that 
 $E \circ (\mathrm{Leg})^{-1}$ is, as one would hope, the familiar Hamiltonian function $H: T^*{\cal Q} \rightarrow {\RR}$ given by $(q,p) \equiv (q^{\alpha}, p_{\alpha}) \mapsto {\dot q}p - {\cal L} \equiv {\dot q}^{\alpha}p_{\alpha} - {\cal L}$, with each ${\dot q}^{\alpha}$ being a function of the $q$s and $p$s.

Thus  the question whether the accelerations are determined by the positions and velocities is translated to the question of whether the momenta $p_\alpha = \frac{\pp  \La}{\pp  \dot q^\alpha}$, are invertible, as functions of the velocities. If $M_{\alpha \beta}=\frac{\pp  p_\alpha}{\pp  \dot q^\beta}$ is not invertible, since $\alpha$ and $\beta$ run over the same indices (i.e. $M$ is a square matrix),  the map from the $\dot q$s to the $p$s, at fixed $q$, is many-to-one. 
That is:  there are constraints among the $p_\beta$ as functions of the velocities. Dropping reference to the velocities, i.e. writing the constraints as functions on phase space, we therefore write equations such as 
\be\label{eq:ctraints}
\Phi^I(q^\alpha, p_\beta)=0,
\ee
where $I$ parametrizes the  constraints.\footnote{Here we assume that the rank of $M_{\alpha\beta}$ is constant, and that the constraints obey regularity conditions. See \cite[Ch. 1.1.2]{HenneauxTeitelboim}.} These conditions are usually called {\em primary constraints}, to emphasize that no equation of motion was used in their derivation. If the constraints are conserved by the equations of motion, then  they correspond, by (the converse of) Noether's theorem, to symmetries of the system: as we will explain in Section \ref{sec:Ham_symp}. 

Now, the constraint surfaces are submanifolds of $T^*\mathcal{Q}$, and it thus follows from \eqref{eq:ctraints} that the inverse transformation, from the momenta $p$ to the velocities $\dot q$, must be  multi-valued, since the dimension of $T\mathcal{Q}$ is the same as that of $T^*\mathcal{Q}$ (viz. $2N$, and so greater than the dimension of  the constraint surfaces).  Thus the inverse image in $T\mathcal{Q}$ of the constraint surface \eqref{eq:ctraints} forms a submanifold.\footnote{The gradients of $\Phi^I$ in the momentum directions along the surface, i.e. the vectors $\frac{\pp \Phi^I}{\pp p_\alpha}$, provide the complete set of null vectors of $M_{\alpha\beta}$, since by \ref{eq:ctraints}
 $$0=\frac{\pp \Phi^I}{\pp \dot q^\beta}=\frac{\pp \Phi^I}{\pp p_\alpha}\frac{\pp p_\alpha}{\pp \dot q^\beta}=\frac{\pp \Phi^I}{\pp p_\alpha}M_{\alpha\beta}.$$
In Section \ref{sec:Ham_symp} we will give a coordinate-free version of these statements.\label{ftnt:M}}   Similarly, if we consider the intersection of the constraint surfaces \eqref{eq:ctraints}, as  $I$ varies (which is often called `{\em the} constraint surface'); and so, the inverse image of that intersection.    

 Correspondingly, within $T\mathcal{Q}$: a non-trivial kernel 
for the matrix $M_{\alpha\beta}$ in \eqref{equ:accel_lagrange} implies that the extrema of the Lagrangian are not isolated: there are 1-parameter families of curves in configuration space that extremize the action functional. In other words, as discussed above: there is indeterminism at the level of the $2N$ variables, $q^\alpha(t), \dot q^\alpha(t)$: (again, compare \cite{Wallace_LagSym}).

 Here is an \textbf{Example} from \citet[Sec. 1.1.1]{HenneauxTeitelboim}, which we will build on in Section \ref{sec:toy}. Let $(q^1, q^2)$ be coordinates for $\mathcal{Q}$, and consider $\mathcal{L}=\frac12(\dot q^1-\dot q^2)^2$. The momenta are $p_1=\dot q^1-\dot q^2$ and $p_2=\dot q^2-\dot q^1$, as is easy to verify. Thus we find the rather simple constraint: ${\cal M}= p_1+p_2=0$. The Legendre transform maps all of $T\mathcal{Q}$  into this constraint surface in $T^*\mathcal{Q}$. Moreover, the entire line $\dot q^2=\dot q^1+c$ is mapped to $p_1=-c=-p_2$. This transformation is therefore neither one-to-one nor onto. 

To render this transformation invertible, we will need to introduce \emph{Lagrange multipliers}, in Section  \ref{sec:Ham_symp}, which can be thought of as coordinates on the manifolds that are the inverse values in $T\mathcal{Q}$ (for \eqref{eq:Leg}) of a given point on the constraint surface lying in $T^*\mathcal{Q}$.\footnote{It is also relatively easy to show that two velocities that lie in the pre-image of the same momentum are related by a linear combination of the null vectors of $M_{\alpha\beta}$, namely $\frac{\pp \Phi^I}{\pp p_\alpha}$ (cf. footnote \ref{ftnt:M}); the coefficients of this linear combination are the Lagrange multipliers. In Section \ref{sec:Ham_symp} we will see a coordinate-free version of these statements. \label{ftnt:symp_flow} }

\section{The Hamiltonian formalism and constraints}\label{sec:Ham_symp}

To understand the basic features of constraints and the symmetries they generate in a manner that will be helpful in what follows, it pays to introduce the symplectic formalism for Hamiltonian mechanics. 

The great advantage of the symplectic formalism is that it treats momentum and configuration variables on a par. By so doing we see phase space $\mathcal{P}$ as a high-dimensional manifold---infinite-dimensional, in field-theory---endowed with a certain geometric structure. 
In the simple example above, $\mathcal{P}$ would be the $2N$-dimensional manifold  whose geometric structure is a symplectic 2-form, given, in  the global coordinates $(q^\alpha , p_\alpha)$, by:
\be \omega=\sum_\alpha \d q^\alpha\wedge\d p_\alpha.
\ee
Though we have given $\omega$ in the specific choice $(q,p)$ of coordinates,\footnote{These are always available locally, thanks to Darboux's theorem: which states, in modern geometric terms, that a manifold equipped with a symplectic 2-form is locally a cotangent bundle. Cf. e.g.  \cite[p. 230-232]{arnold1989}; or for a philosophical introduction,  \cite[ Section 6.6]{Butterfield2006}.} it is a coordinate-independent, differential geometric object on $\mathcal{P}$. 

The role of the symplectic form $\omega$ is to convert a vector field into a one-form, or vice-versa. And  since a scalar function defines a one-form, viz. its gradient, $\omega$ converts a scalar function like the Hamiltonian into a vector field, whose integral curves are a flow in phase space. We construe these curves as the dynamical trajectories of the system;  (i.e. assuming that the given scalar function encodes the forces operative on and in the system). 
Thus we take a Hamiltonian function $H:\mathcal{P}\rightarrow \RR$ to specify the dynamics by assigning to each dynamical state its total energy.   For from any such smooth scalar function, we can obtain a one-form $\d H$, and then   use $\omega$ to define a vector field $X_H$, by: 
\be\label{eq:fund_symp} \omega(X_H, \bullet)=\d H (\bullet).
\ee 
This vector field on phase space   is then integrated to yield  a dynamical trajectory through each point. Compare Figure 1. 

\begin{figure}[h!]\label{fig1}
\center
\includegraphics[width=0.5\textwidth]{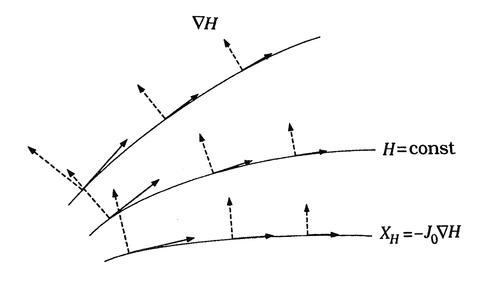}
\caption{An illustration of the relation between a scalar function, its gradient,  and its associated symplectic vector field, using the canonical symplectic form and the Euclidean metric  to identify $\d H$ with $\nabla H$, as a vector orthogonal to the level surfaces $H=$const. Supposing, in the $(q, p)$-coordinates, that $\nabla H=(\pp_q H, \pp _p H)$, then, writing the  inversion map as $J_0$ , we have: $X_H=-J_0(\pp_q H, \pp _p H)=(-\pp_p H, \pp_q H)$, which is orthogonal to $\nabla H$ and thus along $H=$ const.}
\end{figure}

Indeed, in the simple mechanical case without  constraints, we can plug coordinates into this equation to recover a local description of the dynamics, i.e. the familiar form of  Hamilton’s equations. In particular, the relation between Poisson brackets,   defined as usual by
\be\label{eq:Pois}
\{f, h\}:= \frac{\pp f}{\pp q^\alpha}\frac{\pp h}{\pp p_\alpha} - \frac{\pp f}{\pp p_\alpha}\frac{\pp h}{\pp q^\alpha}; \,\; \mbox{so that Hamilton’s equations with $h$ are:}  \,\; 
\frac{df}{dt} = \{f, h\} \, ,
\ee 
and the symplectic form $\omega$ is: 
\be\label{PoisOmeg}
 \{f, h\} = \d f (X_h) = \omega(X_f, X_h),
\ee
for $f, h\in C^\infty(\mathcal{P})$.\footnote{Here, as usual,  $\d f(X)$ is the contraction between 1-forms and vectors; and $\d f(X)$ is equal to $X(f)$ i.e. the directional derivative of a scalar function $f$ along $X$.}

For consistency, since the exterior derivative squares to zero i.e. $\d ^2\equiv0$,  $\omega$ must be closed, i.e. $\d \omega=0$. Moreover, if we would like the dynamical trajectory associated to $H$  to be unique,   $\omega$ must be  non-degenerate. That is: $\omega(v, \bullet)$ must be injective, i.e. have only the zero vector in its kernel. 

Although this last condition is always taken to hold on the full phase space $\mathcal{P}$, it needs to be relaxed in gauge theories, precisely because of  constraints. So although we require $\omega$ to be non-degenerate in $\mathcal{P}$, it does not need to be non-degenerate once we restrict it  to {\em the} constraint surface---meaning now
 the intersection of the constraint surfaces \eqref{eq:ctraints}, as  $I$ varies:
\be\label{eq:interGam}
\Gamma:=\{z\in \mathcal{P}, |\, \Phi^I(z)=0,\quad \text{for all}\,\, I\}.
\ee

 As depicted in Figure 1: the $\Phi^I$,  as scalar functions on phase space (for each $I$), have a (differential geometric) gradient, $\d \Phi^I$, which are in one-one correspondence with  vector fields $X_{\Phi^I} =: X_{I}$ due to the symplectic structure of phase space: namely, through  $\omega(X_I, \bullet)=\d \Phi^I$. The key idea is that, just as the flow specified by the Hamiltonian function conserves energy, these vector fields associated to $\Phi^I$ are tangential to, and so preserve, the intersection of all  the constraint- and energy-surfaces.  That is, in a less geometric (and maybe more familiar) language: they not only commute with the Hamiltonian and conserve energy, but also conserve the charges associated with the constraints.
 
  Here we will only consider  a certain type of constraints, called {\em first-class constraints}. These are defined as constraints whose Poisson bracket with every constraint vanishes on the constraint surface $\Gamma$ (though perhaps not elsewhere). Then the assumption that all the constraints are first-class implies that the flow of each vector field $X_I$, associated to each constraint, is tangent to $\Gamma$. This assumption also   
 means that we need not  concern ourselves with the several steps involved in the {\em Dirac algorithm}.\footnote{\label{algor}{Without the assumption of first-class, we still have an algorithm for finding whether the constraints generate symmetries. This algorithm can be summarised in the more familiar Poisson bracket notation, as follows. Suppose we are given some initial set of constraints $\Gamma_0:=\{z\in \mathcal{P}, |\, {\cal M}^{I_0}(z)=0,\quad \text{for all}\,\, I_0\}$, such that, e.g. for some $I_0, J_0$, we have  $\{{\cal M}^{I_0}, {\cal M}^{J_0}\}_{|\Gamma_0}={\cal M}^{K_1}\neq 0$ where clearly ${\cal M}^{K_1}$ is not included among the original constraints, since it does not vanish on $\Gamma^0$. (Here the restriction to $\Gamma_0$ serves to emphasize that the vector fields need not commute everywhere on phase space, but only on the surface where the constraints vanish.) We would then  add this new constraint to the others, to form a new `surface', $\Gamma_1$, and repeat the test above, until, eventually we get:
$$
 \{ {\cal M}^{I_n}, H\}_{|\Gamma^n}\equiv\omega(X_{I_n}, X_H)_{|\Gamma^n}\equiv X_{H}({\cal M}^{I_n})_{|\Gamma^n}=0,\quad \text{for all}\,\, I_n,
$$
and 
$$
\{{\cal M}^{I_n}, {\cal M}^{J_n}\}_{|\Gamma_n}\equiv \omega(X_{I_n}, X_{J_n})_{|\Gamma_n}=0,\quad \text{for all}\,\, I_n,J_n. 
$$
(It is possible that the iteration yields only the empty set, in which case the system is dynamically inconsistent).}}  In general, these steps are necessary because the  primary constraints \ref{eq:ctraints} that  emerge from the Legendre transform might fail to be preserved by either the Hamiltonian or by other constraints---but this will be ensured if the constraints are first-class.  Thus in general, one must then seek a type of reflective equilibrium: successively imposing further restrictions to  submanifolds of phase space, until the corresponding symplectic flows of all the constraints preserve the Hamiltonian and become tangent to the constraint submanifold. Compare footnote \ref{algor}, and  \cite{Pons2005}; and for a complete account, \cite[Chapter 2]{HenneauxTeitelboim}. 
 
 In the language of symplectic geometry, we define the embedding $\iota:\Gamma\rightarrow \mathcal{P}$ and require that  the pullback of the symplectic form $ \iota^*\omega$ obey:
\be\label{eq:Ham_Gamma}
 \iota^*\omega(X_H, \bullet)=\d (H|_{\Gamma}), \quad \iota^*\omega(X_I, \bullet)|_{\Gamma}=\d ({\cal M}_I{}|_{\Gamma})\equiv 0
\ee
(since ${\cal M}_I{}|_{\Gamma}\equiv 0$). Thus, on the constraint surface, because we assumed constraints to be first-class, the vector fields $X_I$ generated by the constraints are null directions of the symplectic form restricted to the constraint surface. This generalizes, in a  coordinate-free formalism, the content of footnote \ref{ftnt:symp_flow}. 

These directions are \emph{gauge}. 
The set of points of phase space that are reached by the $X_I$'s integral curves, from a given point, is called a \emph{gauge orbit}.  And the elements within each such orbit   are taken to be physically equivalent. Here, physical equivalence of two points of phase space is understood as a matter of any physical quantity, taken as a phase function, i.e. a real-valued function on phase space, having the same value for the two points. That is: a physical quantity must be {\em gauge-invariant}: taken as a phase function, it  must be constant on each gauge orbit.\footnote{\label{ftnt:magic} {Indeed, the null directions of  $i^*(\omega)$ are necessary and sufficient to characterise the generators of gauge symmetry. For suppose that what we know is that  a certain class of vector fields $X_I$ is such that $\omega(X_I, \bullet)=0$.  Since  the exterior derivative $d$ commutes with pullbacks, if $\omega$ is closed, $i^*\omega=:\tilde \omega$ is also closed. Thus using the Cartan Magic formula relating Lie derivatives, contractions $i$ and the exterior derivative $\d $:   
$$
 L_{X_I}\tilde\omega=(\d i_{X_I}+ i_{X_I}\d)\tilde\omega=0; 
$$
i.e. the first term also vanishes because $\tilde\omega(X_I,\bullet)=0$. So $\tilde\omega$ itself is invariant along $X_I$. Moreover, if we take the commutator of $X_I, X_J$, i.e. $[ X_I, X_J]= L_{X_I}X_J$,  contract it with $\tilde\omega$, and remember the formula:
$$L_{X_I}(\tilde\omega(X_J, \bullet))=\tilde\omega( L_{X_I}X_J, \bullet)+  (L_{X_I}\tilde\omega)(X_J, \bullet) \, ,
$$
we obtain that, since both $L_{X_I}(\tilde\omega(X_J, \bullet))=0$  and $L_{X_I}\tilde\omega=0$, it is also the case that $\tilde\omega( [{X_I}, X_J], \bullet)=0$. Thus, by the Frobenius theorem  the kernel of the pullback  $i^*(\omega)$ forms an integrable distribution which  integrates to give  the orbits of the symmetry transformation.  This means, for the discussion below (in Section \ref{sec:off}) of symplectic reduction, in which a Lie group $G$ acts {\em ab initio} on the phase space, that:} we can define a projection operator $\pi:\Gamma\rightarrow \Gamma/G$; and, ultimately the degeneracy of $i^*\omega$ allows one to define a \emph{reduced symplectic form}, $\bar\omega$,  on the space of orbits, given by $\pi^*\bar\omega=i^*\omega$. See \cite[Ch. 1]{Marsden2007}.}

 For simplicity, we have here suppressed  a few important qualifications (that are widely recognized). Firstly: in some systems, the requirement that the primary constraints \eqref{eq:ctraints} be preserved in time implies a new relation between the $q$s and $p$s, independent of these constraints. Such a relation is called a {\em secondary} constraint. Secondary constraints can be first-class: an important example being the Hamiltonian i.e. scalar constraint  (governing time-evolution) in canonical general relativity. And some such secondary first-class constraints are {\em not} gauge generators. But though important, these points do not affect this thesis. For details, compare \cite[Sections 1.1.5, 1.2, 1.6.3 and 3.3]{HenneauxTeitelboim} and \cite{Pitts_Ham}.   A second simplifying assumption---which applies to electromagnetism, and indeed to most familiar physical theories---is that the commutation algebra of the constraints closes irrespectively of the satisfaction of the equations of motion: it forms what is usually called a \emph{closed algebra}. Thirdly, we   have also assumed that the constraints are {\em irreducible}, i.e. that all the constraint equations \ref{eq:ctraints} (for both primary and secondary constraints) are independent of each other: (so that roughly speaking, there are no ``symmetries among the symmetries''  that they generate). And lastly, we   have also assumed that the commutation algebra of the constraints forms a true Lie algebra, i.e. the structure `constants' are true constants, not functions on phase space; (this assumption fails for general relativity).\footnote{\label{Pitts}{We should also note a controversy. \cite{Pitts_electro} claims that, even under these assumptions, and for electromagnetism, the main (and orthodox) idea above---that points in the same gauge orbit, are physically equivalent---fails. He claims that even a first-class constraint can fail to be a gauge generator, i.e. it can generate instead what he calls a `bad physical change'. Our own view is that the main idea holds good. (The dispute turns on the transformation properties of Lagrange multipliers in the canonical Lagrangian; and we think the treatment by e.g. \cite[equations 19.11 and 19.13, as clarified and supported by 3.26 and 3.31]{HenneauxTeitelboim}  answers Pitts’ arguments.)}}

 The existence of null vector fields implies the Hamiltonian flow is not unique: if $X_H$ solves \eqref{eq:Ham_Gamma}, then so does $X_H+a^I X_I$ for any set of  coefficients $a^I$ that are arbitrary functions of time. So, since the dynamics preserves the constraint surface, instead of taking $H$ as the Hamiltonian function generating the dynamics, we may equivalently take the \emph{total Hamiltonian}:
\be\label{eq:Ham_tot}
H_T=H+a^I{\cal M}_I.
\ee
Along the constraint surface the dynamics according to $H_T$  will be indistinguishable from that according to $H$.  Here, by `indistinguishable',  we mean that  two trajectories within $\Gamma$ that start at a common point in $\Gamma$ and that are determined, respectively, by the two choices of Hamiltonian, will at any later (indeed: any earlier!) time, lie in the same gauge orbit as each other---and so will at all times agree on the values of all gauge-invariant quantities.

In sum,  one of the  significant (as well as practical) differences between the Hamiltonian and the Lagrangian   frameworks is that  in the Hamiltonian framework the symmetries are not `guessed'  from the form of the action functional. Instead, they are obtained from the constraints that emerge 
when  the Legendre transformation is applied. That is, the constraints that emerge from the Legendre transformation are associated to vector fields; and, with a few auxiliary assumptions,   the flow of   each of these vector fields conserves the constraints and the Hamiltonian, and   the vector field is thus taken to be the generator of symmetries on ${\Gamma}$. The action of the symmetry on any quantity $Q$ is given by  
\be\label{symquant}
 X_I(Q)=\d Q(X_I)=\omega(X_I, X_Q)=\{\Phi^I, Q\}.
\ee

 Finally let us sum up this discussion of the Lagrangian and Hamiltonian frameworks’ treatments of constraints, by stressing a concordance between them: a concordance despite the many-one mapping from the first to the second, i.e. the fact that at fixed $q$, many $\dot q$ map to a single $p$. 
 Namely: there is a  one-one correspondence between the gauge orbits in the Hamiltonian framework and  degeneracy directions of the Lagrangian in the configuration space; (cf. Sections 3.1-2 in \cite{HenneauxTeitelboim}).

 \end{appendix}

\begin{thebibliography}{}

\bibitem [\protect \citeauthoryear {%
Adams%
}{%
Adams%
}{%
{\protect \APACyear {1979}}%
}]{%
Adams1979}
\APACinsertmetastar {%
Adams1979}%
\begin{APACrefauthors}%
Adams, R\BPBI M.%
\end{APACrefauthors}%
\unskip\
\newblock
\APACrefYearMonthDay{1979}{}{}.
\newblock
{\BBOQ}\APACrefatitle {{Primitive Thisness and Primitive Identity}} {{Primitive
  Thisness and Primitive Identity}}.{\BBCQ}
\newblock
\APACjournalVolNumPages{Journal of Philosophy}{76}{1}{5--26}.
\newblock
\begin{APACrefDOI} \doi{10.2307/2025812} \end{APACrefDOI}
\PrintBackRefs{\CurrentBib}

\bibitem [\protect \citeauthoryear {%
Aharonov%
\ \BBA {} Bohm%
}{%
Aharonov%
\ \BBA {} Bohm%
}{%
{\protect \APACyear {1959}}%
}]{%
aharonovbohm1959}
\APACinsertmetastar {%
aharonovbohm1959}%
\begin{APACrefauthors}%
Aharonov, Y.%
\BCBT {}\ \BBA {} Bohm, D.%
\end{APACrefauthors}%
\unskip\
\newblock
\APACrefYearMonthDay{1959}{Aug}{}.
\newblock
{\BBOQ}\APACrefatitle {Significance of Electromagnetic Potentials in the
  Quantum Theory} {Significance of electromagnetic potentials in the quantum
  theory}.{\BBCQ}
\newblock
\APACjournalVolNumPages{The Physical Review}{115}{}{485--491}.
\newblock
\begin{APACrefURL} \url{https://link.aps.org/doi/10.1103/PhysRev.115.485}
  \end{APACrefURL}
\newblock
\begin{APACrefDOI} \doi{10.1103/PhysRev.115.485} \end{APACrefDOI}
\PrintBackRefs{\CurrentBib}

\bibitem [\protect \citeauthoryear {%
Anandan%
}{%
Anandan%
}{%
{\protect \APACyear {1977}}%
}]{%
Anandan1977}
\APACinsertmetastar {%
Anandan1977}%
\begin{APACrefauthors}%
Anandan, J.%
\end{APACrefauthors}%
\unskip\
\newblock
\APACrefYearMonthDay{1977}{Mar}{}.
\newblock
{\BBOQ}\APACrefatitle {Gravitational and rotational effects in quantum
  interference} {Gravitational and rotational effects in quantum
  interference}.{\BBCQ}
\newblock
\APACjournalVolNumPages{Phys. Rev. D}{15}{}{1448--1457}.
\newblock
\begin{APACrefURL} \url{https://link.aps.org/doi/10.1103/PhysRevD.15.1448}
  \end{APACrefURL}
\newblock
\begin{APACrefDOI} \doi{10.1103/PhysRevD.15.1448} \end{APACrefDOI}
\PrintBackRefs{\CurrentBib}

\bibitem [\protect \citeauthoryear {%
Anandan%
}{%
Anandan%
}{%
{\protect \APACyear {1993}}%
}]{%
Anandan_1993}
\APACinsertmetastar {%
Anandan_1993}%
\begin{APACrefauthors}%
Anandan, J.%
\end{APACrefauthors}%
\unskip\
\newblock
\APACrefYearMonthDay{1993}{}{}.
\newblock
{\BBOQ}\APACrefatitle {{Remarks Concerning the Geometries of Gravity and Gauge
  Fields}} {{Remarks Concerning the Geometries of Gravity and Gauge
  Fields}}.{\BBCQ}
\newblock
\BIn{} B\BPBI L.~Hu, M\BPBI P.~Ryan Jr\BCBL {}\ \BBA {} C\BPBI V.~Vishveshwara\
  (\BEDS), \APACrefbtitle {Directions in General Relativity: Proceedings of the
  1993 International Symposium, Maryland: Papers in Honor of Charles Misner}
  {Directions in general relativity: Proceedings of the 1993 international
  symposium, maryland: Papers in honor of charles misner}\ (\BVOL~1,
  \BPG~10–20).
\newblock
\APACaddressPublisher{}{Cambridge University Press}.
\newblock
\begin{APACrefDOI} \doi{10.1017/CBO9780511628863.005} \end{APACrefDOI}
\PrintBackRefs{\CurrentBib}

\bibitem [\protect \citeauthoryear {%
Arnold%
}{%
Arnold%
}{%
{\protect \APACyear {1989}}%
}]{%
arnold1989}
\APACinsertmetastar {%
arnold1989}%
\begin{APACrefauthors}%
Arnold, V.%
\end{APACrefauthors}%
\unskip\
\newblock
\APACrefYear{1989}.
\newblock
\APACrefbtitle {Mathematical methods of classical mechanics} {Mathematical
  methods of classical mechanics}\ (\BVOL~60).
\newblock
\APACaddressPublisher{}{Springer, Berlin}.
\PrintBackRefs{\CurrentBib}

\bibitem [\protect \citeauthoryear {%
Arnowitt%
, Deser%
\BCBL {}\ \BBA {} Misner%
}{%
Arnowitt%
\ \protect \BOthers {.}}{%
{\protect \APACyear {1962}}%
}]{%
ADM}
\APACinsertmetastar {%
ADM}%
\begin{APACrefauthors}%
Arnowitt, R.%
, Deser, S.%
\BCBL {}\ \BBA {} Misner, C.%
\end{APACrefauthors}%
\unskip\
\newblock
\APACrefYearMonthDay{1962}{}{}.
\newblock
{\BBOQ}\APACrefatitle {The Dynamics of General Relativity, pp. 227-264} {The
  dynamics of general relativity, pp. 227-264}.{\BBCQ}
\newblock
\BIn{} \APACrefbtitle {in Gravitation: an introduction to current research, L.
  Witten, ed.} {in gravitation: an introduction to current research, l. witten,
  ed.}
\newblock
\APACaddressPublisher{}{Wiley, New York}.
\PrintBackRefs{\CurrentBib}

\bibitem [\protect \citeauthoryear {%
Arntzenius%
}{%
Arntzenius%
}{%
{\protect \APACyear {2012}}%
}]{%
Arntzenius_book}
\APACinsertmetastar {%
Arntzenius_book}%
\begin{APACrefauthors}%
Arntzenius, F.%
\end{APACrefauthors}%
\unskip\
\newblock
\APACrefYear{2012}.
\newblock
\APACrefbtitle {Space, Time, and Stuff} {Space, time, and stuff}.
\newblock
\APACaddressPublisher{}{Oxford University Press, Oxford}.
\PrintBackRefs{\CurrentBib}

\bibitem [\protect \citeauthoryear {%
Ashtekar%
}{%
Ashtekar%
}{%
{\protect \APACyear {1987}}%
}]{%
AshtekarNapoli}
\APACinsertmetastar {%
AshtekarNapoli}%
\begin{APACrefauthors}%
Ashtekar, A.%
\end{APACrefauthors}%
\unskip\
\newblock
\APACrefYear{1987}.
\newblock
\APACrefbtitle {{Asymptotic Quantization: Based on 1984 Naples Lectures}}
  {{Asymptotic Quantization: Based on 1984 Naples Lectures}}.
\newblock
\APACaddressPublisher{}{(Monographs and Textbooks in Physical Science Lecture
  Notes, Vol 2)}.
\PrintBackRefs{\CurrentBib}

\bibitem [\protect \citeauthoryear {%
Ashtekar%
, Bonga%
\BCBL {}\ \BBA {} Kesavan%
}{%
Ashtekar%
\ \protect \BOthers {.}}{%
{\protect \APACyear {2014}}%
}]{%
AshtekarBonga_I}
\APACinsertmetastar {%
AshtekarBonga_I}%
\begin{APACrefauthors}%
Ashtekar, A.%
, Bonga, B.%
\BCBL {}\ \BBA {} Kesavan, A.%
\end{APACrefauthors}%
\unskip\
\newblock
\APACrefYearMonthDay{2014}{Dec}{}.
\newblock
{\BBOQ}\APACrefatitle {Asymptotics with a positive cosmological constant: I.
  Basic framework} {Asymptotics with a positive cosmological constant: I. basic
  framework}.{\BBCQ}
\newblock
\APACjournalVolNumPages{Classical and Quantum Gravity}{32}{2}{025004}.
\newblock
\begin{APACrefURL} \url{http://dx.doi.org/10.1088/0264-9381/32/2/025004}
  \end{APACrefURL}
\newblock
\begin{APACrefDOI} \doi{10.1088/0264-9381/32/2/025004} \end{APACrefDOI}
\PrintBackRefs{\CurrentBib}

\bibitem [\protect \citeauthoryear {%
Ashtekar%
\ \BBA {} Hansen%
}{%
Ashtekar%
\ \BBA {} Hansen%
}{%
{\protect \APACyear {1978}}%
}]{%
AshtekarHansen}
\APACinsertmetastar {%
AshtekarHansen}%
\begin{APACrefauthors}%
Ashtekar, A.%
\BCBT {}\ \BBA {} Hansen, R\BPBI O.%
\end{APACrefauthors}%
\unskip\
\newblock
\APACrefYearMonthDay{1978}{}{}.
\newblock
{\BBOQ}\APACrefatitle {{A unified treatment of null and spatial infinity in
  general relativity. I. Universal structure, asymptotic symmetries, and
  conserved quantities at spatial infinity}} {{A unified treatment of null and
  spatial infinity in general relativity. I. Universal structure, asymptotic
  symmetries, and conserved quantities at spatial infinity}}.{\BBCQ}
\newblock
\APACjournalVolNumPages{Journal of Mathematical Physics}{19}{7}{1542-1566}.
\newblock
\begin{APACrefURL} \url{https://doi.org/10.1063/1.523863} \end{APACrefURL}
\newblock
\begin{APACrefDOI} \doi{10.1063/1.523863} \end{APACrefDOI}
\PrintBackRefs{\CurrentBib}

\bibitem [\protect \citeauthoryear {%
Ashtekar%
\ \BBA {} Magnon%
}{%
Ashtekar%
\ \BBA {} Magnon%
}{%
{\protect \APACyear {1984}}%
}]{%
Ashtekar_1984}
\APACinsertmetastar {%
Ashtekar_1984}%
\begin{APACrefauthors}%
Ashtekar, A.%
\BCBT {}\ \BBA {} Magnon, A.%
\end{APACrefauthors}%
\unskip\
\newblock
\APACrefYearMonthDay{1984}{jul}{}.
\newblock
{\BBOQ}\APACrefatitle {{Asymptotically anti-de Sitter space-times}}
  {{Asymptotically anti-de Sitter space-times}}.{\BBCQ}
\newblock
\APACjournalVolNumPages{Classical and Quantum Gravity}{1}{4}{L39--L44}.
\newblock
\begin{APACrefURL} \url{https://doi.org/10.1088/0264-9381/1/4/002}
  \end{APACrefURL}
\newblock
\begin{APACrefDOI} \doi{10.1088/0264-9381/1/4/002} \end{APACrefDOI}
\PrintBackRefs{\CurrentBib}

\bibitem [\protect \citeauthoryear {%
Ashtekar~A.%
}{%
Ashtekar~A.%
}{%
{\protect \APACyear {1981}}%
}]{%
AshtekarStreubel}
\APACinsertmetastar {%
AshtekarStreubel}%
\begin{APACrefauthors}%
Ashtekar~A., S\BPBI M.%
\end{APACrefauthors}%
\unskip\
\newblock
\APACrefYearMonthDay{1981}{}{}.
\newblock
{\BBOQ}\APACrefatitle {Symplectic geometry of radiative modes and conserved
  quantities at null infinity} {Symplectic geometry of radiative modes and
  conserved quantities at null infinity}.{\BBCQ}
\newblock
\APACjournalVolNumPages{Proc. R. Soc. Lond. A, 376}{}{}{}.
\newblock
\begin{APACrefDOI} \doi{10.1098/rspa.1981.0109} \end{APACrefDOI}
\PrintBackRefs{\CurrentBib}

\bibitem [\protect \citeauthoryear {%
Atiyah%
}{%
Atiyah%
}{%
{\protect \APACyear {1957}}%
}]{%
AtiyahLie}
\APACinsertmetastar {%
AtiyahLie}%
\begin{APACrefauthors}%
Atiyah, M.%
\end{APACrefauthors}%
\unskip\
\newblock
\APACrefYearMonthDay{1957}{}{}.
\newblock
{\BBOQ}\APACrefatitle {Complex analytic connections in fibre bundles} {Complex
  analytic connections in fibre bundles}.{\BBCQ}
\newblock
\APACjournalVolNumPages{Transactions of the American Mathematical
  Society}{}{}{}.
\PrintBackRefs{\CurrentBib}

\bibitem [\protect \citeauthoryear {%
Avery%
\ \BBA {} Schwab%
}{%
Avery%
\ \BBA {} Schwab%
}{%
{\protect \APACyear {2016}}%
}]{%
Avery_2016}
\APACinsertmetastar {%
Avery_2016}%
\begin{APACrefauthors}%
Avery, S\BPBI G.%
\BCBT {}\ \BBA {} Schwab, B\BPBI U\BPBI W.%
\end{APACrefauthors}%
\unskip\
\newblock
\APACrefYearMonthDay{2016}{Feb}{}.
\newblock
{\BBOQ}\APACrefatitle {{Noether’s second theorem and Ward identities for
  gauge symmetries}} {{Noether’s second theorem and Ward identities for gauge
  symmetries}}.{\BBCQ}
\newblock
\APACjournalVolNumPages{Journal of High Energy Physics}{2016}{2}{}.
\newblock
\begin{APACrefURL} \url{http://dx.doi.org/10.1007/JHEP02(2016)031}
  \end{APACrefURL}
\newblock
\begin{APACrefDOI} \doi{10.1007/jhep02(2016)031} \end{APACrefDOI}
\PrintBackRefs{\CurrentBib}

\bibitem [\protect \citeauthoryear {%
Baez%
\ \BBA {} Munian%
}{%
Baez%
\ \BBA {} Munian%
}{%
{\protect \APACyear {1994}}%
}]{%
Baez_book}
\APACinsertmetastar {%
Baez_book}%
\begin{APACrefauthors}%
Baez, J.%
\BCBT {}\ \BBA {} Munian, J.%
\end{APACrefauthors}%
\unskip\
\newblock
\APACrefYear{1994}.
\newblock
\APACrefbtitle {{Gauge Fields, Knots and Gravity}} {{Gauge Fields, Knots and
  Gravity}}.
\newblock
\APACaddressPublisher{}{World Scientific}.
\PrintBackRefs{\CurrentBib}

\bibitem [\protect \citeauthoryear {%
Barnich%
\ \BBA {} Brandt%
}{%
Barnich%
\ \BBA {} Brandt%
}{%
{\protect \APACyear {2002}}%
}]{%
BarnichBrandt2003}
\APACinsertmetastar {%
BarnichBrandt2003}%
\begin{APACrefauthors}%
Barnich, G.%
\BCBT {}\ \BBA {} Brandt, F.%
\end{APACrefauthors}%
\unskip\
\newblock
\APACrefYearMonthDay{2002}{}{}.
\newblock
{\BBOQ}\APACrefatitle {{Covariant theory of asymptotic symmetries, conservation
  laws and central charges}} {{Covariant theory of asymptotic symmetries,
  conservation laws and central charges}}.{\BBCQ}
\newblock
\APACjournalVolNumPages{Nucl. Phys.}{B633}{}{3-82}.
\newblock
\begin{APACrefDOI} \doi{10.1016/S0550-3213(02)00251-1} \end{APACrefDOI}
\PrintBackRefs{\CurrentBib}

\bibitem [\protect \citeauthoryear {%
J\BPBI W.~Barrett%
}{%
J\BPBI W.~Barrett%
}{%
{\protect \APACyear {1991}}%
}]{%
Barrett_hol}
\APACinsertmetastar {%
Barrett_hol}%
\begin{APACrefauthors}%
Barrett, J\BPBI W.%
\end{APACrefauthors}%
\unskip\
\newblock
\APACrefYearMonthDay{1991}{Sep}{01}.
\newblock
{\BBOQ}\APACrefatitle {Holonomy and path structures in general relativity and
  Yang-Mills theory} {Holonomy and path structures in general relativity and
  yang-mills theory}.{\BBCQ}
\newblock
\APACjournalVolNumPages{International Journal of Theoretical
  Physics}{30}{9}{1171--1215}.
\newblock
\begin{APACrefURL} \url{https://doi.org/10.1007/BF00671007} \end{APACrefURL}
\newblock
\begin{APACrefDOI} \doi{10.1007/BF00671007} \end{APACrefDOI}
\PrintBackRefs{\CurrentBib}

\bibitem [\protect \citeauthoryear {%
T.~Barrett%
}{%
T.~Barrett%
}{%
{\protect \APACyear {2018}}%
}]{%
Barrett2018_sym}
\APACinsertmetastar {%
Barrett2018_sym}%
\begin{APACrefauthors}%
Barrett, T.%
\end{APACrefauthors}%
\unskip\
\newblock
\APACrefYearMonthDay{2018}{}{}.
\newblock
{\BBOQ}\APACrefatitle {{What Do Symmetries Tell Us about Structure?}} {{What Do
  Symmetries Tell Us about Structure?}}{\BBCQ}
\newblock
\APACjournalVolNumPages{Philosophy of Science}{85}{4}{617-639}.
\newblock
\begin{APACrefURL} \url{https://doi.org/10.1086/699156} \end{APACrefURL}
\newblock
\begin{APACrefDOI} \doi{10.1086/699156} \end{APACrefDOI}
\PrintBackRefs{\CurrentBib}

\bibitem [\protect \citeauthoryear {%
Beig%
\ \BBA {} Chru\ifmmode~\acute{s}\else \'{s}\fi{}ciel%
}{%
Beig%
\ \BBA {} Chru\ifmmode~\acute{s}\else \'{s}\fi{}ciel%
}{%
{\protect \APACyear {2017}}%
}]{%
ChruscielBeig_shield}
\APACinsertmetastar {%
ChruscielBeig_shield}%
\begin{APACrefauthors}%
Beig, R.%
\BCBT {}\ \BBA {} Chru\ifmmode~\acute{s}\else \'{s}\fi{}ciel, P\BPBI T.%
\end{APACrefauthors}%
\unskip\
\newblock
\APACrefYearMonthDay{2017}{Mar}{}.
\newblock
{\BBOQ}\APACrefatitle {Shielding linearized gravity} {Shielding linearized
  gravity}.{\BBCQ}
\newblock
\APACjournalVolNumPages{{Phys. Rev. D}}{95}{}{064063}.
\newblock
\begin{APACrefURL} \url{https://link.aps.org/doi/10.1103/PhysRevD.95.064063}
  \end{APACrefURL}
\newblock
\begin{APACrefDOI} \doi{10.1103/PhysRevD.95.064063} \end{APACrefDOI}
\PrintBackRefs{\CurrentBib}

\bibitem [\protect \citeauthoryear {%
Beig%
\ \BBA {} OMurchadha%
}{%
Beig%
\ \BBA {} OMurchadha%
}{%
{\protect \APACyear {1987}}%
}]{%
BeigOMurchadha}
\APACinsertmetastar {%
BeigOMurchadha}%
\begin{APACrefauthors}%
Beig, R.%
\BCBT {}\ \BBA {} OMurchadha, N.%
\end{APACrefauthors}%
\unskip\
\newblock
\APACrefYearMonthDay{1987}{}{}.
\newblock
{\BBOQ}\APACrefatitle {{The Poincar\'e group as the symmetry group of canonical
  general relativity}} {{The Poincar\'e group as the symmetry group of
  canonical general relativity}}.{\BBCQ}
\newblock
\APACjournalVolNumPages{Annals of Physics}{174}{2}{463 - 498}.
\newblock
\begin{APACrefURL}
  \url{http://www.sciencedirect.com/science/article/pii/0003491687900376}
  \end{APACrefURL}
\newblock
\begin{APACrefDOI} \doi{https://doi.org/10.1016/0003-4916(87)90037-6}
  \end{APACrefDOI}
\PrintBackRefs{\CurrentBib}

\bibitem [\protect \citeauthoryear {%
Belot%
}{%
Belot%
}{%
{\protect \APACyear {1998}}%
}]{%
Belot1998}
\APACinsertmetastar {%
Belot1998}%
\begin{APACrefauthors}%
Belot, G.%
\end{APACrefauthors}%
\unskip\
\newblock
\APACrefYearMonthDay{1998}{}{}.
\newblock
{\BBOQ}\APACrefatitle {{Understanding Electromagnetism}} {{Understanding
  Electromagnetism}}.{\BBCQ}
\newblock
\APACjournalVolNumPages{The British Journal for the Philosophy of
  Science}{49}{4}{531-555}.
\newblock
\begin{APACrefURL} \url{https://doi.org/10.1093/bjps/49.4.531} \end{APACrefURL}
\newblock
\begin{APACrefDOI} \doi{10.1093/bjps/49.4.531} \end{APACrefDOI}
\PrintBackRefs{\CurrentBib}

\bibitem [\protect \citeauthoryear {%
Belot%
}{%
Belot%
}{%
{\protect \APACyear {2003}}%
}]{%
Belot2003}
\APACinsertmetastar {%
Belot2003}%
\begin{APACrefauthors}%
Belot, G.%
\end{APACrefauthors}%
\unskip\
\newblock
\APACrefYearMonthDay{2003}{}{}.
\newblock
{\BBOQ}\APACrefatitle {Symmetry and gauge freedom} {Symmetry and gauge
  freedom}.{\BBCQ}
\newblock
\APACjournalVolNumPages{Studies in History and Philosophy of Science Part B:
  Studies in History and Philosophy of Modern Physics}{34}{2}{189 - 225}.
\newblock
\begin{APACrefURL}
  \url{http://www.sciencedirect.com/science/article/pii/S1355219803000042}
  \end{APACrefURL}
\newblock
\begin{APACrefDOI} \doi{https://doi.org/10.1016/S1355-2198(03)00004-2}
  \end{APACrefDOI}
\PrintBackRefs{\CurrentBib}

\bibitem [\protect \citeauthoryear {%
Belot%
}{%
Belot%
}{%
{\protect \APACyear {2013}}%
}]{%
Belot_sym}
\APACinsertmetastar {%
Belot_sym}%
\begin{APACrefauthors}%
Belot, G.%
\end{APACrefauthors}%
\unskip\
\newblock
\APACrefYearMonthDay{2013}{}{}.
\newblock
{\BBOQ}\APACrefatitle {{Symmetry and Equivalence}} {{Symmetry and
  Equivalence}}.{\BBCQ}
\newblock
\BIn{} \APACrefbtitle {The Oxford Handbook of Philosophy of Physics.} {The
  oxford handbook of philosophy of physics.}
\newblock
\APACaddressPublisher{}{Oxford University Press. Edited by Batterman, R.}
\PrintBackRefs{\CurrentBib}

\bibitem [\protect \citeauthoryear {%
Belot%
}{%
Belot%
}{%
{\protect \APACyear {2018}}%
}]{%
Belot50}
\APACinsertmetastar {%
Belot50}%
\begin{APACrefauthors}%
Belot, G.%
\end{APACrefauthors}%
\unskip\
\newblock
\APACrefYearMonthDay{2018}{}{}.
\newblock
{\BBOQ}\APACrefatitle {Fifty Million Elvis Fans Can't be Wrong} {Fifty million
  elvis fans can't be wrong}.{\BBCQ}
\newblock
\APACjournalVolNumPages{Nous}{52}{4}{946-981}.
\newblock
\begin{APACrefURL}
  \url{https://onlinelibrary.wiley.com/doi/abs/10.1111/nous.12200}
  \end{APACrefURL}
\newblock
\begin{APACrefDOI} \doi{10.1111/nous.12200} \end{APACrefDOI}
\PrintBackRefs{\CurrentBib}

\bibitem [\protect \citeauthoryear {%
Belot%
\ \BBA {} Earman%
}{%
Belot%
\ \BBA {} Earman%
}{%
{\protect \APACyear {1999}}%
}]{%
belot_earman_1999}
\APACinsertmetastar {%
belot_earman_1999}%
\begin{APACrefauthors}%
Belot, G.%
\BCBT {}\ \BBA {} Earman, J.%
\end{APACrefauthors}%
\unskip\
\newblock
\APACrefYearMonthDay{1999}{}{}.
\newblock
{\BBOQ}\APACrefatitle {From metaphysics to physics} {From metaphysics to
  physics}.{\BBCQ}
\newblock
\BIn{} J.~Butterfield\ \BBA {} C.~Pagonis\ (\BEDS), \APACrefbtitle {From
  Physics to Philosophy} {From physics to philosophy}\ (\BPG~166–186).
\newblock
\APACaddressPublisher{}{Cambridge University Press}.
\newblock
\begin{APACrefDOI} \doi{10.1017/CBO9780511597947.009} \end{APACrefDOI}
\PrintBackRefs{\CurrentBib}

\bibitem [\protect \citeauthoryear {%
Belot%
\ \BBA {} Earman%
}{%
Belot%
\ \BBA {} Earman%
}{%
{\protect \APACyear {2001}}%
}]{%
belot_earman_2001}
\APACinsertmetastar {%
belot_earman_2001}%
\begin{APACrefauthors}%
Belot, G.%
\BCBT {}\ \BBA {} Earman, J.%
\end{APACrefauthors}%
\unskip\
\newblock
\APACrefYearMonthDay{2001}{}{}.
\newblock
{\BBOQ}\APACrefatitle {{Pre-Socratic quantum gravity}} {{Pre-Socratic quantum
  gravity}}.{\BBCQ}
\newblock
\BIn{} C.~Callender\ \BBA {} N.~Huggett\ (\BEDS), \APACrefbtitle {Physics Meets
  Philosophy at the Planck Scale: Contemporary Theories in Quantum Gravity}
  {Physics meets philosophy at the planck scale: Contemporary theories in
  quantum gravity}\ (\BPG~213-255).
\newblock
\APACaddressPublisher{}{Cambridge University Press}.
\newblock
\begin{APACrefDOI} \doi{10.1017/CBO9780511612909.011} \end{APACrefDOI}
\PrintBackRefs{\CurrentBib}

\bibitem [\protect \citeauthoryear {%
Belot%
\ \protect \BOthers {.}}{%
Belot%
\ \protect \BOthers {.}}{%
{\protect \APACyear {2009}}%
}]{%
Synopsis_gauge}
\APACinsertmetastar {%
Synopsis_gauge}%
\begin{APACrefauthors}%
Belot, G.%
, Earman, J.%
, Healey, R.%
, Maudlin, T.%
, Nounou, A.%
\BCBL {}\ \BBA {} Struyve, W.%
\end{APACrefauthors}%
\unskip\
\newblock
\APACrefYearMonthDay{2009}{June}{}.
\newblock
\APACrefbtitle {Synopsis and Discussion: Philosophy of Gauge Theory.} {Synopsis
  and discussion: Philosophy of gauge theory.}
\newblock
\begin{APACrefURL} \url{http://philsci-archive.pitt.edu/4728/} \end{APACrefURL}
\PrintBackRefs{\CurrentBib}

\bibitem [\protect \citeauthoryear {%
Berghofer%
\ \protect \BOthers {.}}{%
Berghofer%
\ \protect \BOthers {.}}{%
{\protect \APACyear {2021}}%
}]{%
Elements_gauge}
\APACinsertmetastar {%
Elements_gauge}%
\begin{APACrefauthors}%
Berghofer, P.%
, François, J.%
, Friederich, S.%
, Gomes, H.%
, Hetzroni, G.%
, Maas, A.%
\BCBL {}\ \BBA {} Sondenheimer, R.%
\end{APACrefauthors}%
\unskip\
\newblock
\APACrefYearMonthDay{2021}{}{}.
\newblock
\APACrefbtitle {Gauge Symmetries, Symmetry Breaking, and Gauge-Invariant
  Approaches.} {Gauge symmetries, symmetry breaking, and gauge-invariant
  approaches.}
\PrintBackRefs{\CurrentBib}

\bibitem [\protect \citeauthoryear {%
Bergmann%
\ \BBA {} Komar%
}{%
Bergmann%
\ \BBA {} Komar%
}{%
{\protect \APACyear {1960}}%
}]{%
Bergmann_Komar}
\APACinsertmetastar {%
Bergmann_Komar}%
\begin{APACrefauthors}%
Bergmann, P\BPBI G.%
\BCBT {}\ \BBA {} Komar, A\BPBI B.%
\end{APACrefauthors}%
\unskip\
\newblock
\APACrefYearMonthDay{1960}{Apr}{}.
\newblock
{\BBOQ}\APACrefatitle {{Poisson Brackets Between Locally Defined Observables in
  General Relativity}} {{Poisson Brackets Between Locally Defined Observables
  in General Relativity}}.{\BBCQ}
\newblock
\APACjournalVolNumPages{Physical Review Letters}{4}{}{432--433}.
\newblock
\begin{APACrefURL} \url{https://link.aps.org/doi/10.1103/PhysRevLett.4.432}
  \end{APACrefURL}
\newblock
\begin{APACrefDOI} \doi{10.1103/PhysRevLett.4.432} \end{APACrefDOI}
\PrintBackRefs{\CurrentBib}

\bibitem [\protect \citeauthoryear {%
Black%
}{%
Black%
}{%
{\protect \APACyear {2000}}%
}]{%
Black_quidditism}
\APACinsertmetastar {%
Black_quidditism}%
\begin{APACrefauthors}%
Black, R.%
\end{APACrefauthors}%
\unskip\
\newblock
\APACrefYearMonthDay{2000}{}{}.
\newblock
{\BBOQ}\APACrefatitle {{Against Quidditism}} {{Against Quidditism}}.{\BBCQ}
\newblock
\APACjournalVolNumPages{Australasian Journal of Philosophy}{78}{1}{87--104}.
\newblock
\begin{APACrefDOI} \doi{10.1080/00048400012349371} \end{APACrefDOI}
\PrintBackRefs{\CurrentBib}

\bibitem [\protect \citeauthoryear {%
Bleecker%
}{%
Bleecker%
}{%
{\protect \APACyear {1981}}%
}]{%
Bleecker}
\APACinsertmetastar {%
Bleecker}%
\begin{APACrefauthors}%
Bleecker, D.%
\end{APACrefauthors}%
\unskip\
\newblock
\APACrefYear{1981}.
\newblock
\APACrefbtitle {Gauge Theory and Variational Principles} {Gauge theory and
  variational principles}.
\newblock
\APACaddressPublisher{}{Dover Publications}.
\PrintBackRefs{\CurrentBib}

\bibitem [\protect \citeauthoryear {%
Blohmann%
, Fernandes%
\BCBL {}\ \BBA {} Weinstein%
}{%
Blohmann%
\ \protect \BOthers {.}}{%
{\protect \APACyear {2013}}%
}]{%
Blohmann_2013}
\APACinsertmetastar {%
Blohmann_2013}%
\begin{APACrefauthors}%
Blohmann, C.%
, Fernandes, M\BPBI C\BPBI B.%
\BCBL {}\ \BBA {} Weinstein, A.%
\end{APACrefauthors}%
\unskip\
\newblock
\APACrefYearMonthDay{2013}{Jan}{}.
\newblock
{\BBOQ}\APACrefatitle {Groupoid symmetry and constraints in general relativity}
  {Groupoid symmetry and constraints in general relativity}.{\BBCQ}
\newblock
\APACjournalVolNumPages{Communications in Contemporary
  Mathematics}{15}{01}{1250061}.
\newblock
\begin{APACrefURL} \url{http://dx.doi.org/10.1142/S0219199712500617}
  \end{APACrefURL}
\newblock
\begin{APACrefDOI} \doi{10.1142/s0219199712500617} \end{APACrefDOI}
\PrintBackRefs{\CurrentBib}

\bibitem [\protect \citeauthoryear {%
Bonora%
\ \BBA {} Cotta-Ramusino%
}{%
Bonora%
\ \BBA {} Cotta-Ramusino%
}{%
{\protect \APACyear {1983}}%
}]{%
Bonora1983}
\APACinsertmetastar {%
Bonora1983}%
\begin{APACrefauthors}%
Bonora, L.%
\BCBT {}\ \BBA {} Cotta-Ramusino, P.%
\end{APACrefauthors}%
\unskip\
\newblock
\APACrefYearMonthDay{1983}{dec}{}.
\newblock
{\BBOQ}\APACrefatitle {{Some remarks on BRS transformations, anomalies and the
  cohomology of the Lie algebra of the group of gauge transformations}} {{Some
  remarks on BRS transformations, anomalies and the cohomology of the Lie
  algebra of the group of gauge transformations}}.{\BBCQ}
\newblock
\APACjournalVolNumPages{Communications in Mathematical
  Physics}{87}{4}{589--603}.
\newblock
\begin{APACrefURL} \url{http://link.springer.com/10.1007/BF01208267}
  \end{APACrefURL}
\newblock
\begin{APACrefDOI} \doi{10.1007/BF01208267} \end{APACrefDOI}
\PrintBackRefs{\CurrentBib}

\bibitem [\protect \citeauthoryear {%
Brading%
\ \BBA {} Brown%
}{%
Brading%
\ \BBA {} Brown%
}{%
{\protect \APACyear {2000}}%
{\protect \APACexlab {{\protect \BCnt {1}}}}}]{%
BradingBrown_Noether}
\APACinsertmetastar {%
BradingBrown_Noether}%
\begin{APACrefauthors}%
Brading, K.%
\BCBT {}\ \BBA {} Brown, H\BPBI R.%
\end{APACrefauthors}%
\unskip\
\newblock
\APACrefYearMonthDay{2000{\protect \BCnt {1}}}{}{}.
\newblock
{\BBOQ}\APACrefatitle {{Noether's theorems and gauge symmetries}} {{Noether's
  theorems and gauge symmetries}}.{\BBCQ}
\newblock

\PrintBackRefs{\CurrentBib}

\bibitem [\protect \citeauthoryear {%
Brading%
\ \BBA {} Brown%
}{%
Brading%
\ \BBA {} Brown%
}{%
{\protect \APACyear {2000}}%
{\protect \APACexlab {{\protect \BCnt {2}}}}}]{%
bradingbrown2000n}
\APACinsertmetastar {%
bradingbrown2000n}%
\begin{APACrefauthors}%
Brading, K.%
\BCBT {}\ \BBA {} Brown, H\BPBI R.%
\end{APACrefauthors}%
\unskip\
\newblock
\APACrefYearMonthDay{2000{\protect \BCnt {2}}}{}{}.
\newblock
{\BBOQ}\APACrefatitle {Noether's theorems and gauge symmetries} {Noether's
  theorems and gauge symmetries}.{\BBCQ}
\newblock

\newblock
\APACrefnote{Unpublished manuscript,
  \url{https://arxiv.org/abs/hep-th/0009058}}
\PrintBackRefs{\CurrentBib}

\bibitem [\protect \citeauthoryear {%
Brading%
\ \BBA {} Brown%
}{%
Brading%
\ \BBA {} Brown%
}{%
{\protect \APACyear {2003}}%
}]{%
BradingBrown_chapter}
\APACinsertmetastar {%
BradingBrown_chapter}%
\begin{APACrefauthors}%
Brading, K.%
\BCBT {}\ \BBA {} Brown, H\BPBI R.%
\end{APACrefauthors}%
\unskip\
\newblock
\APACrefYearMonthDay{2003}{}{}.
\newblock
{\BBOQ}\APACrefatitle {Symmetries and Noether's theorems} {Symmetries and
  noether's theorems}.{\BBCQ}
\newblock
\BIn{} K.~Brading\ \BBA {} E.~Castellani\ (\BEDS), \APACrefbtitle {Symmetries
  in Physics: Philosophical Reflections} {Symmetries in physics: Philosophical
  reflections}\ (\BPG~89–109).
\newblock
\APACaddressPublisher{}{Cambridge University Press}.
\newblock
\begin{APACrefDOI} \doi{10.1017/CBO9780511535369.006} \end{APACrefDOI}
\PrintBackRefs{\CurrentBib}

\bibitem [\protect \citeauthoryear {%
Brading%
\ \BBA {} Brown%
}{%
Brading%
\ \BBA {} Brown%
}{%
{\protect \APACyear {2004}}%
}]{%
BradingBrown}
\APACinsertmetastar {%
BradingBrown}%
\begin{APACrefauthors}%
Brading, K.%
\BCBT {}\ \BBA {} Brown, H\BPBI R.%
\end{APACrefauthors}%
\unskip\
\newblock
\APACrefYearMonthDay{2004}{}{}.
\newblock
{\BBOQ}\APACrefatitle {Are Gauge Symmetry Transformations Observable?} {Are
  gauge symmetry transformations observable?}{\BBCQ}
\newblock
\APACjournalVolNumPages{The British Journal for the Philosophy of
  Science}{55}{4}{645--665}.
\newblock
\begin{APACrefURL} \url{http://www.jstor.org/stable/3541620} \end{APACrefURL}
\PrintBackRefs{\CurrentBib}

\bibitem [\protect \citeauthoryear {%
Bradley%
\ \BBA {} Weatherall%
}{%
Bradley%
\ \BBA {} Weatherall%
}{%
{\protect \APACyear {2021}}%
}]{%
BradleyWeatherall_hole}
\APACinsertmetastar {%
BradleyWeatherall_hole}%
\begin{APACrefauthors}%
Bradley, C.%
\BCBT {}\ \BBA {} Weatherall, J\BPBI O.%
\end{APACrefauthors}%
\unskip\
\newblock
\APACrefYearMonthDay{2021}{}{}.
\newblock
\APACrefbtitle {Mathematical Responses to the Hole Argument: Then and Now.}
  {Mathematical responses to the hole argument: Then and now.}
\PrintBackRefs{\CurrentBib}

\bibitem [\protect \citeauthoryear {%
Brighouse%
}{%
Brighouse%
}{%
{\protect \APACyear {1994}}%
}]{%
Brighouse_hole}
\APACinsertmetastar {%
Brighouse_hole}%
\begin{APACrefauthors}%
Brighouse, C.%
\end{APACrefauthors}%
\unskip\
\newblock
\APACrefYearMonthDay{1994}{}{}.
\newblock
{\BBOQ}\APACrefatitle {Spacetime and Holes} {Spacetime and holes}.{\BBCQ}
\newblock
\APACjournalVolNumPages{PSA: Proceedings of the Biennial Meeting of the
  Philosophy of Science Association}{1994}{}{117--125}.
\newblock
\begin{APACrefURL} \url{http://www.jstor.org/stable/193017} \end{APACrefURL}
\PrintBackRefs{\CurrentBib}

\bibitem [\protect \citeauthoryear {%
Brown%
}{%
Brown%
}{%
{\protect \APACyear {1999}}%
}]{%
brown1998a}
\APACinsertmetastar {%
brown1998a}%
\begin{APACrefauthors}%
Brown, H.%
\end{APACrefauthors}%
\unskip\
\newblock
\APACrefYearMonthDay{1999}{}{}.
\newblock
{\BBOQ}\APACrefatitle {{Aspects of Objectivity in Quantum Mechanics}} {{Aspects
  of Objectivity in Quantum Mechanics}}.{\BBCQ}
\newblock
\BIn{} J.~Butterfield\ \BBA {} C.~Pagonis\ (\BEDS), \APACrefbtitle {{From
  Physics to Philosophy}.} {{From Physics to Philosophy}.}
\newblock
\APACaddressPublisher{}{Cambridge: Cambridge University Press}.
\newblock
\APACrefnote{\url{http://philsci-archive.pitt.edu/223/}}
\PrintBackRefs{\CurrentBib}

\bibitem [\protect \citeauthoryear {%
Brown%
}{%
Brown%
}{%
{\protect \APACyear {2006}}%
}]{%
Brown_book}
\APACinsertmetastar {%
Brown_book}%
\begin{APACrefauthors}%
Brown, H.%
\end{APACrefauthors}%
\unskip\
\newblock
\APACrefYear{2006}.
\newblock
\APACrefbtitle {{Physical Relativity: Space-Time Structure from a Dynamical
  Perspective}} {{Physical Relativity: Space-Time Structure from a Dynamical
  Perspective}}.
\newblock
\APACaddressPublisher{}{Oxford University Press}.
\PrintBackRefs{\CurrentBib}

\bibitem [\protect \citeauthoryear {%
Buividovich%
\ \BBA {} Polikarpov%
}{%
Buividovich%
\ \BBA {} Polikarpov%
}{%
{\protect \APACyear {2008}}%
}]{%
Buividovich_2008}
\APACinsertmetastar {%
Buividovich_2008}%
\begin{APACrefauthors}%
Buividovich, P.%
\BCBT {}\ \BBA {} Polikarpov, M.%
\end{APACrefauthors}%
\unskip\
\newblock
\APACrefYearMonthDay{2008}{Dec}{}.
\newblock
{\BBOQ}\APACrefatitle {Entanglement entropy in gauge theories and the
  holographic principle for electric strings} {Entanglement entropy in gauge
  theories and the holographic principle for electric strings}.{\BBCQ}
\newblock
\APACjournalVolNumPages{Physics Letters B}{670}{2}{141–145}.
\newblock
\begin{APACrefURL} \url{http://dx.doi.org/10.1016/j.physletb.2008.10.032}
  \end{APACrefURL}
\newblock
\begin{APACrefDOI} \doi{10.1016/j.physletb.2008.10.032} \end{APACrefDOI}
\PrintBackRefs{\CurrentBib}

\bibitem [\protect \citeauthoryear {%
Butterfield%
}{%
Butterfield%
}{%
{\protect \APACyear {1989}}%
}]{%
Butterfield_hole}
\APACinsertmetastar {%
Butterfield_hole}%
\begin{APACrefauthors}%
Butterfield, J.%
\end{APACrefauthors}%
\unskip\
\newblock
\APACrefYearMonthDay{1989}{03}{}.
\newblock
{\BBOQ}\APACrefatitle {{The Hole Truth}} {{The Hole Truth}}.{\BBCQ}
\newblock
\APACjournalVolNumPages{The British Journal for the Philosophy of
  Science}{40}{1}{1-28}.
\newblock
\begin{APACrefURL} \url{https://doi.org/10.1093/bjps/40.1.1} \end{APACrefURL}
\newblock
\begin{APACrefDOI} \doi{10.1093/bjps/40.1.1} \end{APACrefDOI}
\PrintBackRefs{\CurrentBib}

\bibitem [\protect \citeauthoryear {%
Butterfield%
}{%
Butterfield%
}{%
{\protect \APACyear {2006}}%
{\protect \APACexlab {{\protect \BCnt {1}}}}}]{%
ButterfieldPoint}
\APACinsertmetastar {%
ButterfieldPoint}%
\begin{APACrefauthors}%
Butterfield, J.%
\end{APACrefauthors}%
\unskip\
\newblock
\APACrefYearMonthDay{2006{\protect \BCnt {1}}}{}{}.
\newblock
{\BBOQ}\APACrefatitle {Against Pointillisme About Mechanics} {Against
  pointillisme about mechanics}.{\BBCQ}
\newblock
\APACjournalVolNumPages{British Journal for the Philosophy of
  Science}{57}{4}{709--753}.
\newblock
\begin{APACrefDOI} \doi{10.1093/bjps/axl026} \end{APACrefDOI}
\PrintBackRefs{\CurrentBib}

\bibitem [\protect \citeauthoryear {%
Butterfield%
}{%
Butterfield%
}{%
{\protect \APACyear {2006}}%
{\protect \APACexlab {{\protect \BCnt {2}}}}}]{%
Butterfield2006}
\APACinsertmetastar {%
Butterfield2006}%
\begin{APACrefauthors}%
Butterfield, J.%
\end{APACrefauthors}%
\unskip\
\newblock
\APACrefYearMonthDay{2006{\protect \BCnt {2}}}{}{}.
\newblock
{\BBOQ}\APACrefatitle {On Symmetry and Conserved Quantities in Classical
  Mechanics} {On symmetry and conserved quantities in classical
  mechanics}.{\BBCQ}
\newblock
\BIn{} W.~Demopoulos\ \BBA {} I.~Pitowsky\ (\BEDS), \APACrefbtitle {Physical
  Theory and its Interpretation: Essays in Honor of Jeffrey Bub} {Physical
  theory and its interpretation: Essays in honor of jeffrey bub}\ (\BPGS\
  43--100).
\newblock
\APACaddressPublisher{Dordrecht}{Springer Netherlands}.
\newblock
\begin{APACrefURL} \url{https://doi.org/10.1007/1-4020-4876-9_3}
  \end{APACrefURL}
\newblock
\begin{APACrefDOI} \doi{10.1007/1-4020-4876-9_3} \end{APACrefDOI}
\PrintBackRefs{\CurrentBib}

\bibitem [\protect \citeauthoryear {%
Butterfield%
}{%
Butterfield%
}{%
{\protect \APACyear {2007}}%
}]{%
Butterfield_symp}
\APACinsertmetastar {%
Butterfield_symp}%
\begin{APACrefauthors}%
Butterfield, J.%
\end{APACrefauthors}%
\unskip\
\newblock
\APACrefYearMonthDay{2007}{}{}.
\newblock
{\BBOQ}\APACrefatitle {ON SYMPLECTIC REDUCTION IN CLASSICAL MECHANICS} {On
  symplectic reduction in classical mechanics}.{\BBCQ}
\newblock
\BIn{} J.~Butterfield\ \BBA {} J.~Earman\ (\BEDS), \APACrefbtitle {Philosophy
  of Physics} {Philosophy of physics}\ (\BPG~1 - 131).
\newblock
\APACaddressPublisher{Amsterdam}{North-Holland}.
\newblock
\begin{APACrefURL}
  \url{http://www.sciencedirect.com/science/article/pii/B978044451560550004X}
  \end{APACrefURL}
\newblock
\begin{APACrefDOI} \doi{https://doi.org/10.1016/B978-044451560-5/50004-X}
  \end{APACrefDOI}
\PrintBackRefs{\CurrentBib}

\bibitem [\protect \citeauthoryear {%
Butterfield%
\ \BBA {} Gomes%
}{%
Butterfield%
\ \BBA {} Gomes%
}{%
{\protect \APACyear {2020}}%
}]{%
ButterfieldGomes_1}
\APACinsertmetastar {%
ButterfieldGomes_1}%
\begin{APACrefauthors}%
Butterfield, J.%
\BCBT {}\ \BBA {} Gomes, H.%
\end{APACrefauthors}%
\unskip\
\newblock
\APACrefYearMonthDay{2020}{}{}.
\newblock
{\BBOQ}\APACrefatitle {Functionalism as a species of reduction} {Functionalism
  as a species of reduction}.{\BBCQ}
\newblock
\BIn{} \APACrefbtitle {{Current Debates in Philosophy of Science: In Honor of
  Roberto Torretti}.} {{Current Debates in Philosophy of Science: In Honor of
  Roberto Torretti}.}
\newblock
\APACaddressPublisher{}{Springer, New York.}
\PrintBackRefs{\CurrentBib}

\bibitem [\protect \citeauthoryear {%
Carlotto%
\ \BBA {} Schoen%
}{%
Carlotto%
\ \BBA {} Schoen%
}{%
{\protect \APACyear {2016}}%
}]{%
CarlottoSchoen2016}
\APACinsertmetastar {%
CarlottoSchoen2016}%
\begin{APACrefauthors}%
Carlotto, A.%
\BCBT {}\ \BBA {} Schoen, R.%
\end{APACrefauthors}%
\unskip\
\newblock
\APACrefYearMonthDay{2016}{}{}.
\newblock
{\BBOQ}\APACrefatitle {{Localizing solutions of the Einstein constraint
  equations}} {{Localizing solutions of the Einstein constraint
  equations}}.{\BBCQ}
\newblock
\APACjournalVolNumPages{Inventiones Mathematicae, volume 205}{}{}{}.
\PrintBackRefs{\CurrentBib}

\bibitem [\protect \citeauthoryear {%
Carrozza%
\ \BBA {} Hoehn%
}{%
Carrozza%
\ \BBA {} Hoehn%
}{%
{\protect \APACyear {2021}}%
}]{%
carrozza2021}
\APACinsertmetastar {%
carrozza2021}%
\begin{APACrefauthors}%
Carrozza, S.%
\BCBT {}\ \BBA {} Hoehn, P\BPBI A.%
\end{APACrefauthors}%
\unskip\
\newblock
\APACrefYearMonthDay{2021}{}{}.
\newblock
\APACrefbtitle {Edge modes as reference frames and boundary actions from
  post-selection.} {Edge modes as reference frames and boundary actions from
  post-selection.}
\PrintBackRefs{\CurrentBib}

\bibitem [\protect \citeauthoryear {%
Chasova%
}{%
Chasova%
}{%
{\protect \APACyear {2019}}%
}]{%
Chasova}
\APACinsertmetastar {%
Chasova}%
\begin{APACrefauthors}%
Chasova, V.%
\end{APACrefauthors}%
\unskip\
\newblock
\APACrefYearMonthDay{2019}{}{}.
\newblock
{\BBOQ}\APACrefatitle {Direct Empirical Status of Theoretical Symmetries in
  Physics} {Direct empirical status of theoretical symmetries in
  physics}.{\BBCQ}
\newblock
\APACjournalVolNumPages{PhD thesis}{}{}{}.
\PrintBackRefs{\CurrentBib}

\bibitem [\protect \citeauthoryear {%
Choquet-Bruhat%
}{%
Choquet-Bruhat%
}{%
{\protect \APACyear {2008}}%
}]{%
choquet2008}
\APACinsertmetastar {%
choquet2008}%
\begin{APACrefauthors}%
Choquet-Bruhat, Y.%
\end{APACrefauthors}%
\unskip\
\newblock
\APACrefYear{2008}.
\newblock
\APACrefbtitle {{General Relativity and the Einstein Equations}} {{General
  Relativity and the Einstein Equations}}.
\newblock
\APACaddressPublisher{}{Oxford University Press, Oxford}.
\newblock
\begin{APACrefURL} \url{https://books.google.co.uk/books?id=UjHbm5rfpi8C}
  \end{APACrefURL}
\PrintBackRefs{\CurrentBib}

\bibitem [\protect \citeauthoryear {%
Ciambelli%
\ \BBA {} Leigh%
}{%
Ciambelli%
\ \BBA {} Leigh%
}{%
{\protect \APACyear {2021}}%
}]{%
Ciambelli}
\APACinsertmetastar {%
Ciambelli}%
\begin{APACrefauthors}%
Ciambelli, L.%
\BCBT {}\ \BBA {} Leigh, R\BPBI G.%
\end{APACrefauthors}%
\unskip\
\newblock
\APACrefYearMonthDay{2021}{}{}.
\newblock
\APACrefbtitle {Lie Algebroids and the Geometry of Off-shell BRST.} {Lie
  algebroids and the geometry of off-shell brst.}
\PrintBackRefs{\CurrentBib}

\bibitem [\protect \citeauthoryear {%
Curiel%
}{%
Curiel%
}{%
{\protect \APACyear {{\protect \bibnodate {}}}}%
}]{%
Curiel_priv}
\APACinsertmetastar {%
Curiel_priv}%
\begin{APACrefauthors}%
Curiel, E.%
\end{APACrefauthors}%
\unskip\
\newblock
\APACrefYearMonthDay{{\protect \bibnodate {}}}{}{}.
\newblock
\APACrefbtitle {Private communication, 2021.} {Private communication, 2021.}
\newblock
\APACrefnote{2021}
\PrintBackRefs{\CurrentBib}

\bibitem [\protect \citeauthoryear {%
Curiel%
}{%
Curiel%
}{%
{\protect \APACyear {2018}}%
}]{%
Curiel2018}
\APACinsertmetastar {%
Curiel2018}%
\begin{APACrefauthors}%
Curiel, E.%
\end{APACrefauthors}%
\unskip\
\newblock
\APACrefYearMonthDay{2018}{}{}.
\newblock
{\BBOQ}\APACrefatitle {{On the Existence of Spacetime Structure}} {{On the
  Existence of Spacetime Structure}}.{\BBCQ}
\newblock
\APACjournalVolNumPages{The British Journal for the Philosophy of
  Science}{69}{2}{447-483}.
\newblock
\begin{APACrefURL} \url{https://doi.org/10.1093/bjps/axw014} \end{APACrefURL}
\newblock
\begin{APACrefDOI} \doi{10.1093/bjps/axw014} \end{APACrefDOI}
\PrintBackRefs{\CurrentBib}

\bibitem [\protect \citeauthoryear {%
Dasgupta%
}{%
Dasgupta%
}{%
{\protect \APACyear {2011}}%
}]{%
Dasgupta_bare}
\APACinsertmetastar {%
Dasgupta_bare}%
\begin{APACrefauthors}%
Dasgupta, S.%
\end{APACrefauthors}%
\unskip\
\newblock
\APACrefYearMonthDay{2011}{}{}.
\newblock
{\BBOQ}\APACrefatitle {THE BARE NECESSITIES*} {The bare necessities*}.{\BBCQ}
\newblock
\APACjournalVolNumPages{Philosophical Perspectives}{25}{1}{115-160}.
\newblock
\begin{APACrefURL}
  \url{https://onlinelibrary.wiley.com/doi/abs/10.1111/j.1520-8583.2011.00210.x}
  \end{APACrefURL}
\newblock
\begin{APACrefDOI} \doi{https://doi.org/10.1111/j.1520-8583.2011.00210.x}
  \end{APACrefDOI}
\PrintBackRefs{\CurrentBib}

\bibitem [\protect \citeauthoryear {%
de Almeida%
\ \BBA {} Saa%
}{%
de Almeida%
\ \BBA {} Saa%
}{%
{\protect \APACyear {2006}}%
}]{%
Saa_electron}
\APACinsertmetastar {%
Saa_electron}%
\begin{APACrefauthors}%
de Almeida, C.%
\BCBT {}\ \BBA {} Saa, A.%
\end{APACrefauthors}%
\unskip\
\newblock
\APACrefYearMonthDay{2006}{}{}.
\newblock
{\BBOQ}\APACrefatitle {The radiation of a uniformly accelerated charge is
  beyond the horizon: A simple derivation} {The radiation of a uniformly
  accelerated charge is beyond the horizon: A simple derivation}.{\BBCQ}
\newblock
\APACjournalVolNumPages{American Journal of Physics}{74}{2}{154-158}.
\newblock
\begin{APACrefURL} \url{https://doi.org/10.1119/1.2162548} \end{APACrefURL}
\newblock
\begin{APACrefDOI} \doi{10.1119/1.2162548} \end{APACrefDOI}
\PrintBackRefs{\CurrentBib}

\bibitem [\protect \citeauthoryear {%
de Haro%
}{%
de Haro%
}{%
{\protect \APACyear {2021}}%
}]{%
Seb_Noether}
\APACinsertmetastar {%
Seb_Noether}%
\begin{APACrefauthors}%
de Haro, S.%
\end{APACrefauthors}%
\unskip\
\newblock
\APACrefYearMonthDay{2021}{}{}.
\newblock
{\BBOQ}\APACrefatitle {{Noether's Theorems and Energy in General Relativity}}
  {{Noether's Theorems and Energy in General Relativity}}.{\BBCQ}
\newblock
\APACjournalVolNumPages{Comissioned from Cambridge University Press, for
  special edition celebrating 100 years of Noether's theorems}{}{}{}.
\PrintBackRefs{\CurrentBib}

\bibitem [\protect \citeauthoryear {%
de León%
\ \BBA {} Zajac%
}{%
de León%
\ \BBA {} Zajac%
}{%
{\protect \APACyear {2020}}%
}]{%
LeonZajac}
\APACinsertmetastar {%
LeonZajac}%
\begin{APACrefauthors}%
de León, M.%
\BCBT {}\ \BBA {} Zajac, M.%
\end{APACrefauthors}%
\unskip\
\newblock
\APACrefYearMonthDay{2020}{Jun}{}.
\newblock
{\BBOQ}\APACrefatitle {Hamilton–Jacobi theory for gauge field theories}
  {Hamilton–jacobi theory for gauge field theories}.{\BBCQ}
\newblock
\APACjournalVolNumPages{Journal of Geometry and Physics}{152}{}{103636}.
\newblock
\begin{APACrefURL} \url{http://dx.doi.org/10.1016/j.geomphys.2020.103636}
  \end{APACrefURL}
\newblock
\begin{APACrefDOI} \doi{10.1016/j.geomphys.2020.103636} \end{APACrefDOI}
\PrintBackRefs{\CurrentBib}

\bibitem [\protect \citeauthoryear {%
Dewar%
}{%
Dewar%
}{%
{\protect \APACyear {2017}}%
}]{%
Dewar2017}
\APACinsertmetastar {%
Dewar2017}%
\begin{APACrefauthors}%
Dewar, N.%
\end{APACrefauthors}%
\unskip\
\newblock
\APACrefYearMonthDay{2017}{09}{}.
\newblock
{\BBOQ}\APACrefatitle {{Sophistication about Symmetries}} {{Sophistication
  about Symmetries}}.{\BBCQ}
\newblock
\APACjournalVolNumPages{The British Journal for the Philosophy of
  Science}{70}{2}{485-521}.
\newblock
\begin{APACrefURL} \url{https://doi.org/10.1093/bjps/axx021} \end{APACrefURL}
\newblock
\begin{APACrefDOI} \doi{10.1093/bjps/axx021} \end{APACrefDOI}
\PrintBackRefs{\CurrentBib}

\bibitem [\protect \citeauthoryear {%
Diez%
\ \BBA {} Rudolph%
}{%
Diez%
\ \BBA {} Rudolph%
}{%
{\protect \APACyear {2019}}%
}]{%
Slice_diez}
\APACinsertmetastar {%
Slice_diez}%
\begin{APACrefauthors}%
Diez, T.%
\BCBT {}\ \BBA {} Rudolph, G.%
\end{APACrefauthors}%
\unskip\
\newblock
\APACrefYearMonthDay{2019}{Aug}{}.
\newblock
{\BBOQ}\APACrefatitle {Slice theorem and orbit type stratification in infinite
  dimensions} {Slice theorem and orbit type stratification in infinite
  dimensions}.{\BBCQ}
\newblock
\APACjournalVolNumPages{Differential Geometry and its
  Applications}{65}{}{176–211}.
\newblock
\begin{APACrefURL} \url{http://dx.doi.org/10.1016/j.difgeo.2019.03.005}
  \end{APACrefURL}
\newblock
\begin{APACrefDOI} \doi{10.1016/j.difgeo.2019.03.005} \end{APACrefDOI}
\PrintBackRefs{\CurrentBib}

\bibitem [\protect \citeauthoryear {%
Dirac%
}{%
Dirac%
}{%
{\protect \APACyear {1930}}%
}]{%
Dirac_QM}
\APACinsertmetastar {%
Dirac_QM}%
\begin{APACrefauthors}%
Dirac, P\BPBI A\BPBI M.%
\end{APACrefauthors}%
\unskip\
\newblock
\APACrefYear{1930}.
\newblock
\APACrefbtitle {{The Principles of Quantum Mechanics.}} {{The Principles of
  Quantum Mechanics.}}
\newblock
\APACaddressPublisher{}{Clarendon Press}.
\PrintBackRefs{\CurrentBib}

\bibitem [\protect \citeauthoryear {%
Donnelly%
\ \BBA {} Freidel%
}{%
Donnelly%
\ \BBA {} Freidel%
}{%
{\protect \APACyear {2016}}%
}]{%
DonnellyFreidel}
\APACinsertmetastar {%
DonnellyFreidel}%
\begin{APACrefauthors}%
Donnelly, W.%
\BCBT {}\ \BBA {} Freidel, L.%
\end{APACrefauthors}%
\unskip\
\newblock
\APACrefYearMonthDay{2016}{}{}.
\newblock
{\BBOQ}\APACrefatitle {{Local subsystems in gauge theory and gravity}} {{Local
  subsystems in gauge theory and gravity}}.{\BBCQ}
\newblock
\APACjournalVolNumPages{JHEP}{09}{}{102}.
\newblock
\begin{APACrefDOI} \doi{10.1007/JHEP09(2016)102} \end{APACrefDOI}
\PrintBackRefs{\CurrentBib}

\bibitem [\protect \citeauthoryear {%
Donnelly%
\ \BBA {} Giddings%
}{%
Donnelly%
\ \BBA {} Giddings%
}{%
{\protect \APACyear {2016}}%
}]{%
Donnelly_Giddings}
\APACinsertmetastar {%
Donnelly_Giddings}%
\begin{APACrefauthors}%
Donnelly, W.%
\BCBT {}\ \BBA {} Giddings, S\BPBI B.%
\end{APACrefauthors}%
\unskip\
\newblock
\APACrefYearMonthDay{2016}{Jan}{}.
\newblock
{\BBOQ}\APACrefatitle {Diffeomorphism-invariant observables and their nonlocal
  algebra} {Diffeomorphism-invariant observables and their nonlocal
  algebra}.{\BBCQ}
\newblock
\APACjournalVolNumPages{Physical Review D}{93}{2}{}.
\newblock
\begin{APACrefURL} \url{http://dx.doi.org/10.1103/PhysRevD.93.024030}
  \end{APACrefURL}
\newblock
\begin{APACrefDOI} \doi{10.1103/physrevd.93.024030} \end{APACrefDOI}
\PrintBackRefs{\CurrentBib}

\bibitem [\protect \citeauthoryear {%
Dougherty%
}{%
Dougherty%
}{%
{\protect \APACyear {2019}}%
}]{%
Dougherty_CP}
\APACinsertmetastar {%
Dougherty_CP}%
\begin{APACrefauthors}%
Dougherty, J.%
\end{APACrefauthors}%
\unskip\
\newblock
\APACrefYearMonthDay{2019}{}{}.
\newblock
{\BBOQ}\APACrefatitle {Large Gauge Transformations and the Strong CP Problem}
  {Large gauge transformations and the strong cp problem}.{\BBCQ}
\newblock
\APACjournalVolNumPages{Studies in the History and Philosophy of Modern
  Physics, vol 37, 2020}{}{}{}.
\PrintBackRefs{\CurrentBib}

\bibitem [\protect \citeauthoryear {%
Dougherty%
}{%
Dougherty%
}{%
{\protect \APACyear {2020}}%
}]{%
DoughertyAB}
\APACinsertmetastar {%
DoughertyAB}%
\begin{APACrefauthors}%
Dougherty, J.%
\end{APACrefauthors}%
\unskip\
\newblock
\APACrefYearMonthDay{2020}{}{}.
\newblock
{\BBOQ}\APACrefatitle {{The Non-Ideal Theory of the Aharonov--Bohm Effect}}
  {{The Non-Ideal Theory of the Aharonov--Bohm Effect}}.{\BBCQ}
\newblock
\APACjournalVolNumPages{Synthese}{}{}{}.
\PrintBackRefs{\CurrentBib}

\bibitem [\protect \citeauthoryear {%
Dowker%
}{%
Dowker%
}{%
{\protect \APACyear {1967}}%
}]{%
Dowker1967}
\APACinsertmetastar {%
Dowker1967}%
\begin{APACrefauthors}%
Dowker, J\BPBI S.%
\end{APACrefauthors}%
\unskip\
\newblock
\APACrefYearMonthDay{1967}{}{}.
\newblock
{\BBOQ}\APACrefatitle {{A gravitational Aharonov-Bohm effect}} {{A
  gravitational Aharonov-Bohm effect}}.{\BBCQ}
\newblock
\APACjournalVolNumPages{Il Nuovo Cimento B (1965-1970)}{52}{}{129-135}.
\PrintBackRefs{\CurrentBib}

\bibitem [\protect \citeauthoryear {%
Earman%
}{%
Earman%
}{%
{\protect \APACyear {1986}}%
}]{%
Earman_det}
\APACinsertmetastar {%
Earman_det}%
\begin{APACrefauthors}%
Earman, J.%
\end{APACrefauthors}%
\unskip\
\newblock
\APACrefYear{1986}.
\newblock
\APACrefbtitle {{A Primer on Determinism}} {{A Primer on Determinism}}.
\newblock
\APACaddressPublisher{}{Springer, Netherlands.}
\PrintBackRefs{\CurrentBib}

\bibitem [\protect \citeauthoryear {%
Earman%
}{%
Earman%
}{%
{\protect \APACyear {1987}}%
}]{%
Earman_local}
\APACinsertmetastar {%
Earman_local}%
\begin{APACrefauthors}%
Earman, J.%
\end{APACrefauthors}%
\unskip\
\newblock
\APACrefYearMonthDay{1987}{September}{}.
\newblock
{\BBOQ}\APACrefatitle {Locality, Nonlocality and Action at a Distance: A
  Skeptical Review of Some Philosophical Dogmas} {Locality, nonlocality and
  action at a distance: A skeptical review of some philosophical
  dogmas}.{\BBCQ}
\newblock
\BIn{} R.~Kargon, P.~Achinstein\BCBL {}\ \BBA {} W\BPBI T.~Kelvin\ (\BEDS),
  \APACrefbtitle {Kelvin's Baltimore lectures and modern theoretical physics :
  historical and philosophical perspectives} {Kelvin's baltimore lectures and
  modern theoretical physics : historical and philosophical perspectives}\
  (\BPGS\ 449 -- 490).
\newblock
\APACaddressPublisher{Cambridge}{MIT Press}.
\newblock
\begin{APACrefURL} \url{http://d-scholarship.pitt.edu/12972/} \end{APACrefURL}
\PrintBackRefs{\CurrentBib}

\bibitem [\protect \citeauthoryear {%
Earman%
}{%
Earman%
}{%
{\protect \APACyear {1989}}%
}]{%
Earman_world}
\APACinsertmetastar {%
Earman_world}%
\begin{APACrefauthors}%
Earman, J.%
\end{APACrefauthors}%
\unskip\
\newblock
\APACrefYear{1989}.
\newblock
\APACrefbtitle {World Enough and Spacetime} {World enough and spacetime}.
\newblock
\APACaddressPublisher{}{MIT press}.
\PrintBackRefs{\CurrentBib}

\bibitem [\protect \citeauthoryear {%
Earman%
}{%
Earman%
}{%
{\protect \APACyear {2002}}%
}]{%
Earman_gmatters}
\APACinsertmetastar {%
Earman_gmatters}%
\begin{APACrefauthors}%
Earman, J.%
\end{APACrefauthors}%
\unskip\
\newblock
\APACrefYearMonthDay{2002}{}{}.
\newblock
{\BBOQ}\APACrefatitle {Gauge Matters} {Gauge matters}.{\BBCQ}
\newblock
\APACjournalVolNumPages{Philosophy of Science}{69}{S3}{S209-S220}.
\newblock
\begin{APACrefURL} \url{https://doi.org/10.1086/341847} \end{APACrefURL}
\newblock
\begin{APACrefDOI} \doi{10.1086/341847} \end{APACrefDOI}
\PrintBackRefs{\CurrentBib}

\bibitem [\protect \citeauthoryear {%
Earman%
}{%
Earman%
}{%
{\protect \APACyear {2003}}%
{\protect \APACexlab {{\protect \BCnt {1}}}}}]{%
earman_ode}
\APACinsertmetastar {%
earman_ode}%
\begin{APACrefauthors}%
Earman, J.%
\end{APACrefauthors}%
\unskip\
\newblock
\APACrefYearMonthDay{2003{\protect \BCnt {1}}}{}{}.
\newblock
{\BBOQ}\APACrefatitle {Tracking down gauge: an ode to the constrained
  Hamiltonian formalism} {Tracking down gauge: an ode to the constrained
  hamiltonian formalism}.{\BBCQ}
\newblock
\BIn{} K.~Brading\ \BBA {} E.~Castellani\ (\BEDS), \APACrefbtitle {Symmetries
  in Physics: Philosophical Reflections} {Symmetries in physics: Philosophical
  reflections}\ (\BPG~140–162).
\newblock
\APACaddressPublisher{}{Cambridge University Press}.
\newblock
\begin{APACrefDOI} \doi{10.1017/CBO9780511535369.009} \end{APACrefDOI}
\PrintBackRefs{\CurrentBib}

\bibitem [\protect \citeauthoryear {%
Earman%
}{%
Earman%
}{%
{\protect \APACyear {2003}}%
{\protect \APACexlab {{\protect \BCnt {2}}}}}]{%
Earman2003_ode}
\APACinsertmetastar {%
Earman2003_ode}%
\begin{APACrefauthors}%
Earman, J.%
\end{APACrefauthors}%
\unskip\
\newblock
\APACrefYearMonthDay{2003{\protect \BCnt {2}}}{}{}.
\newblock
{\BBOQ}\APACrefatitle {{Tracking Down Gauge: An Ode to the Constrained
  Hamiltonian Formalism}} {{Tracking Down Gauge: An Ode to the Constrained
  Hamiltonian Formalism}}.{\BBCQ}
\newblock
\BIn{} K.~Brading\ \BBA {} E.~Castellani\ (\BEDS), \APACrefbtitle {Symmetries
  in Physics: Philosophical Reflections} {Symmetries in physics: Philosophical
  reflections}\ (\BPGS\ 140--62).
\newblock
\APACaddressPublisher{}{Cambridge University Press}.
\PrintBackRefs{\CurrentBib}

\bibitem [\protect \citeauthoryear {%
Earman%
}{%
Earman%
}{%
{\protect \APACyear {2004}}%
{\protect \APACexlab {{\protect \BCnt {1}}}}}]{%
Earman2004}
\APACinsertmetastar {%
Earman2004}%
\begin{APACrefauthors}%
Earman, J.%
\end{APACrefauthors}%
\unskip\
\newblock
\APACrefYearMonthDay{2004{\protect \BCnt {1}}}{}{}.
\newblock
{\BBOQ}\APACrefatitle {Curie's Principle and Spontaneous Symmetry Breaking}
  {Curie's principle and spontaneous symmetry breaking}.{\BBCQ}
\newblock
\APACjournalVolNumPages{International Studies in the Philosophy of
  Science}{18}{2 \& 3}{173--198}.
\newblock
\begin{APACrefDOI} \doi{10.1080/0269859042000311299} \end{APACrefDOI}
\PrintBackRefs{\CurrentBib}

\bibitem [\protect \citeauthoryear {%
Earman%
}{%
Earman%
}{%
{\protect \APACyear {2004}}%
{\protect \APACexlab {{\protect \BCnt {2}}}}}]{%
Earman2004b}
\APACinsertmetastar {%
Earman2004b}%
\begin{APACrefauthors}%
Earman, J.%
\end{APACrefauthors}%
\unskip\
\newblock
\APACrefYearMonthDay{2004{\protect \BCnt {2}}}{}{}.
\newblock
{\BBOQ}\APACrefatitle {{Laws, Symmetry, and Symmetry Breaking: Invariance,
  Conservation Principles, and Objectivity}} {{Laws, Symmetry, and Symmetry
  Breaking: Invariance, Conservation Principles, and Objectivity}}.{\BBCQ}
\newblock
\APACjournalVolNumPages{Philosophy of Science}{71}{5}{1227--1241}.
\newblock
\begin{APACrefURL} \url{https://www.jstor.org/stable/10.1086/428016}
  \end{APACrefURL}
\PrintBackRefs{\CurrentBib}

\bibitem [\protect \citeauthoryear {%
Earman%
}{%
Earman%
}{%
{\protect \APACyear {2019}}%
}]{%
Earman2019}
\APACinsertmetastar {%
Earman2019}%
\begin{APACrefauthors}%
Earman, J.%
\end{APACrefauthors}%
\unskip\
\newblock
\APACrefYearMonthDay{2019}{May}{01}.
\newblock
{\BBOQ}\APACrefatitle {{The role of idealizations in the Aharonov--Bohm
  effect}} {{The role of idealizations in the Aharonov--Bohm effect}}.{\BBCQ}
\newblock
\APACjournalVolNumPages{Synthese}{196}{5}{1991--2019}.
\newblock
\begin{APACrefURL} \url{https://doi.org/10.1007/s11229-017-1522-9}
  \end{APACrefURL}
\newblock
\begin{APACrefDOI} \doi{10.1007/s11229-017-1522-9} \end{APACrefDOI}
\PrintBackRefs{\CurrentBib}

\bibitem [\protect \citeauthoryear {%
Earman%
\ \BBA {} Norton%
}{%
Earman%
\ \BBA {} Norton%
}{%
{\protect \APACyear {1987}}%
}]{%
EarmanNorton1987}
\APACinsertmetastar {%
EarmanNorton1987}%
\begin{APACrefauthors}%
Earman, J.%
\BCBT {}\ \BBA {} Norton, J.%
\end{APACrefauthors}%
\unskip\
\newblock
\APACrefYearMonthDay{1987}{12}{}.
\newblock
{\BBOQ}\APACrefatitle {{What Price Spacetime Substantivalism? The Hole Story}}
  {{What Price Spacetime Substantivalism? The Hole Story}}.{\BBCQ}
\newblock
\APACjournalVolNumPages{The British Journal for the Philosophy of
  Science}{38}{4}{515-525}.
\newblock
\begin{APACrefURL} \url{https://doi.org/10.1093/bjps/38.4.515} \end{APACrefURL}
\newblock
\begin{APACrefDOI} \doi{10.1093/bjps/38.4.515} \end{APACrefDOI}
\PrintBackRefs{\CurrentBib}

\bibitem [\protect \citeauthoryear {%
Ebin%
}{%
Ebin%
}{%
{\protect \APACyear {1970}}%
}]{%
Ebin}
\APACinsertmetastar {%
Ebin}%
\begin{APACrefauthors}%
Ebin, D.%
\end{APACrefauthors}%
\unskip\
\newblock
\APACrefYearMonthDay{1970}{}{}.
\newblock
{\BBOQ}\APACrefatitle {The Manifold of Riemmanian Metrics} {The manifold of
  riemmanian metrics}.{\BBCQ}
\newblock
\APACjournalVolNumPages{Symp. Pure Math., AMS, 11}{11,15}{}{}.
\PrintBackRefs{\CurrentBib}

\bibitem [\protect \citeauthoryear {%
Ehlers%
, Pirani%
\BCBL {}\ \BBA {} Schild%
}{%
Ehlers%
\ \protect \BOthers {.}}{%
{\protect \APACyear {2012}}%
}]{%
EPS}
\APACinsertmetastar {%
EPS}%
\begin{APACrefauthors}%
Ehlers, J.%
, Pirani, F.%
\BCBL {}\ \BBA {} Schild, A.%
\end{APACrefauthors}%
\unskip\
\newblock
\APACrefYearMonthDay{2012}{}{}.
\newblock
{\BBOQ}\APACrefatitle {{Republication of: The geometry of free fall and light
  propagation.}} {{Republication of: The geometry of free fall and light
  propagation.}}{\BBCQ}
\newblock
\APACjournalVolNumPages{General Relativity and Gravitation, 44,
  1587–1609}{}{}{}.
\PrintBackRefs{\CurrentBib}

\bibitem [\protect \citeauthoryear {%
Ehrenberg%
\ \BBA {} Siday%
}{%
Ehrenberg%
\ \BBA {} Siday%
}{%
{\protect \APACyear {1949}}%
}]{%
ehrenberg1949refractive}
\APACinsertmetastar {%
ehrenberg1949refractive}%
\begin{APACrefauthors}%
Ehrenberg, W.%
\BCBT {}\ \BBA {} Siday, R\BPBI E.%
\end{APACrefauthors}%
\unskip\
\newblock
\APACrefYearMonthDay{1949}{jan}{}.
\newblock
{\BBOQ}\APACrefatitle {{The Refractive Index in Electron Optics and the
  Principles of Dynamics}} {{The Refractive Index in Electron Optics and the
  Principles of Dynamics}}.{\BBCQ}
\newblock
\APACjournalVolNumPages{Proceedings of the Physical Society. Section
  B}{62}{1}{8--21}.
\newblock
\begin{APACrefURL} \url{https://doi.org/10.1088/0370-1301/62/1/303}
  \end{APACrefURL}
\newblock
\begin{APACrefDOI} \doi{10.1088/0370-1301/62/1/303} \end{APACrefDOI}
\PrintBackRefs{\CurrentBib}

\bibitem [\protect \citeauthoryear {%
Einstein%
}{%
Einstein%
}{%
{\protect \APACyear {1948}}%
}]{%
Einstein}
\APACinsertmetastar {%
Einstein}%
\begin{APACrefauthors}%
Einstein, A.%
\end{APACrefauthors}%
\unskip\
\newblock
\APACrefYearMonthDay{1948}{}{}.
\newblock
{\BBOQ}\APACrefatitle {QUANTUM MECHANICS AND REALITY} {Quantum mechanics and
  reality}.{\BBCQ}
\newblock
\APACjournalVolNumPages{Dialectica}{2}{3-4}{320-324}.
\newblock
\begin{APACrefURL}
  \url{https://onlinelibrary.wiley.com/doi/abs/10.1111/j.1746-8361.1948.tb00704.x}
  \end{APACrefURL}
\newblock
\begin{APACrefDOI} \doi{https://doi.org/10.1111/j.1746-8361.1948.tb00704.x}
  \end{APACrefDOI}
\PrintBackRefs{\CurrentBib}

\bibitem [\protect \citeauthoryear {%
Einstein%
}{%
Einstein%
}{%
{\protect \APACyear {1987}}%
}]{%
CPAE}
\APACinsertmetastar {%
CPAE}%
\begin{APACrefauthors}%
Einstein, A.%
\end{APACrefauthors}%
\unskip\
\newblock
\APACrefYear{1987}.
\newblock
\APACrefbtitle {{The Collected Papers of Albert Einstein. (Cited by volume and
  document number.)}} {{The Collected Papers of Albert Einstein. (Cited by
  volume and document number.)}}.
\newblock
\APACaddressPublisher{}{Princeton University Press}.
\PrintBackRefs{\CurrentBib}

\bibitem [\protect \citeauthoryear {%
Fischer%
\ \BBA {} Marsden%
}{%
Fischer%
\ \BBA {} Marsden%
}{%
{\protect \APACyear {1979}}%
}]{%
fischermarsden}
\APACinsertmetastar {%
fischermarsden}%
\begin{APACrefauthors}%
Fischer, A\BPBI E.%
\BCBT {}\ \BBA {} Marsden, J\BPBI E.%
\end{APACrefauthors}%
\unskip\
\newblock
\APACrefYearMonthDay{1979}{}{}.
\newblock
{\BBOQ}\APACrefatitle {The initial value problem and the dynamical formulation
  of general relativity.} {The initial value problem and the dynamical
  formulation of general relativity.}{\BBCQ}
\newblock
\BIn{} \APACrefbtitle {General relativity : an Einstein centenary survey.
  Cambridge University Press , New York, pp. 138-211.} {General relativity : an
  einstein centenary survey. cambridge university press , new york, pp.
  138-211.}
\PrintBackRefs{\CurrentBib}

\bibitem [\protect \citeauthoryear {%
Fletcher%
}{%
Fletcher%
}{%
{\protect \APACyear {2021}}%
}]{%
Fletcher2018}
\APACinsertmetastar {%
Fletcher2018}%
\begin{APACrefauthors}%
Fletcher, S.%
\end{APACrefauthors}%
\unskip\
\newblock
\APACrefYearMonthDay{2021}{}{}.
\newblock
{\BBOQ}\APACrefatitle {An invitation to approximate symmetry, with three
  applications to intertheoretic relations} {An invitation to approximate
  symmetry, with three applications to intertheoretic relations}.{\BBCQ}
\newblock
\APACjournalVolNumPages{Synthese, vol. 198}{}{}{}.
\PrintBackRefs{\CurrentBib}

\bibitem [\protect \citeauthoryear {%
Ford%
\ \BBA {} Vilenkin%
}{%
Ford%
\ \BBA {} Vilenkin%
}{%
{\protect \APACyear {1981}}%
}]{%
Ford_1981}
\APACinsertmetastar {%
Ford_1981}%
\begin{APACrefauthors}%
Ford, L\BPBI H.%
\BCBT {}\ \BBA {} Vilenkin, A.%
\end{APACrefauthors}%
\unskip\
\newblock
\APACrefYearMonthDay{1981}{sep}{}.
\newblock
{\BBOQ}\APACrefatitle {{A gravitational analogue of the Aharonov-Bohm effect}}
  {{A gravitational analogue of the Aharonov-Bohm effect}}.{\BBCQ}
\newblock
\APACjournalVolNumPages{Journal of Physics A: Mathematical and
  General}{14}{9}{2353--2357}.
\newblock
\begin{APACrefURL} \url{https://doi.org/10.1088/0305-4470/14/9/030}
  \end{APACrefURL}
\newblock
\begin{APACrefDOI} \doi{10.1088/0305-4470/14/9/030} \end{APACrefDOI}
\PrintBackRefs{\CurrentBib}

\bibitem [\protect \citeauthoryear {%
Friederich%
}{%
Friederich%
}{%
{\protect \APACyear {2014}}%
}]{%
Friederich2014}
\APACinsertmetastar {%
Friederich2014}%
\begin{APACrefauthors}%
Friederich, S.%
\end{APACrefauthors}%
\unskip\
\newblock
\APACrefYearMonthDay{2014}{04}{}.
\newblock
{\BBOQ}\APACrefatitle {{Symmetry, Empirical Equivalence, and Identity}}
  {{Symmetry, Empirical Equivalence, and Identity}}.{\BBCQ}
\newblock
\APACjournalVolNumPages{The British Journal for the Philosophy of
  Science}{66}{3}{537-559}.
\newblock
\begin{APACrefURL} \url{https://doi.org/10.1093/bjps/axt046} \end{APACrefURL}
\newblock
\begin{APACrefDOI} \doi{10.1093/bjps/axt046} \end{APACrefDOI}
\PrintBackRefs{\CurrentBib}

\bibitem [\protect \citeauthoryear {%
Friederich%
}{%
Friederich%
}{%
{\protect \APACyear {2017}}%
}]{%
Friederich2017}
\APACinsertmetastar {%
Friederich2017}%
\begin{APACrefauthors}%
Friederich, S.%
\end{APACrefauthors}%
\unskip\
\newblock
\APACrefYearMonthDay{2017}{}{}.
\newblock
{\BBOQ}\APACrefatitle {Symmetries and the Identity of Physical States}
  {Symmetries and the identity of physical states}.{\BBCQ}
\newblock
\BIn{} M.~Massimi, J\BHBI W.~Romeijn\BCBL {}\ \BBA {} G.~Schurz\ (\BEDS),
  \APACrefbtitle {EPSA15 Selected Papers} {Epsa15 selected papers}\ (\BPGS\
  153--165).
\newblock
\APACaddressPublisher{Cham}{Springer International Publishing}.
\PrintBackRefs{\CurrentBib}

\bibitem [\protect \citeauthoryear {%
M.~Fröb%
}{%
M.~Fröb%
}{%
{\protect \APACyear {2018}}%
}]{%
Frob_proj}
\APACinsertmetastar {%
Frob_proj}%
\begin{APACrefauthors}%
Fröb, M.%
\end{APACrefauthors}%
\unskip\
\newblock
\APACrefYearMonthDay{2018}{Feb}{}.
\newblock
{\BBOQ}\APACrefatitle {Gauge-invariant quantum gravitational corrections to
  correlation functions} {Gauge-invariant quantum gravitational corrections to
  correlation functions}.{\BBCQ}
\newblock
\APACjournalVolNumPages{Classical and Quantum Gravity}{35}{5}{055006}.
\newblock
\begin{APACrefURL} \url{http://dx.doi.org/10.1088/1361-6382/aaa74c}
  \end{APACrefURL}
\newblock
\begin{APACrefDOI} \doi{10.1088/1361-6382/aaa74c} \end{APACrefDOI}
\PrintBackRefs{\CurrentBib}

\bibitem [\protect \citeauthoryear {%
M\BPBI B.~Fröb%
\ \BBA {} Lima%
}{%
M\BPBI B.~Fröb%
\ \BBA {} Lima%
}{%
{\protect \APACyear {2021}}%
}]{%
Lima_proj}
\APACinsertmetastar {%
Lima_proj}%
\begin{APACrefauthors}%
Fröb, M\BPBI B.%
\BCBT {}\ \BBA {} Lima, W\BPBI C\BPBI C.%
\end{APACrefauthors}%
\unskip\
\newblock
\APACrefYearMonthDay{2021}{}{}.
\newblock
\APACrefbtitle {Cosmological Perturbations and Invariant Observables in
  Geodesic Lightcone Coordinates.} {Cosmological perturbations and invariant
  observables in geodesic lightcone coordinates.}
\PrintBackRefs{\CurrentBib}

\bibitem [\protect \citeauthoryear {%
Geiller%
}{%
Geiller%
}{%
{\protect \APACyear {2017}}%
}]{%
Geiller:2017xad}
\APACinsertmetastar {%
Geiller:2017xad}%
\begin{APACrefauthors}%
Geiller, M.%
\end{APACrefauthors}%
\unskip\
\newblock
\APACrefYearMonthDay{2017}{}{}.
\newblock
{\BBOQ}\APACrefatitle {{Edge modes and corner ambiguities in 3d Chern–Simons
  theory and gravity}} {{Edge modes and corner ambiguities in 3d Chern–Simons
  theory and gravity}}.{\BBCQ}
\newblock
\APACjournalVolNumPages{Nucl. Phys.}{B924}{}{312-365}.
\newblock
\begin{APACrefDOI} \doi{10.1016/j.nuclphysb.2017.09.010} \end{APACrefDOI}
\PrintBackRefs{\CurrentBib}

\bibitem [\protect \citeauthoryear {%
Geiller%
\ \BBA {} Jai-akson%
}{%
Geiller%
\ \BBA {} Jai-akson%
}{%
{\protect \APACyear {2020}}%
}]{%
Geiller_edge2020}
\APACinsertmetastar {%
Geiller_edge2020}%
\begin{APACrefauthors}%
Geiller, M.%
\BCBT {}\ \BBA {} Jai-akson, P.%
\end{APACrefauthors}%
\unskip\
\newblock
\APACrefYearMonthDay{2020}{Sep}{}.
\newblock
{\BBOQ}\APACrefatitle {Extended actions, dynamics of edge modes, and
  entanglement entropy} {Extended actions, dynamics of edge modes, and
  entanglement entropy}.{\BBCQ}
\newblock
\APACjournalVolNumPages{Journal of High Energy Physics}{2020}{9}{}.
\newblock
\begin{APACrefURL} \url{http://dx.doi.org/10.1007/JHEP09(2020)134}
  \end{APACrefURL}
\newblock
\begin{APACrefDOI} \doi{10.1007/jhep09(2020)134} \end{APACrefDOI}
\PrintBackRefs{\CurrentBib}

\bibitem [\protect \citeauthoryear {%
Geroch%
}{%
Geroch%
}{%
{\protect \APACyear {1970}}%
}]{%
Geroch1970}
\APACinsertmetastar {%
Geroch1970}%
\begin{APACrefauthors}%
Geroch, R.%
\end{APACrefauthors}%
\unskip\
\newblock
\APACrefYearMonthDay{1970}{}{}.
\newblock
{\BBOQ}\APACrefatitle {{Domain of Dependence}} {{Domain of Dependence}}.{\BBCQ}
\newblock
\APACjournalVolNumPages{Journal of Mathematical Physics}{11}{2}{437-449}.
\newblock
\begin{APACrefURL} \url{https://doi.org/10.1063/1.1665157} \end{APACrefURL}
\newblock
\begin{APACrefDOI} \doi{10.1063/1.1665157} \end{APACrefDOI}
\PrintBackRefs{\CurrentBib}

\bibitem [\protect \citeauthoryear {%
Giesel%
, Herzog%
\BCBL {}\ \BBA {} Singh%
}{%
Giesel%
\ \protect \BOthers {.}}{%
{\protect \APACyear {2018}}%
}]{%
Giesel_geom_clocks}
\APACinsertmetastar {%
Giesel_geom_clocks}%
\begin{APACrefauthors}%
Giesel, K.%
, Herzog, A.%
\BCBL {}\ \BBA {} Singh, P.%
\end{APACrefauthors}%
\unskip\
\newblock
\APACrefYearMonthDay{2018}{Jul}{}.
\newblock
{\BBOQ}\APACrefatitle {Gauge invariant variables for cosmological perturbation
  theory using geometrical clocks} {Gauge invariant variables for cosmological
  perturbation theory using geometrical clocks}.{\BBCQ}
\newblock
\APACjournalVolNumPages{Classical and Quantum Gravity}{35}{15}{155012}.
\newblock
\begin{APACrefURL} \url{http://dx.doi.org/10.1088/1361-6382/aacda2}
  \end{APACrefURL}
\newblock
\begin{APACrefDOI} \doi{10.1088/1361-6382/aacda2} \end{APACrefDOI}
\PrintBackRefs{\CurrentBib}

\bibitem [\protect \citeauthoryear {%
Gilbarg%
\ \BBA {} Trudinger%
}{%
Gilbarg%
\ \BBA {} Trudinger%
}{%
{\protect \APACyear {2001}}%
}]{%
Trudinger}
\APACinsertmetastar {%
Trudinger}%
\begin{APACrefauthors}%
Gilbarg, D.%
\BCBT {}\ \BBA {} Trudinger, N.%
\end{APACrefauthors}%
\unskip\
\newblock
\APACrefYear{2001}.
\newblock
\APACrefbtitle {{Elliptic Partial Differential Equations of Second Order}}
  {{Elliptic Partial Differential Equations of Second Order}}.
\newblock
\APACaddressPublisher{}{Springer}.
\PrintBackRefs{\CurrentBib}

\bibitem [\protect \citeauthoryear {%
Giulini%
}{%
Giulini%
}{%
{\protect \APACyear {1995}}%
}]{%
Giulini_asymptotic}
\APACinsertmetastar {%
Giulini_asymptotic}%
\begin{APACrefauthors}%
Giulini, D.%
\end{APACrefauthors}%
\unskip\
\newblock
\APACrefYearMonthDay{1995}{}{}.
\newblock
{\BBOQ}\APACrefatitle {Asymptotic symmetry groups of long-ranged gauge
  configurations} {Asymptotic symmetry groups of long-ranged gauge
  configurations}.{\BBCQ}
\newblock
\APACjournalVolNumPages{Modern Physics Letters A}{10}{28}{2059-2070}.
\newblock
\begin{APACrefURL} \url{https://doi.org/10.1142/S0217732395002210}
  \end{APACrefURL}
\newblock
\begin{APACrefDOI} \doi{10.1142/S0217732395002210} \end{APACrefDOI}
\PrintBackRefs{\CurrentBib}

\bibitem [\protect \citeauthoryear {%
G\"ockeler%
\ \BBA {} Sch\"ucker%
}{%
G\"ockeler%
\ \BBA {} Sch\"ucker%
}{%
{\protect \APACyear {1989}}%
}]{%
gc1989dg}
\APACinsertmetastar {%
gc1989dg}%
\begin{APACrefauthors}%
G\"ockeler, M.%
\BCBT {}\ \BBA {} Sch\"ucker, T.%
\end{APACrefauthors}%
\unskip\
\newblock
\APACrefYear{1989}.
\newblock
\APACrefbtitle {Differential Geometry, Gauge Theories, and Gravity}
  {Differential geometry, gauge theories, and gravity}.
\newblock
\APACaddressPublisher{}{Cambridge: Cambridge University Press}.
\PrintBackRefs{\CurrentBib}

\bibitem [\protect \citeauthoryear {%
Gomes%
}{%
Gomes%
}{%
{\protect \APACyear {2018}}%
}]{%
gomes:aop2018}
\APACinsertmetastar {%
gomes:aop2018}%
\begin{APACrefauthors}%
Gomes, H.%
\end{APACrefauthors}%
\unskip\
\newblock
\APACrefYearMonthDay{2018}{}{}.
\newblock
{\BBOQ}\APACrefatitle {Local gravity theories in conformal superspace} {Local
  gravity theories in conformal superspace}.{\BBCQ}
\newblock
\APACjournalVolNumPages{Annals of Physics}{}{}{}.
\newblock
\begin{APACrefURL}
  \url{http://www.sciencedirect.com/science/article/pii/S0003491618301507}
  \end{APACrefURL}
\newblock
\begin{APACrefDOI} \doi{https://doi.org/10.1016/j.aop.2018.05.014}
  \end{APACrefDOI}
\PrintBackRefs{\CurrentBib}

\bibitem [\protect \citeauthoryear {%
Gomes%
}{%
Gomes%
}{%
{\protect \APACyear {2019}}%
{\protect \APACexlab {{\protect \BCnt {1}}}}}]{%
GomesStudies}
\APACinsertmetastar {%
GomesStudies}%
\begin{APACrefauthors}%
Gomes, H.%
\end{APACrefauthors}%
\unskip\
\newblock
\APACrefYearMonthDay{2019{\protect \BCnt {1}}}{}{}.
\newblock
{\BBOQ}\APACrefatitle {Gauging the boundary in field-space} {Gauging the
  boundary in field-space}.{\BBCQ}
\newblock
\APACjournalVolNumPages{Studies in History and Philosophy of Science Part B:
  Studies in History and Philosophy of Modern Physics}{}{}{}.
\newblock
\begin{APACrefURL}
  \url{http://www.sciencedirect.com/science/article/pii/S1355219818302144}
  \end{APACrefURL}
\newblock
\begin{APACrefDOI} \doi{https://doi.org/10.1016/j.shpsb.2019.04.002}
  \end{APACrefDOI}
\PrintBackRefs{\CurrentBib}

\bibitem [\protect \citeauthoryear {%
Gomes%
}{%
Gomes%
}{%
{\protect \APACyear {2019}}%
{\protect \APACexlab {{\protect \BCnt {2}}}}}]{%
GomesNoether}
\APACinsertmetastar {%
GomesNoether}%
\begin{APACrefauthors}%
Gomes, H.%
\end{APACrefauthors}%
\unskip\
\newblock
\APACrefYearMonthDay{2019{\protect \BCnt {2}}}{}{}.
\newblock
{\BBOQ}\APACrefatitle {Noether charges, gauge-invariance, and non-locality}
  {Noether charges, gauge-invariance, and non-locality}.{\BBCQ}
\newblock
\APACjournalVolNumPages{Comissioned from Cambridge University Press, for
  special edition celebrating 100 years of Noether's theorems}{}{}{}.
\PrintBackRefs{\CurrentBib}

\bibitem [\protect \citeauthoryear {%
Gomes%
}{%
Gomes%
}{%
{\protect \APACyear {2021}}%
{\protect \APACexlab {{\protect \BCnt {1}}}}}]{%
Gomes_new}
\APACinsertmetastar {%
Gomes_new}%
\begin{APACrefauthors}%
Gomes, H.%
\end{APACrefauthors}%
\unskip\
\newblock
\APACrefYearMonthDay{2021{\protect \BCnt {1}}}{}{}.
\newblock
{\BBOQ}\APACrefatitle {Holism as the significance of gauge symmetries} {Holism
  as the significance of gauge symmetries}.{\BBCQ}
\newblock
\APACjournalVolNumPages{European Journal of Philosophy of Science, vol 11,
  87}{}{}{}.
\PrintBackRefs{\CurrentBib}

\bibitem [\protect \citeauthoryear {%
Gomes%
}{%
Gomes%
}{%
{\protect \APACyear {2021}}%
{\protect \APACexlab {{\protect \BCnt {2}}}}}]{%
DES_gf}
\APACinsertmetastar {%
DES_gf}%
\begin{APACrefauthors}%
Gomes, H.%
\end{APACrefauthors}%
\unskip\
\newblock
\APACrefYearMonthDay{2021{\protect \BCnt {2}}}{}{}.
\newblock
{\BBOQ}\APACrefatitle {The role of representational conventions in assessing
  the empirical signicance of symmetries} {The role of representational
  conventions in assessing the empirical signicance of symmetries}.{\BBCQ}
\newblock
\APACjournalVolNumPages{(to appear in Studies in History and Philosophy of
  Modern Physics)}{}{}{}.
\PrintBackRefs{\CurrentBib}

\bibitem [\protect \citeauthoryear {%
Gomes%
}{%
Gomes%
}{%
{\protect \APACyear {2021}}%
{\protect \APACexlab {{\protect \BCnt {3}}}}}]{%
Samediff_1}
\APACinsertmetastar {%
Samediff_1}%
\begin{APACrefauthors}%
Gomes, H.%
\end{APACrefauthors}%
\unskip\
\newblock
\APACrefYearMonthDay{2021{\protect \BCnt {3}}}{}{}.
\newblock
{\BBOQ}\APACrefatitle {{Same-diff? Part I: Conceptual similarities (and one
  difference) between gauge transformations and diffeomorphisms}} {{Same-diff?
  Part I: Conceptual similarities (and one difference) between gauge
  transformations and diffeomorphisms}}.{\BBCQ}
\newblock
\APACjournalVolNumPages{Arxiv: 2110.07203. Submitted.}{}{}{}.
\PrintBackRefs{\CurrentBib}

\bibitem [\protect \citeauthoryear {%
Gomes%
}{%
Gomes%
}{%
{\protect \APACyear {2021}}%
{\protect \APACexlab {{\protect \BCnt {4}}}}}]{%
Samediff_2}
\APACinsertmetastar {%
Samediff_2}%
\begin{APACrefauthors}%
Gomes, H.%
\end{APACrefauthors}%
\unskip\
\newblock
\APACrefYearMonthDay{2021{\protect \BCnt {4}}}{}{}.
\newblock
{\BBOQ}\APACrefatitle {{Same-diff? Part II: A compendium of similarities
  between gauge transformations and diffeomorphisms}} {{Same-diff? Part II: A
  compendium of similarities between gauge transformations and
  diffeomorphisms}}.{\BBCQ}
\newblock
\APACjournalVolNumPages{Arxiv: 2110.07204. Submitted.}{}{}{}.
\PrintBackRefs{\CurrentBib}

\bibitem [\protect \citeauthoryear {%
Gomes%
\ \BBA {} Butterfield%
}{%
Gomes%
\ \BBA {} Butterfield%
}{%
{\protect \APACyear {2021}}%
{\protect \APACexlab {{\protect \BCnt {1}}}}}]{%
GomesButterfield_hole}
\APACinsertmetastar {%
GomesButterfield_hole}%
\begin{APACrefauthors}%
Gomes, H.%
\BCBT {}\ \BBA {} Butterfield, J.%
\end{APACrefauthors}%
\unskip\
\newblock
\APACrefYearMonthDay{2021{\protect \BCnt {1}}}{}{}.
\newblock
{\BBOQ}\APACrefatitle {Assessing the hole argument} {Assessing the hole
  argument}.{\BBCQ}
\newblock
\APACjournalVolNumPages{In preparation}{}{}{}.
\PrintBackRefs{\CurrentBib}

\bibitem [\protect \citeauthoryear {%
Gomes%
\ \BBA {} Butterfield%
}{%
Gomes%
\ \BBA {} Butterfield%
}{%
{\protect \APACyear {2021}}%
{\protect \APACexlab {{\protect \BCnt {2}}}}}]{%
GomesButterfield_counter}
\APACinsertmetastar {%
GomesButterfield_counter}%
\begin{APACrefauthors}%
Gomes, H.%
\BCBT {}\ \BBA {} Butterfield, J.%
\end{APACrefauthors}%
\unskip\
\newblock
\APACrefYearMonthDay{2021{\protect \BCnt {2}}}{}{}.
\newblock
{\BBOQ}\APACrefatitle {Counterpart relations in general relativity
  (forthcoming)} {Counterpart relations in general relativity
  (forthcoming)}.{\BBCQ}
\newblock

\PrintBackRefs{\CurrentBib}

\bibitem [\protect \citeauthoryear {%
Gomes%
\ \BBA {} Butterfield%
}{%
Gomes%
\ \BBA {} Butterfield%
}{%
{\protect \APACyear {2021}}%
{\protect \APACexlab {{\protect \BCnt {3}}}}}]{%
GomesButterfield_glimpse}
\APACinsertmetastar {%
GomesButterfield_glimpse}%
\begin{APACrefauthors}%
Gomes, H.%
\BCBT {}\ \BBA {} Butterfield, J.%
\end{APACrefauthors}%
\unskip\
\newblock
\APACrefYearMonthDay{2021{\protect \BCnt {3}}}{}{}.
\newblock
{\BBOQ}\APACrefatitle {A glimpse of symplectic reduction} {A glimpse of
  symplectic reduction}.{\BBCQ}
\newblock
\APACjournalVolNumPages{In preparation}{}{}{}.
\PrintBackRefs{\CurrentBib}

\bibitem [\protect \citeauthoryear {%
Gomes%
\ \BBA {} Butterfield%
}{%
Gomes%
\ \BBA {} Butterfield%
}{%
{\protect \APACyear {2021}}%
{\protect \APACexlab {{\protect \BCnt {4}}}}}]{%
GomesButterfield_electro}
\APACinsertmetastar {%
GomesButterfield_electro}%
\begin{APACrefauthors}%
Gomes, H.%
\BCBT {}\ \BBA {} Butterfield, J.%
\end{APACrefauthors}%
\unskip\
\newblock
\APACrefYearMonthDay{2021{\protect \BCnt {4}}}{}{}.
\newblock
{\BBOQ}\APACrefatitle {How to choose a gauge: the example of electromagnetism}
  {How to choose a gauge: the example of electromagnetism}.{\BBCQ}
\newblock
\APACjournalVolNumPages{In preparation}{}{}{}.
\PrintBackRefs{\CurrentBib}

\bibitem [\protect \citeauthoryear {%
Gomes%
\ \BBA {} Gryb%
}{%
Gomes%
\ \BBA {} Gryb%
}{%
{\protect \APACyear {2021}}%
}]{%
GomesGryb_KK}
\APACinsertmetastar {%
GomesGryb_KK}%
\begin{APACrefauthors}%
Gomes, H.%
\BCBT {}\ \BBA {} Gryb, S.%
\end{APACrefauthors}%
\unskip\
\newblock
\APACrefYearMonthDay{2021}{}{}.
\newblock
{\BBOQ}\APACrefatitle {{Angular momentum without rotation: Turbocharging
  relationalism}} {{Angular momentum without rotation: Turbocharging
  relationalism}}.{\BBCQ}
\newblock
\APACjournalVolNumPages{Studies in History and Philosophy of Science Part
  A}{88}{}{138-155}.
\newblock
\begin{APACrefURL}
  \url{https://www.sciencedirect.com/science/article/pii/S0039368121000704}
  \end{APACrefURL}
\newblock
\begin{APACrefDOI} \doi{https://doi.org/10.1016/j.shpsa.2021.05.006}
  \end{APACrefDOI}
\PrintBackRefs{\CurrentBib}

\bibitem [\protect \citeauthoryear {%
Gomes%
, Gryb%
\BCBL {}\ \BBA {} Koslowski%
}{%
Gomes%
\ \protect \BOthers {.}}{%
{\protect \APACyear {2011}}%
}]{%
SD_first}
\APACinsertmetastar {%
SD_first}%
\begin{APACrefauthors}%
Gomes, H.%
, Gryb, S.%
\BCBL {}\ \BBA {} Koslowski, T.%
\end{APACrefauthors}%
\unskip\
\newblock
\APACrefYearMonthDay{2011}{}{}.
\newblock
{\BBOQ}\APACrefatitle {{Einstein gravity as a 3D conformally invariant theory}}
  {{Einstein gravity as a 3D conformally invariant theory}}.{\BBCQ}
\newblock
\APACjournalVolNumPages{Class. Quant. Grav.}{28}{}{045005}.
\newblock
\begin{APACrefDOI} \doi{10.1088/0264-9381/28/4/045005} \end{APACrefDOI}
\PrintBackRefs{\CurrentBib}

\bibitem [\protect \citeauthoryear {%
Gomes%
, Hopfmüller%
\BCBL {}\ \BBA {} Riello%
}{%
Gomes%
\ \protect \BOthers {.}}{%
{\protect \APACyear {2019}}%
}]{%
GomesHopfRiello}
\APACinsertmetastar {%
GomesHopfRiello}%
\begin{APACrefauthors}%
Gomes, H.%
, Hopfmüller, F.%
\BCBL {}\ \BBA {} Riello, A.%
\end{APACrefauthors}%
\unskip\
\newblock
\APACrefYearMonthDay{2019}{}{}.
\newblock
{\BBOQ}\APACrefatitle {A unified geometric framework for boundary charges and
  dressings: Non-Abelian theory and matter} {A unified geometric framework for
  boundary charges and dressings: Non-abelian theory and matter}.{\BBCQ}
\newblock
\APACjournalVolNumPages{Nuclear Physics B}{941}{}{249 - 315}.
\newblock
\begin{APACrefURL}
  \url{http://www.sciencedirect.com/science/article/pii/S0550321319300483}
  \end{APACrefURL}
\newblock
\begin{APACrefDOI} \doi{https://doi.org/10.1016/j.nuclphysb.2019.02.020}
  \end{APACrefDOI}
\PrintBackRefs{\CurrentBib}

\bibitem [\protect \citeauthoryear {%
Gomes%
\ \BBA {} Koslowski%
}{%
Gomes%
\ \BBA {} Koslowski%
}{%
{\protect \APACyear {2012}}%
}]{%
SD_linking}
\APACinsertmetastar {%
SD_linking}%
\begin{APACrefauthors}%
Gomes, H.%
\BCBT {}\ \BBA {} Koslowski, T.%
\end{APACrefauthors}%
\unskip\
\newblock
\APACrefYearMonthDay{2012}{}{}.
\newblock
{\BBOQ}\APACrefatitle {{The Link between General Relativity and Shape
  Dynamics}} {{The Link between General Relativity and Shape Dynamics}}.{\BBCQ}
\newblock
\APACjournalVolNumPages{Class.Quant.Grav.}{29}{}{075009}.
\newblock
\begin{APACrefDOI} \doi{10.1088/0264-9381/29/7/075009} \end{APACrefDOI}
\PrintBackRefs{\CurrentBib}

\bibitem [\protect \citeauthoryear {%
Gomes%
\ \BBA {} Riello%
}{%
Gomes%
\ \BBA {} Riello%
}{%
{\protect \APACyear {2017}}%
}]{%
GomesRiello2016}
\APACinsertmetastar {%
GomesRiello2016}%
\begin{APACrefauthors}%
Gomes, H.%
\BCBT {}\ \BBA {} Riello, A.%
\end{APACrefauthors}%
\unskip\
\newblock
\APACrefYearMonthDay{2017}{}{}.
\newblock
{\BBOQ}\APACrefatitle {{The observer’s ghost: notes on a field space
  connection}} {{The observer’s ghost: notes on a field space
  connection}}.{\BBCQ}
\newblock
\APACjournalVolNumPages{Journal of High Energy Physics (JHEP)}{05}{}{017}.
\newblock
\begin{APACrefURL}
  \url{https://link.springer.com/article/10.1007%2FJHEP05%282017%29017}
  \end{APACrefURL}
\newblock
\begin{APACrefDOI} \doi{10.1007/JHEP05(2017)017} \end{APACrefDOI}
\PrintBackRefs{\CurrentBib}

\bibitem [\protect \citeauthoryear {%
Gomes%
\ \BBA {} Riello%
}{%
Gomes%
\ \BBA {} Riello%
}{%
{\protect \APACyear {2018}}%
}]{%
GomesRiello2018}
\APACinsertmetastar {%
GomesRiello2018}%
\begin{APACrefauthors}%
Gomes, H.%
\BCBT {}\ \BBA {} Riello, A.%
\end{APACrefauthors}%
\unskip\
\newblock
\APACrefYearMonthDay{2018}{Jul}{}.
\newblock
{\BBOQ}\APACrefatitle {Unified geometric framework for boundary charges and
  particle dressings} {Unified geometric framework for boundary charges and
  particle dressings}.{\BBCQ}
\newblock
\APACjournalVolNumPages{Physical Review D}{98}{}{025013}.
\newblock
\begin{APACrefURL} \url{https://link.aps.org/doi/10.1103/PhysRevD.98.025013}
  \end{APACrefURL}
\newblock
\begin{APACrefDOI} \doi{10.1103/PhysRevD.98.025013} \end{APACrefDOI}
\PrintBackRefs{\CurrentBib}

\bibitem [\protect \citeauthoryear {%
Gomes%
\ \BBA {} Riello%
}{%
Gomes%
\ \BBA {} Riello%
}{%
{\protect \APACyear {2020}}%
}]{%
GomesRiello_theta}
\APACinsertmetastar {%
GomesRiello_theta}%
\begin{APACrefauthors}%
Gomes, H.%
\BCBT {}\ \BBA {} Riello, A.%
\end{APACrefauthors}%
\unskip\
\newblock
\APACrefYearMonthDay{2020}{}{}.
\newblock
{\BBOQ}\APACrefatitle {{Eliminativism and the QCD-theta term: What gauge
  transformations cannot do.}} {{Eliminativism and the QCD-theta term: What
  gauge transformations cannot do.}}{\BBCQ}
\newblock
\APACjournalVolNumPages{Submitted}{}{}{}.
\PrintBackRefs{\CurrentBib}

\bibitem [\protect \citeauthoryear {%
Gomes%
\ \BBA {} Riello%
}{%
Gomes%
\ \BBA {} Riello%
}{%
{\protect \APACyear {2021}}%
}]{%
GomesRiello_new}
\APACinsertmetastar {%
GomesRiello_new}%
\begin{APACrefauthors}%
Gomes, H.%
\BCBT {}\ \BBA {} Riello, A.%
\end{APACrefauthors}%
\unskip\
\newblock
\APACrefYearMonthDay{2021}{}{}.
\newblock
{\BBOQ}\APACrefatitle {{The quasilocal degrees of freedom of Yang-Mills
  theory}} {{The quasilocal degrees of freedom of Yang-Mills theory}}.{\BBCQ}
\newblock
\APACjournalVolNumPages{SciPost Phys.}{10}{}{130}.
\newblock
\begin{APACrefURL} \url{https://scipost.org/10.21468/SciPostPhys.10.6.130}
  \end{APACrefURL}
\newblock
\begin{APACrefDOI} \doi{10.21468/SciPostPhys.10.6.130} \end{APACrefDOI}
\PrintBackRefs{\CurrentBib}

\bibitem [\protect \citeauthoryear {%
Gomes%
, Roberts%
\BCBL {}\ \BBA {} Butterfield%
}{%
Gomes%
\ \protect \BOthers {.}}{%
{\protect \APACyear {2021}}%
}]{%
GomesButterfieldRoberts}
\APACinsertmetastar {%
GomesButterfieldRoberts}%
\begin{APACrefauthors}%
Gomes, H.%
, Roberts, B.%
\BCBL {}\ \BBA {} Butterfield, J.%
\end{APACrefauthors}%
\unskip\
\newblock
\APACrefYearMonthDay{2021}{}{}.
\newblock
{\BBOQ}\APACrefatitle {{The Gauge Argument: a Noether Reason}} {{The Gauge
  Argument: a Noether Reason}}.{\BBCQ}
\newblock
\APACjournalVolNumPages{Forthcoming in The Physics and Philosophy of Noether's
  Theorems, Edited by Read, Roberts and Teh, Cambridge University Press}{}{}{}.
\PrintBackRefs{\CurrentBib}

\bibitem [\protect \citeauthoryear {%
Greaves%
\ \BBA {} Wallace%
}{%
Greaves%
\ \BBA {} Wallace%
}{%
{\protect \APACyear {2014}}%
}]{%
GreavesWallace}
\APACinsertmetastar {%
GreavesWallace}%
\begin{APACrefauthors}%
Greaves, H.%
\BCBT {}\ \BBA {} Wallace, D.%
\end{APACrefauthors}%
\unskip\
\newblock
\APACrefYearMonthDay{2014}{}{}.
\newblock
{\BBOQ}\APACrefatitle {Empirical Consequences of Symmetries} {Empirical
  consequences of symmetries}.{\BBCQ}
\newblock
\APACjournalVolNumPages{British Journal for the Philosophy of
  Science}{65}{1}{59--89}.
\PrintBackRefs{\CurrentBib}

\bibitem [\protect \citeauthoryear {%
Gribov%
}{%
Gribov%
}{%
{\protect \APACyear {1978}}%
}]{%
Gribov:1977wm}
\APACinsertmetastar {%
Gribov:1977wm}%
\begin{APACrefauthors}%
Gribov, V\BPBI N.%
\end{APACrefauthors}%
\unskip\
\newblock
\APACrefYearMonthDay{1978}{}{}.
\newblock
{\BBOQ}\APACrefatitle {{Quantization of Nonabelian Gauge Theories}}
  {{Quantization of Nonabelian Gauge Theories}}.{\BBCQ}
\newblock
\APACjournalVolNumPages{Nucl. Phys.}{B139}{}{1}.
\newblock
\APACrefnote{[,1(1977)]}
\newblock
\begin{APACrefDOI} \doi{10.1016/0550-3213(78)90175-X} \end{APACrefDOI}
\PrintBackRefs{\CurrentBib}

\bibitem [\protect \citeauthoryear {%
Harlow%
\ \BBA {} Wu%
}{%
Harlow%
\ \BBA {} Wu%
}{%
{\protect \APACyear {2019}}%
}]{%
Harlow_cov}
\APACinsertmetastar {%
Harlow_cov}%
\begin{APACrefauthors}%
Harlow, D.%
\BCBT {}\ \BBA {} Wu, J\BHBI Q.%
\end{APACrefauthors}%
\unskip\
\newblock
\APACrefYearMonthDay{2019}{}{}.
\newblock
{\BBOQ}\APACrefatitle {{Covariant phase space with boundaries}} {{Covariant
  phase space with boundaries}}.{\BBCQ}
\newblock

\PrintBackRefs{\CurrentBib}

\bibitem [\protect \citeauthoryear {%
Harlow%
\ \BBA {} Wu%
}{%
Harlow%
\ \BBA {} Wu%
}{%
{\protect \APACyear {2021}}%
}]{%
Harlow_JT}
\APACinsertmetastar {%
Harlow_JT}%
\begin{APACrefauthors}%
Harlow, D.%
\BCBT {}\ \BBA {} Wu, J\BHBI Q.%
\end{APACrefauthors}%
\unskip\
\newblock
\APACrefYearMonthDay{2021}{}{}.
\newblock
\APACrefbtitle {Algebra of diffeomorphism-invariant observables in
  Jackiw-Teitelboim Gravity.} {Algebra of diffeomorphism-invariant observables
  in jackiw-teitelboim gravity.}
\PrintBackRefs{\CurrentBib}

\bibitem [\protect \citeauthoryear {%
Hawking%
\ \BBA {} Ellis%
}{%
Hawking%
\ \BBA {} Ellis%
}{%
{\protect \APACyear {1975}}%
}]{%
HawkingEllis}
\APACinsertmetastar {%
HawkingEllis}%
\begin{APACrefauthors}%
Hawking, S\BPBI W.%
\BCBT {}\ \BBA {} Ellis, G\BPBI F\BPBI R.%
\end{APACrefauthors}%
\unskip\
\newblock
\APACrefYear{1975}.
\newblock
\APACrefbtitle {{The Large Scale Structure of Space-Time (Cambridge Monographs
  on Mathematical Physics)}} {{The Large Scale Structure of Space-Time
  (Cambridge Monographs on Mathematical Physics)}}.
\newblock
\APACaddressPublisher{}{Cambridge University Press}.
\newblock
\begin{APACrefURL}
  \url{http://www.amazon.com/Structure-Space-Time-Cambridge-Monographs-Mathematical/dp/0521099064}
  \end{APACrefURL}
\PrintBackRefs{\CurrentBib}

\bibitem [\protect \citeauthoryear {%
Hayward%
}{%
Hayward%
}{%
{\protect \APACyear {2013}}%
}]{%
Hayward_book}
\APACinsertmetastar {%
Hayward_book}%
\begin{APACrefauthors}%
Hayward, S\BPBI A.%
\end{APACrefauthors}%
\unskip\
\newblock
\APACrefYear{2013}.
\newblock
\APACrefbtitle {{Black Holes}} {{Black Holes}}.
\newblock
\APACaddressPublisher{}{WORLD SCIENTIFIC}.
\newblock
\begin{APACrefURL} \url{https://www.worldscientific.com/doi/abs/10.1142/8604}
  \end{APACrefURL}
\newblock
\begin{APACrefDOI} \doi{10.1142/8604} \end{APACrefDOI}
\PrintBackRefs{\CurrentBib}

\bibitem [\protect \citeauthoryear {%
Healey%
}{%
Healey%
}{%
{\protect \APACyear {1997}}%
}]{%
healey1997ab}
\APACinsertmetastar {%
healey1997ab}%
\begin{APACrefauthors}%
Healey, R.%
\end{APACrefauthors}%
\unskip\
\newblock
\APACrefYearMonthDay{1997}{}{}.
\newblock
{\BBOQ}\APACrefatitle {Nonlocality and the {A}haronov-{B}ohm effect}
  {Nonlocality and the {A}haronov-{B}ohm effect}.{\BBCQ}
\newblock
\APACjournalVolNumPages{Philosophy of Science}{64}{1}{18--41}.
\PrintBackRefs{\CurrentBib}

\bibitem [\protect \citeauthoryear {%
Healey%
}{%
Healey%
}{%
{\protect \APACyear {1999}}%
}]{%
healey1999ab}
\APACinsertmetastar {%
healey1999ab}%
\begin{APACrefauthors}%
Healey, R.%
\end{APACrefauthors}%
\unskip\
\newblock
\APACrefYearMonthDay{1999}{}{}.
\newblock
{\BBOQ}\APACrefatitle {Quantum analogies: {A} reply to {M}audlin} {Quantum
  analogies: {A} reply to {M}audlin}.{\BBCQ}
\newblock
\APACjournalVolNumPages{Philosophy of Science}{66}{3}{440--447}.
\PrintBackRefs{\CurrentBib}

\bibitem [\protect \citeauthoryear {%
Healey%
}{%
Healey%
}{%
{\protect \APACyear {2004}}%
}]{%
Healey2004}
\APACinsertmetastar {%
Healey2004}%
\begin{APACrefauthors}%
Healey, R.%
\end{APACrefauthors}%
\unskip\
\newblock
\APACrefYearMonthDay{2004}{}{}.
\newblock
{\BBOQ}\APACrefatitle {Gauge theories and holisms} {Gauge theories and
  holisms}.{\BBCQ}
\newblock
\APACjournalVolNumPages{Studies in History and Philosophy of Science Part B:
  Studies in History and Philosophy of Modern Physics}{35}{4}{619-642}.
\newblock
\begin{APACrefURL}
  \url{https://www.sciencedirect.com/science/article/pii/S1355219804000553}
  \end{APACrefURL}
\newblock
\begin{APACrefDOI} \doi{https://doi.org/10.1016/j.shpsb.2004.07.003}
  \end{APACrefDOI}
\PrintBackRefs{\CurrentBib}

\bibitem [\protect \citeauthoryear {%
Healey%
}{%
Healey%
}{%
{\protect \APACyear {2007}}%
}]{%
Healey_book}
\APACinsertmetastar {%
Healey_book}%
\begin{APACrefauthors}%
Healey, R.%
\end{APACrefauthors}%
\unskip\
\newblock
\APACrefYear{2007}.
\newblock
\APACrefbtitle {{Gauging What's Real: The Conceptual Foundations of Gauge
  Theories}} {{Gauging What's Real: The Conceptual Foundations of Gauge
  Theories}}.
\newblock
\APACaddressPublisher{}{Oxford University Press}.
\PrintBackRefs{\CurrentBib}

\bibitem [\protect \citeauthoryear {%
Healey%
}{%
Healey%
}{%
{\protect \APACyear {2009}}%
}]{%
Healey2009}
\APACinsertmetastar {%
Healey2009}%
\begin{APACrefauthors}%
Healey, R.%
\end{APACrefauthors}%
\unskip\
\newblock
\APACrefYearMonthDay{2009}{08}{}.
\newblock
{\BBOQ}\APACrefatitle {{Perfect Symmetries}} {{Perfect Symmetries}}.{\BBCQ}
\newblock
\APACjournalVolNumPages{The British Journal for the Philosophy of
  Science}{60}{4}{697-720}.
\newblock
\begin{APACrefURL} \url{https://doi.org/10.1093/bjps/axp033} \end{APACrefURL}
\newblock
\begin{APACrefDOI} \doi{10.1093/bjps/axp033} \end{APACrefDOI}
\PrintBackRefs{\CurrentBib}

\bibitem [\protect \citeauthoryear {%
Heisenberg%
}{%
Heisenberg%
}{%
{\protect \APACyear {1971}}%
}]{%
Heisenberg_dial}
\APACinsertmetastar {%
Heisenberg_dial}%
\begin{APACrefauthors}%
Heisenberg, W.%
\end{APACrefauthors}%
\unskip\
\newblock
\APACrefYear{1971}.
\newblock
\APACrefbtitle {{Physics and Beyond}} {{Physics and Beyond}}\ (t.~Arnold
  J.~Pomerans, \BED{}).
\newblock
\APACaddressPublisher{}{Harper, New York}.
\PrintBackRefs{\CurrentBib}

\bibitem [\protect \citeauthoryear {%
Henneaux%
\ \BBA {} Teitelboim%
}{%
Henneaux%
\ \BBA {} Teitelboim%
}{%
{\protect \APACyear {1985}}%
}]{%
Asymp_ads}
\APACinsertmetastar {%
Asymp_ads}%
\begin{APACrefauthors}%
Henneaux, M.%
\BCBT {}\ \BBA {} Teitelboim, C.%
\end{APACrefauthors}%
\unskip\
\newblock
\APACrefYearMonthDay{1985}{}{}.
\newblock
{\BBOQ}\APACrefatitle {{Asymptotically anti-de Sitter spaces}} {{Asymptotically
  anti-de Sitter spaces}}.{\BBCQ}
\newblock
\APACjournalVolNumPages{Communications in Mathematical Physics, 98}{}{}{}.
\PrintBackRefs{\CurrentBib}

\bibitem [\protect \citeauthoryear {%
Henneaux%
\ \BBA {} Teitelboim%
}{%
Henneaux%
\ \BBA {} Teitelboim%
}{%
{\protect \APACyear {1992}}%
}]{%
HenneauxTeitelboim}
\APACinsertmetastar {%
HenneauxTeitelboim}%
\begin{APACrefauthors}%
Henneaux, M.%
\BCBT {}\ \BBA {} Teitelboim, C.%
\end{APACrefauthors}%
\unskip\
\newblock
\APACrefYear{1992}.
\newblock
\APACrefbtitle {Quantization of Gauge Systems} {Quantization of gauge systems}.
\newblock
\APACaddressPublisher{}{Princeton University Press}.
\PrintBackRefs{\CurrentBib}

\bibitem [\protect \citeauthoryear {%
Henneaux%
\ \BBA {} Troessaert%
}{%
Henneaux%
\ \BBA {} Troessaert%
}{%
{\protect \APACyear {2018}}%
}]{%
Henneaux:2018gfi}
\APACinsertmetastar {%
Henneaux:2018gfi}%
\begin{APACrefauthors}%
Henneaux, M.%
\BCBT {}\ \BBA {} Troessaert, C.%
\end{APACrefauthors}%
\unskip\
\newblock
\APACrefYearMonthDay{2018}{}{}.
\newblock
{\BBOQ}\APACrefatitle {{Asymptotic symmetries of electromagnetism at spatial
  infinity}} {{Asymptotic symmetries of electromagnetism at spatial
  infinity}}.{\BBCQ}
\newblock
\APACjournalVolNumPages{JHEP}{05}{}{137}.
\newblock
\begin{APACrefDOI} \doi{10.1007/JHEP05(2018)137} \end{APACrefDOI}
\PrintBackRefs{\CurrentBib}

\bibitem [\protect \citeauthoryear {%
Henneaux%
\ \BBA {} Troessaert%
}{%
Henneaux%
\ \BBA {} Troessaert%
}{%
{\protect \APACyear {2019}}%
}]{%
Henneaux_rev}
\APACinsertmetastar {%
Henneaux_rev}%
\begin{APACrefauthors}%
Henneaux, M.%
\BCBT {}\ \BBA {} Troessaert, C.%
\end{APACrefauthors}%
\unskip\
\newblock
\APACrefYearMonthDay{2019}{}{}.
\newblock
{\BBOQ}\APACrefatitle {{The asymptotic structure of gravity at spatial infinity
  in four spacetime dimensions}} {{The asymptotic structure of gravity at
  spatial infinity in four spacetime dimensions}}.{\BBCQ}
\newblock

\PrintBackRefs{\CurrentBib}

\bibitem [\protect \citeauthoryear {%
Hetzroni%
}{%
Hetzroni%
}{%
{\protect \APACyear {2021}}%
}]{%
Guy_ghosts}
\APACinsertmetastar {%
Guy_ghosts}%
\begin{APACrefauthors}%
Hetzroni, G.%
\end{APACrefauthors}%
\unskip\
\newblock
\APACrefYearMonthDay{2021}{}{}.
\newblock
{\BBOQ}\APACrefatitle {{Gauge and Ghosts}} {{Gauge and Ghosts}}.{\BBCQ}
\newblock
\APACjournalVolNumPages{The British Journal for the Philosophy of
  Science}{72}{3}{773-796}.
\newblock
\begin{APACrefURL} \url{https://doi.org/10.1093/bjps/axz021} \end{APACrefURL}
\newblock
\begin{APACrefDOI} \doi{10.1093/bjps/axz021} \end{APACrefDOI}
\PrintBackRefs{\CurrentBib}

\bibitem [\protect \citeauthoryear {%
Hiley%
}{%
Hiley%
}{%
{\protect \APACyear {2013}}%
}]{%
hiley2013ABeffect}
\APACinsertmetastar {%
hiley2013ABeffect}%
\begin{APACrefauthors}%
Hiley, B.%
\end{APACrefauthors}%
\unskip\
\newblock
\APACrefYearMonthDay{2013}{}{}.
\newblock
{\BBOQ}\APACrefatitle {{The early history of the Aharonov-Bohm effect}} {{The
  early history of the Aharonov-Bohm effect}}.{\BBCQ}
\newblock
\APACjournalVolNumPages{arXiv preprint arXiv:1304.4736}{}{}{}.
\PrintBackRefs{\CurrentBib}

\bibitem [\protect \citeauthoryear {%
Hoefer%
}{%
Hoefer%
}{%
{\protect \APACyear {1996}}%
}]{%
Hoefer_hole}
\APACinsertmetastar {%
Hoefer_hole}%
\begin{APACrefauthors}%
Hoefer, C.%
\end{APACrefauthors}%
\unskip\
\newblock
\APACrefYearMonthDay{1996}{}{}.
\newblock
{\BBOQ}\APACrefatitle {{The Metaphysics of Space-Time Substantivalism}} {{The
  Metaphysics of Space-Time Substantivalism}}.{\BBCQ}
\newblock
\APACjournalVolNumPages{The Journal of Philosophy}{93}{1}{5--27}.
\newblock
\begin{APACrefURL} \url{http://www.jstor.org/stable/2941016} \end{APACrefURL}
\PrintBackRefs{\CurrentBib}

\bibitem [\protect \citeauthoryear {%
Iftime%
\ \BBA {} Stachel%
}{%
Iftime%
\ \BBA {} Stachel%
}{%
{\protect \APACyear {2006}}%
}]{%
Stachel_Iftime_short}
\APACinsertmetastar {%
Stachel_Iftime_short}%
\begin{APACrefauthors}%
Iftime, M.%
\BCBT {}\ \BBA {} Stachel, J.%
\end{APACrefauthors}%
\unskip\
\newblock
\APACrefYearMonthDay{2006}{}{}.
\newblock
{\BBOQ}\APACrefatitle {The hole argument for covariant theories} {The hole
  argument for covariant theories}.{\BBCQ}
\newblock
\APACjournalVolNumPages{General Relativity and Gravitation \bf{38},
  1241–1252}{}{}{}.
\PrintBackRefs{\CurrentBib}

\bibitem [\protect \citeauthoryear {%
Isenberg%
\ \BBA {} Marsden%
}{%
Isenberg%
\ \BBA {} Marsden%
}{%
{\protect \APACyear {1982}}%
}]{%
isenberg1982slice}
\APACinsertmetastar {%
isenberg1982slice}%
\begin{APACrefauthors}%
Isenberg, J.%
\BCBT {}\ \BBA {} Marsden, J\BPBI E.%
\end{APACrefauthors}%
\unskip\
\newblock
\APACrefYearMonthDay{1982}{}{}.
\newblock
{\BBOQ}\APACrefatitle {A slice theorem for the space of solutions of Einstein's
  equations} {A slice theorem for the space of solutions of einstein's
  equations}.{\BBCQ}
\newblock
\APACjournalVolNumPages{Physics Reports}{89}{2}{179--222}.
\PrintBackRefs{\CurrentBib}

\bibitem [\protect \citeauthoryear {%
Isham%
}{%
Isham%
}{%
{\protect \APACyear {1992}}%
}]{%
Isham_POT}
\APACinsertmetastar {%
Isham_POT}%
\begin{APACrefauthors}%
Isham, C\BPBI J.%
\end{APACrefauthors}%
\unskip\
\newblock
\APACrefYearMonthDay{1992}{}{}.
\newblock
{\BBOQ}\APACrefatitle {{Canonical quantum gravity and the problem of time}}
  {{Canonical quantum gravity and the problem of time}}.{\BBCQ}
\newblock
\BIn{} \APACrefbtitle {{19th International Colloquium on Group Theoretical
  Methods in Physics (GROUP 19) Salamanca, Spain, June 29-July 5, 1992}.}
  {{19th International Colloquium on Group Theoretical Methods in Physics
  (GROUP 19) Salamanca, Spain, June 29-July 5, 1992}.}
\PrintBackRefs{\CurrentBib}

\bibitem [\protect \citeauthoryear {%
Itzykson%
\ \BBA {} Zuber%
}{%
Itzykson%
\ \BBA {} Zuber%
}{%
{\protect \APACyear {1980}}%
}]{%
ItzyksonZuber}
\APACinsertmetastar {%
ItzyksonZuber}%
\begin{APACrefauthors}%
Itzykson, C.%
\BCBT {}\ \BBA {} Zuber, J\BHBI B.%
\end{APACrefauthors}%
\unskip\
\newblock
\APACrefYear{1980}.
\newblock
\APACrefbtitle {{Quantum Field Theory}} {{Quantum Field Theory}}.
\newblock
\APACaddressPublisher{}{McGraw-Hill Inc.}
\PrintBackRefs{\CurrentBib}

\bibitem [\protect \citeauthoryear {%
Jackson%
}{%
Jackson%
}{%
{\protect \APACyear {1975}}%
}]{%
JacksonBook}
\APACinsertmetastar {%
JacksonBook}%
\begin{APACrefauthors}%
Jackson, J\BPBI D.%
\end{APACrefauthors}%
\unskip\
\newblock
\APACrefYear{1975}.
\newblock
\APACrefbtitle {{Classical electrodynamics; 2nd ed.}} {{Classical
  electrodynamics; 2nd ed.}}
\newblock
\APACaddressPublisher{New York, NY}{Wiley}.
\newblock
\begin{APACrefURL} \url{https://cds.cern.ch/record/100964} \end{APACrefURL}
\PrintBackRefs{\CurrentBib}

\bibitem [\protect \citeauthoryear {%
Jacobs%
}{%
Jacobs%
}{%
{\protect \APACyear {2021}}%
{\protect \APACexlab {{\protect \BCnt {1}}}}}]{%
Jacobs_Inv}
\APACinsertmetastar {%
Jacobs_Inv}%
\begin{APACrefauthors}%
Jacobs, C.%
\end{APACrefauthors}%
\unskip\
\newblock
\APACrefYearMonthDay{2021{\protect \BCnt {1}}}{}{}.
\newblock
{\BBOQ}\APACrefatitle {{Invariance or Equivalence: A Tale of Two Principles}}
  {{Invariance or Equivalence: A Tale of Two Principles}}.{\BBCQ}
\newblock
\APACjournalVolNumPages{Synthese}{}{}{1--21}.
\newblock
\begin{APACrefDOI} \doi{10.1007/s11229-021-03205-5} \end{APACrefDOI}
\PrintBackRefs{\CurrentBib}

\bibitem [\protect \citeauthoryear {%
Jacobs%
}{%
Jacobs%
}{%
{\protect \APACyear {2021}}%
{\protect \APACexlab {{\protect \BCnt {2}}}}}]{%
Jacobs_thesis}
\APACinsertmetastar {%
Jacobs_thesis}%
\begin{APACrefauthors}%
Jacobs, C.%
\end{APACrefauthors}%
\unskip\
\newblock
\APACrefYear{2021{\protect \BCnt {2}}}.
\unskip\
\newblock
\APACrefbtitle {{Symmetries as a Guide to theStructure of Physical Quantities}}
  {{Symmetries as a Guide to theStructure of Physical Quantities}}\
  \APACtypeAddressSchool {\BUPhD}{}{}.
\unskip\
\newblock
\APACaddressSchool {}{University of Oxford}.
\PrintBackRefs{\CurrentBib}

\bibitem [\protect \citeauthoryear {%
Jacobson%
\ \BBA {} Nguyen%
}{%
Jacobson%
\ \BBA {} Nguyen%
}{%
{\protect \APACyear {2019}}%
}]{%
Jacobson_2019}
\APACinsertmetastar {%
Jacobson_2019}%
\begin{APACrefauthors}%
Jacobson, T.%
\BCBT {}\ \BBA {} Nguyen, P.%
\end{APACrefauthors}%
\unskip\
\newblock
\APACrefYearMonthDay{2019}{Aug}{}.
\newblock
{\BBOQ}\APACrefatitle {Diffeomorphism invariance and the black hole information
  paradox} {Diffeomorphism invariance and the black hole information
  paradox}.{\BBCQ}
\newblock
\APACjournalVolNumPages{Physical Review D}{100}{4}{}.
\newblock
\begin{APACrefURL} \url{http://dx.doi.org/10.1103/PhysRevD.100.046002}
  \end{APACrefURL}
\newblock
\begin{APACrefDOI} \doi{10.1103/physrevd.100.046002} \end{APACrefDOI}
\PrintBackRefs{\CurrentBib}

\bibitem [\protect \citeauthoryear {%
Janssen%
\ \BBA {} Renn%
}{%
Janssen%
\ \BBA {} Renn%
}{%
{\protect \APACyear {2015}}%
}]{%
Janssen}
\APACinsertmetastar {%
Janssen}%
\begin{APACrefauthors}%
Janssen, M.%
\BCBT {}\ \BBA {} Renn, J.%
\end{APACrefauthors}%
\unskip\
\newblock
\APACrefYearMonthDay{2015}{}{}.
\newblock
{\BBOQ}\APACrefatitle {{Arch and scaffold: How Einstein found his field
  equations}} {{Arch and scaffold: How Einstein found his field
  equations}}.{\BBCQ}
\newblock
\APACjournalVolNumPages{Physics Today}{68}{11}{30-36}.
\newblock
\begin{APACrefURL} \url{https://doi.org/10.1063/PT.3.2979} \end{APACrefURL}
\newblock
\begin{APACrefDOI} \doi{10.1063/PT.3.2979} \end{APACrefDOI}
\PrintBackRefs{\CurrentBib}

\bibitem [\protect \citeauthoryear {%
Kaluza%
}{%
Kaluza%
}{%
{\protect \APACyear {1921}}%
}]{%
KaluzaKlein}
\APACinsertmetastar {%
KaluzaKlein}%
\begin{APACrefauthors}%
Kaluza, T.%
\end{APACrefauthors}%
\unskip\
\newblock
\APACrefYearMonthDay{1921}{{\APACmonth{01}}}{}.
\newblock
{\BBOQ}\APACrefatitle {{Zum Unit{\"a}tsproblem der Physik}} {{Zum
  Unit{\"a}tsproblem der Physik}}.{\BBCQ}
\newblock
\APACjournalVolNumPages{Sitzungsberichte der K{\"o}niglich Preu{\ss}ischen
  Akademie der Wissenschaften (Berlin}{}{}{966-972}.
\PrintBackRefs{\CurrentBib}

\bibitem [\protect \citeauthoryear {%
Kelly%
\ \BBA {} Marolf%
}{%
Kelly%
\ \BBA {} Marolf%
}{%
{\protect \APACyear {2012}}%
}]{%
Kelly_2012}
\APACinsertmetastar {%
Kelly_2012}%
\begin{APACrefauthors}%
Kelly, W.%
\BCBT {}\ \BBA {} Marolf, D.%
\end{APACrefauthors}%
\unskip\
\newblock
\APACrefYearMonthDay{2012}{sep}{}.
\newblock
{\BBOQ}\APACrefatitle {{Phase spaces for asymptotically de Sitter cosmologies}}
  {{Phase spaces for asymptotically de Sitter cosmologies}}.{\BBCQ}
\newblock
\APACjournalVolNumPages{Classical and Quantum Gravity}{29}{20}{205013}.
\newblock
\begin{APACrefURL} \url{https://doi.org/10.1088/0264-9381/29/20/205013}
  \end{APACrefURL}
\newblock
\begin{APACrefDOI} \doi{10.1088/0264-9381/29/20/205013} \end{APACrefDOI}
\PrintBackRefs{\CurrentBib}

\bibitem [\protect \citeauthoryear {%
F.~Klein%
}{%
F.~Klein%
}{%
{\protect \APACyear {1893}}%
}]{%
Klein_erlangen}
\APACinsertmetastar {%
Klein_erlangen}%
\begin{APACrefauthors}%
Klein, F.%
\end{APACrefauthors}%
\unskip\
\newblock
\APACrefYearMonthDay{1893}{}{}.
\newblock
{\BBOQ}\APACrefatitle {{Vergleichende Betrachtungen über neuere geometrische
  Forschungen}} {{Vergleichende Betrachtungen über neuere geometrische
  Forschungen}}.{\BBCQ}
\newblock
\APACjournalVolNumPages{Math. Ann. 43}{}{}{}.
\PrintBackRefs{\CurrentBib}

\bibitem [\protect \citeauthoryear {%
O.~Klein%
}{%
O.~Klein%
}{%
{\protect \APACyear {1986}}%
}]{%
Klein1938}
\APACinsertmetastar {%
Klein1938}%
\begin{APACrefauthors}%
Klein, O.%
\end{APACrefauthors}%
\unskip\
\newblock
\APACrefYearMonthDay{1986}{}{}.
\newblock
{\BBOQ}\APACrefatitle {On the theory of charged fields} {On the theory of
  charged fields}.{\BBCQ}
\newblock
\APACjournalVolNumPages{Surveys in High Energy Physics}{5}{3}{269-285}.
\newblock
\begin{APACrefURL} \url{https://doi.org/10.1080/01422418608228775}
  \end{APACrefURL}
\newblock
\begin{APACrefDOI} \doi{10.1080/01422418608228775} \end{APACrefDOI}
\PrintBackRefs{\CurrentBib}

\bibitem [\protect \citeauthoryear {%
Kobayaschi%
}{%
Kobayaschi%
}{%
{\protect \APACyear {1957}}%
}]{%
Kobayaschi_bundle}
\APACinsertmetastar {%
Kobayaschi_bundle}%
\begin{APACrefauthors}%
Kobayaschi, S.%
\end{APACrefauthors}%
\unskip\
\newblock
\APACrefYearMonthDay{1957}{}{}.
\newblock
{\BBOQ}\APACrefatitle {Theory of connections} {Theory of connections}.{\BBCQ}
\newblock
\APACjournalVolNumPages{Annali di Matematica 43, 119–194}{}{}{}.
\PrintBackRefs{\CurrentBib}

\bibitem [\protect \citeauthoryear {%
Kobayashi%
\ \BBA {} Nomizu%
}{%
Kobayashi%
\ \BBA {} Nomizu%
}{%
{\protect \APACyear {1963}}%
}]{%
kobayashivol1}
\APACinsertmetastar {%
kobayashivol1}%
\begin{APACrefauthors}%
Kobayashi, S.%
\BCBT {}\ \BBA {} Nomizu, K.%
\end{APACrefauthors}%
\unskip\
\newblock
\APACrefYear{1963}.
\newblock
\APACrefbtitle {Foundations of differential geometry. {V}ol {I}} {Foundations
  of differential geometry. {V}ol {I}}.
\newblock
\APACaddressPublisher{}{Interscience Publishers, a division of John Wiley \&
  Sons, New York-Lond on}.
\PrintBackRefs{\CurrentBib}

\bibitem [\protect \citeauthoryear {%
Kolar%
, Michor%
\BCBL {}\ \BBA {} Slovak%
}{%
Kolar%
\ \protect \BOthers {.}}{%
{\protect \APACyear {1993}}%
}]{%
Kolar_book}
\APACinsertmetastar {%
Kolar_book}%
\begin{APACrefauthors}%
Kolar, I.%
, Michor, P.%
\BCBL {}\ \BBA {} Slovak, J.%
\end{APACrefauthors}%
\unskip\
\newblock
\APACrefYear{1993}.
\newblock
\APACrefbtitle {{Natural Operations in Differential Geometry}} {{Natural
  Operations in Differential Geometry}}.
\newblock
\APACaddressPublisher{}{Springer}.
\PrintBackRefs{\CurrentBib}

\bibitem [\protect \citeauthoryear {%
Komar%
}{%
Komar%
}{%
{\protect \APACyear {1958}}%
}]{%
Komar_inv}
\APACinsertmetastar {%
Komar_inv}%
\begin{APACrefauthors}%
Komar, A.%
\end{APACrefauthors}%
\unskip\
\newblock
\APACrefYearMonthDay{1958}{Aug}{}.
\newblock
{\BBOQ}\APACrefatitle {{Construction of a Complete Set of Independent
  Observables in the General Theory of Relativity}} {{Construction of a
  Complete Set of Independent Observables in the General Theory of
  Relativity}}.{\BBCQ}
\newblock
\APACjournalVolNumPages{Physical Review}{111}{}{1182--1187}.
\newblock
\begin{APACrefURL} \url{https://link.aps.org/doi/10.1103/PhysRev.111.1182}
  \end{APACrefURL}
\newblock
\begin{APACrefDOI} \doi{10.1103/PhysRev.111.1182} \end{APACrefDOI}
\PrintBackRefs{\CurrentBib}

\bibitem [\protect \citeauthoryear {%
Kondracki%
\ \BBA {} Rogulski%
}{%
Kondracki%
\ \BBA {} Rogulski%
}{%
{\protect \APACyear {1983}}%
}]{%
kondracki1983}
\APACinsertmetastar {%
kondracki1983}%
\begin{APACrefauthors}%
Kondracki, W.%
\BCBT {}\ \BBA {} Rogulski, J.%
\end{APACrefauthors}%
\unskip\
\newblock
\APACrefYear{1983}.
\newblock
\APACrefbtitle {On the Stratification of the Orbit Space for the Action of
  Automorphisms on Connections. On Conjugacy Classes of Closed Subgroups. On
  the Notion of Stratification} {On the stratification of the orbit space for
  the action of automorphisms on connections. on conjugacy classes of closed
  subgroups. on the notion of stratification}.
\newblock
\APACaddressPublisher{}{Inst., Acad.}
\newblock
\begin{APACrefURL} \url{https://books.google.co.uk/books?id=LK0JrgEACAAJ}
  \end{APACrefURL}
\PrintBackRefs{\CurrentBib}

\bibitem [\protect \citeauthoryear {%
Kosmann-Schwarzbach%
}{%
Kosmann-Schwarzbach%
}{%
{\protect \APACyear {2011}}%
}]{%
ks2011noether}
\APACinsertmetastar {%
ks2011noether}%
\begin{APACrefauthors}%
Kosmann-Schwarzbach, Y.%
\end{APACrefauthors}%
\unskip\
\newblock
\APACrefYear{2011}.
\newblock
\APACrefbtitle {{The Noether Theorems: Invariance and Conservation Laws in the
  Twentieth Century}} {{The Noether Theorems: Invariance and Conservation Laws
  in the Twentieth Century}}.
\newblock
\APACaddressPublisher{}{New York: Springer Science+Business Media, LLC}.
\newblock
\APACrefnote{Translated by Bertram E. Schwarzbach}
\PrintBackRefs{\CurrentBib}

\bibitem [\protect \citeauthoryear {%
Kosso%
}{%
Kosso%
}{%
{\protect \APACyear {2000}}%
}]{%
Kosso}
\APACinsertmetastar {%
Kosso}%
\begin{APACrefauthors}%
Kosso, P.%
\end{APACrefauthors}%
\unskip\
\newblock
\APACrefYearMonthDay{2000}{}{}.
\newblock
{\BBOQ}\APACrefatitle {The Empirical Status of Symmetries in Physics} {The
  empirical status of symmetries in physics}.{\BBCQ}
\newblock
\APACjournalVolNumPages{The British Journal for the Philosophy of
  Science}{51}{1}{81--98}.
\newblock
\begin{APACrefURL} \url{http://www.jstor.org/stable/3541749} \end{APACrefURL}
\PrintBackRefs{\CurrentBib}

\bibitem [\protect \citeauthoryear {%
Kragh%
}{%
Kragh%
}{%
{\protect \APACyear {1999}}%
}]{%
Kragh1999}
\APACinsertmetastar {%
Kragh1999}%
\begin{APACrefauthors}%
Kragh, H.%
\end{APACrefauthors}%
\unskip\
\newblock
\APACrefYear{1999}.
\newblock
\APACrefbtitle {{Quantum Generations}} {{Quantum Generations}}.
\newblock
\APACaddressPublisher{}{Princeton University Press}.
\PrintBackRefs{\CurrentBib}

\bibitem [\protect \citeauthoryear {%
Kretschmann%
}{%
Kretschmann%
}{%
{\protect \APACyear {1918}}%
}]{%
Kretschmann_inv}
\APACinsertmetastar {%
Kretschmann_inv}%
\begin{APACrefauthors}%
Kretschmann, E.%
\end{APACrefauthors}%
\unskip\
\newblock
\APACrefYearMonthDay{1918}{}{}.
\newblock
{\BBOQ}\APACrefatitle {{Über den physikalischen Sinn der
  Relativitätspostulate, A. Einsteins neue und seine ursprüngliche
  Relativitätstheorie}} {{Über den physikalischen Sinn der
  Relativitätspostulate, A. Einsteins neue und seine ursprüngliche
  Relativitätstheorie}}.{\BBCQ}
\newblock
\APACjournalVolNumPages{Annalen der Physik}{358}{16}{575-614}.
\newblock
\begin{APACrefURL}
  \url{https://onlinelibrary.wiley.com/doi/abs/10.1002/andp.19183581602}
  \end{APACrefURL}
\newblock
\begin{APACrefDOI} \doi{https://doi.org/10.1002/andp.19183581602}
  \end{APACrefDOI}
\PrintBackRefs{\CurrentBib}

\bibitem [\protect \citeauthoryear {%
Kriegl%
\ \BBA {} Michor%
}{%
Kriegl%
\ \BBA {} Michor%
}{%
{\protect \APACyear {1997}}%
}]{%
Michor}
\APACinsertmetastar {%
Michor}%
\begin{APACrefauthors}%
Kriegl, A.%
\BCBT {}\ \BBA {} Michor, P\BPBI W.%
\end{APACrefauthors}%
\unskip\
\newblock
\APACrefYear{1997}.
\newblock
\APACrefbtitle {{The Convenient Setting of Global Analysis}} {{The Convenient
  Setting of Global Analysis}}.
\newblock
\APACaddressPublisher{}{Mathematical Surveys and Monographs}.
\newblock
\begin{APACrefURL}
  \url{https://www.google.com/url?sa=t{\&}rct=j{\&}q={\&}esrc=s{\&}source=web{\&}cd=1{\&}cad=rja{\&}uact=8{\&}ved=0ahUKEwishv{\_}6seTOAhUElR4KHZkXAq8QFggmMAA{\&}url=http{\%}3A{\%}2F{\%}2Fwww.mat.univie.ac.at{\%}2F{~}michor{\%}2Fapbookh-ams.pdf{\&}usg=AFQjCNGBbW7PYwbWLS329sZJtvIQXI8Gfw{\&}sig2=HeiuQD2v23zbfVXcyU3zIA}
  \end{APACrefURL}
\newblock
\begin{APACrefDOI} \doi{http://dx.doi.org/10.1090/surv/053} \end{APACrefDOI}
\PrintBackRefs{\CurrentBib}

\bibitem [\protect \citeauthoryear {%
Kripke%
}{%
Kripke%
}{%
{\protect \APACyear {1982}}%
}]{%
Kripke_naming}
\APACinsertmetastar {%
Kripke_naming}%
\begin{APACrefauthors}%
Kripke, S.%
\end{APACrefauthors}%
\unskip\
\newblock
\APACrefYear{1982}.
\newblock
\APACrefbtitle {{Naming and Necessity}} {{Naming and Necessity}}.
\newblock
\APACaddressPublisher{}{Wiley-Blackwell}.
\PrintBackRefs{\CurrentBib}

\bibitem [\protect \citeauthoryear {%
Kucha\v{r}%
}{%
Kucha\v{r}%
}{%
{\protect \APACyear {2011}}%
}]{%
Kuchar_Time}
\APACinsertmetastar {%
Kuchar_Time}%
\begin{APACrefauthors}%
Kucha\v{r}, K.%
\end{APACrefauthors}%
\unskip\
\newblock
\APACrefYearMonthDay{2011}{}{}.
\newblock
{\BBOQ}\APACrefatitle {TIME AND INTERPRETATIONS OF QUANTUM GRAVITY} {Time and
  interpretations of quantum gravity}.{\BBCQ}
\newblock
\APACjournalVolNumPages{International Journal of Modern Physics
  D}{20}{supp01}{3-86}.
\newblock
\begin{APACrefURL}
  \url{http://www.worldscientific.com/doi/abs/10.1142/S0218271811019347}
  \end{APACrefURL}
\newblock
\begin{APACrefDOI} \doi{10.1142/S0218271811019347} \end{APACrefDOI}
\PrintBackRefs{\CurrentBib}

\bibitem [\protect \citeauthoryear {%
Ladyman%
}{%
Ladyman%
}{%
{\protect \APACyear {2015}}%
}]{%
Ladyman_DES}
\APACinsertmetastar {%
Ladyman_DES}%
\begin{APACrefauthors}%
Ladyman, J.%
\end{APACrefauthors}%
\unskip\
\newblock
\APACrefYearMonthDay{2015}{}{}.
\newblock
{\BBOQ}\APACrefatitle {Representation and Symmetry in Physics} {Representation
  and symmetry in physics}.{\BBCQ}
\newblock
\APACjournalVolNumPages{unpublished}{}{}{}.
\PrintBackRefs{\CurrentBib}

\bibitem [\protect \citeauthoryear {%
Landsman%
}{%
Landsman%
}{%
{\protect \APACyear {2021}}%
}]{%
Landsman_GR}
\APACinsertmetastar {%
Landsman_GR}%
\begin{APACrefauthors}%
Landsman, K.%
\end{APACrefauthors}%
\unskip\
\newblock
\APACrefYear{2021}.
\newblock
\APACrefbtitle {{Foundations of General Relativity}} {{Foundations of General
  Relativity}}.
\newblock
\APACaddressPublisher{}{Radboud University Press}.
\PrintBackRefs{\CurrentBib}

\bibitem [\protect \citeauthoryear {%
Lang%
}{%
Lang%
}{%
{\protect \APACyear {1999}}%
}]{%
Lang_book}
\APACinsertmetastar {%
Lang_book}%
\begin{APACrefauthors}%
Lang, S.%
\end{APACrefauthors}%
\unskip\
\newblock
\APACrefYear{1999}.
\newblock
\APACrefbtitle {{Fundamentals of differential geometry. Graduate Texts in
  Mathematics,191.}} {{Fundamentals of differential geometry. Graduate Texts in
  Mathematics,191.}}
\newblock
\APACaddressPublisher{}{Springer, New York}.
\PrintBackRefs{\CurrentBib}

\bibitem [\protect \citeauthoryear {%
Lange%
}{%
Lange%
}{%
{\protect \APACyear {2007}}%
}]{%
Lange_meta}
\APACinsertmetastar {%
Lange_meta}%
\begin{APACrefauthors}%
Lange, M.%
\end{APACrefauthors}%
\unskip\
\newblock
\APACrefYearMonthDay{2007}{}{}.
\newblock
{\BBOQ}\APACrefatitle {Laws and meta-laws of nature: Conservation laws and
  symmetries} {Laws and meta-laws of nature: Conservation laws and
  symmetries}.{\BBCQ}
\newblock
\APACjournalVolNumPages{Studies in History and Philosophy of Science Part B:
  Studies in History and Philosophy of Modern Physics}{38}{3}{457-481}.
\newblock
\begin{APACrefURL}
  \url{https://www.sciencedirect.com/science/article/pii/S1355219806000943}
  \end{APACrefURL}
\newblock
\begin{APACrefDOI} \doi{https://doi.org/10.1016/j.shpsb.2006.08.003}
  \end{APACrefDOI}
\PrintBackRefs{\CurrentBib}

\bibitem [\protect \citeauthoryear {%
Lee%
\ \BBA {} Wald%
}{%
Lee%
\ \BBA {} Wald%
}{%
{\protect \APACyear {1990}}%
}]{%
Lee:1990nz}
\APACinsertmetastar {%
Lee:1990nz}%
\begin{APACrefauthors}%
Lee, J.%
\BCBT {}\ \BBA {} Wald, R\BPBI M.%
\end{APACrefauthors}%
\unskip\
\newblock
\APACrefYearMonthDay{1990}{}{}.
\newblock
{\BBOQ}\APACrefatitle {{Local symmetries and constraints}} {{Local symmetries
  and constraints}}.{\BBCQ}
\newblock
\APACjournalVolNumPages{J. Math. Phys.}{31}{}{725-743}.
\newblock
\begin{APACrefDOI} \doi{10.1063/1.528801} \end{APACrefDOI}
\PrintBackRefs{\CurrentBib}

\bibitem [\protect \citeauthoryear {%
Lehmkuhl%
}{%
Lehmkuhl%
}{%
{\protect \APACyear {2018}}%
}]{%
Lehmkuhl2018}
\APACinsertmetastar {%
Lehmkuhl2018}%
\begin{APACrefauthors}%
Lehmkuhl, D.%
\end{APACrefauthors}%
\unskip\
\newblock
\APACrefYearMonthDay{2018}{}{}.
\newblock
{\BBOQ}\APACrefatitle {{The Metaphysics of Super-Substantivalism}} {{The
  Metaphysics of Super-Substantivalism}}.{\BBCQ}
\newblock
\APACjournalVolNumPages{Noûs}{52}{1}{24-46}.
\newblock
\begin{APACrefURL}
  \url{https://onlinelibrary.wiley.com/doi/abs/10.1111/nous.12163}
  \end{APACrefURL}
\newblock
\begin{APACrefDOI} \doi{https://doi.org/10.1111/nous.12163} \end{APACrefDOI}
\PrintBackRefs{\CurrentBib}

\bibitem [\protect \citeauthoryear {%
D.~Lewis%
}{%
D.~Lewis%
}{%
{\protect \APACyear {1970}}%
}]{%
Lewis_defT}
\APACinsertmetastar {%
Lewis_defT}%
\begin{APACrefauthors}%
Lewis, D.%
\end{APACrefauthors}%
\unskip\
\newblock
\APACrefYearMonthDay{1970}{}{}.
\newblock
{\BBOQ}\APACrefatitle {{How to Define Theoretical Terms}} {{How to Define
  Theoretical Terms}}.{\BBCQ}
\newblock
\APACjournalVolNumPages{Journal of Philosophy, {\bf 67}, {\em
  pp.~427-446.}}{}{}{}.
\PrintBackRefs{\CurrentBib}

\bibitem [\protect \citeauthoryear {%
D.~Lewis%
}{%
D.~Lewis%
}{%
{\protect \APACyear {1972}}%
}]{%
Lewis_func}
\APACinsertmetastar {%
Lewis_func}%
\begin{APACrefauthors}%
Lewis, D.%
\end{APACrefauthors}%
\unskip\
\newblock
\APACrefYearMonthDay{1972}{}{}.
\newblock
{\BBOQ}\APACrefatitle {Psychophysical and theoretical identifications}
  {Psychophysical and theoretical identifications}.{\BBCQ}
\newblock
\APACjournalVolNumPages{Australasian Journal of Philosophy}{}{}{}.
\PrintBackRefs{\CurrentBib}

\bibitem [\protect \citeauthoryear {%
D\BPBI K.~Lewis%
}{%
D\BPBI K.~Lewis%
}{%
{\protect \APACyear {1986}}%
}]{%
Lewis1986}
\APACinsertmetastar {%
Lewis1986}%
\begin{APACrefauthors}%
Lewis, D\BPBI K.%
\end{APACrefauthors}%
\unskip\
\newblock
\APACrefYear{1986}.
\newblock
\APACrefbtitle {On the Plurality of Worlds} {On the plurality of worlds}.
\newblock
\APACaddressPublisher{}{Blackwell Publishers}.
\PrintBackRefs{\CurrentBib}

\bibitem [\protect \citeauthoryear {%
D\BPBI K.~Lewis%
}{%
D\BPBI K.~Lewis%
}{%
{\protect \APACyear {2009}}%
}]{%
LewisRamsey}
\APACinsertmetastar {%
LewisRamsey}%
\begin{APACrefauthors}%
Lewis, D\BPBI K.%
\end{APACrefauthors}%
\unskip\
\newblock
\APACrefYearMonthDay{2009}{}{}.
\newblock
{\BBOQ}\APACrefatitle {Ramseyan Humility} {Ramseyan humility}.{\BBCQ}
\newblock
\BIn{} D.~Braddon{-}Mitchell\ \BBA {} R.~Nola\ (\BEDS), \APACrefbtitle
  {Conceptual Analysis and Philosophical Naturalism} {Conceptual analysis and
  philosophical naturalism}\ (\BPGS\ 203--222).
\newblock
\APACaddressPublisher{}{MIT Press}.
\PrintBackRefs{\CurrentBib}

\bibitem [\protect \citeauthoryear {%
Lifshitz%
}{%
Lifshitz%
}{%
{\protect \APACyear {1987}}%
}]{%
LandauLifschitzVol2}
\APACinsertmetastar {%
LandauLifschitzVol2}%
\begin{APACrefauthors}%
Lifshitz, L\BPBI D\BPBI L\BPBI E.%
\end{APACrefauthors}%
\unskip\
\newblock
\APACrefYear{1987}.
\newblock
\APACrefbtitle {Course of thoretical physics, Vol 2: The Classical Theory of
  Fields} {Course of thoretical physics, vol 2: The classical theory of
  fields}.
\newblock
\APACaddressPublisher{}{Butterworth-Heinemann}.
\PrintBackRefs{\CurrentBib}

\bibitem [\protect \citeauthoryear {%
Linnemann%
\ \BBA {} Read%
}{%
Linnemann%
\ \BBA {} Read%
}{%
{\protect \APACyear {2021}}%
}]{%
LinnemannRead}
\APACinsertmetastar {%
LinnemannRead}%
\begin{APACrefauthors}%
Linnemann, N.%
\BCBT {}\ \BBA {} Read, J.%
\end{APACrefauthors}%
\unskip\
\newblock
\APACrefYearMonthDay{2021}{}{}.
\newblock
\APACrefbtitle {Constructive Axiomatics in Spacetime Physics Part I:
  Walkthrough to the Ehlers-Pirani-Schild Axiomatisation.} {Constructive
  axiomatics in spacetime physics part i: Walkthrough to the
  ehlers-pirani-schild axiomatisation.}
\PrintBackRefs{\CurrentBib}

\bibitem [\protect \citeauthoryear {%
Lyre%
}{%
Lyre%
}{%
{\protect \APACyear {2001}}%
}]{%
Lyre_gauge}
\APACinsertmetastar {%
Lyre_gauge}%
\begin{APACrefauthors}%
Lyre, H.%
\end{APACrefauthors}%
\unskip\
\newblock
\APACrefYearMonthDay{2001}{}{}.
\newblock
{\BBOQ}\APACrefatitle {The Principles of Gauging} {The principles of
  gauging}.{\BBCQ}
\newblock
\APACjournalVolNumPages{Philosophy of Science}{68}{3}{S371--S381}.
\newblock
\begin{APACrefURL} \url{http://www.jstor.org/stable/3080959} \end{APACrefURL}
\PrintBackRefs{\CurrentBib}

\bibitem [\protect \citeauthoryear {%
Lyre%
}{%
Lyre%
}{%
{\protect \APACyear {2009}}%
}]{%
lyre2009ab}
\APACinsertmetastar {%
lyre2009ab}%
\begin{APACrefauthors}%
Lyre, H.%
\end{APACrefauthors}%
\unskip\
\newblock
\APACrefYearMonthDay{2009}{}{}.
\newblock
{\BBOQ}\APACrefatitle {{Aharonov-Bohm Effect}} {{Aharonov-Bohm Effect}}.{\BBCQ}
\newblock
\BIn{} D.~Greenberger, K.~Hentschel\BCBL {}\ \BBA {} F.~Weinert\ (\BEDS),
  \APACrefbtitle {Compendium of Quantum Physics: Concepts, Experiments, History
  and Philosophy} {Compendium of quantum physics: Concepts, experiments,
  history and philosophy}\ (\BPG~1-3).
\newblock
\APACaddressPublisher{}{Berlin Heidelberg: Springer-Verlag}.
\PrintBackRefs{\CurrentBib}

\bibitem [\protect \citeauthoryear {%
Mackenzie%
}{%
Mackenzie%
}{%
{\protect \APACyear {2005}}%
}]{%
mackenzie_2005}
\APACinsertmetastar {%
mackenzie_2005}%
\begin{APACrefauthors}%
Mackenzie, K\BPBI C\BPBI H.%
\end{APACrefauthors}%
\unskip\
\newblock
\APACrefYear{2005}.
\newblock
\APACrefbtitle {{General Theory of Lie Groupoids and Lie Algebroids}} {{General
  Theory of Lie Groupoids and Lie Algebroids}}.
\newblock
\APACaddressPublisher{}{Cambridge University Press}.
\newblock
\begin{APACrefDOI} \doi{10.1017/CBO9781107325883} \end{APACrefDOI}
\PrintBackRefs{\CurrentBib}

\bibitem [\protect \citeauthoryear {%
Marsden%
}{%
Marsden%
}{%
{\protect \APACyear {2007}}%
}]{%
Marsden2007}
\APACinsertmetastar {%
Marsden2007}%
\begin{APACrefauthors}%
Marsden, J.%
\end{APACrefauthors}%
\unskip\
\newblock
\APACrefYearMonthDay{2007}{}{}.
\newblock
{\BBOQ}\APACrefatitle {{Symplectic Reduction}} {{Symplectic Reduction}}.{\BBCQ}
\newblock
\BIn{} \APACrefbtitle {Hamiltonian Reduction by Stages} {Hamiltonian reduction
  by stages}\ (\BPGS\ 3--42).
\newblock
\APACaddressPublisher{Berlin, Heidelberg}{Springer Berlin Heidelberg}.
\newblock
\begin{APACrefURL} \url{https://doi.org/10.1007/978-3-540-72470-4_1}
  \end{APACrefURL}
\newblock
\begin{APACrefDOI} \doi{10.1007/978-3-540-72470-4_1} \end{APACrefDOI}
\PrintBackRefs{\CurrentBib}

\bibitem [\protect \citeauthoryear {%
Martens%
\ \BBA {} Read%
}{%
Martens%
\ \BBA {} Read%
}{%
{\protect \APACyear {2020}}%
}]{%
ReadMartens}
\APACinsertmetastar {%
ReadMartens}%
\begin{APACrefauthors}%
Martens, N\BPBI C.%
\BCBT {}\ \BBA {} Read, J.%
\end{APACrefauthors}%
\unskip\
\newblock
\APACrefYearMonthDay{2020}{}{}.
\newblock
{\BBOQ}\APACrefatitle {Sophistry about symmetries?} {Sophistry about
  symmetries?}{\BBCQ}
\newblock
\APACjournalVolNumPages{Synthese}{}{}{}.
\newblock
\begin{APACrefURL} \url{http://philsci-archive.pitt.edu/17184/}
  \end{APACrefURL}
\PrintBackRefs{\CurrentBib}

\bibitem [\protect \citeauthoryear {%
Martin%
}{%
Martin%
}{%
{\protect \APACyear {2002}}%
}]{%
martin2002g}
\APACinsertmetastar {%
martin2002g}%
\begin{APACrefauthors}%
Martin, C\BPBI A.%
\end{APACrefauthors}%
\unskip\
\newblock
\APACrefYearMonthDay{2002}{}{}.
\newblock
{\BBOQ}\APACrefatitle {Gauge principles, gauge arguments and the logic of
  nature} {Gauge principles, gauge arguments and the logic of nature}.{\BBCQ}
\newblock
\APACjournalVolNumPages{Philosophy of Science}{69}{S3}{S221--S234}.
\PrintBackRefs{\CurrentBib}

\bibitem [\protect \citeauthoryear {%
Mattingly%
}{%
Mattingly%
}{%
{\protect \APACyear {2006}}%
}]{%
Mattingly_gauge}
\APACinsertmetastar {%
Mattingly_gauge}%
\begin{APACrefauthors}%
Mattingly, J.%
\end{APACrefauthors}%
\unskip\
\newblock
\APACrefYearMonthDay{2006}{}{}.
\newblock
{\BBOQ}\APACrefatitle {Which gauge matters?} {Which gauge matters?}{\BBCQ}
\newblock
\APACjournalVolNumPages{Studies in History and Philosophy of Science Part B:
  Studies in History and Philosophy of Modern Physics}{37}{2}{243-262}.
\newblock
\begin{APACrefURL}
  \url{https://www.sciencedirect.com/science/article/pii/S1355219806000037}
  \end{APACrefURL}
\newblock
\begin{APACrefDOI} \doi{https://doi.org/10.1016/j.shpsb.2005.08.001}
  \end{APACrefDOI}
\PrintBackRefs{\CurrentBib}

\bibitem [\protect \citeauthoryear {%
Maudlin%
}{%
Maudlin%
}{%
{\protect \APACyear {1998}}%
}]{%
Maudlin_response}
\APACinsertmetastar {%
Maudlin_response}%
\begin{APACrefauthors}%
Maudlin, T.%
\end{APACrefauthors}%
\unskip\
\newblock
\APACrefYearMonthDay{1998}{}{}.
\newblock
{\BBOQ}\APACrefatitle {Healey on the Aharonov-Bohm Effect} {Healey on the
  aharonov-bohm effect}.{\BBCQ}
\newblock
\APACjournalVolNumPages{Philosophy of Science}{65}{2}{361--368}.
\newblock
\begin{APACrefURL} \url{http://www.jstor.org/stable/188266} \end{APACrefURL}
\PrintBackRefs{\CurrentBib}

\bibitem [\protect \citeauthoryear {%
Maudlin%
}{%
Maudlin%
}{%
{\protect \APACyear {2015}}%
}]{%
Maudlin_book}
\APACinsertmetastar {%
Maudlin_book}%
\begin{APACrefauthors}%
Maudlin, T.%
\end{APACrefauthors}%
\unskip\
\newblock
\APACrefYear{2015}.
\newblock
\APACrefbtitle {{Philosophy of Physics: Space and Time (Princeton Foundations
  of Contemporary Philosophy, 5)}} {{Philosophy of Physics: Space and Time
  (Princeton Foundations of Contemporary Philosophy, 5)}}.
\newblock
\APACaddressPublisher{}{Princeton University Press}.
\PrintBackRefs{\CurrentBib}

\bibitem [\protect \citeauthoryear {%
Maudlin%
}{%
Maudlin%
}{%
{\protect \APACyear {2018}}%
}]{%
Maudlin_ontology}
\APACinsertmetastar {%
Maudlin_ontology}%
\begin{APACrefauthors}%
Maudlin, T.%
\end{APACrefauthors}%
\unskip\
\newblock
\APACrefYearMonthDay{2018}{}{}.
\newblock
{\BBOQ}\APACrefatitle {Ontological Clarity via Canonical Presentation:
  Electromagnetism and the Aharonov–Bohm Effect} {Ontological clarity via
  canonical presentation: Electromagnetism and the aharonov–bohm
  effect}.{\BBCQ}
\newblock
\APACjournalVolNumPages{Entropy}{20}{6}{}.
\newblock
\begin{APACrefURL} \url{https://www.mdpi.com/1099-4300/20/6/465}
  \end{APACrefURL}
\newblock
\begin{APACrefDOI} \doi{10.3390/e20060465} \end{APACrefDOI}
\PrintBackRefs{\CurrentBib}

\bibitem [\protect \citeauthoryear {%
Mercati%
}{%
Mercati%
}{%
{\protect \APACyear {2017}}%
}]{%
Flavio_tutorial}
\APACinsertmetastar {%
Flavio_tutorial}%
\begin{APACrefauthors}%
Mercati, F.%
\end{APACrefauthors}%
\unskip\
\newblock
\APACrefYearMonthDay{2017}{}{}.
\newblock
{\BBOQ}\APACrefatitle {Shape Dynamics: Relativity and Relationalism} {Shape
  dynamics: Relativity and relationalism}.{\BBCQ}
\newblock
\APACjournalVolNumPages{Oxford University Press}{}{}{}.
\PrintBackRefs{\CurrentBib}

\bibitem [\protect \citeauthoryear {%
Misner%
\ \BBA {} Wheeler%
}{%
Misner%
\ \BBA {} Wheeler%
}{%
{\protect \APACyear {1957}}%
}]{%
MisnerWheeler1957}
\APACinsertmetastar {%
MisnerWheeler1957}%
\begin{APACrefauthors}%
Misner, C.%
\BCBT {}\ \BBA {} Wheeler, J.%
\end{APACrefauthors}%
\unskip\
\newblock
\APACrefYearMonthDay{1957}{}{}.
\newblock
{\BBOQ}\APACrefatitle {Classical physics as geometry} {Classical physics as
  geometry}.{\BBCQ}
\newblock
\APACjournalVolNumPages{Annals of Physics}{2}{6}{525-603}.
\newblock
\begin{APACrefURL}
  \url{https://www.sciencedirect.com/science/article/pii/0003491657900490}
  \end{APACrefURL}
\newblock
\begin{APACrefDOI} \doi{https://doi.org/10.1016/0003-4916(57)90049-0}
  \end{APACrefDOI}
\PrintBackRefs{\CurrentBib}

\bibitem [\protect \citeauthoryear {%
Mitter%
\ \BBA {} Viallet%
}{%
Mitter%
\ \BBA {} Viallet%
}{%
{\protect \APACyear {1981}}%
}]{%
Mitter:1979un}
\APACinsertmetastar {%
Mitter:1979un}%
\begin{APACrefauthors}%
Mitter, P\BPBI K.%
\BCBT {}\ \BBA {} Viallet, C\BPBI M.%
\end{APACrefauthors}%
\unskip\
\newblock
\APACrefYearMonthDay{1981}{}{}.
\newblock
{\BBOQ}\APACrefatitle {{On the Bundle of Connections and the Gauge Orbit
  Manifold in {Yang-Mills} Theory}} {{On the Bundle of Connections and the
  Gauge Orbit Manifold in {Yang-Mills} Theory}}.{\BBCQ}
\newblock
\APACjournalVolNumPages{Commun. Math. Phys.}{79}{}{457}.
\newblock
\begin{APACrefDOI} \doi{10.1007/BF01209307} \end{APACrefDOI}
\PrintBackRefs{\CurrentBib}

\bibitem [\protect \citeauthoryear {%
Morita%
}{%
Morita%
}{%
{\protect \APACyear {2001}}%
}]{%
Morita_book}
\APACinsertmetastar {%
Morita_book}%
\begin{APACrefauthors}%
Morita, S.%
\end{APACrefauthors}%
\unskip\
\newblock
\APACrefYear{2001}.
\newblock
\APACrefbtitle {{Translations of Mathematical MonographsIwanami Series in
  Modern Mathematics: Geometry of Differential forms}} {{Translations of
  Mathematical MonographsIwanami Series in Modern Mathematics: Geometry of
  Differential forms}}.
\newblock
\APACaddressPublisher{}{American Mathematical Society}.
\PrintBackRefs{\CurrentBib}

\bibitem [\protect \citeauthoryear {%
Mulder%
}{%
Mulder%
}{%
{\protect \APACyear {2021}}%
}]{%
Mulder_AB}
\APACinsertmetastar {%
Mulder_AB}%
\begin{APACrefauthors}%
Mulder, R.%
\end{APACrefauthors}%
\unskip\
\newblock
\APACrefYearMonthDay{2021}{}{}.
\newblock
{\BBOQ}\APACrefatitle {{Gauge-Underdetermination and Shades of Locality in the
  Aharonov–Bohm Effect}} {{Gauge-Underdetermination and Shades of Locality in
  the Aharonov–Bohm Effect}}.{\BBCQ}
\newblock
\APACjournalVolNumPages{Foundations of Physics}{}{}{}.
\PrintBackRefs{\CurrentBib}

\bibitem [\protect \citeauthoryear {%
Muller%
}{%
Muller%
}{%
{\protect \APACyear {2011}}%
}]{%
Muller_abysmal}
\APACinsertmetastar {%
Muller_abysmal}%
\begin{APACrefauthors}%
Muller, F\BPBI A.%
\end{APACrefauthors}%
\unskip\
\newblock
\APACrefYearMonthDay{2011}{}{}.
\newblock
{\BBOQ}\APACrefatitle {How to Defeat Wüthrich’s Abysmal Embarrassment
  Argument against Space-Time Structuralism} {How to defeat wüthrich’s
  abysmal embarrassment argument against space-time structuralism}.{\BBCQ}
\newblock
\APACjournalVolNumPages{Philosophy of Science}{78}{5}{1046--1057}.
\newblock
\begin{APACrefURL} \url{http://www.jstor.org/stable/10.1086/662634}
  \end{APACrefURL}
\PrintBackRefs{\CurrentBib}

\bibitem [\protect \citeauthoryear {%
Mundy%
}{%
Mundy%
}{%
{\protect \APACyear {1992}}%
}]{%
Mundy1992}
\APACinsertmetastar {%
Mundy1992}%
\begin{APACrefauthors}%
Mundy, B.%
\end{APACrefauthors}%
\unskip\
\newblock
\APACrefYearMonthDay{1992}{}{}.
\newblock
{\BBOQ}\APACrefatitle {{Space-Time and Isomorphism}} {{Space-Time and
  Isomorphism}}.{\BBCQ}
\newblock
\APACjournalVolNumPages{PSA: Proceedings of the Biennial Meeting of the
  Philosophy of Science Association}{1992}{Volume One: Contributed
  Papers}{515--527}.
\PrintBackRefs{\CurrentBib}

\bibitem [\protect \citeauthoryear {%
Myrvold%
}{%
Myrvold%
}{%
{\protect \APACyear {2011}}%
}]{%
Myrvold2010}
\APACinsertmetastar {%
Myrvold2010}%
\begin{APACrefauthors}%
Myrvold, W\BPBI C.%
\end{APACrefauthors}%
\unskip\
\newblock
\APACrefYearMonthDay{2011}{}{}.
\newblock
{\BBOQ}\APACrefatitle {{Nonseparability, Classical, and Quantum}}
  {{Nonseparability, Classical, and Quantum}}.{\BBCQ}
\newblock
\APACjournalVolNumPages{The British Journal for the Philosophy of
  Science}{62}{2}{417-432}.
\newblock
\begin{APACrefURL} \url{https://doi.org/10.1093/bjps/axq036} \end{APACrefURL}
\newblock
\begin{APACrefDOI} \doi{10.1093/bjps/axq036} \end{APACrefDOI}
\PrintBackRefs{\CurrentBib}

\bibitem [\protect \citeauthoryear {%
Møller-Nielsen%
}{%
Møller-Nielsen%
}{%
{\protect \APACyear {2017}}%
}]{%
Moller}
\APACinsertmetastar {%
Moller}%
\begin{APACrefauthors}%
Møller-Nielsen, T.%
\end{APACrefauthors}%
\unskip\
\newblock
\APACrefYearMonthDay{2017}{}{}.
\newblock
{\BBOQ}\APACrefatitle {Invariance, Interpretation, and Motivation} {Invariance,
  interpretation, and motivation}.{\BBCQ}
\newblock
\APACjournalVolNumPages{Philosophy of Science}{84}{5}{1253-1264}.
\newblock
\begin{APACrefURL} \url{https://doi.org/10.1086/694087} \end{APACrefURL}
\newblock
\begin{APACrefDOI} \doi{10.1086/694087} \end{APACrefDOI}
\PrintBackRefs{\CurrentBib}

\bibitem [\protect \citeauthoryear {%
Noether%
}{%
Noether%
}{%
{\protect \APACyear {1918}}%
}]{%
noether1918a}
\APACinsertmetastar {%
noether1918a}%
\begin{APACrefauthors}%
Noether, E.%
\end{APACrefauthors}%
\unskip\
\newblock
\APACrefYearMonthDay{1918}{}{}.
\newblock
{\BBOQ}\APACrefatitle {Invariante {V}ariationsprobleme} {Invariante
  {V}ariationsprobleme}.{\BBCQ}
\newblock
\APACjournalVolNumPages{Nachr. D. König. Gesellsch. D. Wiss. Zu Göttingen,
  Math-phys. Klasse}{}{}{235–257}.
\newblock
\APACrefnote{English translation by M. A. Tavel:
  \url{https://arxiv.org/abs/physics/0503066}}
\PrintBackRefs{\CurrentBib}

\bibitem [\protect \citeauthoryear {%
Nounou%
}{%
Nounou%
}{%
{\protect \APACyear {2003}}%
}]{%
nounou2003ab}
\APACinsertmetastar {%
nounou2003ab}%
\begin{APACrefauthors}%
Nounou, A.%
\end{APACrefauthors}%
\unskip\
\newblock
\APACrefYearMonthDay{2003}{}{}.
\newblock
{\BBOQ}\APACrefatitle {A fourth way to the {A}haronov–{B}ohm effect.} {A
  fourth way to the {A}haronov–{B}ohm effect.}{\BBCQ}
\newblock
\BIn{} K.~Brading\ \BBA {} E.~Castellani\ (\BEDS), \APACrefbtitle {{Symmetries
  in Physics: Philosophical Reflections}} {{Symmetries in Physics:
  Philosophical Reflections}}\ (\BPG~174-2000).
\newblock
\APACaddressPublisher{}{Cambridge: Cambridge University Press}.
\PrintBackRefs{\CurrentBib}

\bibitem [\protect \citeauthoryear {%
Nozick%
}{%
Nozick%
}{%
{\protect \APACyear {2001}}%
}]{%
Nozick}
\APACinsertmetastar {%
Nozick}%
\begin{APACrefauthors}%
Nozick, R.%
\end{APACrefauthors}%
\unskip\
\newblock
\APACrefYear{2001}.
\newblock
\APACrefbtitle {{Invariances: The Structure of the Objective World. }}
  {{Invariances: The Structure of the Objective World. }}.
\newblock
\APACaddressPublisher{}{Harvard University Press. (Cambridge, Mass.)}.
\PrintBackRefs{\CurrentBib}

\bibitem [\protect \citeauthoryear {%
Olver%
}{%
Olver%
}{%
{\protect \APACyear {1986}}%
}]{%
Olver_book}
\APACinsertmetastar {%
Olver_book}%
\begin{APACrefauthors}%
Olver, P.%
\end{APACrefauthors}%
\unskip\
\newblock
\APACrefYear{1986}.
\newblock
\APACrefbtitle {Applications of Lie Groups to Differential Equations}
  {Applications of lie groups to differential equations}.
\newblock
\APACaddressPublisher{}{Springer-Verlag New York}.
\PrintBackRefs{\CurrentBib}

\bibitem [\protect \citeauthoryear {%
O'Neill%
}{%
O'Neill%
}{%
{\protect \APACyear {1983}}%
}]{%
Oneill}
\APACinsertmetastar {%
Oneill}%
\begin{APACrefauthors}%
O'Neill, B.%
\end{APACrefauthors}%
\unskip\
\newblock
\APACrefYear{1983}.
\newblock
\APACrefbtitle {{Semi-Riemannian Geometry With Applications to Relativity}}
  {{Semi-Riemannian Geometry With Applications to Relativity}}.
\newblock
\APACaddressPublisher{}{Academic Press}.
\PrintBackRefs{\CurrentBib}

\bibitem [\protect \citeauthoryear {%
O’Raifertaigh%
}{%
O’Raifertaigh%
}{%
{\protect \APACyear {1997}}%
}]{%
Dawning_book}
\APACinsertmetastar {%
Dawning_book}%
\begin{APACrefauthors}%
O’Raifertaigh, L.%
\end{APACrefauthors}%
\unskip\
\newblock
\APACrefYear{1997}.
\newblock
\APACrefbtitle {{The Dawning of Gauge Theory}} {{The Dawning of Gauge Theory}}.
\newblock
\APACaddressPublisher{}{Princeton University Press}.
\newblock
\begin{APACrefURL} \url{http://www.jstor.org/stable/j.ctv10vm2qt}
  \end{APACrefURL}
\PrintBackRefs{\CurrentBib}

\bibitem [\protect \citeauthoryear {%
Palais%
}{%
Palais%
}{%
{\protect \APACyear {1961}}%
}]{%
Palais}
\APACinsertmetastar {%
Palais}%
\begin{APACrefauthors}%
Palais, R.%
\end{APACrefauthors}%
\unskip\
\newblock
\APACrefYearMonthDay{1961}{}{}.
\newblock
{\BBOQ}\APACrefatitle {On the existence of slices for the actions of
  non-compact groups} {On the existence of slices for the actions of
  non-compact groups}.{\BBCQ}
\newblock
\APACjournalVolNumPages{Ann. of Math.}{73}{}{295-322}.
\PrintBackRefs{\CurrentBib}

\bibitem [\protect \citeauthoryear {%
Penrose%
}{%
Penrose%
}{%
{\protect \APACyear {1982}}%
}]{%
Penrose_gr_problems}
\APACinsertmetastar {%
Penrose_gr_problems}%
\begin{APACrefauthors}%
Penrose, R.%
\end{APACrefauthors}%
\unskip\
\newblock
\APACrefYearMonthDay{1982}{}{}.
\newblock
{\BBOQ}\APACrefatitle {{Some Unsolved Problems in Classical General
  Relativity}} {{Some Unsolved Problems in Classical General
  Relativity}}.{\BBCQ}
\newblock
\APACjournalVolNumPages{In S.-T. Yau (ed), Seminar on Differential Geometry.
  Princeton: Princeton University Press, 631–668.}{}{}{}.
\PrintBackRefs{\CurrentBib}

\bibitem [\protect \citeauthoryear {%
{Penrose}%
}{%
{Penrose}%
}{%
{\protect \APACyear {1996}}%
}]{%
Penrose_gravi_collapse}
\APACinsertmetastar {%
Penrose_gravi_collapse}%
\begin{APACrefauthors}%
{Penrose}, R.%
\end{APACrefauthors}%
\unskip\
\newblock
\APACrefYearMonthDay{1996}{{\APACmonth{05}}}{}.
\newblock
{\BBOQ}\APACrefatitle {{On Gravity's role in Quantum State Reduction}} {{On
  Gravity's role in Quantum State Reduction}}.{\BBCQ}
\newblock
\APACjournalVolNumPages{General Relativity and Gravitation}{28}{5}{581-600}.
\newblock
\begin{APACrefDOI} \doi{10.1007/BF02105068} \end{APACrefDOI}
\PrintBackRefs{\CurrentBib}

\bibitem [\protect \citeauthoryear {%
Pitts%
}{%
Pitts%
}{%
{\protect \APACyear {2014}}%
{\protect \APACexlab {{\protect \BCnt {1}}}}}]{%
Pitts_Ham}
\APACinsertmetastar {%
Pitts_Ham}%
\begin{APACrefauthors}%
Pitts, J\BPBI B.%
\end{APACrefauthors}%
\unskip\
\newblock
\APACrefYearMonthDay{2014{\protect \BCnt {1}}}{}{}.
\newblock
{\BBOQ}\APACrefatitle {{Change in Hamiltonian general relativity from the lack
  of a time-like Killing vector field}} {{Change in Hamiltonian general
  relativity from the lack of a time-like Killing vector field}}.{\BBCQ}
\newblock
\APACjournalVolNumPages{Studies in History and Philosophy of Science Part B:
  Studies in History and Philosophy of Modern Physics}{47}{}{68-89}.
\newblock
\begin{APACrefURL}
  \url{https://www.sciencedirect.com/science/article/pii/S1355219814000586}
  \end{APACrefURL}
\newblock
\begin{APACrefDOI} \doi{https://doi.org/10.1016/j.shpsb.2014.05.007}
  \end{APACrefDOI}
\PrintBackRefs{\CurrentBib}

\bibitem [\protect \citeauthoryear {%
Pitts%
}{%
Pitts%
}{%
{\protect \APACyear {2014}}%
{\protect \APACexlab {{\protect \BCnt {2}}}}}]{%
Pitts_electro}
\APACinsertmetastar {%
Pitts_electro}%
\begin{APACrefauthors}%
Pitts, J\BPBI B.%
\end{APACrefauthors}%
\unskip\
\newblock
\APACrefYearMonthDay{2014{\protect \BCnt {2}}}{Dec}{}.
\newblock
{\BBOQ}\APACrefatitle {A first class constraint generates not a gauge
  transformation, but a bad physical change: The case of electromagnetism} {A
  first class constraint generates not a gauge transformation, but a bad
  physical change: The case of electromagnetism}.{\BBCQ}
\newblock
\APACjournalVolNumPages{Annals of Physics}{351}{}{382–406}.
\newblock
\begin{APACrefURL} \url{http://dx.doi.org/10.1016/j.aop.2014.08.014}
  \end{APACrefURL}
\newblock
\begin{APACrefDOI} \doi{10.1016/j.aop.2014.08.014} \end{APACrefDOI}
\PrintBackRefs{\CurrentBib}

\bibitem [\protect \citeauthoryear {%
Pons%
}{%
Pons%
}{%
{\protect \APACyear {2005}}%
}]{%
Pons2005}
\APACinsertmetastar {%
Pons2005}%
\begin{APACrefauthors}%
Pons, J\BPBI M.%
\end{APACrefauthors}%
\unskip\
\newblock
\APACrefYearMonthDay{2005}{}{}.
\newblock
{\BBOQ}\APACrefatitle {{On Dirac's incomplete analysis of gauge
  transformations}} {{On Dirac's incomplete analysis of gauge
  transformations}}.{\BBCQ}
\newblock
\APACjournalVolNumPages{Studies in History and Philosophy of Science Part B:
  Studies in History and Philosophy of Modern Physics}{36}{3}{491-518}.
\newblock
\begin{APACrefURL}
  \url{https://www.sciencedirect.com/science/article/pii/S1355219805000456}
  \end{APACrefURL}
\newblock
\begin{APACrefDOI} \doi{https://doi.org/10.1016/j.shpsb.2005.04.004}
  \end{APACrefDOI}
\PrintBackRefs{\CurrentBib}

\bibitem [\protect \citeauthoryear {%
Pooley%
}{%
Pooley%
}{%
{\protect \APACyear {2013}}%
}]{%
Pooley_rel}
\APACinsertmetastar {%
Pooley_rel}%
\begin{APACrefauthors}%
Pooley, O.%
\end{APACrefauthors}%
\unskip\
\newblock
\APACrefYearMonthDay{2013}{}{}.
\newblock
{\BBOQ}\APACrefatitle {{'Substantivalist and Relationalist Approaches to
  Spacetime'. In The Oxford Handbook of Philosophy of Physics.}}
  {{'Substantivalist and Relationalist Approaches to Spacetime'. In The Oxford
  Handbook of Philosophy of Physics.}}{\BBCQ}
\newblock
\BIn{} R.~Batterman\ (\BED), (\BCHAP~15).
\newblock
\APACaddressPublisher{}{Oxford University Press}.
\PrintBackRefs{\CurrentBib}

\bibitem [\protect \citeauthoryear {%
Pooley%
}{%
Pooley%
}{%
{\protect \APACyear {{\protect \BIP {}}}}%
}]{%
Pooley_routledge}
\APACinsertmetastar {%
Pooley_routledge}%
\begin{APACrefauthors}%
Pooley, O.%
\end{APACrefauthors}%
\unskip\
\newblock
\APACrefYearMonthDay{{\protect \BIP {}}}{}{}.
\newblock
{\BBOQ}\APACrefatitle {{The Hole Argument}} {{The Hole Argument}}.{\BBCQ}
\newblock
\BIn{} E.~Knox\ \BBA {} A.~Wilson\ (\BEDS), \APACrefbtitle {The Routledge
  Companion to the Philosophy of Physics.} {The routledge companion to the
  philosophy of physics.}
\newblock
\APACaddressPublisher{}{Routledge}.
\PrintBackRefs{\CurrentBib}

\bibitem [\protect \citeauthoryear {%
Pooley%
\ \BBA {} Read%
}{%
Pooley%
\ \BBA {} Read%
}{%
{\protect \APACyear {2022}}%
}]{%
Pooley_Read}
\APACinsertmetastar {%
Pooley_Read}%
\begin{APACrefauthors}%
Pooley, O.%
\BCBT {}\ \BBA {} Read, J.%
\end{APACrefauthors}%
\unskip\
\newblock
\APACrefYearMonthDay{2022}{}{}.
\newblock
{\BBOQ}\APACrefatitle {{On the Mathematics and Metaphysics of the Hole
  Argument}} {{On the Mathematics and Metaphysics of the Hole
  Argument}}.{\BBCQ}
\newblock
\APACjournalVolNumPages{The British Journal for the Philosophy of
  Science}{}{}{}.
\newblock
\begin{APACrefDOI} \doi{10.1086/718274} \end{APACrefDOI}
\PrintBackRefs{\CurrentBib}

\bibitem [\protect \citeauthoryear {%
Putnam%
}{%
Putnam%
}{%
{\protect \APACyear {1975}}%
}]{%
Putnam_analytic}
\APACinsertmetastar {%
Putnam_analytic}%
\begin{APACrefauthors}%
Putnam, H.%
\end{APACrefauthors}%
\unskip\
\newblock
\APACrefYearMonthDay{1975}{}{}.
\newblock
{\BBOQ}\APACrefatitle {{The Analytic and Synthetic}} {{The Analytic and
  Synthetic}}.{\BBCQ}
\newblock
\BIn{} \APACrefbtitle {Mind, Language and Reality: Philosophical Papers} {Mind,
  language and reality: Philosophical papers}\ (\BPGS\ 33--69).
\newblock
\APACaddressPublisher{}{Cambridge University Press}.
\PrintBackRefs{\CurrentBib}

\bibitem [\protect \citeauthoryear {%
S.~Ramirez%
\ \BBA {} Teh%
}{%
S.~Ramirez%
\ \BBA {} Teh%
}{%
{\protect \APACyear {2019}}%
}]{%
Teh_abandon}
\APACinsertmetastar {%
Teh_abandon}%
\begin{APACrefauthors}%
Ramirez, S.%
\BCBT {}\ \BBA {} Teh, N.%
\end{APACrefauthors}%
\unskip\
\newblock
\APACrefYearMonthDay{2019}{}{}.
\newblock
{\BBOQ}\APACrefatitle {Abandoning Galileo's Ship: The quest for non-relational
  empirical signicance} {Abandoning galileo's ship: The quest for
  non-relational empirical signicance}.{\BBCQ}
\newblock
\APACjournalVolNumPages{preprint}{}{}{}.
\PrintBackRefs{\CurrentBib}

\bibitem [\protect \citeauthoryear {%
S\BPBI M.~Ramirez%
}{%
S\BPBI M.~Ramirez%
}{%
{\protect \APACyear {2019}}%
}]{%
Seb_ramirez}
\APACinsertmetastar {%
Seb_ramirez}%
\begin{APACrefauthors}%
Ramirez, S\BPBI M.%
\end{APACrefauthors}%
\unskip\
\newblock
\APACrefYearMonthDay{2019}{October}{}.
\newblock
{\BBOQ}\APACrefatitle {A puzzle concerning local symmetries and their empirical
  significance} {A puzzle concerning local symmetries and their empirical
  significance}.{\BBCQ}
\newblock
\begin{APACrefURL} \url{http://philsci-archive.pitt.edu/16509/}
  \end{APACrefURL}
\PrintBackRefs{\CurrentBib}

\bibitem [\protect \citeauthoryear {%
Read%
\ \BBA {} Møller-Nielsen%
}{%
Read%
\ \BBA {} Møller-Nielsen%
}{%
{\protect \APACyear {2020}}%
}]{%
ReadMoller}
\APACinsertmetastar {%
ReadMoller}%
\begin{APACrefauthors}%
Read, J.%
\BCBT {}\ \BBA {} Møller-Nielsen, T.%
\end{APACrefauthors}%
\unskip\
\newblock
\APACrefYearMonthDay{2020}{}{}.
\newblock
{\BBOQ}\APACrefatitle {Redundant epistemic symmetries} {Redundant epistemic
  symmetries}.{\BBCQ}
\newblock
\APACjournalVolNumPages{Studies in History and Philosophy of Science Part B:
  Studies in History and Philosophy of Modern Physics}{70}{}{88-97}.
\newblock
\begin{APACrefURL}
  \url{https://www.sciencedirect.com/science/article/pii/S1355219819301649}
  \end{APACrefURL}
\newblock
\begin{APACrefDOI} \doi{https://doi.org/10.1016/j.shpsb.2020.03.002}
  \end{APACrefDOI}
\PrintBackRefs{\CurrentBib}

\bibitem [\protect \citeauthoryear {%
Reck%
\ \BBA {} Schiemer%
}{%
Reck%
\ \BBA {} Schiemer%
}{%
{\protect \APACyear {2020}}%
}]{%
sep-structuralism-mathematics}
\APACinsertmetastar {%
sep-structuralism-mathematics}%
\begin{APACrefauthors}%
Reck, E.%
\BCBT {}\ \BBA {} Schiemer, G.%
\end{APACrefauthors}%
\unskip\
\newblock
\APACrefYearMonthDay{2020}{}{}.
\newblock
{\BBOQ}\APACrefatitle {{Structuralism in the Philosophy of Mathematics}}
  {{Structuralism in the Philosophy of Mathematics}}.{\BBCQ}
\newblock
\BIn{} E\BPBI N.~Zalta\ (\BED), \APACrefbtitle {The {Stanford} Encyclopedia of
  Philosophy} {The {Stanford} encyclopedia of philosophy}\
  (\PrintOrdinal{{S}pring 2020}\ \BEd).
\newblock
\APACaddressPublisher{}{Metaphysics Research Lab, Stanford University}.
\newblock
\APAChowpublished
  {\url{https://plato.stanford.edu/archives/spr2020/entries/structuralism-mathematics/}}.
\PrintBackRefs{\CurrentBib}

\bibitem [\protect \citeauthoryear {%
Regge%
\ \BBA {} Teitelboim%
}{%
Regge%
\ \BBA {} Teitelboim%
}{%
{\protect \APACyear {1974}}%
}]{%
ReggeTeitelboim1974}
\APACinsertmetastar {%
ReggeTeitelboim1974}%
\begin{APACrefauthors}%
Regge, T.%
\BCBT {}\ \BBA {} Teitelboim, C.%
\end{APACrefauthors}%
\unskip\
\newblock
\APACrefYearMonthDay{1974}{}{}.
\newblock
{\BBOQ}\APACrefatitle {{Role of Surface Integrals in the Hamiltonian
  Formulation of General Relativity}} {{Role of Surface Integrals in the
  Hamiltonian Formulation of General Relativity}}.{\BBCQ}
\newblock
\APACjournalVolNumPages{Annals Phys.}{88}{}{286}.
\newblock
\begin{APACrefDOI} \doi{10.1016/0003-4916(74)90404-7} \end{APACrefDOI}
\PrintBackRefs{\CurrentBib}

\bibitem [\protect \citeauthoryear {%
Resnik%
}{%
Resnik%
}{%
{\protect \APACyear {1981}}%
}]{%
Resnik1981}
\APACinsertmetastar {%
Resnik1981}%
\begin{APACrefauthors}%
Resnik, M\BPBI D.%
\end{APACrefauthors}%
\unskip\
\newblock
\APACrefYearMonthDay{1981}{}{}.
\newblock
{\BBOQ}\APACrefatitle {{Mathematics as a Science of Patterns: Ontology and
  Reference}} {{Mathematics as a Science of Patterns: Ontology and
  Reference}}.{\BBCQ}
\newblock
\APACjournalVolNumPages{No\^us}{15}{4}{529--550}.
\newblock
\begin{APACrefDOI} \doi{10.2307/2214851} \end{APACrefDOI}
\PrintBackRefs{\CurrentBib}

\bibitem [\protect \citeauthoryear {%
Riello%
}{%
Riello%
}{%
{\protect \APACyear {2020}}%
}]{%
RielloSoft}
\APACinsertmetastar {%
RielloSoft}%
\begin{APACrefauthors}%
Riello, A.%
\end{APACrefauthors}%
\unskip\
\newblock
\APACrefYearMonthDay{2020}{}{}.
\newblock
{\BBOQ}\APACrefatitle {{Soft charges from the geometry of field space}} {{Soft
  charges from the geometry of field space}}.{\BBCQ}
\newblock
\APACjournalVolNumPages{JHEP}{}{}{}.
\PrintBackRefs{\CurrentBib}

\bibitem [\protect \citeauthoryear {%
Riello%
}{%
Riello%
}{%
{\protect \APACyear {2021}}%
{\protect \APACexlab {{\protect \BCnt {1}}}}}]{%
Riello_new}
\APACinsertmetastar {%
Riello_new}%
\begin{APACrefauthors}%
Riello, A.%
\end{APACrefauthors}%
\unskip\
\newblock
\APACrefYearMonthDay{2021{\protect \BCnt {1}}}{}{}.
\newblock
{\BBOQ}\APACrefatitle {Edge modes without edge modes} {Edge modes without edge
  modes}.{\BBCQ}
\newblock
\APACjournalVolNumPages{forthcoming}{}{}{}.
\PrintBackRefs{\CurrentBib}

\bibitem [\protect \citeauthoryear {%
Riello%
}{%
Riello%
}{%
{\protect \APACyear {2021}}%
{\protect \APACexlab {{\protect \BCnt {2}}}}}]{%
Riello_symp}
\APACinsertmetastar {%
Riello_symp}%
\begin{APACrefauthors}%
Riello, A.%
\end{APACrefauthors}%
\unskip\
\newblock
\APACrefYearMonthDay{2021{\protect \BCnt {2}}}{}{}.
\newblock
{\BBOQ}\APACrefatitle {{Symplectic reduction of Yang-Mills theory with
  boundaries: from superselection sectors to edge modes, and back}}
  {{Symplectic reduction of Yang-Mills theory with boundaries: from
  superselection sectors to edge modes, and back}}.{\BBCQ}
\newblock
\APACjournalVolNumPages{SciPost Phys.}{10}{}{125}.
\newblock
\begin{APACrefURL} \url{https://scipost.org/10.21468/SciPostPhys.10.6.125}
  \end{APACrefURL}
\newblock
\begin{APACrefDOI} \doi{10.21468/SciPostPhys.10.6.125} \end{APACrefDOI}
\PrintBackRefs{\CurrentBib}

\bibitem [\protect \citeauthoryear {%
Ringstr\"om%
}{%
Ringstr\"om%
}{%
{\protect \APACyear {2021}}%
}]{%
Ringstrom_book}
\APACinsertmetastar {%
Ringstrom_book}%
\begin{APACrefauthors}%
Ringstr\"om, H.%
\end{APACrefauthors}%
\unskip\
\newblock
\APACrefYear{2021}.
\newblock
\APACrefbtitle {{On the Topology and Future Stability of the Universe}} {{On
  the Topology and Future Stability of the Universe}}.
\newblock
\APACaddressPublisher{}{Oxford University Press}.
\PrintBackRefs{\CurrentBib}

\bibitem [\protect \citeauthoryear {%
Robb%
}{%
Robb%
}{%
{\protect \APACyear {1936}}%
}]{%
Robb1936}
\APACinsertmetastar {%
Robb1936}%
\begin{APACrefauthors}%
Robb, A\BPBI A.%
\end{APACrefauthors}%
\unskip\
\newblock
\APACrefYear{1936}.
\newblock
\APACrefbtitle {{Geometry of Time and Space}} {{Geometry of Time and Space}}.
\newblock
\APACaddressPublisher{}{Cambridge University Press}.
\PrintBackRefs{\CurrentBib}

\bibitem [\protect \citeauthoryear {%
Rosenstock%
\ \BBA {} Weatherall%
}{%
Rosenstock%
\ \BBA {} Weatherall%
}{%
{\protect \APACyear {2016}}%
{\protect \APACexlab {{\protect \BCnt {1}}}}}]{%
RosenstockWeatherall2016c}
\APACinsertmetastar {%
RosenstockWeatherall2016c}%
\begin{APACrefauthors}%
Rosenstock, S.%
\BCBT {}\ \BBA {} Weatherall, J\BPBI O.%
\end{APACrefauthors}%
\unskip\
\newblock
\APACrefYearMonthDay{2016{\protect \BCnt {1}}}{}{}.
\newblock
{\BBOQ}\APACrefatitle {A categorical equivalence between generalized holonomy
  maps on a connected manifold and principal connections on bundles over that
  manifold} {A categorical equivalence between generalized holonomy maps on a
  connected manifold and principal connections on bundles over that
  manifold}.{\BBCQ}
\newblock
\APACjournalVolNumPages{Journal of Mathematical Physics}{57}{10}{102902}.
\newblock
\APACrefnote{\url{http://philsci-archive.pitt.edu/11904/}}
\PrintBackRefs{\CurrentBib}

\bibitem [\protect \citeauthoryear {%
Rosenstock%
\ \BBA {} Weatherall%
}{%
Rosenstock%
\ \BBA {} Weatherall%
}{%
{\protect \APACyear {2016}}%
{\protect \APACexlab {{\protect \BCnt {2}}}}}]{%
Weatherall_holonomy}
\APACinsertmetastar {%
Weatherall_holonomy}%
\begin{APACrefauthors}%
Rosenstock, S.%
\BCBT {}\ \BBA {} Weatherall, J\BPBI O.%
\end{APACrefauthors}%
\unskip\
\newblock
\APACrefYearMonthDay{2016{\protect \BCnt {2}}}{Oct}{}.
\newblock
{\BBOQ}\APACrefatitle {A categorical equivalence between generalized holonomy
  maps on a connected manifold and principal connections on bundles over that
  manifold} {A categorical equivalence between generalized holonomy maps on a
  connected manifold and principal connections on bundles over that
  manifold}.{\BBCQ}
\newblock
\APACjournalVolNumPages{Journal of Mathematical Physics}{57}{10}{102902}.
\newblock
\begin{APACrefURL} \url{http://dx.doi.org/10.1063/1.4965445} \end{APACrefURL}
\newblock
\begin{APACrefDOI} \doi{10.1063/1.4965445} \end{APACrefDOI}
\PrintBackRefs{\CurrentBib}

\bibitem [\protect \citeauthoryear {%
Rosenstock%
\ \BBA {} Weatherall%
}{%
Rosenstock%
\ \BBA {} Weatherall%
}{%
{\protect \APACyear {2018}}%
}]{%
rosenstockweatherall2018e}
\APACinsertmetastar {%
rosenstockweatherall2018e}%
\begin{APACrefauthors}%
Rosenstock, S.%
\BCBT {}\ \BBA {} Weatherall, J\BPBI O.%
\end{APACrefauthors}%
\unskip\
\newblock
\APACrefYearMonthDay{2018}{}{}.
\newblock
{\BBOQ}\APACrefatitle {Erratum: ``A categorical equivalence between generalized
  holonomy maps on a connected manifold and principal connections on bundles
  over that manifold'' [J. Math. Phys. 57, 102902 (2016)]} {Erratum: ``a
  categorical equivalence between generalized holonomy maps on a connected
  manifold and principal connections on bundles over that manifold'' [j. math.
  phys. 57, 102902 (2016)]}.{\BBCQ}
\newblock
\APACjournalVolNumPages{Journal of Mathematical Physics}{59}{2}{029901}.
\PrintBackRefs{\CurrentBib}

\bibitem [\protect \citeauthoryear {%
Rovelli%
}{%
Rovelli%
}{%
{\protect \APACyear {2002}}%
}]{%
Rovelli_partial}
\APACinsertmetastar {%
Rovelli_partial}%
\begin{APACrefauthors}%
Rovelli, C.%
\end{APACrefauthors}%
\unskip\
\newblock
\APACrefYearMonthDay{2002}{Jun}{}.
\newblock
{\BBOQ}\APACrefatitle {Partial observables} {Partial observables}.{\BBCQ}
\newblock
\APACjournalVolNumPages{Physical Review D}{65}{12}{}.
\newblock
\begin{APACrefURL} \url{http://dx.doi.org/10.1103/PhysRevD.65.124013}
  \end{APACrefURL}
\newblock
\begin{APACrefDOI} \doi{10.1103/physrevd.65.124013} \end{APACrefDOI}
\PrintBackRefs{\CurrentBib}

\bibitem [\protect \citeauthoryear {%
Rovelli%
}{%
Rovelli%
}{%
{\protect \APACyear {2007}}%
}]{%
Rovelli_book}
\APACinsertmetastar {%
Rovelli_book}%
\begin{APACrefauthors}%
Rovelli, C.%
\end{APACrefauthors}%
\unskip\
\newblock
\APACrefYear{2007}.
\newblock
\APACrefbtitle {{Quantum Gravity}} {{Quantum Gravity}}.
\newblock
\APACaddressPublisher{}{Cambridge University Press}.
\PrintBackRefs{\CurrentBib}

\bibitem [\protect \citeauthoryear {%
Rovelli%
}{%
Rovelli%
}{%
{\protect \APACyear {2014}}%
}]{%
RovelliGauge2013}
\APACinsertmetastar {%
RovelliGauge2013}%
\begin{APACrefauthors}%
Rovelli, C.%
\end{APACrefauthors}%
\unskip\
\newblock
\APACrefYearMonthDay{2014}{}{}.
\newblock
{\BBOQ}\APACrefatitle {{Why Gauge?}} {{Why Gauge?}}{\BBCQ}
\newblock
\APACjournalVolNumPages{Found. Phys.}{44}{1}{91-104}.
\newblock
\begin{APACrefDOI} \doi{10.1007/s10701-013-9768-7} \end{APACrefDOI}
\PrintBackRefs{\CurrentBib}

\bibitem [\protect \citeauthoryear {%
Ryder%
}{%
Ryder%
}{%
{\protect \APACyear {1996}}%
}]{%
ryder1996qft}
\APACinsertmetastar {%
ryder1996qft}%
\begin{APACrefauthors}%
Ryder, L\BPBI H.%
\end{APACrefauthors}%
\unskip\
\newblock
\APACrefYear{1996}.
\newblock
\APACrefbtitle {{Quantum Field Theory}} {{Quantum Field Theory}}.
\newblock
\APACaddressPublisher{}{Cambridge: Cambridge University Press}.
\PrintBackRefs{\CurrentBib}

\bibitem [\protect \citeauthoryear {%
Sardanashvily%
}{%
Sardanashvily%
}{%
{\protect \APACyear {2009}}%
}]{%
sardanashvily2009fibre}
\APACinsertmetastar {%
sardanashvily2009fibre}%
\begin{APACrefauthors}%
Sardanashvily, G.%
\end{APACrefauthors}%
\unskip\
\newblock
\APACrefYearMonthDay{2009}{}{}.
\newblock
\APACrefbtitle {Fibre Bundles, Jet Manifolds and Lagrangian Theory. Lectures
  for Theoreticians.} {Fibre bundles, jet manifolds and lagrangian theory.
  lectures for theoreticians.}
\PrintBackRefs{\CurrentBib}

\bibitem [\protect \citeauthoryear {%
Saunders%
}{%
Saunders%
}{%
{\protect \APACyear {2013}}%
}]{%
saunders2013}
\APACinsertmetastar {%
saunders2013}%
\begin{APACrefauthors}%
Saunders, S.%
\end{APACrefauthors}%
\unskip\
\newblock
\APACrefYearMonthDay{2013}{}{}.
\newblock
{\BBOQ}\APACrefatitle {{Rethinking Newton’s Principia}} {{Rethinking
  Newton’s Principia}}.{\BBCQ}
\newblock
\APACjournalVolNumPages{Philosophy of Science}{80}{1}{22--48}.
\PrintBackRefs{\CurrentBib}

\bibitem [\protect \citeauthoryear {%
Schutz%
}{%
Schutz%
}{%
{\protect \APACyear {1980}}%
}]{%
schutz1980g}
\APACinsertmetastar {%
schutz1980g}%
\begin{APACrefauthors}%
Schutz, B\BPBI F.%
\end{APACrefauthors}%
\unskip\
\newblock
\APACrefYear{1980}.
\newblock
\APACrefbtitle {{Geometric Methods of Mathematical Physics}} {{Geometric
  Methods of Mathematical Physics}}.
\newblock
\APACaddressPublisher{}{Cambridge: Cambridge University Press}.
\PrintBackRefs{\CurrentBib}

\bibitem [\protect \citeauthoryear {%
Shech%
}{%
Shech%
}{%
{\protect \APACyear {2018}}%
}]{%
ShechAB}
\APACinsertmetastar {%
ShechAB}%
\begin{APACrefauthors}%
Shech, E.%
\end{APACrefauthors}%
\unskip\
\newblock
\APACrefYearMonthDay{2018}{}{}.
\newblock
{\BBOQ}\APACrefatitle {{Idealizations, Essential Self-Adjointness, and Minimal
  Model Explanation in the Aharonov--Bohm Effect}} {{Idealizations, Essential
  Self-Adjointness, and Minimal Model Explanation in the Aharonov--Bohm
  Effect}}.{\BBCQ}
\newblock
\APACjournalVolNumPages{Synthese}{195}{11}{4839--4863}.
\newblock
\begin{APACrefDOI} \doi{10.1007/s11229-017-1428-6} \end{APACrefDOI}
\PrintBackRefs{\CurrentBib}

\bibitem [\protect \citeauthoryear {%
Shulman%
}{%
Shulman%
}{%
{\protect \APACyear {2017}}%
}]{%
Shulman_hole}
\APACinsertmetastar {%
Shulman_hole}%
\begin{APACrefauthors}%
Shulman, M.%
\end{APACrefauthors}%
\unskip\
\newblock
\APACrefYearMonthDay{2017}{}{}.
\newblock
{\BBOQ}\APACrefatitle {{Homotopy Type Theory: A Synthetic Approach to Higher
  Equalities. In Categories for the Working Philosopher.}} {{Homotopy Type
  Theory: A Synthetic Approach to Higher Equalities. In Categories for the
  Working Philosopher.}}{\BBCQ}
\newblock
\BIn{} (\BCHAP~3).
\newblock
\APACaddressPublisher{}{Oxford University Press}.
\PrintBackRefs{\CurrentBib}

\bibitem [\protect \citeauthoryear {%
Singer%
}{%
Singer%
}{%
{\protect \APACyear {1978}}%
}]{%
Singer:1978dk}
\APACinsertmetastar {%
Singer:1978dk}%
\begin{APACrefauthors}%
Singer, I\BPBI M.%
\end{APACrefauthors}%
\unskip\
\newblock
\APACrefYearMonthDay{1978}{}{}.
\newblock
{\BBOQ}\APACrefatitle {{Some Remarks on the Gribov Ambiguity}} {{Some Remarks
  on the Gribov Ambiguity}}.{\BBCQ}
\newblock
\APACjournalVolNumPages{Commun. Math. Phys.}{60}{}{7-12}.
\newblock
\begin{APACrefDOI} \doi{10.1007/BF01609471} \end{APACrefDOI}
\PrintBackRefs{\CurrentBib}

\bibitem [\protect \citeauthoryear {%
Stachel%
}{%
Stachel%
}{%
{\protect \APACyear {2014}}%
}]{%
Stachel_lrr}
\APACinsertmetastar {%
Stachel_lrr}%
\begin{APACrefauthors}%
Stachel, J.%
\end{APACrefauthors}%
\unskip\
\newblock
\APACrefYearMonthDay{2014}{}{}.
\newblock
{\BBOQ}\APACrefatitle {{The Hole Argument and Some Physical and Philosophical
  Implications.}} {{The Hole Argument and Some Physical and Philosophical
  Implications.}}{\BBCQ}
\newblock
\APACjournalVolNumPages{Living Reviews of Relativity, 17.}{}{}{}.
\PrintBackRefs{\CurrentBib}

\bibitem [\protect \citeauthoryear {%
Strocchi%
}{%
Strocchi%
}{%
{\protect \APACyear {2013}}%
}]{%
Strocchi_book}
\APACinsertmetastar {%
Strocchi_book}%
\begin{APACrefauthors}%
Strocchi, F.%
\end{APACrefauthors}%
\unskip\
\newblock
\APACrefYear{2013}.
\newblock
\APACrefbtitle {An Introduction to Non-Perturbative Foundations of Quantum
  Field Theory} {An introduction to non-perturbative foundations of quantum
  field theory}.
\newblock
\APACaddressPublisher{}{Oxford Universtiy Press}.
\PrintBackRefs{\CurrentBib}

\bibitem [\protect \citeauthoryear {%
Strocchi%
}{%
Strocchi%
}{%
{\protect \APACyear {2015}}%
}]{%
Strocchi_phil}
\APACinsertmetastar {%
Strocchi_phil}%
\begin{APACrefauthors}%
Strocchi, F.%
\end{APACrefauthors}%
\unskip\
\newblock
\APACrefYearMonthDay{2015}{}{}.
\newblock
{\BBOQ}\APACrefatitle {{Symmetries, Symmetry Breaking, Gauge Symmetries}}
  {{Symmetries, Symmetry Breaking, Gauge Symmetries}}.{\BBCQ}
\newblock

\PrintBackRefs{\CurrentBib}

\bibitem [\protect \citeauthoryear {%
Strominger%
}{%
Strominger%
}{%
{\protect \APACyear {2018}}%
}]{%
strominger2018lectures}
\APACinsertmetastar {%
strominger2018lectures}%
\begin{APACrefauthors}%
Strominger, A.%
\end{APACrefauthors}%
\unskip\
\newblock
\APACrefYear{2018}.
\newblock
\APACrefbtitle {Lectures on the infrared structure of gravity and gauge theory}
  {Lectures on the infrared structure of gravity and gauge theory}.
\newblock
\APACaddressPublisher{}{Princeton University Press}.
\PrintBackRefs{\CurrentBib}

\bibitem [\protect \citeauthoryear {%
Swanson%
}{%
Swanson%
}{%
{\protect \APACyear {2019}}%
}]{%
swanson2019ostrogradsky}
\APACinsertmetastar {%
swanson2019ostrogradsky}%
\begin{APACrefauthors}%
Swanson, N.%
\end{APACrefauthors}%
\unskip\
\newblock
\APACrefYearMonthDay{2019}{}{}.
\newblock
{\BBOQ}\APACrefatitle {{On the Ostrogradski Instability, or, Why Physics Really
  Uses Second Derivatives}} {{On the Ostrogradski Instability, or, Why Physics
  Really Uses Second Derivatives}}.{\BBCQ}
\newblock
\APACjournalVolNumPages{The British Journal for the Philosophy of
  Science}{}{}{}.
\newblock
\APACrefnote{\url{http://philsci-archive.pitt.edu/15932/}}
\PrintBackRefs{\CurrentBib}

\bibitem [\protect \citeauthoryear {%
Teh%
}{%
Teh%
}{%
{\protect \APACyear {2016}}%
}]{%
Teh_emp}
\APACinsertmetastar {%
Teh_emp}%
\begin{APACrefauthors}%
Teh, N\BPBI J.%
\end{APACrefauthors}%
\unskip\
\newblock
\APACrefYearMonthDay{2016}{}{}.
\newblock
{\BBOQ}\APACrefatitle {Galileo’s Gauge: Understanding the Empirical
  Significance of Gauge Symmetry} {Galileo’s gauge: Understanding the
  empirical significance of gauge symmetry}.{\BBCQ}
\newblock
\APACjournalVolNumPages{Philosophy of Science}{83}{1}{93-118}.
\newblock
\begin{APACrefURL} \url{https://doi.org/10.1086/684196} \end{APACrefURL}
\newblock
\begin{APACrefDOI} \doi{10.1086/684196} \end{APACrefDOI}
\PrintBackRefs{\CurrentBib}

\bibitem [\protect \citeauthoryear {%
Teitelboim%
}{%
Teitelboim%
}{%
{\protect \APACyear {1973}}%
}]{%
Teitelboim1973}
\APACinsertmetastar {%
Teitelboim1973}%
\begin{APACrefauthors}%
Teitelboim, C.%
\end{APACrefauthors}%
\unskip\
\newblock
\APACrefYearMonthDay{1973}{}{}.
\newblock
{\BBOQ}\APACrefatitle {How commutators of constraints reflect the spacetime
  structure} {How commutators of constraints reflect the spacetime
  structure}.{\BBCQ}
\newblock
\APACjournalVolNumPages{Annals of Physics}{79}{2}{542 - 557}.
\newblock
\begin{APACrefURL}
  \url{http://www.sciencedirect.com/science/article/pii/0003491673900961}
  \end{APACrefURL}
\newblock
\begin{APACrefDOI} \doi{https://doi.org/10.1016/0003-4916(73)90096-1}
  \end{APACrefDOI}
\PrintBackRefs{\CurrentBib}

\bibitem [\protect \citeauthoryear {%
Teller%
}{%
Teller%
}{%
{\protect \APACyear {1997}}%
}]{%
teller1997m}
\APACinsertmetastar {%
teller1997m}%
\begin{APACrefauthors}%
Teller, P.%
\end{APACrefauthors}%
\unskip\
\newblock
\APACrefYearMonthDay{1997}{}{}.
\newblock
{\BBOQ}\APACrefatitle {A metaphysics for contemporary field theories} {A
  metaphysics for contemporary field theories}.{\BBCQ}
\newblock
\APACjournalVolNumPages{Studies in History and Philosophy of Modern
  Physics}{28}{4}{507--522}.
\PrintBackRefs{\CurrentBib}

\bibitem [\protect \citeauthoryear {%
Teller%
}{%
Teller%
}{%
{\protect \APACyear {2000}}%
}]{%
teller2000gauge}
\APACinsertmetastar {%
teller2000gauge}%
\begin{APACrefauthors}%
Teller, P.%
\end{APACrefauthors}%
\unskip\
\newblock
\APACrefYearMonthDay{2000}{}{}.
\newblock
{\BBOQ}\APACrefatitle {The gauge argument} {The gauge argument}.{\BBCQ}
\newblock
\APACjournalVolNumPages{Philosophy of Science}{67}{}{S466--S481}.
\PrintBackRefs{\CurrentBib}

\bibitem [\protect \citeauthoryear {%
Thiemann%
}{%
Thiemann%
}{%
{\protect \APACyear {2003}}%
}]{%
Thiemann_2003}
\APACinsertmetastar {%
Thiemann_2003}%
\begin{APACrefauthors}%
Thiemann, T.%
\end{APACrefauthors}%
\unskip\
\newblock
\APACrefYearMonthDay{2003}{}{}.
\newblock
{\BBOQ}\APACrefatitle {Lectures on Loop Quantum Gravity} {Lectures on loop
  quantum gravity}.{\BBCQ}
\newblock
\APACjournalVolNumPages{Lecture Notes in Physics}{}{}{41–135}.
\newblock
\begin{APACrefURL} \url{http://dx.doi.org/10.1007/978-3-540-45230-0_3}
  \end{APACrefURL}
\newblock
\begin{APACrefDOI} \doi{10.1007/978-3-540-45230-0_3} \end{APACrefDOI}
\PrintBackRefs{\CurrentBib}

\bibitem [\protect \citeauthoryear {%
Thierry-Mieg%
}{%
Thierry-Mieg%
}{%
{\protect \APACyear {1980}}%
}]{%
Thierry-MiegJMP}
\APACinsertmetastar {%
Thierry-MiegJMP}%
\begin{APACrefauthors}%
Thierry-Mieg, J.%
\end{APACrefauthors}%
\unskip\
\newblock
\APACrefYearMonthDay{1980}{}{}.
\newblock
{\BBOQ}\APACrefatitle {Geometrical reinterpretation of Faddeev-Popov ghost
  particles and BRS transformations} {Geometrical reinterpretation of
  faddeev-popov ghost particles and brs transformations}.{\BBCQ}
\newblock
\APACjournalVolNumPages{Journal of Mathematical Physics}{21}{12}{2834-2838}.
\newblock
\begin{APACrefURL} \url{https://doi.org/10.1063/1.524385} \end{APACrefURL}
\newblock
\begin{APACrefDOI} \doi{10.1063/1.524385} \end{APACrefDOI}
\PrintBackRefs{\CurrentBib}

\bibitem [\protect \citeauthoryear {%
't Hooft%
}{%
't Hooft%
}{%
{\protect \APACyear {1980}}%
}]{%
thooft}
\APACinsertmetastar {%
thooft}%
\begin{APACrefauthors}%
't Hooft, G.%
\end{APACrefauthors}%
\unskip\
\newblock
\APACrefYearMonthDay{1980}{}{}.
\newblock
{\BBOQ}\APACrefatitle {{Gauge Theories and the Forces Between Elementary
  Particles}} {{Gauge Theories and the Forces Between Elementary
  Particles}}.{\BBCQ}
\newblock
\APACjournalVolNumPages{Scientific American, 242, pp. 90-166}{}{}{}.
\PrintBackRefs{\CurrentBib}

\bibitem [\protect \citeauthoryear {%
Tong%
}{%
Tong%
}{%
{\protect \APACyear {2018}}%
}]{%
Tong_gt}
\APACinsertmetastar {%
Tong_gt}%
\begin{APACrefauthors}%
Tong, D.%
\end{APACrefauthors}%
\unskip\
\newblock
\APACrefYear{2018}.
\newblock
\APACrefbtitle {Lecture notes on gauge theory} {Lecture notes on gauge theory}.
\newblock
\APACrefnote{Available online at
  \href{http://www.damtp.cam.ac.uk/user/tong/gaugetheory.html}{http://www.damtp.cam.ac.uk/user/tong/gaugetheory.html}}
\PrintBackRefs{\CurrentBib}

\bibitem [\protect \citeauthoryear {%
Torre%
}{%
Torre%
}{%
{\protect \APACyear {1993}}%
}]{%
Torre_local}
\APACinsertmetastar {%
Torre_local}%
\begin{APACrefauthors}%
Torre, C\BPBI G.%
\end{APACrefauthors}%
\unskip\
\newblock
\APACrefYearMonthDay{1993}{Sep}{}.
\newblock
{\BBOQ}\APACrefatitle {Gravitational observables and local symmetries}
  {Gravitational observables and local symmetries}.{\BBCQ}
\newblock
\APACjournalVolNumPages{Phys. Rev. D}{48}{}{R2373--R2376}.
\newblock
\begin{APACrefURL} \url{https://link.aps.org/doi/10.1103/PhysRevD.48.R2373}
  \end{APACrefURL}
\newblock
\begin{APACrefDOI} \doi{10.1103/PhysRevD.48.R2373} \end{APACrefDOI}
\PrintBackRefs{\CurrentBib}

\bibitem [\protect \citeauthoryear {%
Voisin%
}{%
Voisin%
}{%
{\protect \APACyear {2002}}%
}]{%
voisin}
\APACinsertmetastar {%
voisin}%
\begin{APACrefauthors}%
Voisin, C.%
\end{APACrefauthors}%
\unskip\
\newblock
\APACrefYear{2002}.
\newblock
\APACrefbtitle {{Hodge Theory and Complex Algebraic Geometry I}} {{Hodge Theory
  and Complex Algebraic Geometry I}}\ (\BVOL~1; L.~Schneps, \BED{}).
\newblock
\APACaddressPublisher{}{Cambridge University Press}.
\newblock
\begin{APACrefDOI} \doi{10.1017/CBO9780511615344} \end{APACrefDOI}
\PrintBackRefs{\CurrentBib}

\bibitem [\protect \citeauthoryear {%
Wald%
}{%
Wald%
}{%
{\protect \APACyear {1984}}%
}]{%
Wald_book}
\APACinsertmetastar {%
Wald_book}%
\begin{APACrefauthors}%
Wald, R\BPBI M.%
\end{APACrefauthors}%
\unskip\
\newblock
\APACrefYear{1984}.
\newblock
\APACrefbtitle {{General Relativity}} {{General Relativity}}.
\newblock
\APACaddressPublisher{}{University of Chicago Press}.
\PrintBackRefs{\CurrentBib}

\bibitem [\protect \citeauthoryear {%
Wallace%
}{%
Wallace%
}{%
{\protect \APACyear {2002}}%
}]{%
Wallace_LagSym}
\APACinsertmetastar {%
Wallace_LagSym}%
\begin{APACrefauthors}%
Wallace, D.%
\end{APACrefauthors}%
\unskip\
\newblock
\APACrefYearMonthDay{2002}{}{}.
\newblock
{\BBOQ}\APACrefatitle {{Time-Dependent Symmetries: The Link Between Gauge
  Symmetries and Indeterminism}} {{Time-Dependent Symmetries: The Link Between
  Gauge Symmetries and Indeterminism}}.{\BBCQ}
\newblock
\BIn{} K.~Brading\ \BBA {} E.~Castellani\ (\BEDS), \APACrefbtitle {Symmetries
  in Physics: Philosophical Reflections} {Symmetries in physics: Philosophical
  reflections}\ (\BPGS\ 163--173).
\newblock
\APACaddressPublisher{}{Cambridge University Press}.
\PrintBackRefs{\CurrentBib}

\bibitem [\protect \citeauthoryear {%
Wallace%
}{%
Wallace%
}{%
{\protect \APACyear {2009}}%
}]{%
wallace2009anti}
\APACinsertmetastar {%
wallace2009anti}%
\begin{APACrefauthors}%
Wallace, D.%
\end{APACrefauthors}%
\unskip\
\newblock
\APACrefYearMonthDay{2009}{}{}.
\newblock
\APACrefbtitle {{QFT, Antimatter and Symmetry}.} {{QFT, Antimatter and
  Symmetry}.}
\newblock
\APACrefnote{Unpublished Manuscript, \url{http://arxiv.org/abs/0903.3018}}
\PrintBackRefs{\CurrentBib}

\bibitem [\protect \citeauthoryear {%
Wallace%
}{%
Wallace%
}{%
{\protect \APACyear {2014}}%
}]{%
Wallace_deflating}
\APACinsertmetastar {%
Wallace_deflating}%
\begin{APACrefauthors}%
Wallace, D.%
\end{APACrefauthors}%
\unskip\
\newblock
\APACrefYearMonthDay{2014}{}{}.
\newblock
{\BBOQ}\APACrefatitle {{Deflating the Aharonov-Bohm Effect}} {{Deflating the
  Aharonov-Bohm Effect}}.{\BBCQ}
\newblock
\APACjournalVolNumPages{arxiv: 1407.5073}{}{}{}.
\PrintBackRefs{\CurrentBib}

\bibitem [\protect \citeauthoryear {%
Wallace%
}{%
Wallace%
}{%
{\protect \APACyear {2019}}%
{\protect \APACexlab {{\protect \BCnt {1}}}}}]{%
Wallace2019a}
\APACinsertmetastar {%
Wallace2019a}%
\begin{APACrefauthors}%
Wallace, D.%
\end{APACrefauthors}%
\unskip\
\newblock
\APACrefYearMonthDay{2019{\protect \BCnt {1}}}{}{}.
\newblock
\APACrefbtitle {Isolated Systems and their Symmetries, Part I: General
  Framework and Particle-Mechanics Examples.} {Isolated systems and their
  symmetries, part i: General framework and particle-mechanics examples.}
\newblock
\begin{APACrefURL} \url{http://philsci-archive.pitt.edu/16623/}
  \end{APACrefURL}
\PrintBackRefs{\CurrentBib}

\bibitem [\protect \citeauthoryear {%
Wallace%
}{%
Wallace%
}{%
{\protect \APACyear {2019}}%
{\protect \APACexlab {{\protect \BCnt {2}}}}}]{%
Wallace2019b}
\APACinsertmetastar {%
Wallace2019b}%
\begin{APACrefauthors}%
Wallace, D.%
\end{APACrefauthors}%
\unskip\
\newblock
\APACrefYearMonthDay{2019{\protect \BCnt {2}}}{}{}.
\newblock
{\BBOQ}\APACrefatitle {{Isolated systems and their symmetries, part II: local
  and global symmetries of field theories}} {{Isolated systems and their
  symmetries, part II: local and global symmetries of field theories}}.{\BBCQ}
\newblock
\begin{APACrefURL} \url{http://philsci-archive.pitt.edu/16624/}
  \end{APACrefURL}
\PrintBackRefs{\CurrentBib}

\bibitem [\protect \citeauthoryear {%
Wallace%
}{%
Wallace%
}{%
{\protect \APACyear {2019}}%
{\protect \APACexlab {{\protect \BCnt {3}}}}}]{%
Wallace2019}
\APACinsertmetastar {%
Wallace2019}%
\begin{APACrefauthors}%
Wallace, D.%
\end{APACrefauthors}%
\unskip\
\newblock
\APACrefYearMonthDay{2019{\protect \BCnt {3}}}{}{}.
\newblock
{\BBOQ}\APACrefatitle {Observability, redundancy and modality for dynamical
  symmetry transformations} {Observability, redundancy and modality for
  dynamical symmetry transformations}.{\BBCQ}
\newblock
\APACjournalVolNumPages{Forthcoming}{}{}{}.
\newblock
\begin{APACrefURL} \url{http://philsci-archive.pitt.edu/18813/}
  \end{APACrefURL}
\newblock
\APACrefnote{Revised 3/2021 to correct a few typos and add a section on
  Noether's Theorem.}
\PrintBackRefs{\CurrentBib}

\bibitem [\protect \citeauthoryear {%
Wallace%
}{%
Wallace%
}{%
{\protect \APACyear {2019}}%
{\protect \APACexlab {{\protect \BCnt {4}}}}}]{%
Wallace_coords}
\APACinsertmetastar {%
Wallace_coords}%
\begin{APACrefauthors}%
Wallace, D.%
\end{APACrefauthors}%
\unskip\
\newblock
\APACrefYearMonthDay{2019{\protect \BCnt {4}}}{}{}.
\newblock
{\BBOQ}\APACrefatitle {{Who's Afraid of Coordinate Systems? An Essay on
  Representation of Spacetime Structure}} {{Who's Afraid of Coordinate Systems?
  An Essay on Representation of Spacetime Structure}}.{\BBCQ}
\newblock
\APACjournalVolNumPages{Studies in History and Philosophy of Science Part B:
  Studies in History and Philosophy of Modern Physics}{67}{}{125--136}.
\newblock
\begin{APACrefDOI} \doi{10.1016/j.shpsb.2017.07.002} \end{APACrefDOI}
\PrintBackRefs{\CurrentBib}

\bibitem [\protect \citeauthoryear {%
Weatherall%
}{%
Weatherall%
}{%
{\protect \APACyear {2016}}%
}]{%
Weatherall2016_YMGR}
\APACinsertmetastar {%
Weatherall2016_YMGR}%
\begin{APACrefauthors}%
Weatherall, J.%
\end{APACrefauthors}%
\unskip\
\newblock
\APACrefYearMonthDay{2016}{}{}.
\newblock
{\BBOQ}\APACrefatitle {{Fiber bundles, Yang--Mills theory, and general
  relativity}} {{Fiber bundles, Yang--Mills theory, and general
  relativity}}.{\BBCQ}
\newblock
\APACjournalVolNumPages{Synthese}{193}{8}{2389--2425}.
\newblock
\APACrefnote{\url{http://philsci-archive.pitt.edu/11481/}}
\PrintBackRefs{\CurrentBib}

\bibitem [\protect \citeauthoryear {%
Weatherall%
}{%
Weatherall%
}{%
{\protect \APACyear {2018}}%
}]{%
Weatherall_hole}
\APACinsertmetastar {%
Weatherall_hole}%
\begin{APACrefauthors}%
Weatherall, J.%
\end{APACrefauthors}%
\unskip\
\newblock
\APACrefYearMonthDay{2018}{}{}.
\newblock
{\BBOQ}\APACrefatitle {{Regarding the ‘Hole Argument’}} {{Regarding the
  ‘Hole Argument’}}.{\BBCQ}
\newblock
\APACjournalVolNumPages{The British Journal for the Philosophy of
  Science}{69}{2}{329-350}.
\newblock
\begin{APACrefURL} \url{https://doi.org/10.1093/bjps/axw012} \end{APACrefURL}
\newblock
\begin{APACrefDOI} \doi{10.1093/bjps/axw012} \end{APACrefDOI}
\PrintBackRefs{\CurrentBib}

\bibitem [\protect \citeauthoryear {%
Weinberg%
}{%
Weinberg%
}{%
{\protect \APACyear {1964}}%
}]{%
Weinberg_equiv}
\APACinsertmetastar {%
Weinberg_equiv}%
\begin{APACrefauthors}%
Weinberg, S.%
\end{APACrefauthors}%
\unskip\
\newblock
\APACrefYearMonthDay{1964}{}{}.
\newblock
{\BBOQ}\APACrefatitle {Derivation of gauge invariance and the equivalence
  principle from Lorentz invariance of the S- matrix} {Derivation of gauge
  invariance and the equivalence principle from lorentz invariance of the s-
  matrix}.{\BBCQ}
\newblock
\APACjournalVolNumPages{Physics Letters}{9}{4}{357-359}.
\newblock
\begin{APACrefURL}
  \url{https://www.sciencedirect.com/science/article/pii/0031916364903968}
  \end{APACrefURL}
\newblock
\begin{APACrefDOI} \doi{https://doi.org/10.1016/0031-9163(64)90396-8}
  \end{APACrefDOI}
\PrintBackRefs{\CurrentBib}

\bibitem [\protect \citeauthoryear {%
Weinberg%
}{%
Weinberg%
}{%
{\protect \APACyear {2005}}%
{\protect \APACexlab {{\protect \BCnt {1}}}}}]{%
WeinbergQFT1}
\APACinsertmetastar {%
WeinbergQFT1}%
\begin{APACrefauthors}%
Weinberg, S.%
\end{APACrefauthors}%
\unskip\
\newblock
\APACrefYear{2005{\protect \BCnt {1}}}.
\newblock
\APACrefbtitle {{The Quantum Theory of Fields. Vol. 1: Foundations}} {{The
  Quantum Theory of Fields. Vol. 1: Foundations}}.
\newblock
\APACaddressPublisher{}{Cambridge University Press, Cambridge}.
\PrintBackRefs{\CurrentBib}

\bibitem [\protect \citeauthoryear {%
Weinberg%
}{%
Weinberg%
}{%
{\protect \APACyear {2005}}%
{\protect \APACexlab {{\protect \BCnt {2}}}}}]{%
WeinbergQFT2}
\APACinsertmetastar {%
WeinbergQFT2}%
\begin{APACrefauthors}%
Weinberg, S.%
\end{APACrefauthors}%
\unskip\
\newblock
\APACrefYear{2005{\protect \BCnt {2}}}.
\newblock
\APACrefbtitle {The Quantum Theory of Fields. Volume 2. Modern Applications}
  {The quantum theory of fields. volume 2. modern applications}.
\newblock
\APACaddressPublisher{}{Cambridge Univ. Press}.
\PrintBackRefs{\CurrentBib}

\bibitem [\protect \citeauthoryear {%
Weyl%
}{%
Weyl%
}{%
{\protect \APACyear {1929}}%
}]{%
weyl1929g}
\APACinsertmetastar {%
weyl1929g}%
\begin{APACrefauthors}%
Weyl, H.%
\end{APACrefauthors}%
\unskip\
\newblock
\APACrefYearMonthDay{1929}{}{}.
\newblock
{\BBOQ}\APACrefatitle {Gravitation and the electron} {Gravitation and the
  electron}.{\BBCQ}
\newblock
\APACjournalVolNumPages{{Proceedings of the National Academy of Sciences of the
  United States of America}}{15}{4}{323-334}.
\newblock
\APACrefnote{\url{https://www.ncbi.nlm.nih.gov/pmc/articles/PMC522457/}}
\PrintBackRefs{\CurrentBib}

\bibitem [\protect \citeauthoryear {%
Wilkins%
}{%
Wilkins%
}{%
{\protect \APACyear {1989}}%
}]{%
YangMillsSlice}
\APACinsertmetastar {%
YangMillsSlice}%
\begin{APACrefauthors}%
Wilkins, D\BPBI R.%
\end{APACrefauthors}%
\unskip\
\newblock
\APACrefYearMonthDay{1989}{}{}.
\newblock
{\BBOQ}\APACrefatitle {Slice Theorems in Gauge Theory} {Slice theorems in gauge
  theory}.{\BBCQ}
\newblock
\APACjournalVolNumPages{Proceedings of the Royal Irish Academy. Section A:
  Mathematical and Physical Sciences}{89A}{1}{13--34}.
\newblock
\begin{APACrefURL} \url{http://www.jstor.org/stable/20489307} \end{APACrefURL}
\PrintBackRefs{\CurrentBib}

\bibitem [\protect \citeauthoryear {%
W\"uthrich%
}{%
W\"uthrich%
}{%
{\protect \APACyear {2009}}%
}]{%
Wuthrich_abysmal}
\APACinsertmetastar {%
Wuthrich_abysmal}%
\begin{APACrefauthors}%
W\"uthrich, C.%
\end{APACrefauthors}%
\unskip\
\newblock
\APACrefYearMonthDay{2009}{}{}.
\newblock
{\BBOQ}\APACrefatitle {{Challenging the Spacetime Structuralist}} {{Challenging
  the Spacetime Structuralist}}.{\BBCQ}
\newblock
\APACjournalVolNumPages{Philosophy of Science. \textbf{Volume 76}, Number
  5}{}{}{}.
\PrintBackRefs{\CurrentBib}

\bibitem [\protect \citeauthoryear {%
Yang%
\ \BBA {} Mills%
}{%
Yang%
\ \BBA {} Mills%
}{%
{\protect \APACyear {1954}}%
}]{%
YangMills}
\APACinsertmetastar {%
YangMills}%
\begin{APACrefauthors}%
Yang, C\BPBI N.%
\BCBT {}\ \BBA {} Mills, R\BPBI L.%
\end{APACrefauthors}%
\unskip\
\newblock
\APACrefYearMonthDay{1954}{Oct}{}.
\newblock
{\BBOQ}\APACrefatitle {Conservation of Isotopic Spin and Isotopic Gauge
  Invariance} {Conservation of isotopic spin and isotopic gauge
  invariance}.{\BBCQ}
\newblock
\APACjournalVolNumPages{Phys. Rev.}{96}{}{191--195}.
\newblock
\begin{APACrefURL} \url{https://link.aps.org/doi/10.1103/PhysRev.96.191}
  \end{APACrefURL}
\newblock
\begin{APACrefDOI} \doi{10.1103/PhysRev.96.191} \end{APACrefDOI}
\PrintBackRefs{\CurrentBib}

\bibitem [\protect \citeauthoryear {%
York%
}{%
York%
}{%
{\protect \APACyear {1971}}%
}]{%
York}
\APACinsertmetastar {%
York}%
\begin{APACrefauthors}%
York, J\BPBI W.%
\end{APACrefauthors}%
\unskip\
\newblock
\APACrefYearMonthDay{1971}{}{}.
\newblock
{\BBOQ}\APACrefatitle {Gravitational degrees of freedom and the initial-value
  problem} {Gravitational degrees of freedom and the initial-value
  problem}.{\BBCQ}
\newblock
\APACjournalVolNumPages{Phys. Rev. Lett.}{26}{}{1656--1658}.
\newblock
\begin{APACrefDOI} \doi{10.1103/PhysRevLett.26.1656} \end{APACrefDOI}
\PrintBackRefs{\CurrentBib}

\end{thebibliography}

\end{document}